\documentclass[journal, onecolumn, draftclsnofoot]{IEEEtran}

\usepackage{color,array,amsthm}
\usepackage{graphicx}
\usepackage{bbm}
\usepackage{amsmath}
\usepackage{amsfonts}
\usepackage{amssymb}
\usepackage{graphicx}
\usepackage{csquotes}
\usepackage{placeins}
\usepackage{titlesec}
\usepackage{newfloat}
\usepackage{csquotes}
\usepackage{soul}

\usepackage{todonotes}
\usepackage{color}

\usepackage{caption}
\usepackage{subcaption}
\newcommand{\nn}{\nonumber}
\newcommand{\e}{{\mathrm e}}

\begin{document}

\title{Communication protocol for a~satellite-swarm interferometer}

\author{OLIVER NAGY$^1$, MANISH PANDEY$^2$, GEORGIOS EXARCHAKOS$^2$,\\ MARK BENTUM$^{2,3}$, REMCO van der HOFSTAD$^2$}

\markboth{NAGY ET AL.}{COMMUNICATION PROTOCOL FOR A SATELLITE-SWARM INTERFEROMETER}
\maketitle
\vspace{-2cm}
\begin{center}
    ${}^1$Leiden University, Leiden, The Netherlands\\
    ${}^2$Eindhoven University of Technology, Eindhoven\\
    ${}^3$Netherlands Foundation for Research in Radio Astronomy (ASTRON), Dwingeloo, The Netherlands
\end{center}
\vspace{1cm}

\begin{abstract}
Orbiting low frequency antennas for radio astronomy (OLFAR) that capture cosmic signals in the frequency range below 30MHz could provide valuable insights on our Universe. These wireless swarms of satellites form a connectivity graph that allows data exchange between most pairs of satellites. Since this swarm acts as an interferometer, the aim is to compute the cross-correlations between most pairs of satellites. 

We propose a $k$-nearest-neighbour communication protocol, and investigate the minimum neighbourhood size of each satellite that ensures connectivity of at least 95\% of the swarm. We describe the proportion of cross-correlations that can be computed in our method  given an energy budget per satellite. Despite the method's apparent simplicity, it allows us to gain insight into the requirements for such satellite swarms. In particular, we give specific advice on the energy requirements to have sufficient coverage of the relevant baselines.
\end{abstract}


\section{Introduction}\label{sec:intro}

\subsection{Motivation and setting}\label{sec:intro:setting}

Long-wavelength radio astronomy (also called Low-frequency radio astronomy) is targeting for an instrument for observing the Universe at frequencies below 30 MHz. For frequencies above 30 MHz several instruments have been implemented in the past two centuries, like LOFAR (LOw Frequency ARray) \cite{lofar} in the Netherlands and its European extension ILT, the International LOFAR Telescope. However, at frequencies below 30 MHz, Earth-based observations are limited due to a combination of severe ionospheric distortions, almost full reflection of radio waves below 10 MHz, solar eruptions and the radio frequency interference (RFI) of human-made signals. Scientifically this frequency band is extremely interesting, for instance providing information from the faint signals from the Hydrogen in the Cosmological Dark Ages and Cosmic Dawn, the study of Solar activity and space weather at low frequencies, the measure of the auroral radio emission from the large planets in our Solar system, the determination of the radio background spectrum at the Earth-Moon L2 point, the creation of a new low-frequency map of the radio sky, the study of the Earth’s ionosphere, and the detection of bright pulsars and other radio transient phenomena at very low frequencies \cite{jester}.

Many paper studies have been performed in the last decade, to open up this last, virtually unexplored frequency domain in the electromagnetic spectrum \cite{olfar,noire}.

The basic idea is to form a swarm of satellites, each sampling the astronomical signals. Together, they work as an interferometer -- in practice, this means that all the useful information is obtained only after \emph{cross-correlation} of a pair of measurements from 2 different satellites.

In an ideal setting, one would carry out all the possible cross-correlations; in practice, there are limitations coming from a finite energy budget, data routing problems and others.

\subsection{Modelling assumptions}\label{sec:intro:assump}

The problem under consideration is characterized by its inherent complexity, encompassing a multitude of complications stemming from various sources, including the intricate dynamics of the system, challenges associated with data transmission, equipment malfunctions, and other factors. It is important to emphasize that our primary objective is not to achieve a high degree of realism in the modelling process. Instead, our focus lies in conducting a feasibility study under simplified, yet realistic, assumptions. By delineating these assumptions, we aim to establish the extent of applicability of our investigation.

\paragraph{Number of satellites}
Throughout this paper, we denote the number of satellites in our swarm by $n$, and we think of $n$ as being large. In our analysis, we thus make the explicit assumption that the number of satellites involved  is large, yet remains fixed throughout the duration of the study and does not undergo any changes over time. We acknowledge, however, that allowing for dynamic changes in the satellite constellation could introduce valuable insights into the effects of various scenarios, such as catastrophic malfunctions or the reinforcement of the satellite swarm through the deployment of additional satellites.

By considering the possibility of catastrophic malfunctions, we could model the impact of severe failures within the satellite network. These malfunctions could include critical subsystem failures, orbital anomalies, or unexpected events leading to the loss of functionality of one or more satellites. Incorporating such dynamic changes would allow us to assess the system's resilience and the overall robustness of the network in the face of unforeseen disruptions.

Furthermore, considering the deployment of additional satellites to reinforce the existing swarm introduces an important aspect of scalability and adaptability. By accounting for the potential deployment of extra satellites, we can evaluate the system's ability to respond to increasing demands, expand coverage, or mitigate the effects of satellite failures. This consideration becomes particularly relevant when exploring strategies for enhancing network reliability, improving signal coverage, or addressing future capacity requirements.

\paragraph{Position and movement of satellites} 
We adopt the assumption that the initial distribution of satellites is uniformly random within a unit cube. This means that each satellite's position is independently and uniformly chosen within the interval [0,1] for each of its three Cartesian coordinates.

The decision to employ this uniform random distribution can be interpreted as a deliberate choice not to pursue an optimization approach for satellite deployment in our study. Instead, we aim to investigate communication networks governed by {\em spatial} randomness. This approach allows us to capture the inherent unpredictability and diversity present in real-world deployments.

It is important to note that the assumption of static satellites imposes limitations on the applicability of our results. Our findings are most suitable for scenarios where the assumption of static satellite positions is a justifiable approximation. For instance, an illustrative example could be a satellite swarm deployed near one of the stable Lagrange points in the Earth-Sun system. In such cases, the gravitational forces and orbital dynamics may cause satellites to maintain relatively fixed positions over extended periods, rendering the static assumption reasonable. We do wish to stress that the uniformity assumption could arise as the {\em stationary distribution} of dynamical swarms of satellites, and thus presents a snapshot of the dynamical system. Further, our approach can be straightforwardly adapted should such a stationary distribution be non-uniform.

\paragraph{Structure of the communication network} 
In addition to the previously mentioned assumptions, another key assumption we make in our study is that the communication network between the satellites can be represented as a directed graph. Specifically, we construct this graph by connecting each vertex to its $k$ nearest neighbours based on Euclidean distance, where $k$ is chosen appropriately to strike a balance between network connectivity and energy consumption. This type of graph, where each vertex has directed edges connecting it to its $k$ nearest neighbours, is commonly referred to as a $k$-nearest-neighbour ($k$-NN) graph.

The choice to model the communication network as a $k$-NN graph may initially appear arbitrary. However, our research findings demonstrate that this network representation possesses numerous desirable properties when $k\geq4$. By examining the network characteristics and performance metrics associated with $k$-NN graphs, we can gain insights into the behaviour and functionality of the communication system under study.

While there exist alternative network models that could be considered within this setting, such as the random geometric graph, we have intentionally limited our investigation to focus solely on $k$-NN graphs 
due to its simplicity combined with the valuable insights that it provides.

\paragraph{Energy expenditure}
We introduce a significant simplification regarding the energy consumption of satellites. Specifically, we assume that there are only two primary ways in which a satellite can consume energy: communication-related activities and the computation of cross-correlations based on acquired measurements. This assumption, while simplifying the analysis, entails the exclusion of other potential energy expenditure factors that may exist in real-world scenarios. This simplified approach allows us to isolate and examine the energy requirements directly linked to these essential functions and evaluate their effects on system performance and efficiency.

We investigate a setting in which the total energy spent on communication is comparable to the total energy spent on computation, as this is the most interesting setting. Indeed, when either of the two energies is negligible compared to the other, we can ignore that aspect, making the problem significantly simpler. Let $c$ denote the computation costs per cross-correlation, and $E_{\rm max}$ the total amount of energy at the disposal of each satellite. Furthermore, we desire that the total amount of computation power is, on average, a $\beta$-fraction of the total energy, i.e.,
    \begin{equation}\label{eq:assumption energy expenditure}
        c{\binom{n}{2}}=n\beta E_{\rm max}.
    \end{equation}
Without loss of generality, we may work in units such that $c=1$. The parameters $E_{\rm max}$ and $\beta$ will then serve as the key tuning parameters in the paper.

\paragraph{Data flows in the network}
We make the assumption that all data is readily available \enquote{on-demand} as long as a viable path exists between the data source and the destination. This assumption implies that once a valid communication path is established between two satellites, the necessary data can be efficiently transmitted and accessed without delay.

As our analysis progresses, we will demonstrate that, for appropriate values of $k$, there exists a path connecting nearly all pairs of satellites in the network. This observation underscores the connectivity and accessibility of the communication infrastructure we consider.

However, it is important to note that our assumptions disregard various factors related to data flow within the satellite network. Specifically, we neglect the effects of routing, network capacity limitations, finite-speed propagation delays, noisy channels, and the need for retransmissions, among others.

\subsection{Previous works}\label{sec:intro:literature}

\subsubsection{Engineering applications of $k$-NN graphs} 
There is a plethora of challenges on wireless networks to which $k$-NN models have been applied. This is mostly because of its simplicity to implement, its decentralized nature, and its good performance. However, each challenge requires a slight modification of the algorithm. Here, we review $k$-NN based wireless topologies and transmission power control, as well as node placement, as they are the most relevant to the LOFAR connectivity problem we study.

Blough et al. \cite{10.1145/778415.778433}, in one of the earliest studies on $k$-NN-based topology control, ensure that every node in a mesh wireless network is connected to at least $k$ other neighbours. Each node adjusts the transmission range to maintain $k$ bidirectional links. Supported by other studies (e.g., \cite{Blough2002, 10.1145/513800.513803}), the authors argue that the overhead generated on the routing layer to maintain a connected topology graph (a.k.a.\ large strongly connected component) using unidirectional links outweighs the benefits. The size of the routing tables that need to be maintained increases with unidirectional links, since any pair of nodes needs two paths to exchange information back and forth. 
Therefore, the vast majority of protocols in the L2 and L3 OSI layers assume bidirectional links. However, the OLFAR application does not require bidirectional exchange of interferometric observations between any pair of satellites. The cross-correlation of signals can happen at any aggregation point; hence, not all pairs of satellites need to exchange data.

Wireless coverage is another challenge to which $k$-NN algorithms have been applied. For instance, \cite{8808108} and \cite{1205192} devise algorithms to adjust the heading of mobile wireless sensors such that all sensors maintain a neighbourhood of size $k$ indicating convergence of the group of sensors to the same direction. In a relevant problem, node placement and localization, the authors of \cite{OH201891, 7386596} use a fingerprinting method to localize a wireless device based on the received signal strength of the $k$ nearest neighbours. Extending the same idea, but using the channel state information (CSI) instead of the signal strength, has been proposed by \cite{9149443}.

Finally, in resource allocation problems of 6G mobile telecommunication networks, $k$-NN has been used as a classification method to cluster data.
In \cite{8329631}, $k$-NN is used to dynamically allocate the radiating elements of an antenna array to $k$ users based on their spatial patterns. A set of classes of quality of service and channel state information are defined, and best resource allocation is devised. $k$-NN is used to assign the current channel and quality conditions of each user to one of those classes and, based on that, to allocate the radiating elements to form the necessary beams in a massive multiple-input-multiple-output (MIMO) setup. The expected response time of radio resource allocation decisions is in the range of $< 1~\text{ms}$ \cite{9846950}. In the OLFAR communications scenario, where satellites continuously change their position, the relevant resource allocation decisions need to also be performed sufficiently fast. Though we currently focus on static scenarios, the aforementioned study indicates the benefits of $k$-NN even in more dynamic settings. The authors of \cite{9285223} have used $k$-NN graphs to reduce a wireless node's complexity and calculation overhead of the best time schedule of the communication to their neighbours.

Previous studies confirm the wide adoption of $k$-NN graphs in various problems in wireless networks. Yet, these studies view wireless networks as undirected graphs; an unnecessary assumption for OLFAR. We extend those efforts with topology control of unidirectional orbiting swarms for OLFAR applications.

\subsubsection{Connectivity properties of $k$-NN graphs}
The connectivity properties of $k$-NN graph have received substantial attention, both in engineering and mathematical literature \cite{alamgir2012shortest, balister2005connectivity, balister2009critical, balister2013percolation, eppstein1997nearest}. In \cite{alamgir2012shortest}, the behaviour of the shortest path distance in weighted $k$-NN graphs is studied. 
In \cite{eppstein1997nearest}, probabilistic properties, like the expected number of connected components of the $k$-NN graph for a random set of points are discussed.

In contrast to wired networks, wireless networks usually do not come with a fixed set of links between nodes; furthermore, these wireless connections need not be bidirectional. In a situation when one is given a network of wireless nodes distributed at random, e.g., in a way modelled by a Poisson process, it is natural to ask under what conditions is it reasonable to expect a \enquote{fully connected} network of nodes. Here we will consider nodes broadcasting with a fixed finite range of transmission, but we allow for these ranges to vary between the nodes. Naturally, the answer depends on the nature of links (unidirectional vs. bidirectional) and the precise meaning of the term fully connected.

The studies on connectivity properties of $k$-NN graphs carried out by the engineering community focus on a model where each of the links allows for bidirectional communication and these links are established between $k$-nearest neighbours. Originally, the research revolved around \enquote{magic numbers}, that is, values of $k$ which with high probability lead to a network connected in a sense that there exists a path between any two nodes (see e.g.\ \cite{kleinrock1978,ni1994} and many others). It has been suggested that these \enquote{magic numbers} can be any of the integers between three and eight. A breakthrough came in the article \cite{xue2004}, where the authors realized that to maintain connectivity as the area of the square in which we place the nodes (denoted by $A$) tends to infinity, while the intensity of the Poisson process governing the nodes remains constant,  each node needs to be connected to a number of neighbours that is of the order $\log A$. This marks a departure away from the idea of a universal \enquote{magic number}. Furthermore, the same article provides some bounds for the critical value, multiplying $\log{A}$,  describing the zero-one law related to this notion of connectivity.

These results have stimulated a rigorous mathematical investigation. In a series of articles \cite{balister2005connectivity, balister2009critical}, the authors formalize the idea that, given the setting of \cite{xue2004}, connectivity is indeed obtained when $k$ is of the order $\log A$, and first provide a much tighter bound on the critical multiplication constant (0.3043 $\leq c^\star \leq$ 0.5139). Later, they invent a method to compute this critical constant precisely. Further results in these directions are for example \cite{balister2013percolation}.

All of these results concern the case with bidirectional links, which is \emph{not} our setting. For a directed version of this problem, where connectivity is given by an existence of a \emph{directed} path between any two nodes, the article \cite{balister2005connectivity} presents behaviour similar to the undirected model, that is connectivity threshold at $k=c^\star \log A$, with $0.7209 \leq c^\star \leq 0.9967$. It should be noted that these results concern asymptotic behaviour when $n$ grows very large, which limits their direct applicability.

\section{Communication network within the swarm}\label{sec:commnet}

\subsection{Connectivity \& strongly connected components}
As outlined in Section~\ref{sec:intro}.\ref{sec:intro:assump}, we assume that the communication network inside the swarm is modelled as a directed graph. Ideally, one would have a network that allows for communication between \emph{any} two satellites. In mathematical terms, this would mean that the \emph{largest strongly connected components} (LSCC for short) is the entire $k$-NN graph; i.e., there exists a (directed) path between any two vertices. On the other hand, such a strong guarantee in a spatially random network could be quite energetically demanding. 

\begin{figure}[htbp]
    \centerline{\includegraphics[scale=.60]{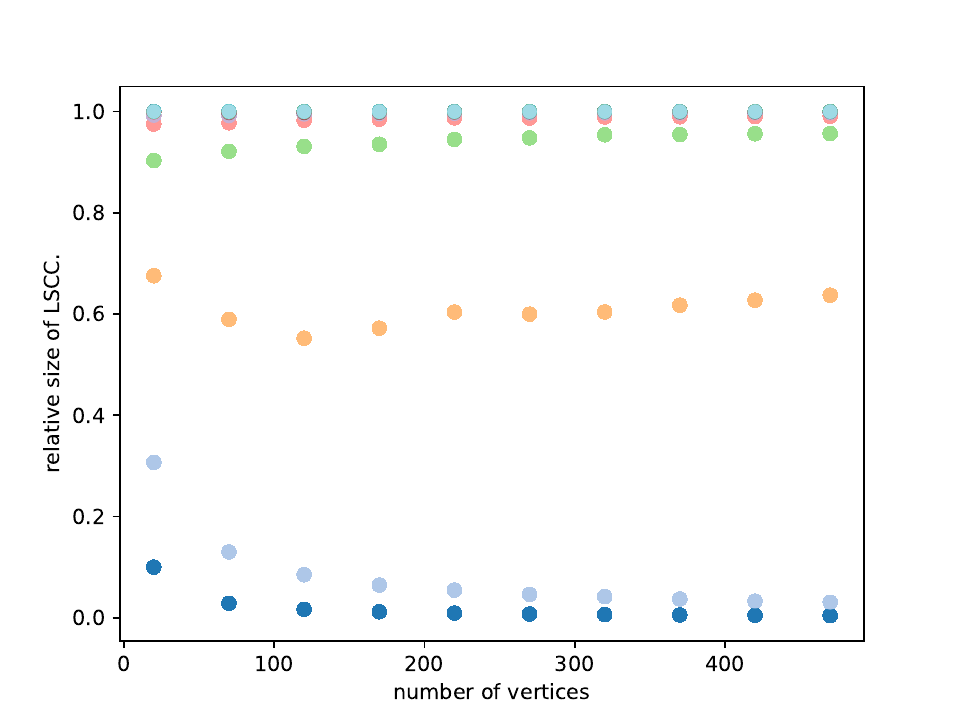}}
    \caption{Mean fraction of vertices in the LSCC for different sizes of the graph and $k\in \{2, \ldots12\}$.}\label{fig:conn}
\end{figure}

In this section, we show simulations that shed light on the effect of the choice of the parameter $k$ on the size of the LSCC. Our results are plotted in Figure~\ref{fig:conn}. This data shows that, in the setting of this study, it makes little sense to consider values of $k\leq 3$, since there the connectivity is simply too low. Starting with $k=4$, we see that the LSCC typically spans at least $\sim 90\%$ of vertices in the graph. While this choice already provides a network that behaves in the required manner, it appears that the optimal choice of $k$ would be $k=5$, since, on the one hand, there is little to no improvement in the typical size of the LSCC for larger values of $k$, while larger values of $k$ lead to higher energy requirements for maintaining communication.

\subsection{Empirical distribution of baseline lengths}
To obtain high-quality data from an interferometric observation, it is desirable to have target observations made with various baselines. In this section, we study the length of the longest baseline within the LSCC and the distribution of baseline lengths. Note that given our assumption about the geometry, the longest possible baseline has length $\sqrt{3}\approx 1.73$ -- this corresponds to the diagonal across the unit cube. In our results, we confine ourselves to the parameter regime $n=0-1,000$, and $k\in\{4, 5, 6 \}$.

\begin{figure}[h]
    \centering
     \begin{subfigure}[b]{0.44\textwidth}
         \centering\includegraphics[width=\textwidth]{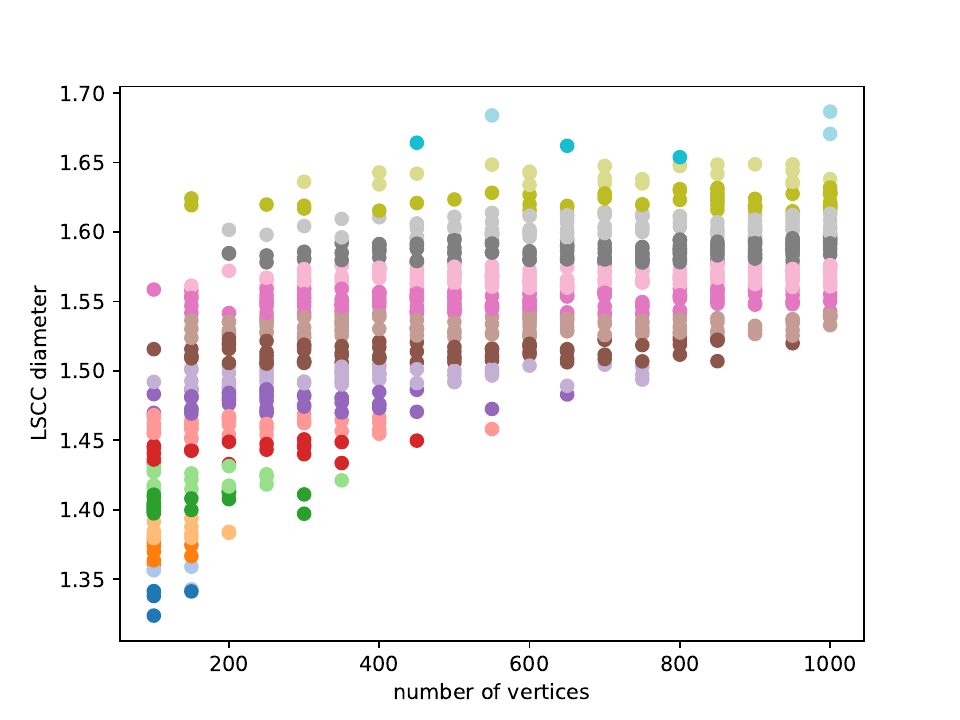}
         \caption{Scatter plots of maximal baseline lengths for $k=5$}
         \label{fig:maxbaseline}
     \end{subfigure}
     \hfill
     \begin{subfigure}[b]{0.44\textwidth}
         \centering\includegraphics[width=\textwidth]{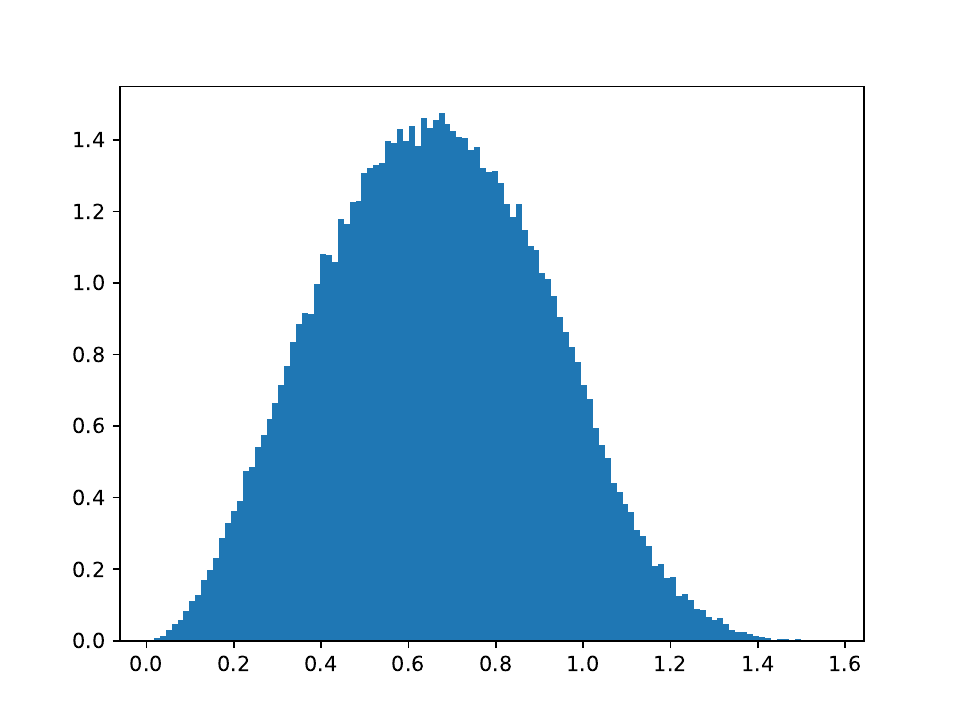}
         \caption{Histogram of  baseline lengths for $n=560$, $k=5$.}
         \label{fig:baseline:i-ex}
     \end{subfigure}
     \caption{Illustrative results related to baseline distribution.}
     \label{fig:baselines}
\end{figure}

In Figure~\ref{fig:maxbaseline}, one can see the scatter plot of the lengths of longest baselines within the LSCC for $k=5$. In the regime of small $n$'s ($n\lesssim 200$), irrespective of the value of~$k$, the length of the maximal baseline within the LSCC varies considerably. On the other hand, for swarms with at least $\sim 200$ satellites, we observe less variance around the mean. Scatter plots for $k=4, 6$ can be found in the Supplementary material. 

Not only the maximal baseline is important for observation, but also their distribution. Our simulations show that the distribution of baseline lengths within the LSCC is well spread-out and spans all the way from very short to almost maximal possible lengths. A representative example of this distribution can be seen in Figure~\ref{fig:baseline:i-ex}. More histograms can be found in the supplementary material.


\subsection{Distribution function for communication costs}
\label{sec:commnet:pdf}
Recall from Section \ref{sec:intro:assump} that we have $n$ satellites distributed uniformly in the cube $[0,1]^3$. Let $U$ be a uniformly chosen satellite. The locations of the satellites close to $U$ can be well approximated by those in a homogeneous Poisson Point Process (PPP) of intensity measure $n$ times the Lebesgue measure. For such a PPP, the number of satellites in a region $A$ of volume ${\rm Vol}(A)$ is Poisson with parameter $n {\rm Vol}(A)$, while the number of satellites in disjoint regions is independent. Indeed, for a PPP, the total number of points is Poisson with parameter $n$, rather than equal to $n$, while the point locations are independently uniformly distributed. Since the Poisson distribution is highly concentrated, a Poisson distribution with parameter $n$ is quite close to $n$, so  it makes little difference to work with a Poisson number of parameter $n$ or with precisely $n$. When rescaling distances by $n^{1/3}$, the PPP becomes of unit intensity, allowing to describe the local environment of the uniform satellite $U$. While this accurately describes the local environment of satellites in the interior of the cube, for satellites closer to the boundary, this obviously creates some inaccuracies that we aim to describe using finite-size corrections.

Let the cost of transmission be $P_{0,t}(k)$, defined as 
\begin{equation}
    P_{0,t}(k) := R^2_k,
\end{equation}
where $R_k$ is the distance to the $k$th nearest neighbour. Let $X_r = |B_r(U)|$ denote the number of points in the ball of radius $r$ around the point $U$. By definition of a Poisson point process, $X_r$ is distributed as a Poisson random variable with parameter equal to the volume of a ball times a parameter which depends on $n$ as $4\pi r^3/{3} n$.

By this observation, we can compute the probability that $P_{0,t}(k)>c$, for $c>0$, as
\begin{align}
    \mathbb{P}(P_{0,t}(k)>c) &= \mathbb{P}( X_{\sqrt{c}}\leq k-1) = \sum_{j=0}^{k-1}\mathbb{P}(X_{\sqrt{c}} = j)\nn\\
    &=\sum_{j=0}^{k-1} \e^{-\lambda(c)}\frac{\lambda(c)^j}{j!}, 
\end{align}
where $\lambda(c) = 4 \pi n c^{\frac{3}{2}}/3$.\\

It follows that the cumulative distribution function (CDF) of $P_{0,t}(k)$ is given by
\begin{equation}\label{eq:Power of transmission: CDF}
    \mathbb{P}(P_{0,t}(k)\leq c) =1 - \e^{-\lambda(c)}\sum_{j=0}^{k-1} \frac{\lambda(c)^j}{j!}.
\end{equation}
To obtain the probability density function, we differentiate \eqref{eq:Power of transmission: CDF} with respect to $c$, to obtain
\begin{align} f_{P_{0,t}(k)}(c) &= \e^{-\lambda(c)}\sum_{j=0}^{k-1} \frac{\lambda(c)^j}{j!}  - \e^{-\lambda(c)}\sum_{j=0}^{k-2} \frac{\lambda(c)^j}{j!}\nn\\
&= \frac{\lambda(c)^{k-1}}{(k-1)!}\e^{-\lambda(c)}\lambda'(c).
\end{align}
Substituting $\lambda(c) = 4 \pi  n c^{\frac{3}{2}}/3$ gives
\begin{equation}\label{eq:Density of power of transmission}
    f_{P_{0,t}(k)}(c) = \frac{3}{2}\left(\frac{4\pi n}{3}\right)^{k}\frac{c^{\frac{3k}{2}-1}}{(k-1)!}\e^{-\frac{4 \pi n c^{3/2}}{3}}.
\end{equation}
Recall that the probability distribution function of the \emph{generalized} gamma distribution is given by
\begin{align}\label{eq:gengamma}
    f_{GG}(x,a,d,p) = \frac{\left( \frac{p}{a^d}\right) x^{d-1} \e^{-\left( \frac{x}{a}\right)^p}}{\Gamma\left(\frac{d}{p}\right)},
\end{align}
where $d,p >0$ are so-called shape parameters, and $a>0$ is the scale parameter. Comparing \eqref{eq:Density of power of transmission} with \eqref{eq:gengamma}, we see that the transmission costs are distributed as a generalized gamma distribution  with parameters
\begin{equation}
\label{eq:GGn}
    f_{P_{0,t}(k)}(c) = f_{GG}\left(c, \left( \frac{4\pi n}{3}\right)^{-2/3}, \frac{3k}{2}, \frac{3}{2}\right).
\end{equation}
This is the approximate transmission cost density for large numbers of satellites, to which we can compare our simulations that involve finite-size corrections, as discussed in more detail in the next section.
In particular, \eqref{eq:GGn}  implies that the transmission cost of a satellite $P_{0,t}(k)$ scales as
\begin{equation}
\label{eq:limit-law-power}
    n^{2/3} P_{0,t}(k)\stackrel{d}{\rightarrow} P_k,
\end{equation}
where $P_k$ is distributed according to the generalized gamma distribution in \eqref{eq:GGn} with $n=1.$

After having established the distribution of the powers of the swarm of satellites, observe that the maximal energy of the satellites is given by $E_{\max}$, which is a finite constant. The generalized gamma distribution, however, has an unbounded support, which means that not all connections present in the $k$-NN graph can actually be established by the swarm. In practice, the empirical distribution of transmission costs will therefore be supported only on the finite interval $[0, E_{\max}]$ and one could approximate it with a truncated distribution which has density 
\begin{align}
     f^{T}_{P_{0,t}(k)}(c) = f_{P_{0,t}(k)}(c)
\end{align}
for $c\in [0,E_{\max}]$, and is equal to $E_{\max}$ with probability $\int_0^{E_{\max}}  f_{P_{0,t}(k)}(s)\,\mathrm{d}s$.

In Section \ref{sec:comp-alloc}, where we study the performance of our communication and computation system, we investigate the LSCC of the realized connections of the swarm as a function of the key tuning parameters $\beta$ and $E_{\rm max}$.

\subsection{Finite-size corrections}\label{sec:commnet:finite-size-corrections}
The distribution function for communication costs derived in the previous section was obtained in the limit of large numbers of satellites. Effectively, this ignores the fact that the cube has a {\em boundary}. This is in contrast with the setting at hand, where we assume that all the satellites are restricted to be within a unit cube, thus introducing \emph{finite-size corrections}. Unfortunately, the error introduced by these finite-size corrections is significant when considering hundreds of deployed satellites.

To remedy this error, we argue that these finite-size corrections will predominantly appear in \emph{one} of the parameters of the class of generalized gamma distributions. Furthermore, we estimate this parameter numerically and provide a simple empirical formula that captures these corrections in the studied regime.

\begin{figure}
     \centering
     \begin{subfigure}[b]{0.44\textwidth}
         \centering
         \includegraphics[width=\textwidth]{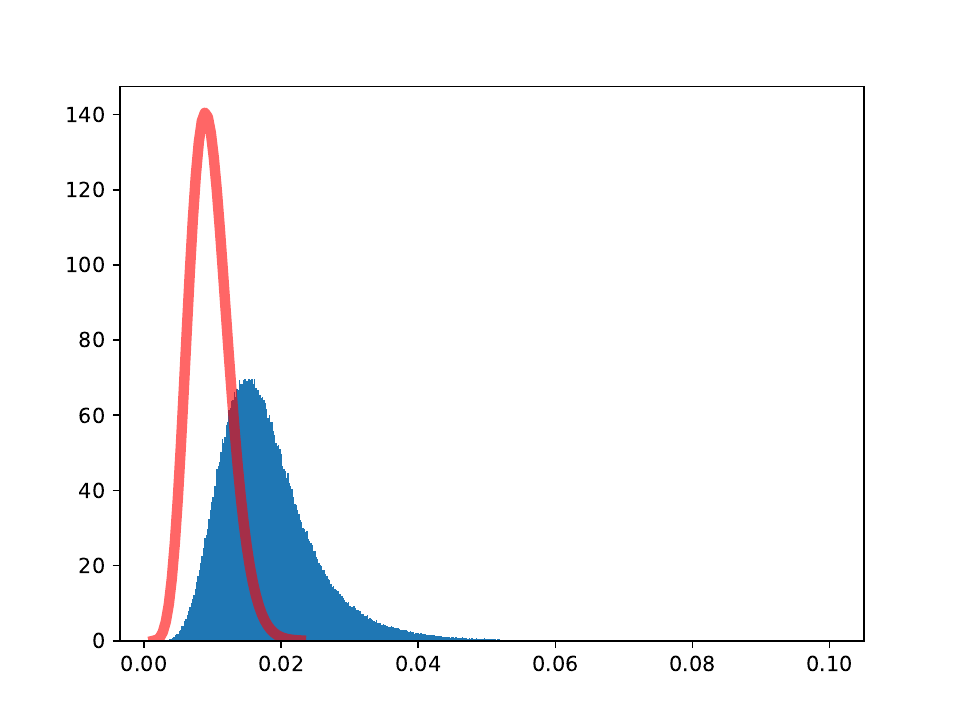}
             \label{fig:hist:uncorr}
     \end{subfigure}
     \hfill
     \begin{subfigure}[b]{0.44\textwidth}
         \centering
         \includegraphics[width=\textwidth]{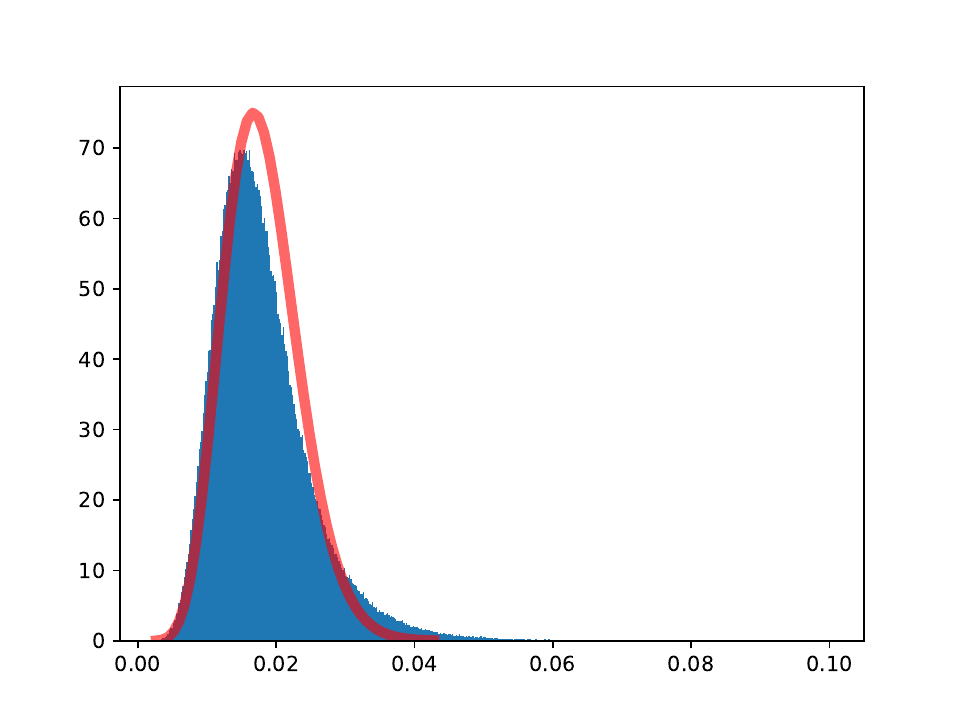}
             \label{fig:hist:corr}
     \end{subfigure}
     \hfill
     \vspace{-0.5cm}
    \caption{Comparison of the simulated transmission cost histograms with the generalized gamma distribution; (a) with and (b) without finite- size corrections, for $n=600$ and $k=5$.}
    \label{fig:hist:corr_uncorr}
\end{figure}

As one can see in Figure~\ref{fig:hist:corr_uncorr}, 
 empirical distributions obtained via simulation are qualitatively similar to the ones obtained in the previous section, but there is a significant quantitative disagreement.

By examining the parametrization of the distributions obtained in the previous section, it is intuitive that the finite-size corrections have to be applied predominantly to the parameter~$a$, since it reflects the intensity of the considered PPP. In the case of configurations with a low number of satellites, one could argue that each of the satellites has a different \emph{effective} intensity of the Poisson random variable describing its neighbourhood. While a direct symbolic computation seems infeasible, it is possible to \emph{estimate} the parameter $a$ (and hence also the effective intensity) by fitting the family of generalized gamma distribution with all parameters except for $a$ fixed to values obtained in Section~\ref{sec:commnet:pdf}, and estimating the parameter~$a$. This way, we obtain a correction in the form
\begin{equation}
    a_{\rm corr} = (0.685 \pm 0.002) n^{-0.73\pm0.01},
\end{equation}
which is valid in the regime of $n\in [100, 1,000]$, $k\in \{4, 5, 6 \}$.
By comparing generalized gamma distributions with $a_{\rm corr}$ as the scale parameter with empirical distributions obtained from simulations (see Figure~\ref{fig:hist:corr_uncorr}), 
we see that this simple correction leads to much better approximation of empirical results.

\section{Applications to computation allocation}
\label{sec:comp-alloc}

\subsection{Cross-correlations in the LSCC}\label{sec:comp-alloc:cross-corr-LSCC}
In this section, we describe an algorithm to divide the computation of the cross-correlations of the satellites in the largest strongly connected component (LSCC).
First, we investigate how the satellites can decide whether they are in the LSCC. After this, we determine how the satellites can divide the computations of the cross-correlations. Due to the choice of $k$ in our $k$-NN graph, the LSCC contains at least $95\%$ of the satellites.

By the definition of a strongly connected component, all the satellites in the LSCC have the same satellites in their in-component, and thus have the same in-component sizes. The satellites outside the LSCC will either have more satellites in their in-component compared to the LSCC in-component (for those in the out-component of the LSCC) or much less (less than $5\%$). When they are in the out-component of the LSCC, their in-component is strictly larger than the LSCC. Otherwise, their in-component will be at most the complement of the LSCC. Thus, the mode of the in-component sizes (the one with the highest frequency) can be used to identify which satellites are in the LSCC. Based on their in-component sizes, each of the satellites can thus be classified into three types:
\begin{enumerate}
    \item \textbf{Satellites in the LSCC:} These are the satellites whose in-component size equals the mode of the in-component sizes.
    \item \textbf{Satellites in out-component of the LSCC minus the LSCC:} These are satellites whose in-component size is strictly larger than this mode.
    \item \textbf{Other satellites:} The size of in-component is less than $5\%$.
\end{enumerate}

When the LSCC is large, the satellites of the type $1$ and $2$ are the most significant, as they have the most data. We perform the cross-correlations between all pairs of these satellites. The number of satellites of type $2$ and $3$ is at most $5\%$ of the total satellites because they lie outside the LSCC. 

After identifying the LSCC, we next investigate how to assign the cross-correlation computations optimally to the satellites, by proposing a strategy based on the energy distribution of the satellites. For this, we \emph{label} the satellites. 

The idea is that each satellite will communicate their energy surplus and their labels to the other satellites. Based on this information, the first labelled satellite gets the first few cross-correlations, the second labelled satellite gets the cross-correlations after the point where the first satellite finished, etc.

We denote the LSCC by $C_{(1)}$. We write the labels of the satellites in the LSCC as $\{1, 2, \ldots, m\}$, where $m = |C_{(1)}|$. We let the surplus energy of the $i$th satellite be $E_i=E_{\rm max}-P_i$, where $E_{\max}$ is the initial energy given to each satellite and $P_i$ is the power needed by the $i$th satellite to communicate to its $k$ nearest neighbours. 

We denote the out-component and in-component of the LSCC by $C_{(1)}^{+}$ and $C_{(1)}^{-}$, respectively. We index the set of all $M= \binom{|C_{(1)}^{-}|}{2}$ cross-correlation pairs lexicographically as
$$C = \{c_1, c_2, \ldots, c_{M}\}.$$ 
Each element of $C$ is an ordered pair denoting a cross-correlation.
In the LSCC, we compute these cross-correlations in ascending order. The satellite with the smallest label in the LSCC does the initial cross-correlations with all its residual energy. The second smallest labelled satellite now performs the cross-correlations from where the first satellite left, and so on. Each satellite computes with all its residual energy.

In more detail, we assign the first $\lfloor E_1\rfloor$ cross-correlations, that is, $\{c_1, c_2, \ldots, c_{\lfloor E_1\rfloor}\}$ to the satellite with label 1. Satellite 2 performs the next $\lfloor E_2\rfloor$ cross-correlations, that is, $\{c_{\lfloor E_1\rfloor +1}, \ldots, c_{\lfloor E_1\rfloor +\lfloor E_2\rfloor}\}$ and so on. This goes on until either all the cross-correlations are computed, or all satellites have done their computations. 

When there is surplus energy remaining after the cross-correlations between the satellites in the in-component of the LSCC have all been computed, we use the surplus energy of the satellites in the out-component to compute their cross-correlations with the satellites in the LSCC's in-component. We write the labels for the satellites in the out-component of LSCC minus LSCC as $\{m+1, m+2, \ldots, l\}$, where $l= |C_{(1)}^{+}|$. Then the total coverage of the cross-correlations $T(n)$ satisfies
\begin{equation}\label{eq: Total cross-correlation count disconnected}
    T(n) = \sum_{i=1}^{|C_{(1)}^{+}|}\left \lfloor E_i\right\rfloor \wedge \binom{|C_{(1)}^{-}|}{2} , 
\end{equation}
and the proportion of cross-correlations $\alpha$ satisfies
\begin{align}
    \alpha &= \frac{1}{\binom{n}{2}}\sum_{i=1}^{|C_{(1)}^{+}|}\left \lfloor E_i\right\rfloor \wedge  \frac{\binom{|C_{(1)}^{-}|}{2}}{\binom{n}{2}}.
\end{align}

By Section \ref{sec:commnet}.\ref{sec:commnet:pdf},
\begin{equation}\label{eq: communication energy}
   P_i \sim  GG\left(\left(\frac{4\pi n}{3}\right)^{-2/3}, \frac{3k}{2}, \frac{3}{2}\right), 
\end{equation} 
where $GG(a, d, p)$ is the generalised gamma distribution with density given by \eqref{eq:gengamma}. 
Since the empirical distribution of the transmission costs converges, for large $n$, we expect that
\begin{equation}\label{eq: proportion cross-correlation count disconnected}
    \alpha \approx \frac{2\mathbb{E}[E]}{n} \times \frac{|C_{(1)}^{+}|}{n}  \wedge  \left(\frac{|C_{(1)}^{-}|}{n}\right)^{2}, 
\end{equation}
where $E$ has the same probability distribution as $E_i = \max\{E_{\max}-P_i, 0\}$, given by
\begin{equation}\label{eq: residue energy}
   E_i \overset{d}{=} \left(E_{\max} - GG\left(\left(\frac{4\pi n}{3}\right)^{-2/3}, \frac{3k}{2}, \frac{3}{2}\right)\right)_{+},
\end{equation} 
where $x_+ = \max\{x, 0\}$.

We can rewrite \eqref{eq: proportion cross-correlation count disconnected} as 
\begin{equation}\label{eq: reparametrized proportion cross-correlation count disconnected}
    \alpha \approx \eta_{+}\frac{2\mathbb{E}[E]}{n}  \wedge  \eta_{-}^{2}, 
\end{equation}
where $\eta_+$ and $\eta_-$ are the proportions of satellites in the out- and in-component of the LSCC, respectively. This gives us our final formula for the total proportion of cross-correlations that can be computed in our communication and computation network, and is the main performance parameter of our system.

\subsection{Simulation study of cross-correlation coverage}
To study the cross-correlation coverage in the $k$-NN network, and to demonstrate the utility of results derived in Sections~\ref{sec:commnet}, we simulated the following scenario:
\begin{enumerate}
    \item Initially, all the $n$ satellites are assigned an energy budget $E_{\rm max} = q_{GG}(p)$, where $q_{GG}(p)$ is the quantile function of the density in \eqref{eq:limit-law-power} in Section~\ref{sec:commnet}.\ref{sec:commnet:pdf} with finite-size correction described in Section~\ref{sec:commnet}.\ref{sec:commnet:finite-size-corrections}, and $p$ is a simulation parameter.
    \item We establish a \emph{pruned} $k$-NN communication network, which is created from $k$-NN graph between the satellites with edges removed when their existence causes the satellite's power consumption to exceed the energy budget $E_{\rm max}$. We iteratively remove the most energetically expensive edges, until the edges no longer deplete the power budget.
    \item We set the energy per one cross-correlation computation as in \eqref{eq:assumption energy expenditure}, and assign computation jobs as described in Section~\ref{sec:comp-alloc}.\ref{sec:comp-alloc:cross-corr-LSCC}.
\end{enumerate}

The choice of $E_{\rm max}$ and $\beta$ strongly influences the outcome of the simulation. For large $n$, one needs to choose $E_{\rm max}$ unrealistically large to avoid all pruning. Thus, the quantile function $ q_{GG}(p)$ allows us to pick $E_{\rm max}$ in a way that will very probably induce network pruning. The intensity of this effect can be effectively controlled by the parameter $p$ (where we note that pruning is independent of $\beta$). The parameter $\beta$ influences the intensity of \emph{competition} for energy between the transmission and computational functions of satellites. 

In our simulations, we focus on three  quantities:
\begin{enumerate}
    \item LSCC reduction factor $\rho_L = \frac{|\text{LSCC post-pruning}|}{|\text{LSCC pre-pruning}|}$,
    \item LSCC coverage factor $\alpha_L = \frac{|\text{jobs assigned within LSCC}|}{|\text{jobs available within LSCC}|}$,
    \item Coverage factor $\alpha = \frac{|\text{jobs assigned within LSCC}|}{|\text{jobs available within swarm}|}$.
\end{enumerate}

Our results indicate that these 3 quantities are highly sensitive to the choice of parameters, with some regimes exhibiting results concentrated in a small interval and others dominated by inherent randomness. Some of these results are shown below, see Figures~\ref{fig:red-all}, \ref{fig:cLSCC-all} and \ref{fig:cALL-all}. For a more detailed discussion, see Supplementary material.

\begin{figure*}[t]
\begin{minipage}[b]{0.44\linewidth}
\centering
\begin{subfigure}[b]{0.49\textwidth}
         \centering
         \includegraphics[width=\textwidth]{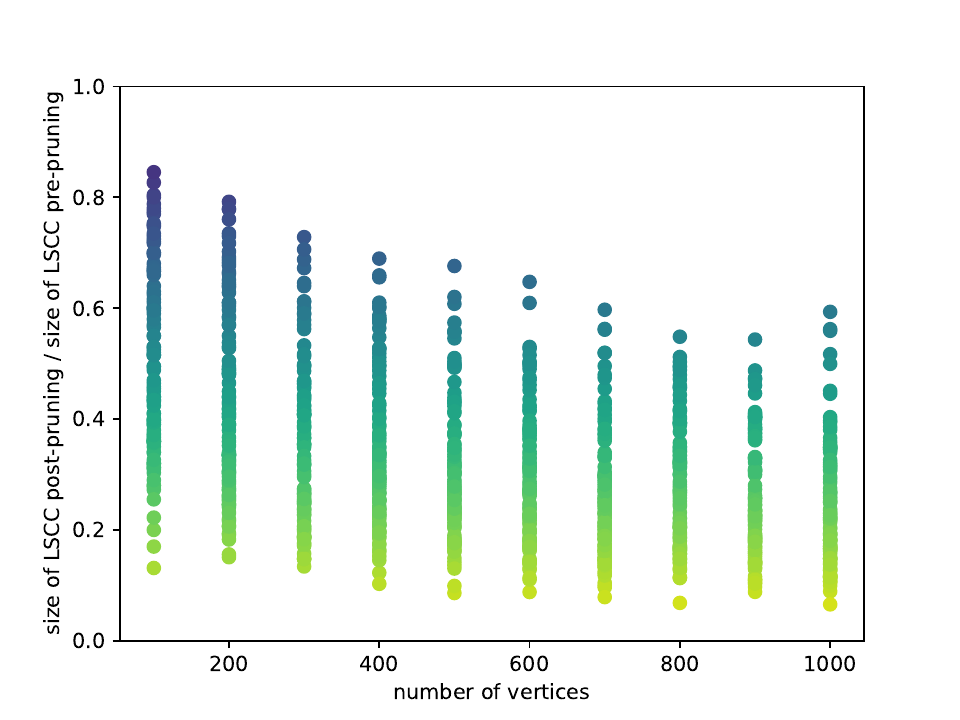}
         \subcaption{$p=0.1$}
         \label{fig:red1}
     \end{subfigure}
     \hfill
    \begin{subfigure}[b]{0.49\textwidth}
         \centering
         \includegraphics[width=\textwidth]{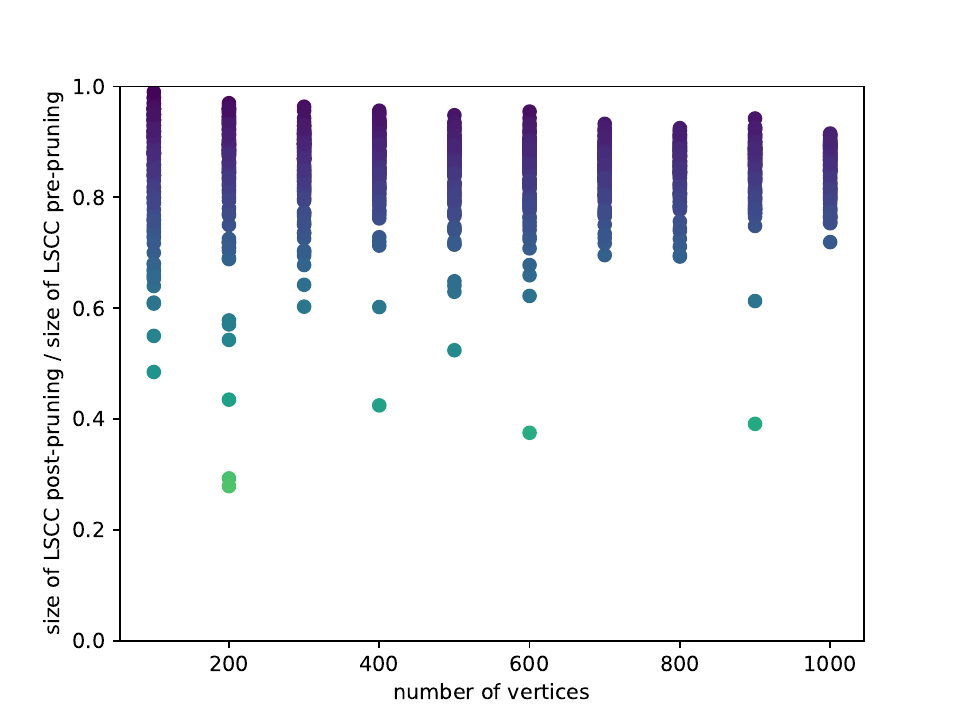}
         \subcaption{$p=0.3$}
         \label{fig:red2}
     \end{subfigure}
     \hfill
     \begin{subfigure}[b]{0.49\textwidth}
         \centering
         \includegraphics[width=\textwidth]{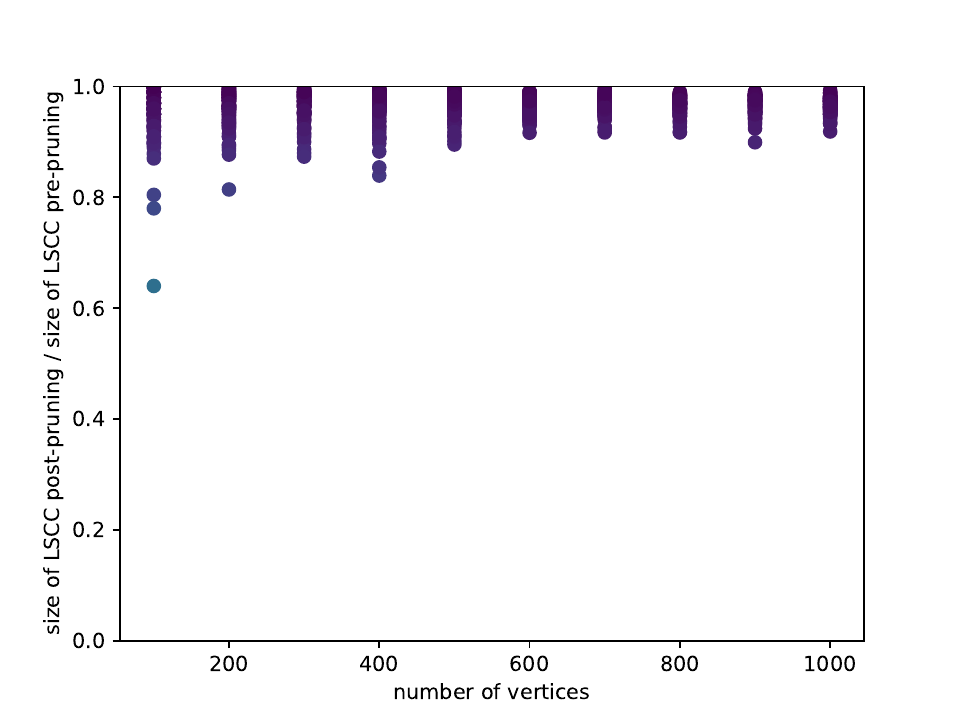}
         \subcaption{$p=0.6$}
         \label{fig:red3}
     \end{subfigure}
     \hfill
     \begin{subfigure}[b]{0.49\textwidth}
         \centering
         \includegraphics[width=\textwidth]{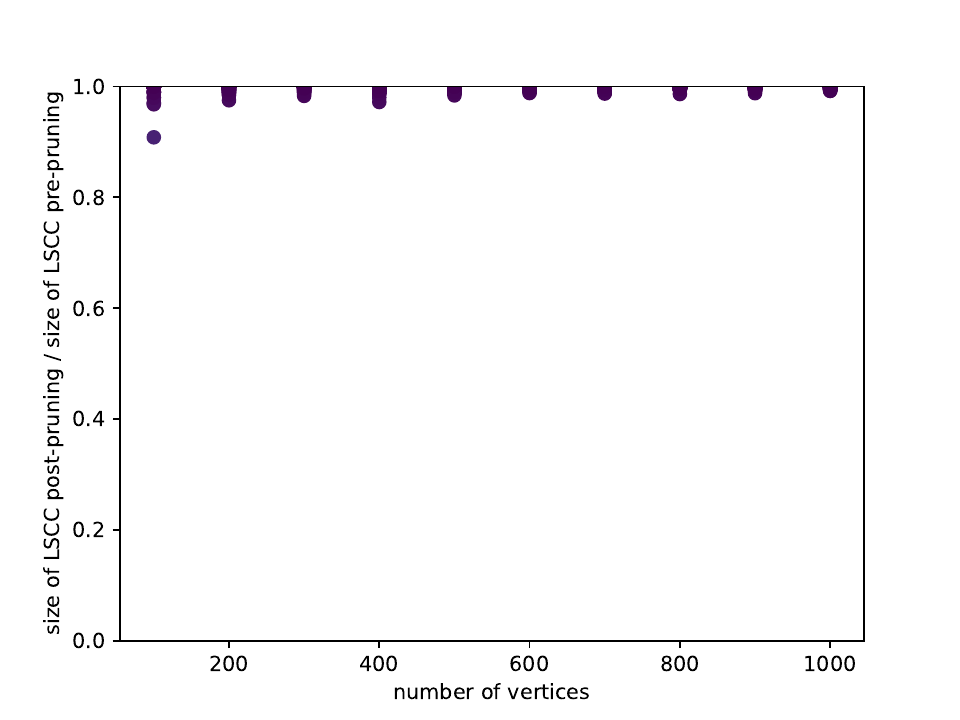}
         \subcaption{$p=0.99$}
         \label{fig:red4}
     \end{subfigure}
\caption{Examples of observed values of the LSCC reduction factor $\rho_L$ for $k=5$, $n\in(100, 1000)$, $p$ varying.}
        \label{fig:red-all}
\end{minipage}
\hfill
\begin{minipage}[b]{0.44\linewidth}
\centering
\begin{subfigure}[b]{0.49\textwidth}
         \centering
         \includegraphics[width=\textwidth]{sats_res/fig_fracLSCC_5-0.100-0.100-Oct5.pdf}
         \subcaption{$p=0.1$}
         \label{fig:red5}
     \end{subfigure}
     \hfill
    \begin{subfigure}[b]{0.49\textwidth}
         \centering
         \includegraphics[width=\textwidth]{sats_res/fig_fracLSCC_5-0.300-0.600-Oct5.pdf}
         \subcaption{$p=0.3$}
         \label{fig:red6}
     \end{subfigure}
     \hfill
     \begin{subfigure}[b]{0.49\textwidth}
         \centering
         \includegraphics[width=\textwidth]{sats_res/fig_fracLSCC_5-0.600-0.700-Oct5.pdf}
         \subcaption{$p=0.6$}
         \label{fig:red7}
     \end{subfigure}
     \hfill
     \begin{subfigure}[b]{0.49\textwidth}
         \centering
         \includegraphics[width=\textwidth]{sats_res/fig_fracLSCC_5-0.990-0.900-Oct5.pdf}
         \subcaption{$p=0.99$}
         \label{fig:red8}
     \end{subfigure}
        \caption{Examples of observed values of the LSCC coverage factor $\alpha_L$ for $k=5$, $p, \beta$ varying.} 
        \label{fig:cLSCC-all}
\end{minipage}
\end{figure*}

\begin{figure*}[h]
     \centering
     \begin{subfigure}[b]{0.24\textwidth}
         \centering\includegraphics[width=\textwidth]{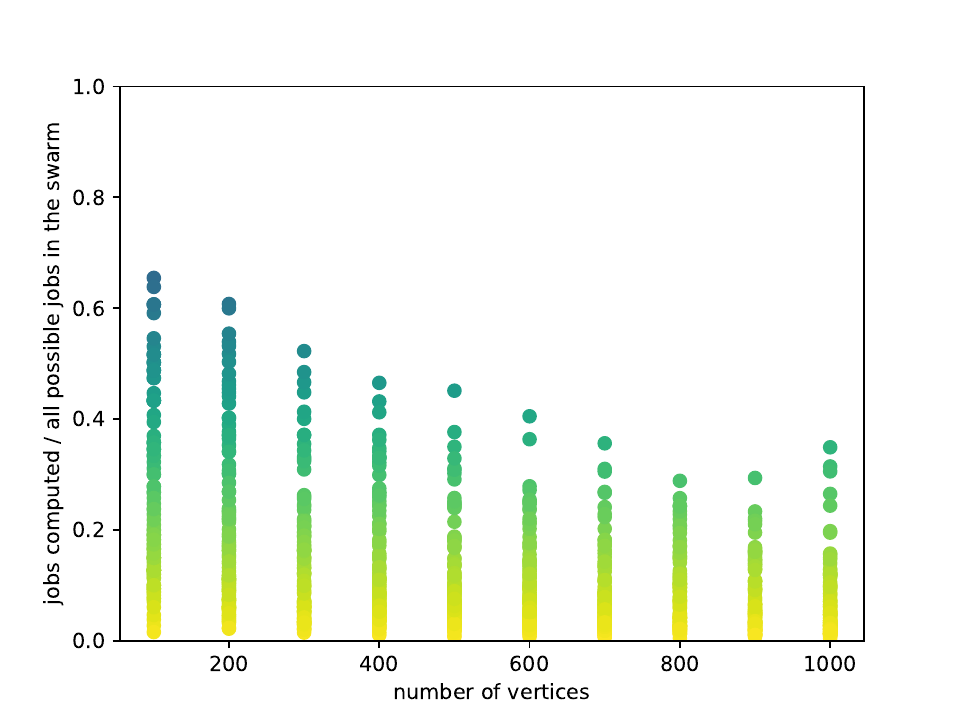}
         \caption{ $p=0.1$, $\beta=0.1$}
         \label{fig:LSCCcov.1.1-5}
     \end{subfigure}
     \begin{subfigure}[b]{0.24\textwidth}
         \centering\includegraphics[width=\textwidth]{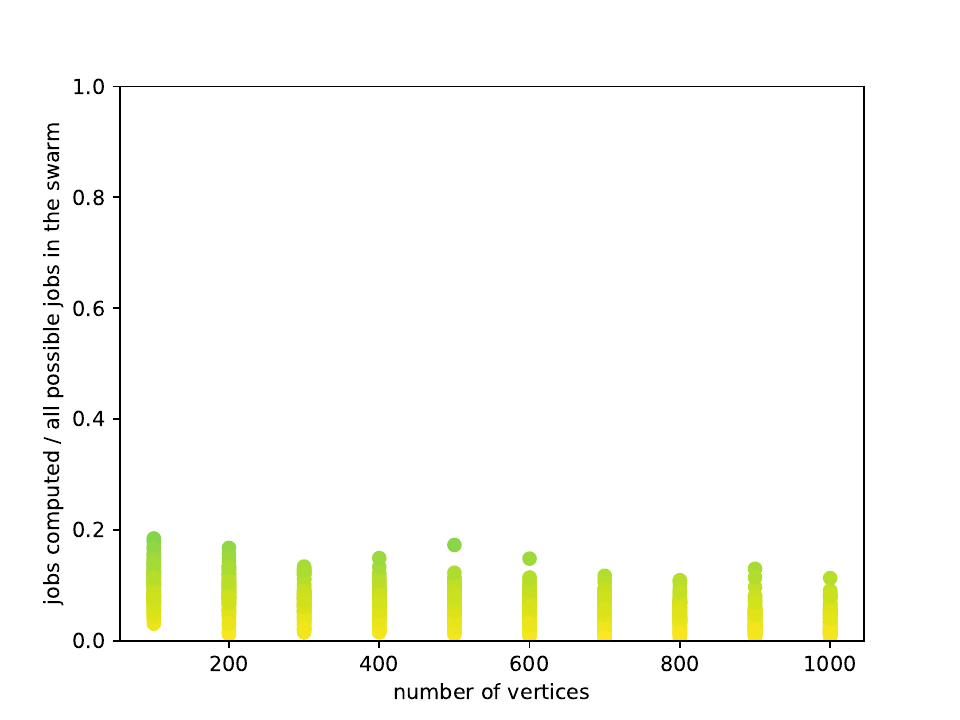}
         \caption{$p=0.1$, $\beta=1.0$}
         \label{fig:LSCCcov.1.0-5}
     \end{subfigure}
     \hfill
         \begin{subfigure}[b]{0.24\textwidth}
         \centering\includegraphics[width=\textwidth]{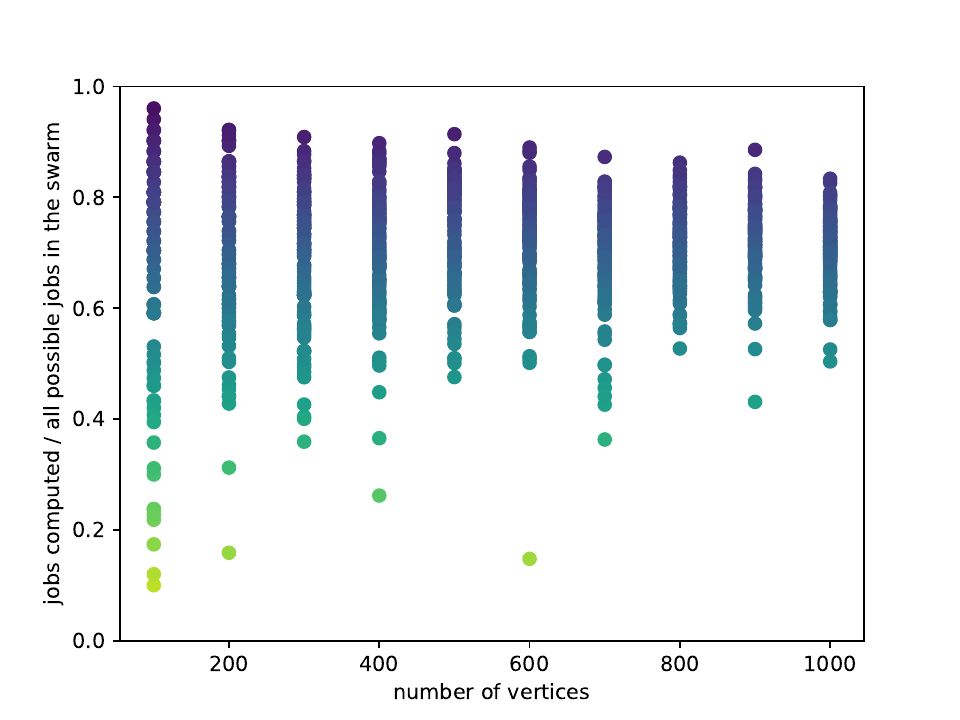}
         \caption{$p=0.3$, $\beta=0.1$}
         \label{fig:LSCCcov.3.1-5}
     \end{subfigure}
     \hfill
     \begin{subfigure}[b]{0.24\textwidth}
         \centering\includegraphics[width=\textwidth]{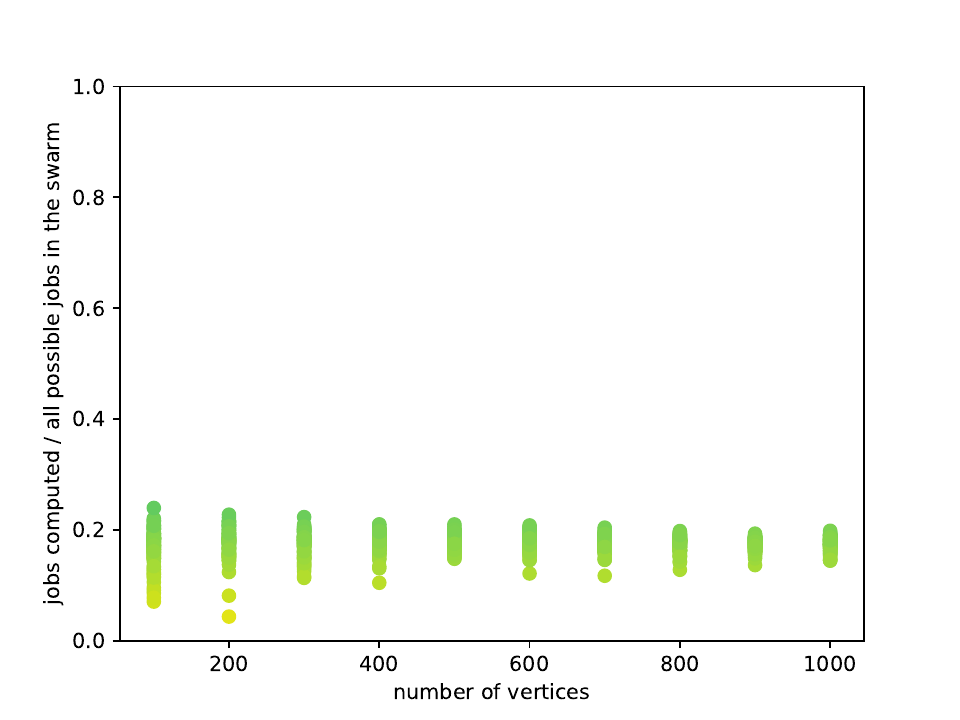}
         \caption{$p=0.3$, $\beta=1.0$}
         \label{fig:LSCCcov.3.0-5}
     \end{subfigure}
     \hfill
          \begin{subfigure}[b]{0.24\textwidth}
         \centering\includegraphics[width=\textwidth]{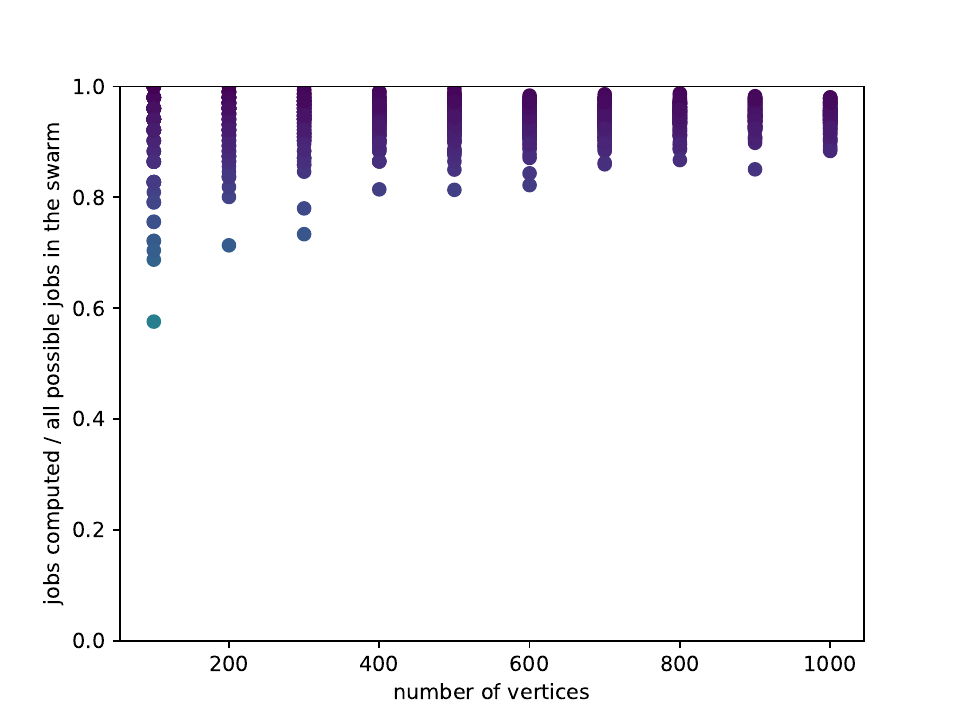}
         \caption{$p=0.7$, $\beta=0.1$}
         \label{fig:LSCCcov.7.1-5}
     \end{subfigure}
     \hfill
     \begin{subfigure}[b]{0.24\textwidth}
         \centering\includegraphics[width=\textwidth]{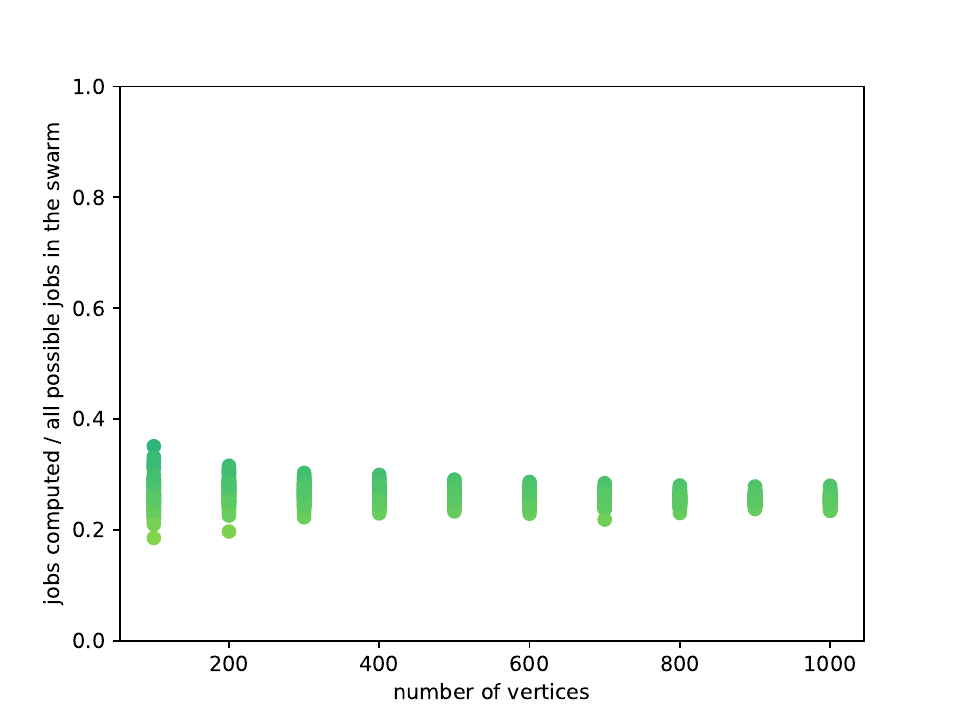}
         \caption{$p=0.7$, $\beta=1.0$}
         \label{fig:LSCCcov.7.0-5}
     \end{subfigure}
     \hfill
          \begin{subfigure}[b]{0.24\textwidth}
         \centering\includegraphics[width=\textwidth]{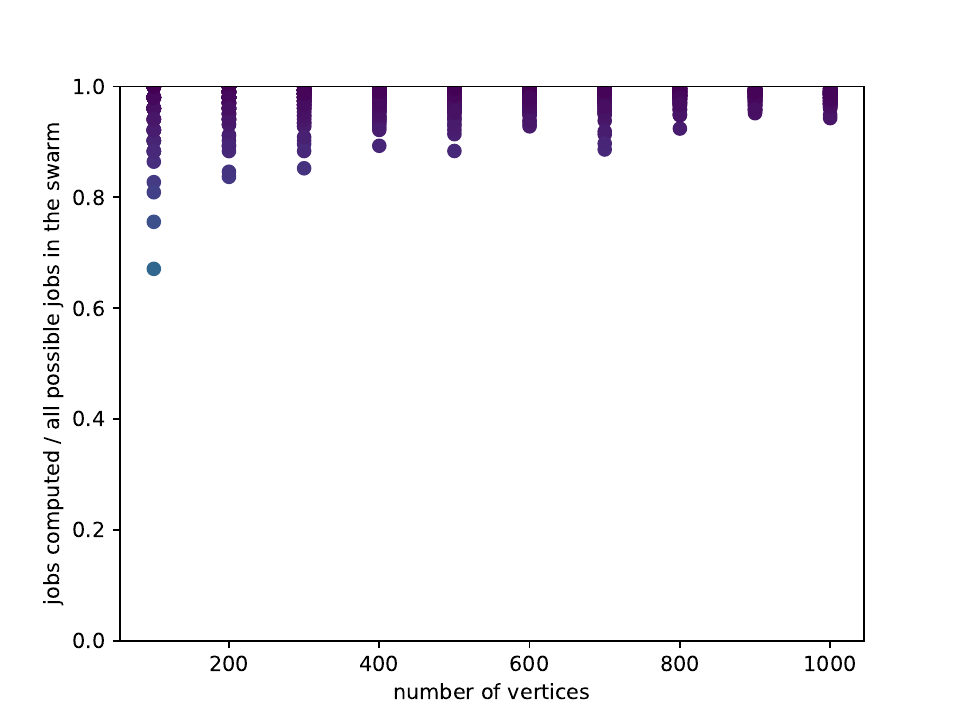}
         \caption{$p=0.99$, $\beta=0.4$}
         \label{fig:LSCCcov.99.4-5}
     \end{subfigure}
     \hfill
          \begin{subfigure}[b]{0.24\textwidth}
         \centering\includegraphics[width=\textwidth]{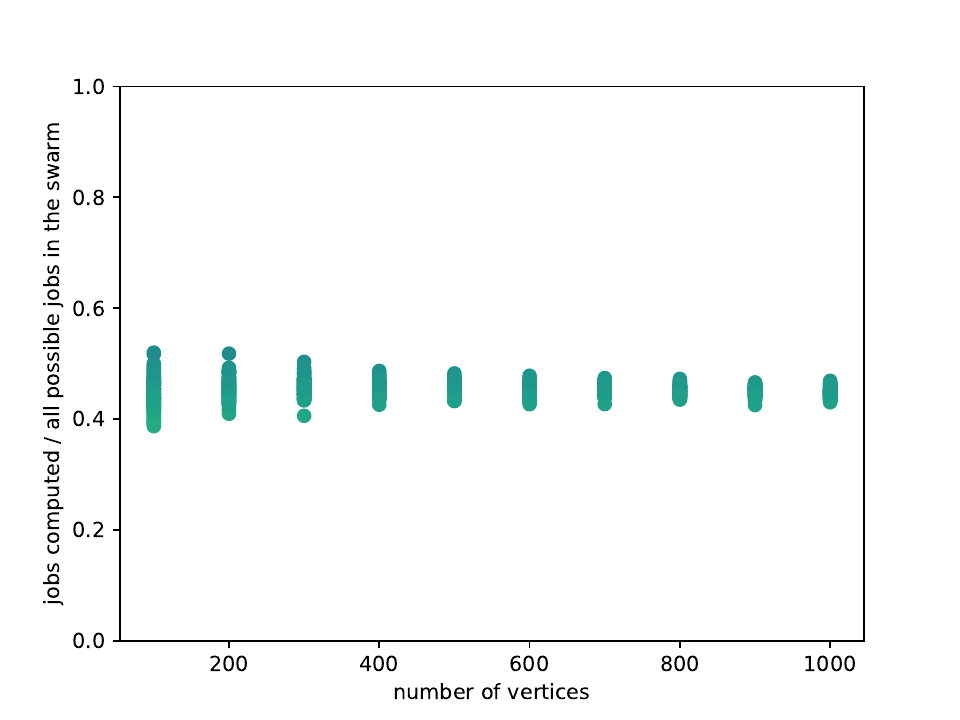}
         \caption{$p=0.99$, $\beta=1.0$}
         \label{fig:LSCCcov.99.0-5}
     \end{subfigure}
     \hfill
     \caption{Examples of observed values of the LSCC coverage factor $\alpha_L$ for $k=5$, $p, \beta$ varying. }
     \label{fig:cALL-all}
\end{figure*}


In engineering terms, the results lead to an algorithm for the parameter choices. We assume that $n$, the size of the satellite swarm, is known. We pick $k$ such that the LSCC is sufficiently large. For $k=5$, the LSCC contains at least 95\% of the vertices, while it is at least 98\% for $k=6$. With this $k$, we can then compute the power distributions over the swarm (including finite-size corrections). Further, the value of the computation cost $c$ (in this paper rescaled to be 1) fixes the product $\beta E_{\max}$ through \eqref{eq:assumption energy expenditure}. In turn, a value of $E_{\max}$ directly translates into a value of $p$ through the relation $E_{\rm max} = q_{GG}(p)$, where $q_{GG}(p)$ is the quantile function associated to the density in \eqref{eq:limit-law-power}. We then need to choose the pair $\beta$ and $p$ such that the desired coverage $\alpha$ is reached. Our analysis provides insight into the relation between the performance $\alpha$ and these parameters.

\section{Conclusion and future work}
\label{sec:conc-disc}
We proposed a communication protocol for satellites, in which satellites choose their power so that they can communicate with their $k$ nearest neighbours. Already for $k=5$, the largest strongly connected component (LSCC) contains more than 95\% of the satellites, thus ensuring sufficient coverage of the baselines needed. We computed the asymptotic distribution of the powers needed by the swarm of satellites, which is described by a generalized gamma distribution. This approximation follows by describing the local neighbourhoods of satellites in terms of a homogeneous Poisson process. Simulations confirm that this approximation already holds reasonably well, even when only a few hundred of satellites are considered after applying an appropriate finite-size corrections.  

Based on the approximation of the communication costs, we proposed an algorithm to guarantee that a large proportion of cross-correlations is computed. This algorithm relied on the assumption that all satellites have access to all signals in the strongly connected component. We computed the {\em coverage} of the cross-correlations that can be computed under the assumption that the total computation costs are a proportion $\beta$ of the total energy available to the satellites.

Some extensions of our work, for future research, are as follows:
\begin{enumerate}
    \item We compute the power that satellites need to communicate with their $k$ nearest-neighbour satellites, 
    Should the satellites be able to signal that they are not in the SCC, then they could increase their power so that they connect to more servers. While this increases their power level, this enhances connectivity. Since the satellites outside the (in-component of) the SCC are not heard at all, it may be worth for them to raise their power. Unfortunately, this does not help the satellites in the out-component of the SCC (as they hear all the signals in the SCC), but it may help satellites that are even outside of that. 
    \item What buffer allocation is needed in order for all $\binom{n}{2}$ cross-correlated pairs to be computed with high probability? Determining this threshold is probably mathematically challenging.
\end{enumerate}

\paragraph*{Acknowledgements.}
The work of ON, MP and RvdH was supported by the Netherlands Organisation for Scientific Research (NWO) through Gravitation-grant NETWORKS-024.002.003. The work of GE and MJB is part of the project ADAPTOR: Autonomous Distribution Architecture on Progressing Topologies and Optimization of Resources, with project number 18651 of the research Open Technology Program, (partly) financed by the Netherlands Organisation for Scientific Research (NWO).
\pagebreak

\bibliographystyle{IEEEtaes.bst}






\bibliography{refs}

\vspace{-2cm}
\begin{IEEEbiography}[{\includegraphics[width=1in,height=1.25in,clip,keepaspectratio]{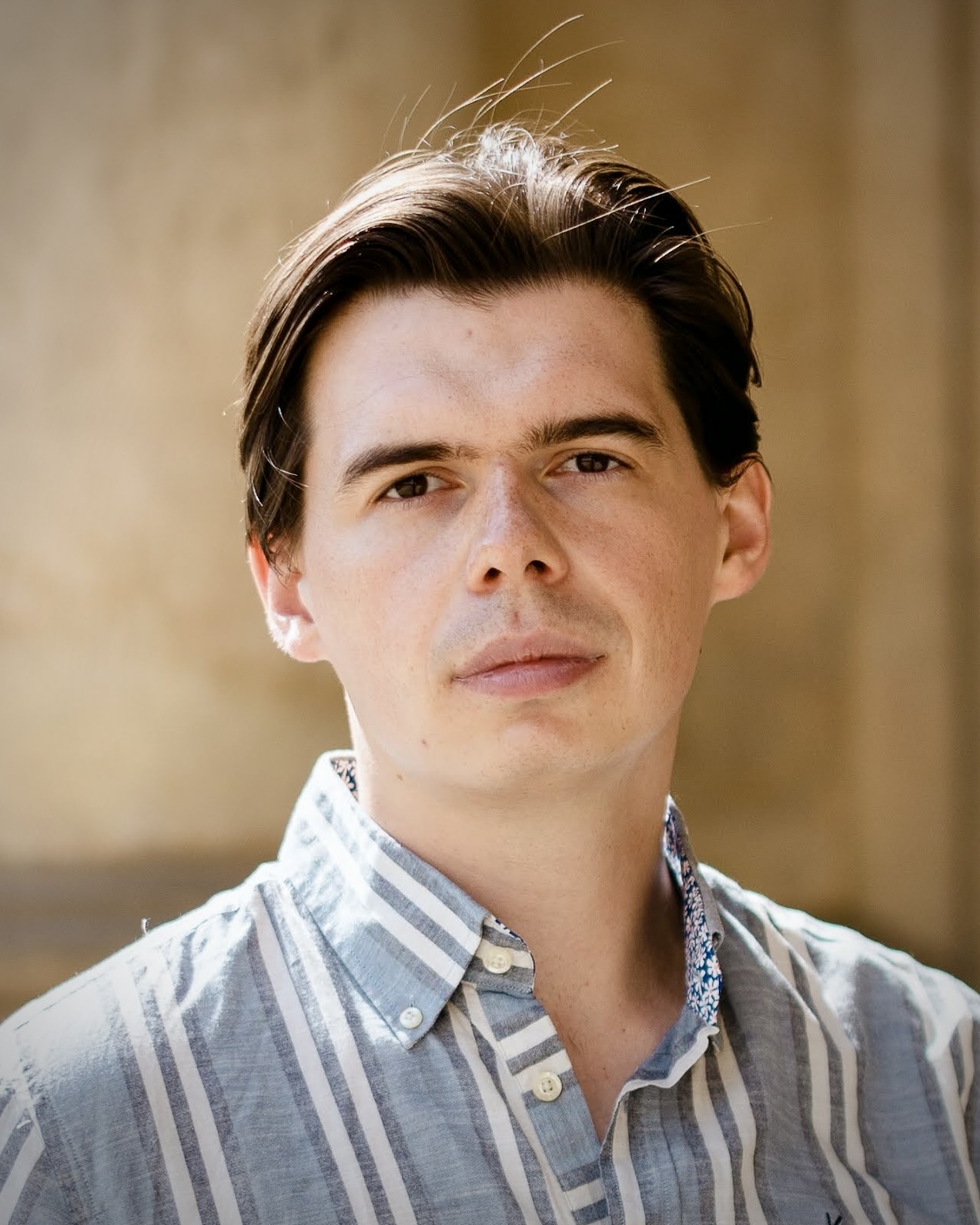}}]{Oliver Nagy}{\space} was born in Bratislava, Slovakia, in 1994. He received his bachelor's degree in physics in February 2017 from Charles University, Prague, Czech Republic. In February 2020, he received a master's degree in theoretical physics from the same university. Also in February 2020, he started his PhD studies in mathematics at Leiden University, Leiden, Netherlands.

He is employed as a doctoral candidate at the Mathematical Institute of Leiden University, Leiden, The Netherlands. He is currently interested in mixing properties of Markov chains in dynamic discrete random environments, and applications of random graphs. He was previously interested in Monte Carlo simulations and mathematical aspects of statistical physics.
\end{IEEEbiography}

\begin{IEEEbiography}[{\includegraphics[width=1in,height=1.25in,clip,keepaspectratio]{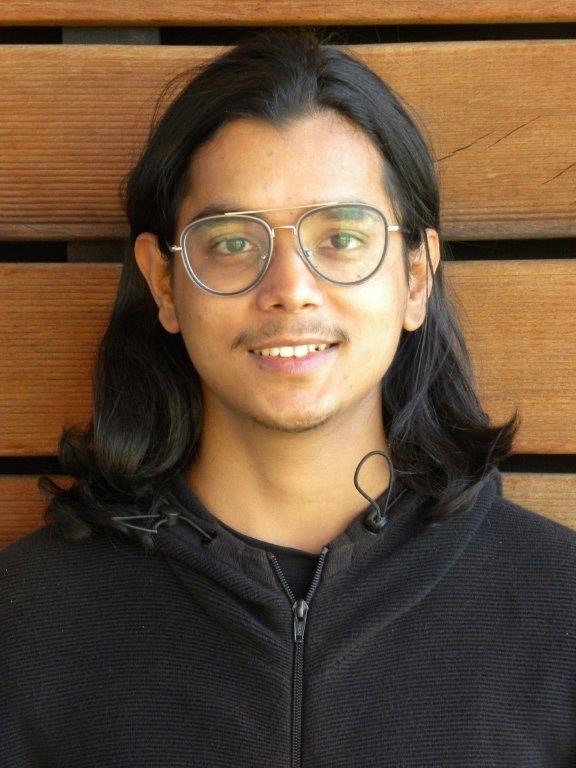}}]{Manish Pandey}{\space} was born in Almora, Uttarakhand, India in 1997. He received his bachelor's degree in mathematics with honours in 2018 from Indian Statistical Institute, Bangalore, Karnataka, India. In 2020, he received his master's degree in statistics with probability specialisation from Indian Statistical Institute, Kolkata, West Bengal, India.

He is currently employed as a doctoral candidate at the Department of Mathematics and Computer Science, Eindhoven University of Technology, Eindhoven, Netherlands. He is interested in centrality measures and connectivity properties in random graphs and other applications of stochastic processes. During his masters, he did a project on the topic `Length of stationary Gaussian excursions' which later published in the Proceedings of the American Mathematical Society. Joining as an INTERN in State Street Global Advisors, Bangalore, Karnataka, India in the group Active Quantitative Equity (AQE), May 2019- July 2019, he worked on portfolio optimization problems. He worked as an INTERN in Petrabytes Corp, Bangalore, Karnataka, India where he studied and implemented machine learning algorithms, Dec 2017- Jan 2018.
\end{IEEEbiography}

\begin{IEEEbiography}[{\includegraphics[width=1in,height=1.25in,clip,keepaspectratio]{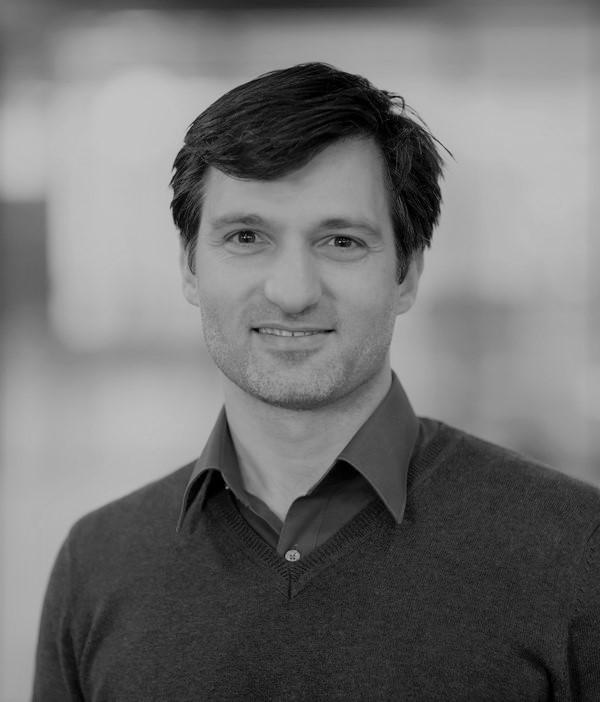}}]{Georgios Exarchakos}{\space} was born in Sparti, Greece, in 1982. He received the MSc degree in Department of Computing from the Imperial College London, United Kingdom, in 2005. He received his PhD degree from the Department of Computer Science in 2009 for his thesis "Hoverlay: a peer-to-peer system for on-demand sharing of capacity across network applications".

From 2009 until 2011, he worked as a Postdoctoral Researcher at the Eindhoven University of Technology on network quality of service and quality of experience. He is now an Assistant Professor on in-network intelligence for reliable network architectures, heading the Advanced Networking Lab of Electrical Engineering. Since 2019, he has been coordinating the research program on IoT Communications of the Center for Wireless Technology, TU/e. He has co-authored the book \textit{Networks for Pervasive Services: Six ways to upgrade the internet} (Springer, 2011) and edited 4 other books. His research interests are on resource allocation in wired/wireless networks, scheduling, in-network intelligence and network architectures.

Dr. Exarchakos has served as TPC member and reviewer for various international conferences and journals including IEEE Access, IEEE Sensors, IEEE IoT, IEEE WF-IoT and more.
\end{IEEEbiography}

\begin{IEEEbiography}[{\includegraphics[width=1in,height=1.25in,clip,keepaspectratio]{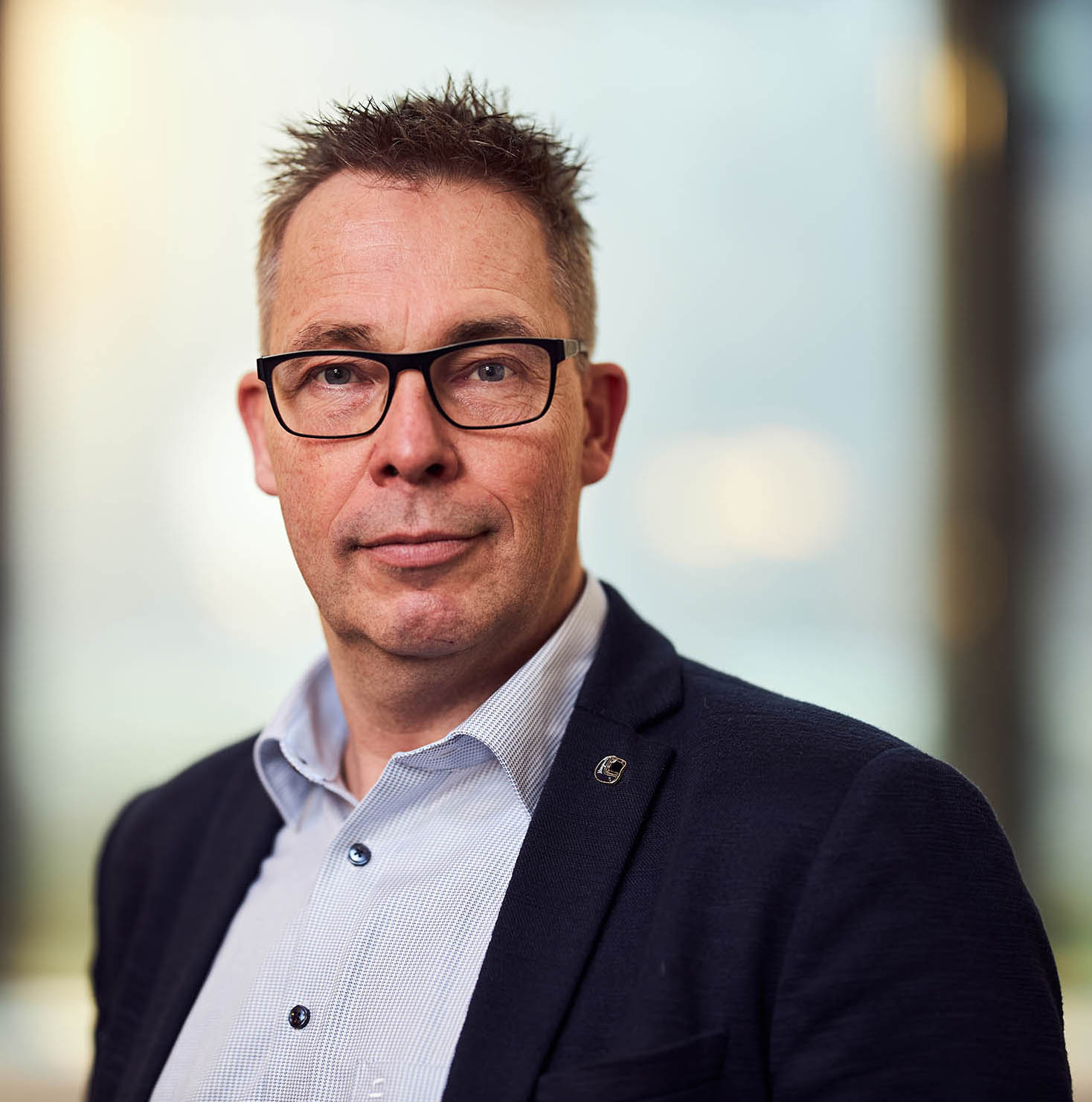}}]{Mark Bentum} (S'92, M'95, SM'09) was born in Smilde, The Netherlands, in 1967. He received the MSc degree in Electrical Engineering (with honours) from the University of Twente, Enschede, The Netherlands, in August 1991. In December 1995, he received the PhD degree for his thesis “Interactive Visualization of Volume Data” also from the University of Twente.

From December 1995 to June 1996, he was a research assistant at the University of Twente in the field of signal processing for mobile telecommunications and medical data processing. In June 1996, he joined the Netherlands Foundation for Research in Astronomy (ASTRON). From 2005 to 2008, he was responsible for the construction of the first software radio telescope in the world, LOFAR (Low Frequency Array). In 2008, he became an Associate Professor in the Telecommunication Engineering Group at the University of Twente. From December 2013 until September 2017, he was also the program director of Electrical Engineering at the University of Twente. In 2017, he became a Full Professor in Radio Science at Eindhoven University of Technology. Since 2023 he is the dean of the Electrical Engineering department of Eindhoven University of Technology. He is now involved with research and education in radio science. His current research interests are radio astronomy, short-range radio communications, novel receiver technologies, channel modelling, interference mitigation, sensor networks and aerospace.
\end{IEEEbiography}

\begin{IEEEbiography}[{\includegraphics[width=1in,height=1.35in,clip,keepaspectratio]{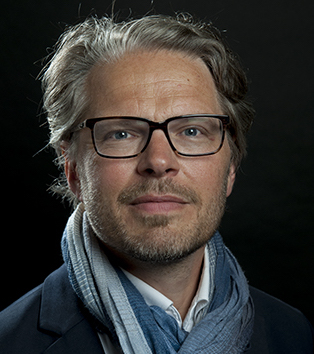}}]{Remco van der Hofstad}{\space} was born in Eindhoven, The Netherlands, in 1971. He received his M.Sc.\  (1993, with honours) and PhD degree (1997) at the Department of Mathematics of Utrecht University, the Netherlands.

He was a Post-Doc at McMaster University in Hamilton, Canada, and worked at Microsoft Research in Redmond, U.S.A. He became  Assistant Professor at Delft University of Technology, the Netherlands, in 1998 and Associate Professor at Eindhoven University of Technology (TU/e), the Netherlands, in 2002. Since 2005, he has been full professor in probability at TU/e. 
His research interests include random graphs and complex networks, stochastic processes and random walks, and interacting statistical mechanics.

Prof.\ van der Hofstad received the Prix Henri Poincaré 2003 (with Gordon Slade), the Rollo Davidson Prize 2007, and is a laureate of the \enquote{Innovative Research VIDI Scheme} 2003 and \enquote{Innovative Research VICI Scheme} 2008. He is one of the 11 co-applicants of the 2013 Gravitation program NETWORKS. In 2018, Prof.\ van der Hofstad was elected to the Royal Academy of Science and Arts (KNAW), where he currently is the chair of the Mathematics Section and a member of the Board for Natural and Technical Sciences.  
\end{IEEEbiography}
\end{document}


\title{Communication protocol for a\\
satellite-swarm interferometer:\\Supplementary material} 
\author{OLIVER NAGY, MANISH PANDEY, GEORGIOS EXARCHAKOS,\\ MARK BENTUM and REMCO van der HOFSTAD}
\markboth{NAGY ET AL. -- Supplementary material~\SMnumber}{}

\maketitle

This document contains a more detailed discussion of some of our results and additional plots omitted from the main paper.

References prefixed with "S\SMnumber" are pointing towards this document. Other references are pointing towards the contents of the main paper. 

\tableofcontents
\listoffigures

\section{Scatter plots of longest baselines}
In Section II.B we have shown the scatter plots of the observed longest baselines for $k=5$. Here we present scatter plots for all $k=4, 5, 6$.

\begin{figure}[h]
     \centering
     \begin{subfigure}[b]{0.32\textwidth}
         \centering
         \includegraphics[width=\textwidth]{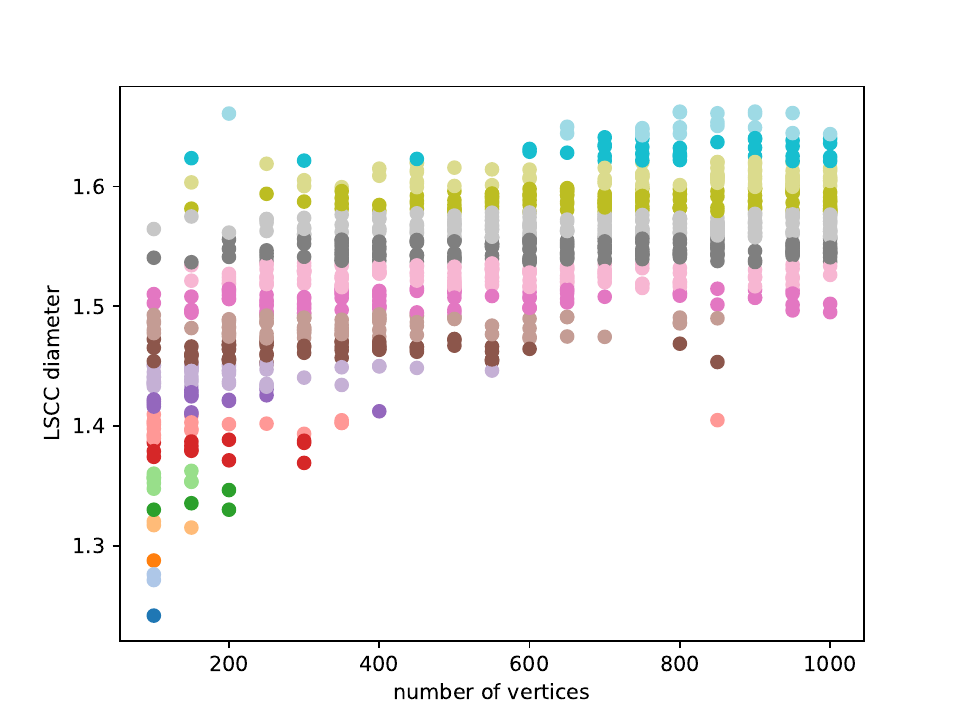}
         \subcaption{$k=4$}
         \label{fig:MB4}
     \end{subfigure}
     \hfill
     \begin{subfigure}[b]{0.32\textwidth}
         \centering
         \includegraphics[width=\textwidth]{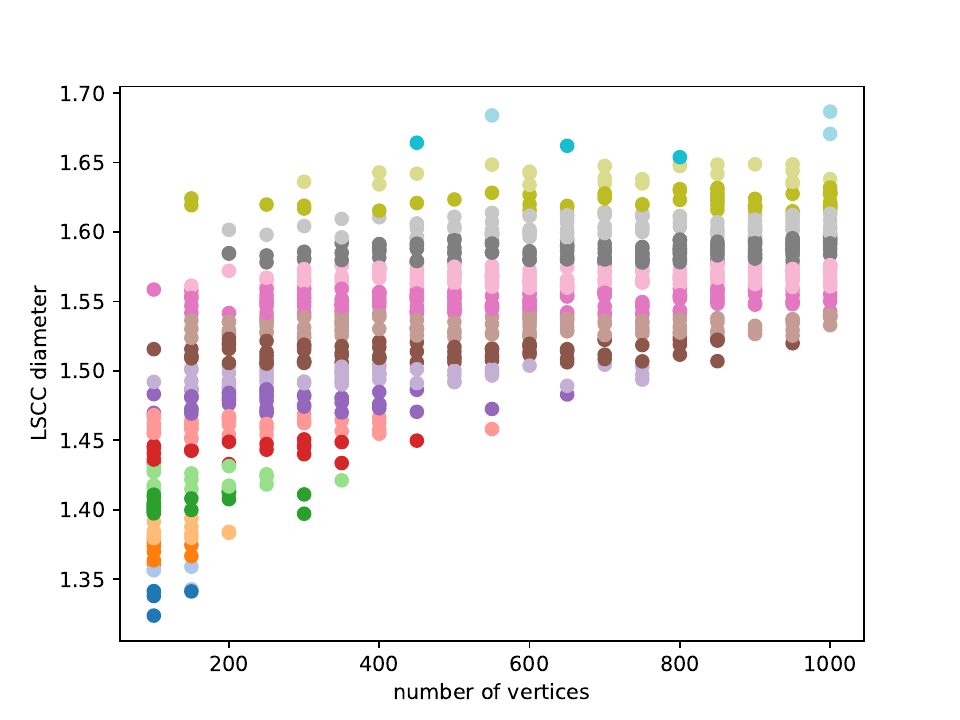}
         \subcaption{$k=5$}
         \label{fig:MB5}
     \end{subfigure}
     \hfill
     \begin{subfigure}[b]{0.32\textwidth}
         \centering
         \includegraphics[width=\textwidth]{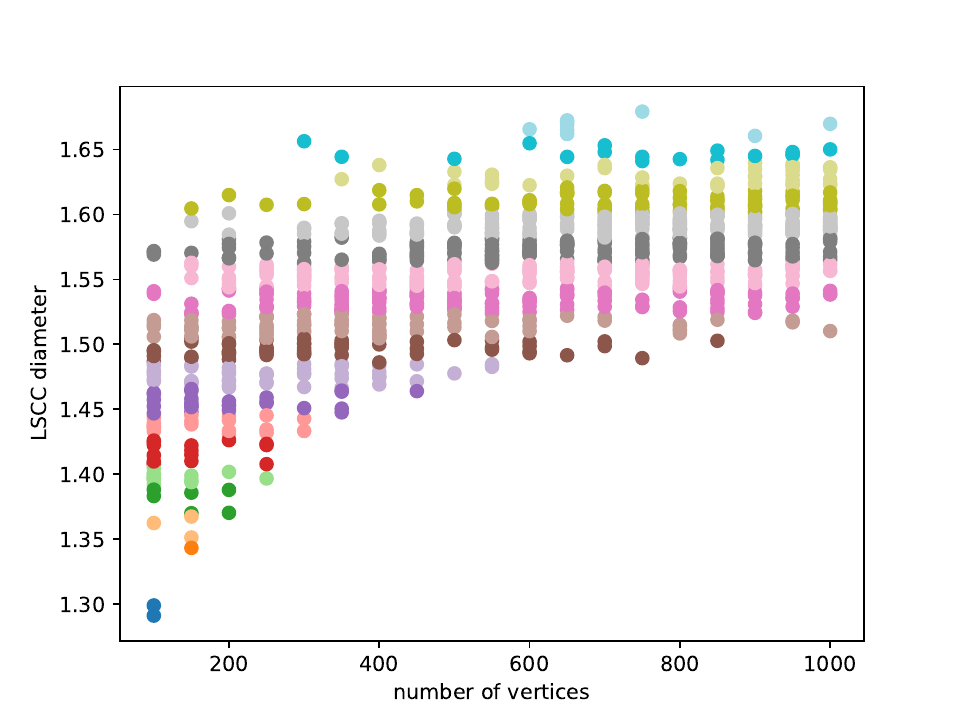}
         \subcaption{$k=6$}
         \label{fig:MB6}
     \end{subfigure}
        \caption{Scatter plot of lengths of longest baselines for $n=100-1000$, $k = \{4, 5, 6\}$.} 
        \label{fig:maxbaseline}
\end{figure}

\section{Histograms of empirical distributions of baseline lengths}

\begin{figure}[p]
     \centering
     \begin{subfigure}[b]{0.325\textwidth}
         \centering
         \includegraphics[width=\textwidth]{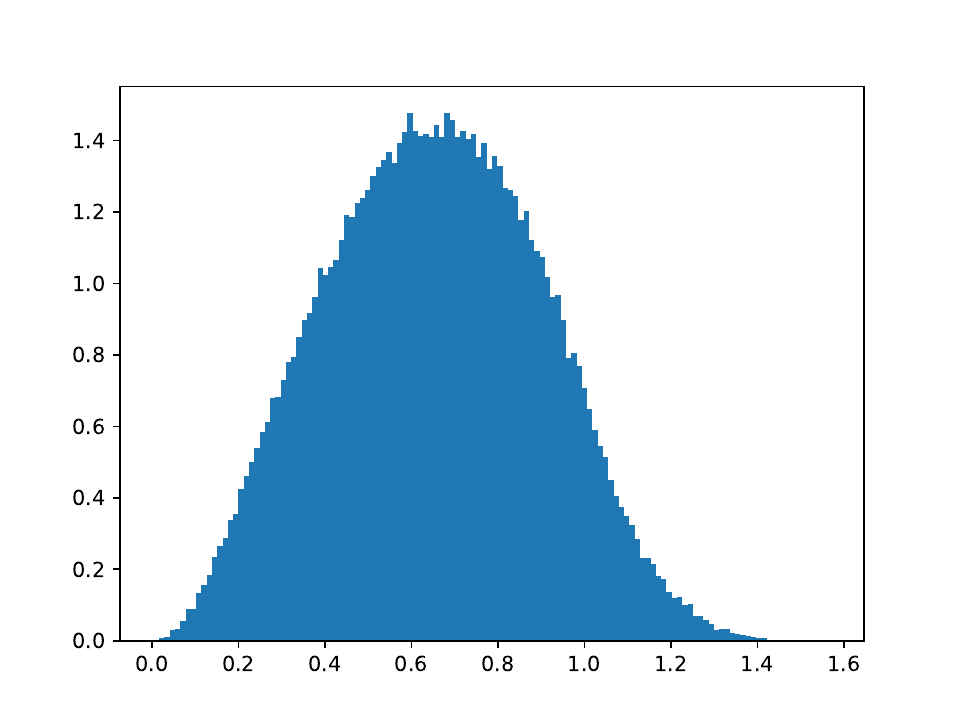}
         \caption{$n=50$}
         \label{fig:hist-EDB-50-4}
     \end{subfigure}
     \hfill
     \begin{subfigure}[b]{0.325\textwidth}
         \centering
         \includegraphics[width=\textwidth]{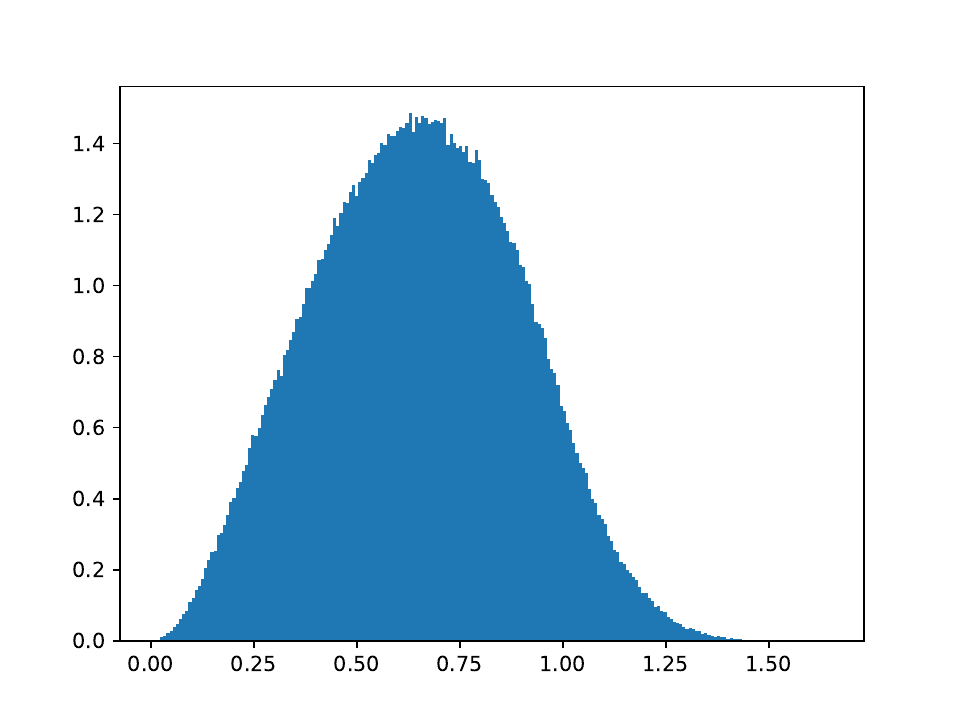}
         \caption{$n=100$}
         \label{fig:hist-EDB-100-4}
     \end{subfigure}
     \hfill
     \begin{subfigure}[b]{0.325\textwidth}
         \centering
         \includegraphics[width=\textwidth]{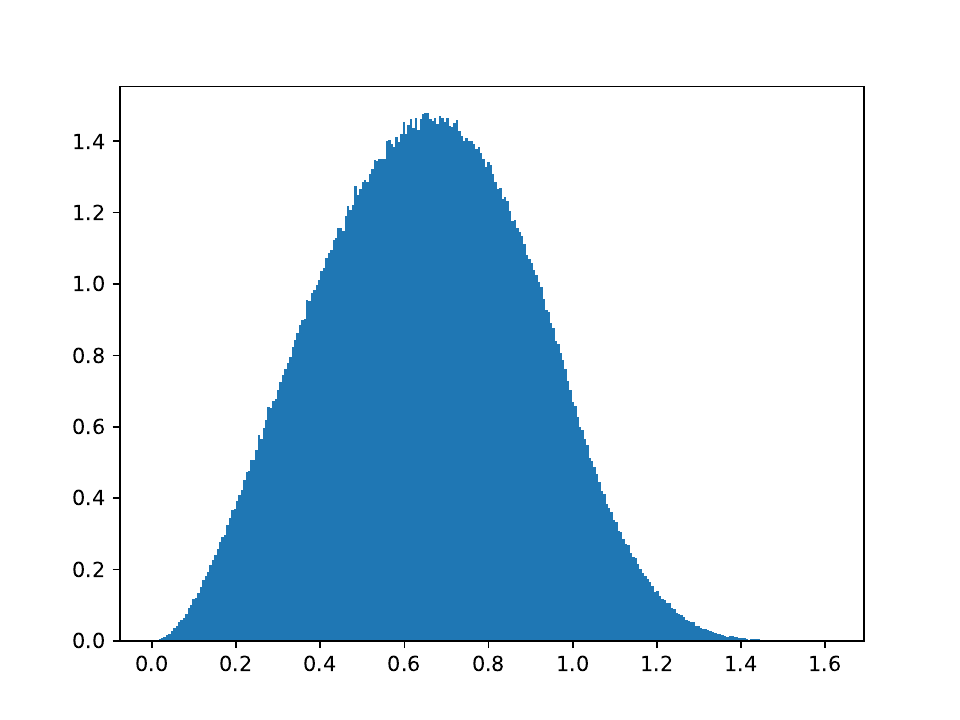}
         \caption{$n=150$}
         \label{fig:hist-EDB-150-4}
     \end{subfigure}
     \hfill
     \begin{subfigure}[b]{0.325\textwidth}
         \centering
         \includegraphics[width=\textwidth]{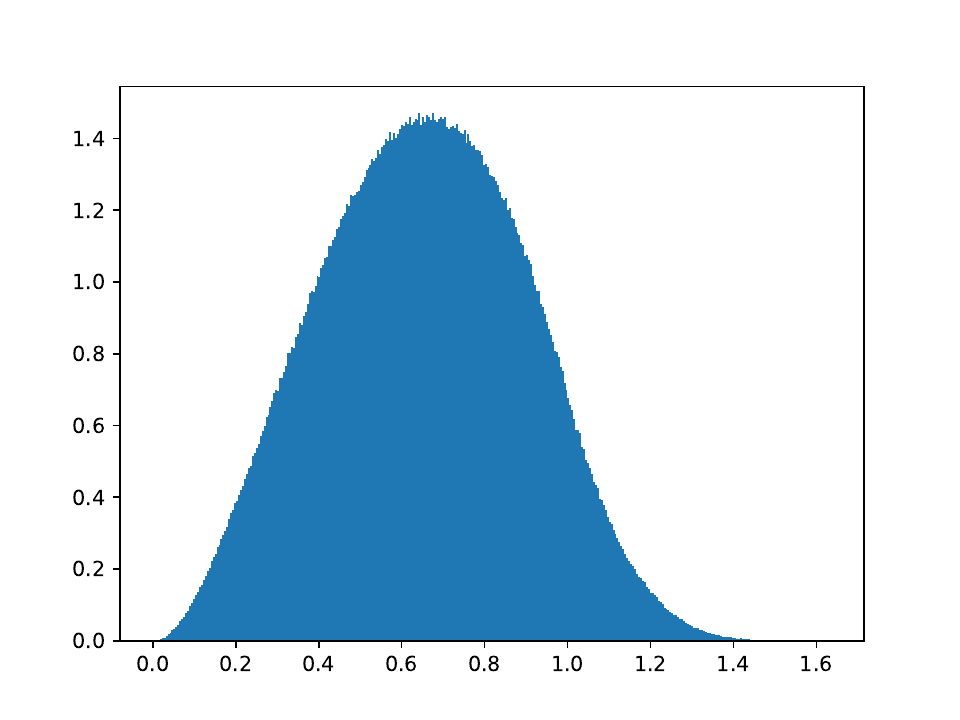}
         \caption{$n=200$}
         \label{fig:hist-EDB-200-4}
     \end{subfigure}
     \hfill
     \begin{subfigure}[b]{0.325\textwidth}
         \centering
         \includegraphics[width=\textwidth]{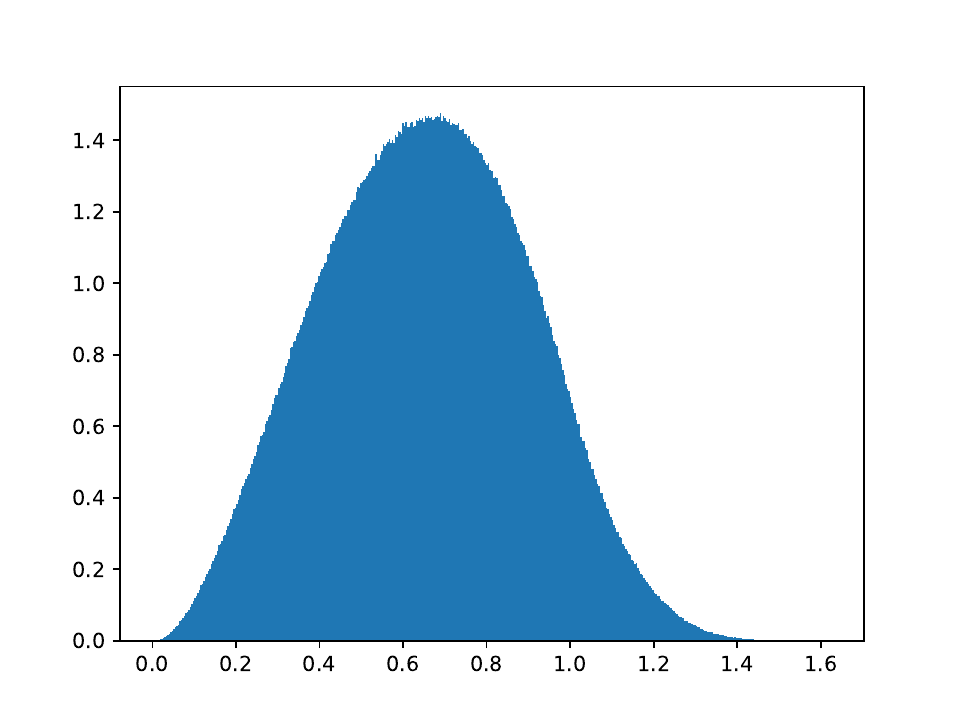}
         \caption{$n=300$}
         \label{fig:hist-EDB-300-4}
     \end{subfigure}
     \hfill
     \begin{subfigure}[b]{0.325\textwidth}
         \centering
         \includegraphics[width=\textwidth]{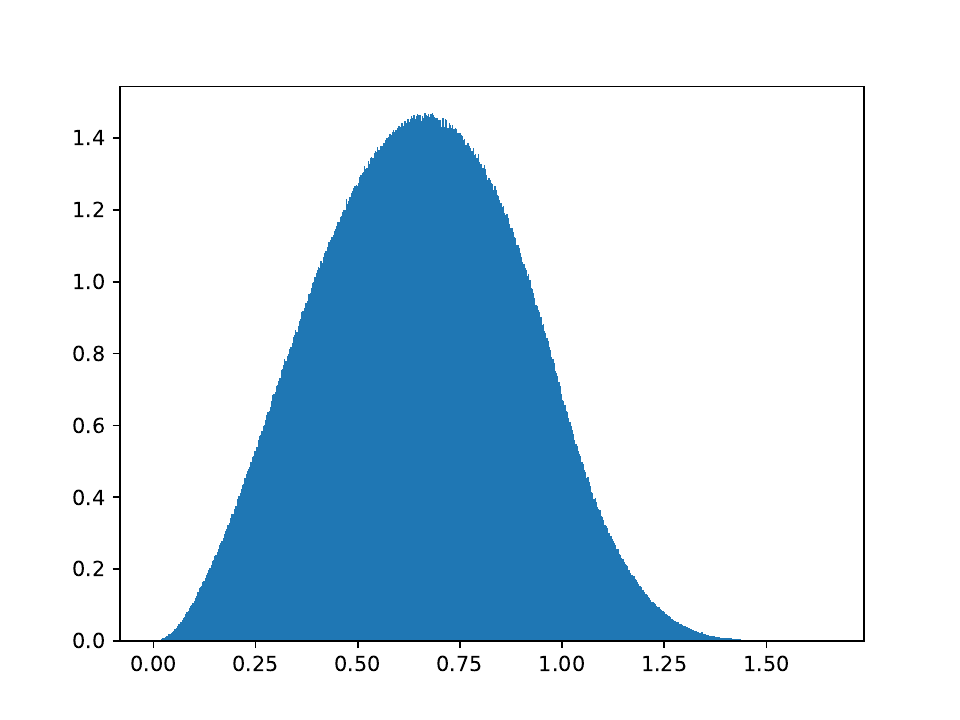}
         \caption{$n=400$}
         \label{fig:hist-EDB-400-4}
     \end{subfigure}
     \hfill
          \begin{subfigure}[b]{0.325\textwidth}
         \centering
         \includegraphics[width=\textwidth]{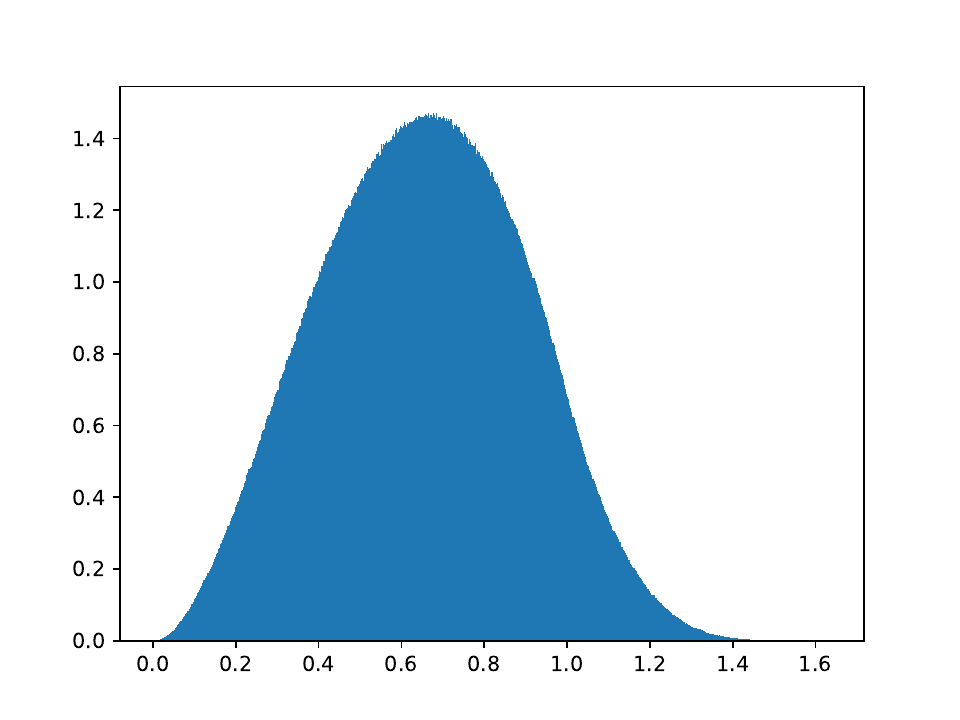}
         \caption{$n=500$}
         \label{fig:hist-EDB-500-4}
     \end{subfigure}
     \hfill
          \begin{subfigure}[b]{0.325\textwidth}
         \centering
         \includegraphics[width=\textwidth]{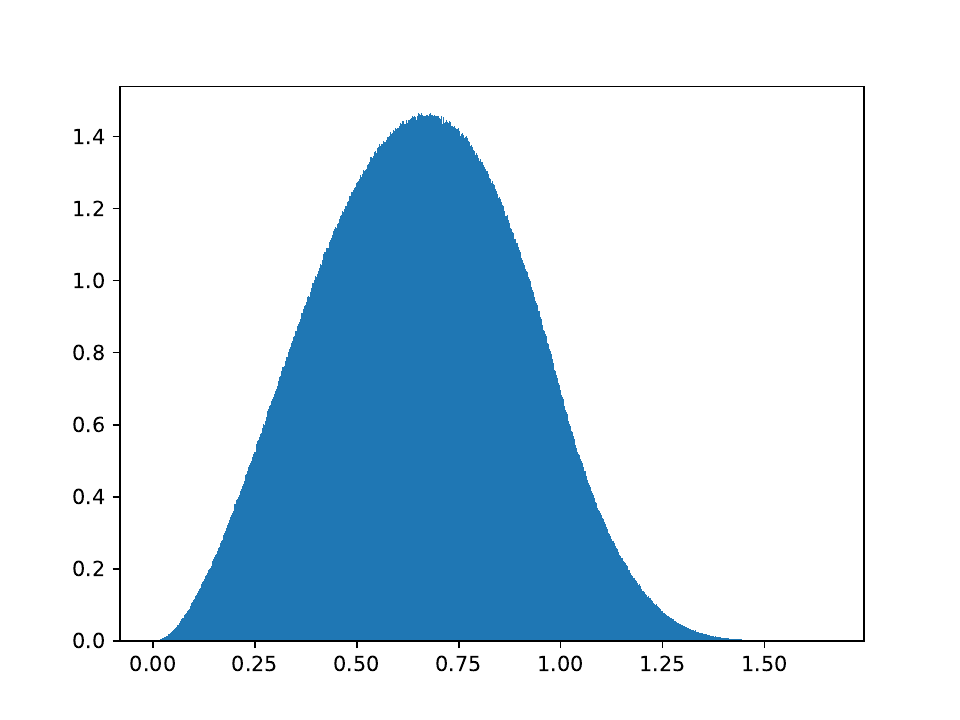}
         \caption{$n=600$}
         \label{fig:hist-EDB-600-4}
     \end{subfigure}
     \hfill
          \begin{subfigure}[b]{0.325\textwidth}
         \centering
         \includegraphics[width=\textwidth]{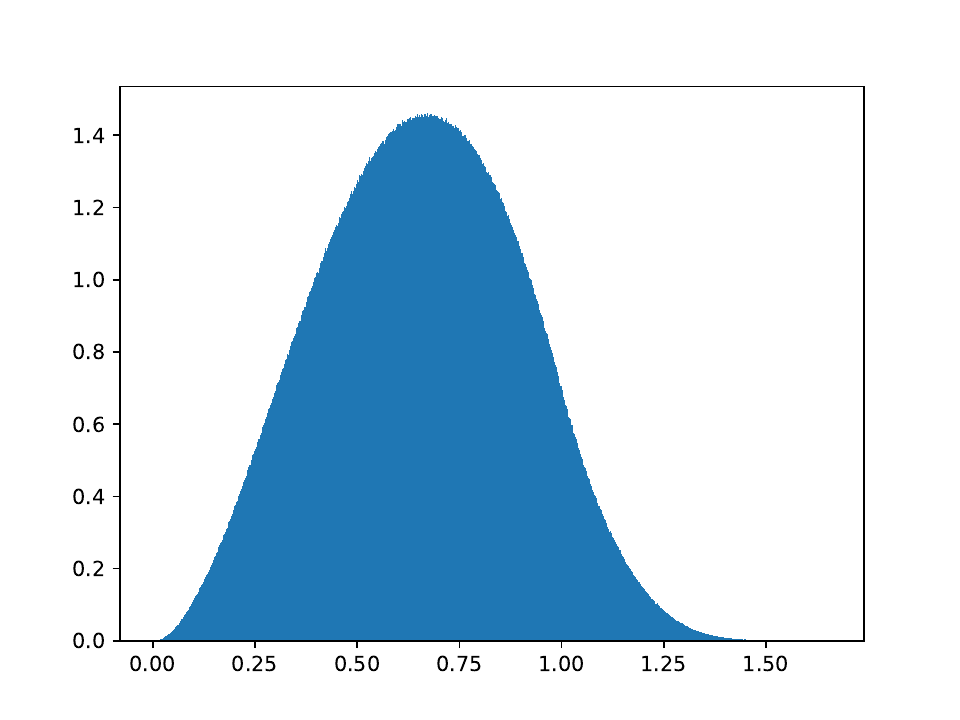}
         \caption{$n=700$}
         \label{fig:hist-EDB-700-4}
     \end{subfigure}
     \hfill
          \begin{subfigure}[b]{0.325\textwidth}
         \centering
         \includegraphics[width=\textwidth]{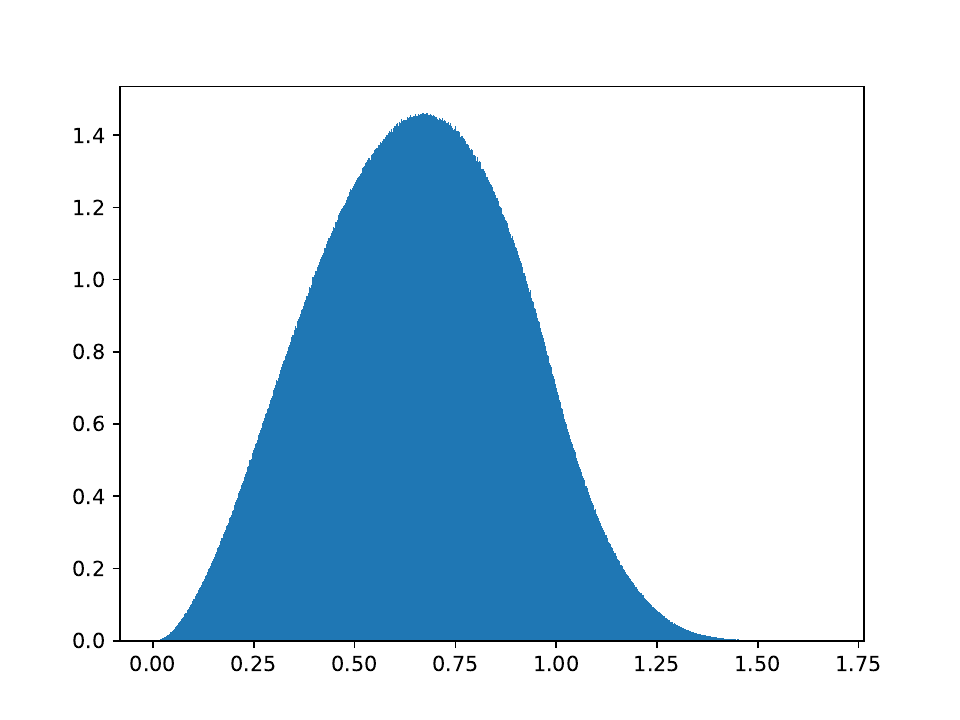}
         \caption{$n=800$}
         \label{fig:hist-EDB-800-4}
     \end{subfigure}
     \hfill
          \begin{subfigure}[b]{0.325\textwidth}
         \centering
         \includegraphics[width=\textwidth]{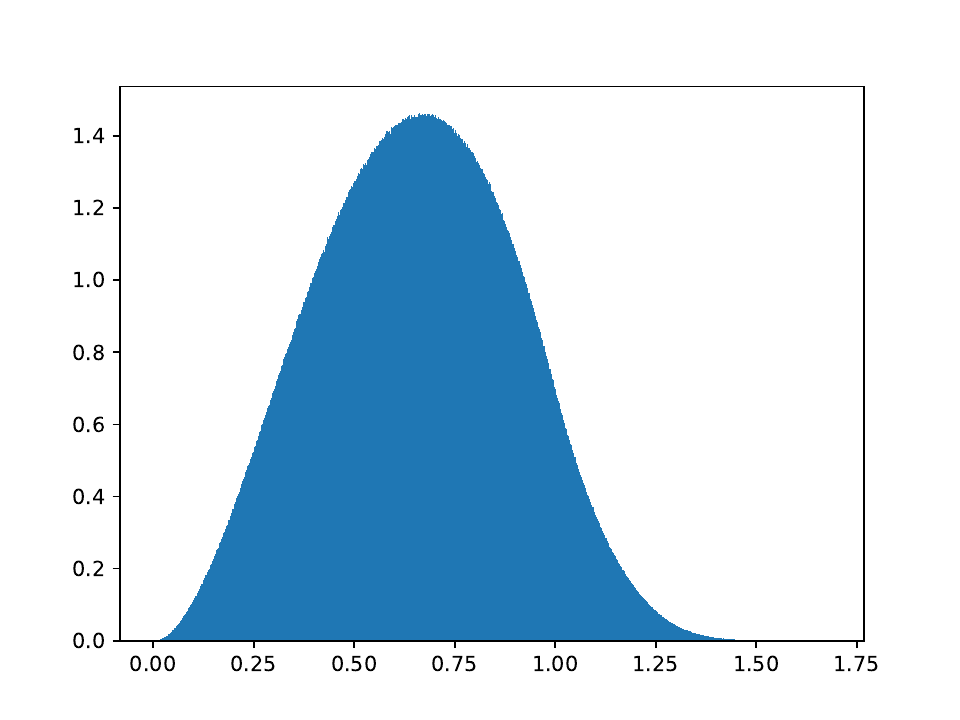}
         \caption{$n=900$}
         \label{fig:hist-EDB-900-4}
     \end{subfigure}
     \hfill
          \begin{subfigure}[b]{0.325\textwidth}
         \centering
         \includegraphics[width=\textwidth]{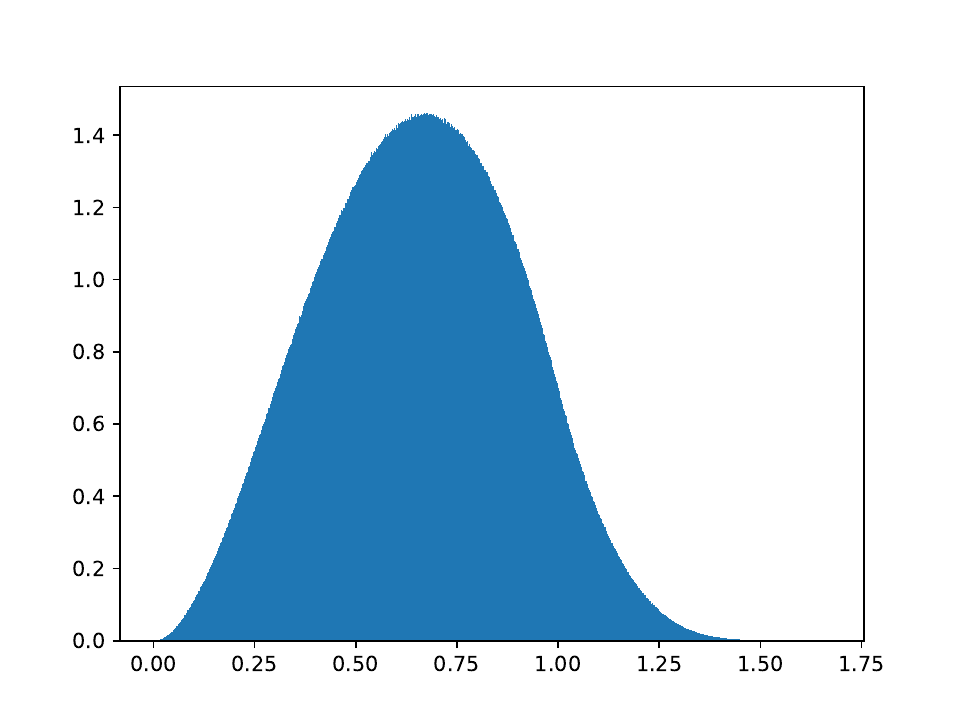}
         \caption{$n=1000$}
         \label{fig:hist-EDB-1000-4}
     \end{subfigure}
         \caption{Histograms of the empirical densities of baseline lengths for various $n$ and $k=4$.}
    \label{fig:ALL-EDBhist-4}
\end{figure}

\begin{figure}[p]
     \centering
     \begin{subfigure}[b]{0.325\textwidth}
         \centering
         \includegraphics[width=\textwidth]{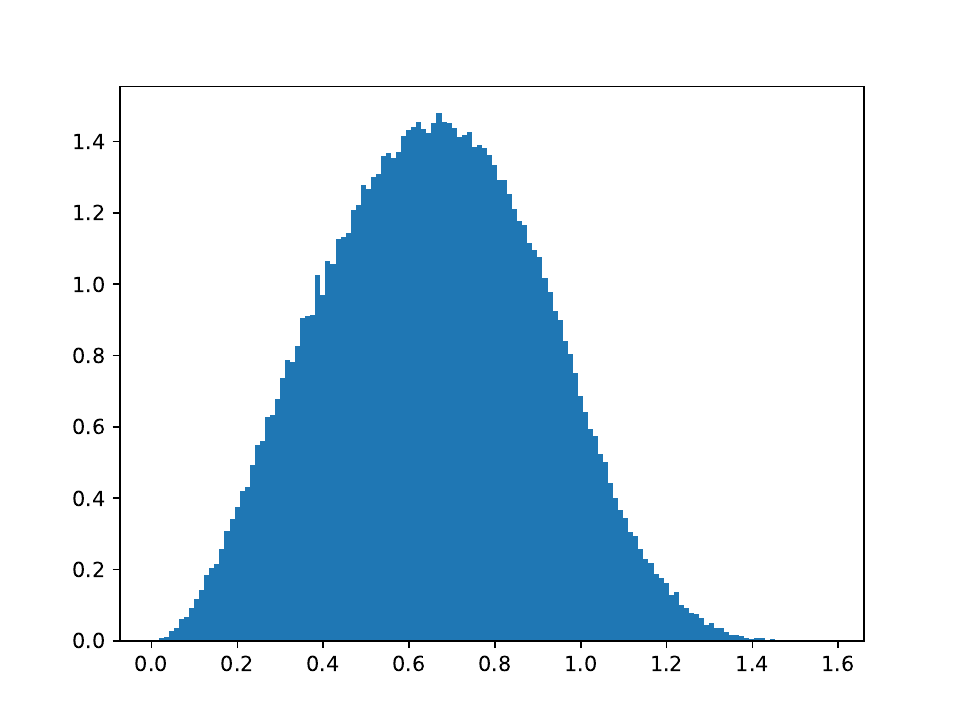}
         \caption{$n=50$}
         \label{fig:hist-EDB-50-5}
     \end{subfigure}
     \hfill
     \begin{subfigure}[b]{0.325\textwidth}
         \centering
         \includegraphics[width=\textwidth]{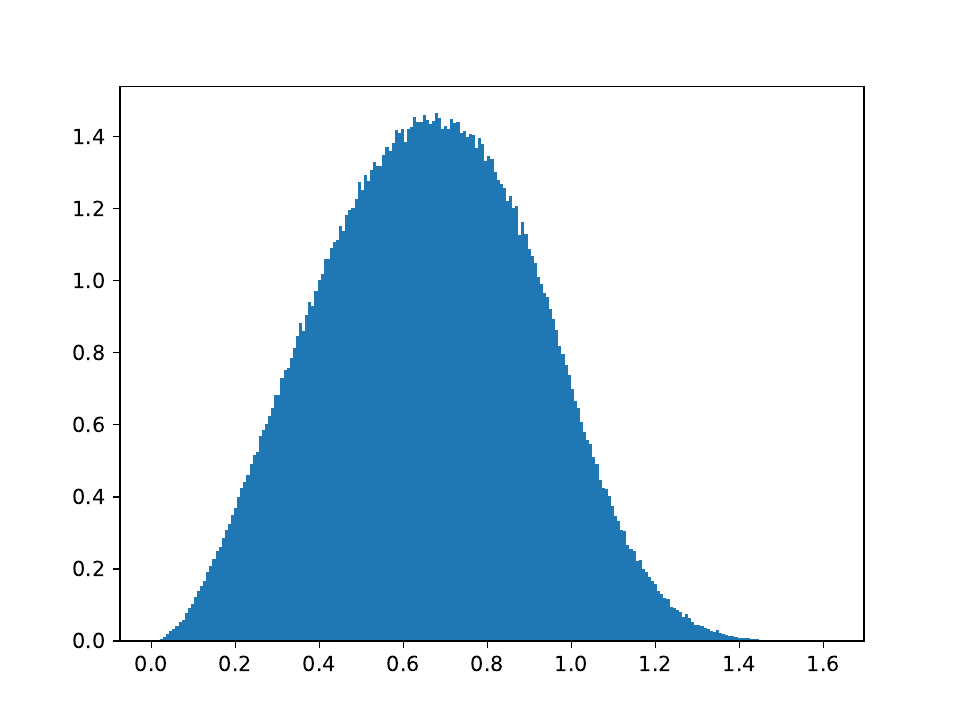}
         \caption{$n=100$}
         \label{fig:hist-EDB-100-5}
     \end{subfigure}
     \hfill
     \begin{subfigure}[b]{0.325\textwidth}
         \centering
         \includegraphics[width=\textwidth]{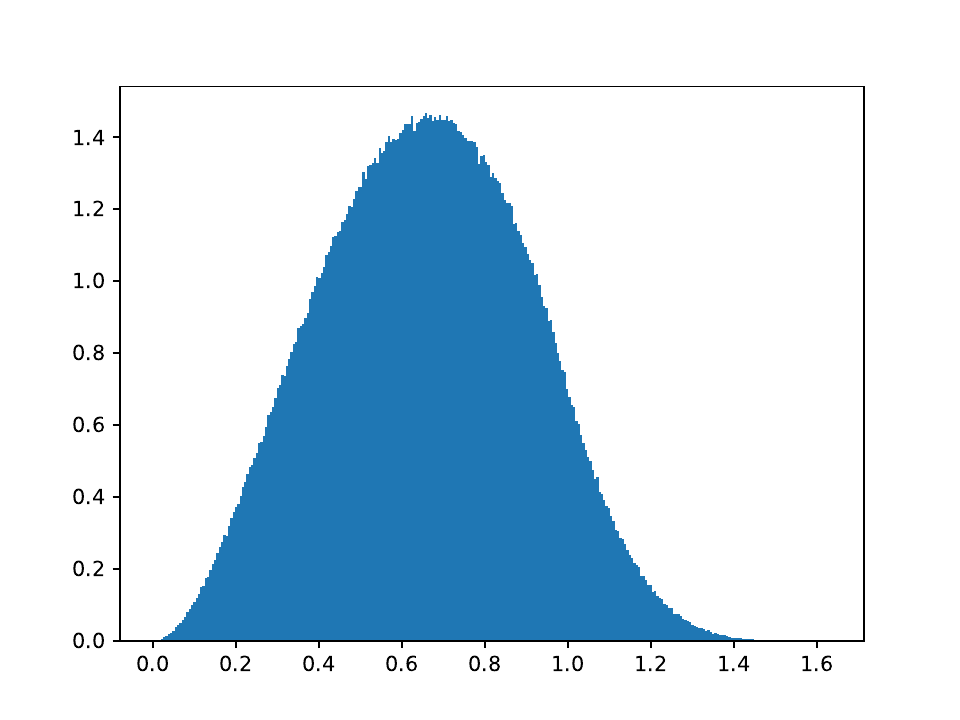}
         \caption{$n=150$}
         \label{fig:hist-EDB-150-5}
     \end{subfigure}
     \hfill
     \begin{subfigure}[b]{0.325\textwidth}
         \centering
         \includegraphics[width=\textwidth]{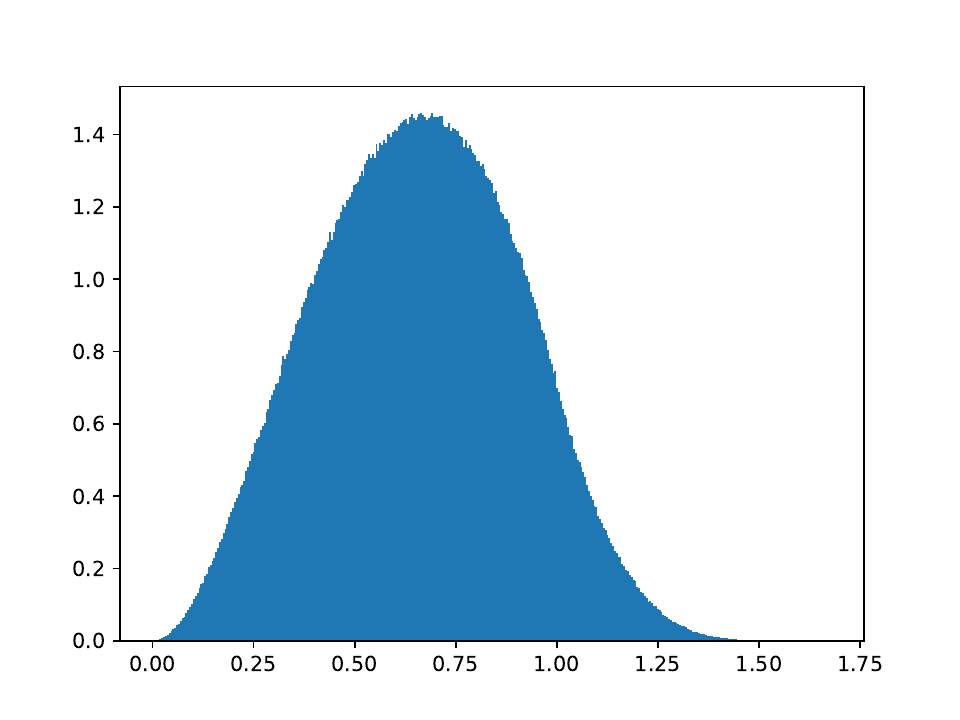}
         \caption{$n=200$}
         \label{fig:hist-EDB-200-5}
     \end{subfigure}
     \hfill
     \begin{subfigure}[b]{0.325\textwidth}
         \centering
         \includegraphics[width=\textwidth]{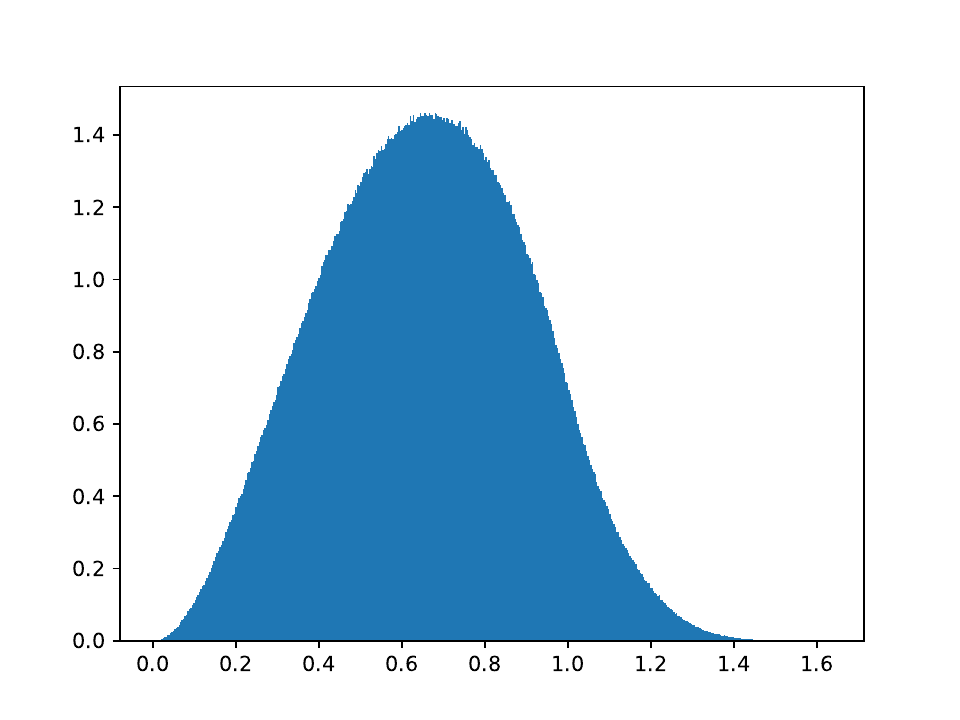}
         \caption{$n=300$}
         \label{fig:hist-EDB-300-5}
     \end{subfigure}
     \hfill
     \begin{subfigure}[b]{0.325\textwidth}
         \centering
         \includegraphics[width=\textwidth]{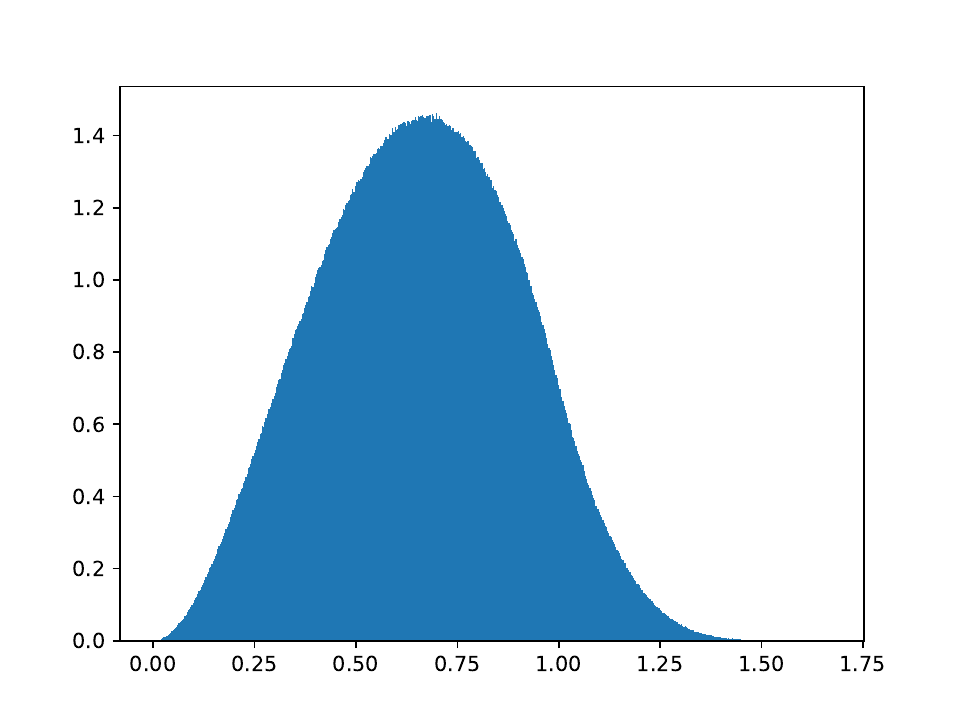}
         \caption{$n=400$}
         \label{fig:hist-EDB-400-5}
     \end{subfigure}
     \hfill
          \begin{subfigure}[b]{0.325\textwidth}
         \centering
         \includegraphics[width=\textwidth]{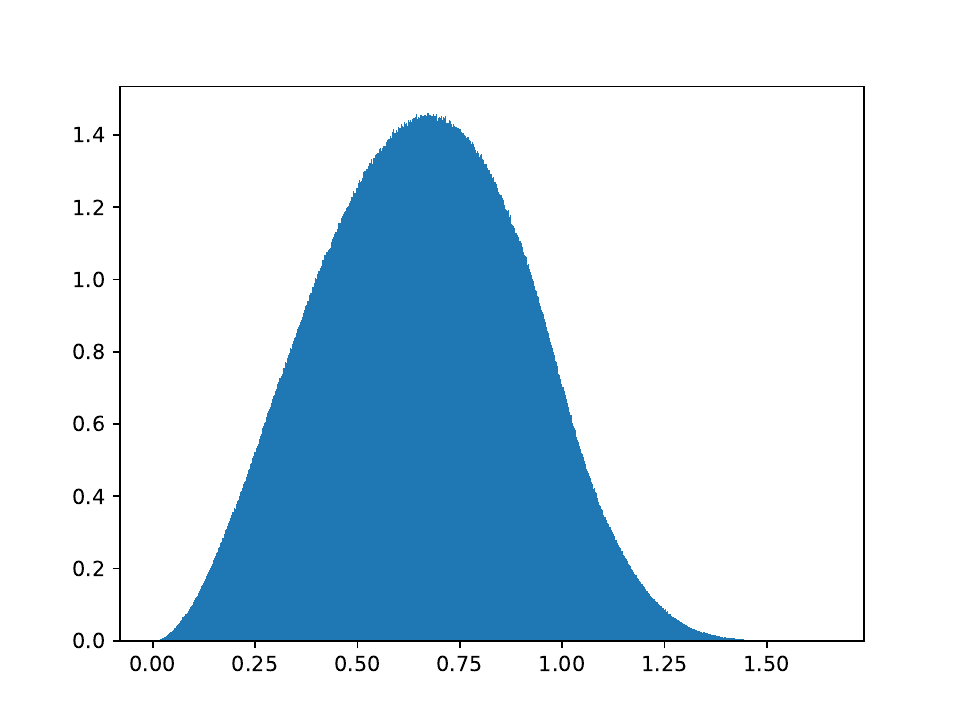}
         \caption{$n=500$}
         \label{fig:hist-EDB-500-5}
     \end{subfigure}
     \hfill
          \begin{subfigure}[b]{0.325\textwidth}
         \centering
         \includegraphics[width=\textwidth]{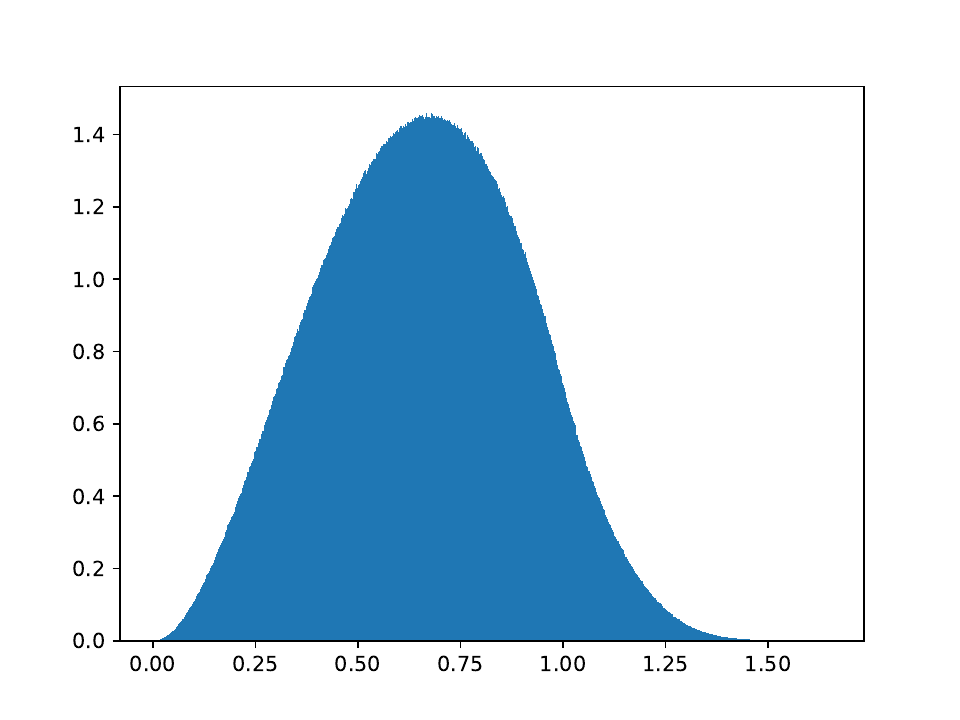}
         \caption{$n=600$}
         \label{fig:hist-EDB-600-5}
     \end{subfigure}
     \hfill
          \begin{subfigure}[b]{0.325\textwidth}
         \centering
         \includegraphics[width=\textwidth]{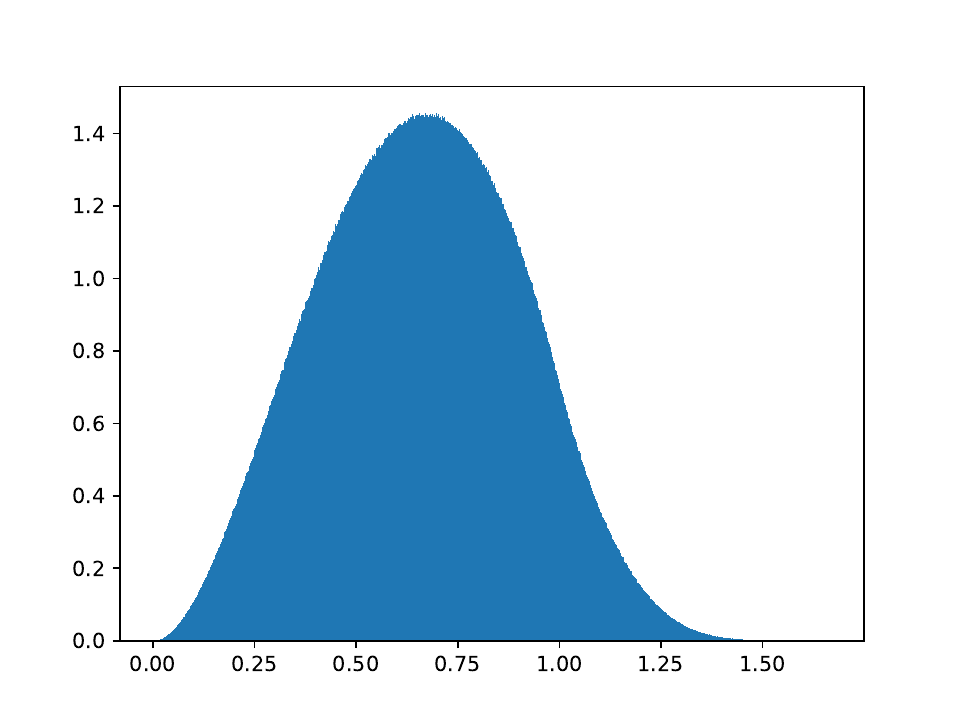}
         \caption{$n=700$}
         \label{fig:hist-EDB-700-5}
     \end{subfigure}
     \hfill
          \begin{subfigure}[b]{0.325\textwidth}
         \centering
         \includegraphics[width=\textwidth]{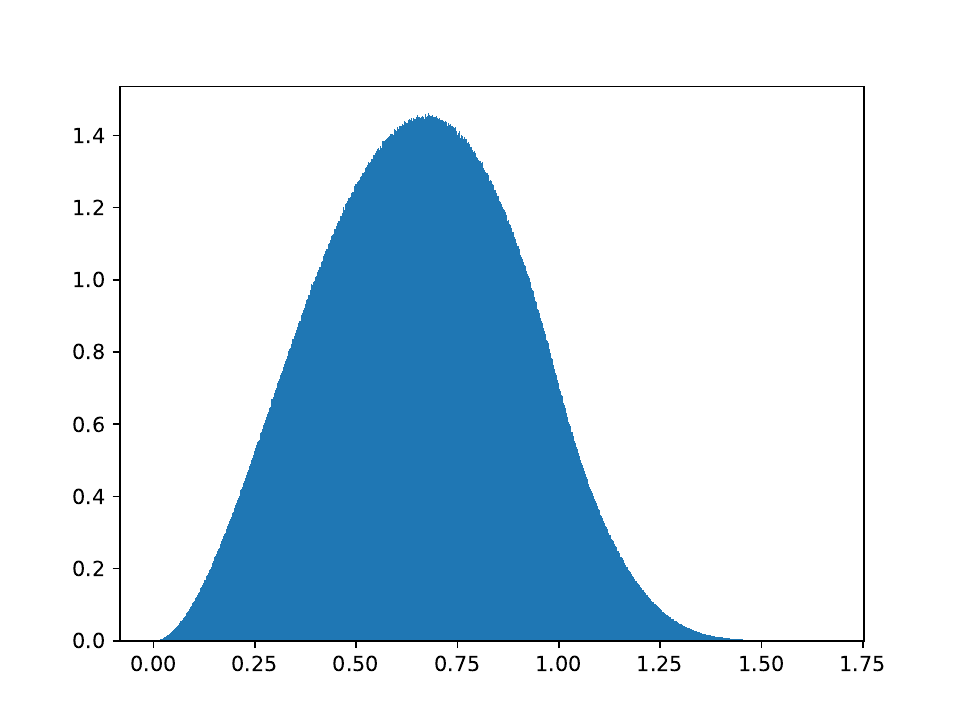}
         \caption{$n=800$}
         \label{fig:hist-EDB-800-5}
     \end{subfigure}
     \hfill
          \begin{subfigure}[b]{0.325\textwidth}
         \centering
         \includegraphics[width=\textwidth]{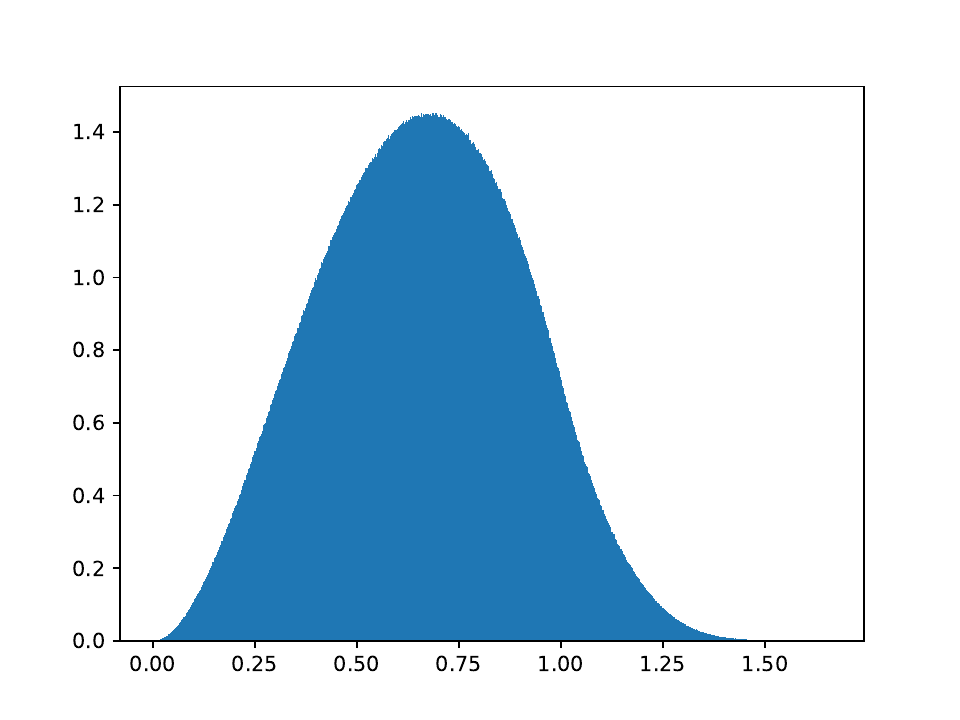}
         \caption{$n=900$}
         \label{fig:hist-EDB-900-5}
     \end{subfigure}
     \hfill
          \begin{subfigure}[b]{0.325\textwidth}
         \centering
         \includegraphics[width=\textwidth]{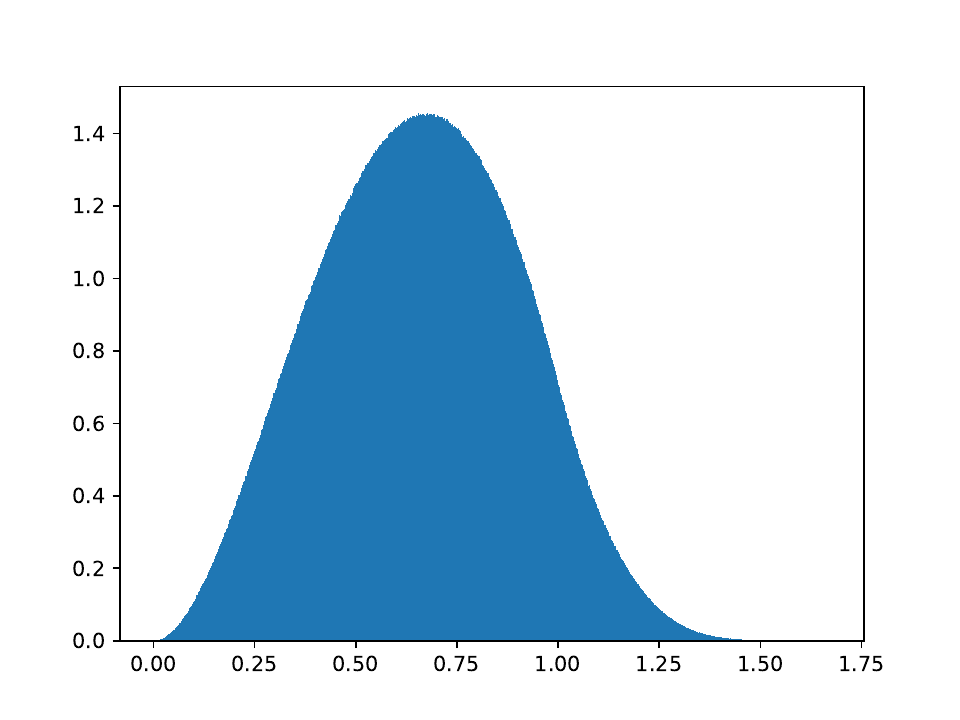}
         \caption{$n=1000$}
         \label{fig:hist-EDB-1000-5}
     \end{subfigure}
         \caption{Histograms of the empirical densities of baseline lengths for various $n$ and $k=5$.}
    \label{fig:ALL-EDBhist-5}
\end{figure}

\begin{figure}[p]
     \centering
     \begin{subfigure}[b]{0.325\textwidth}
         \centering
         \includegraphics[width=\textwidth]{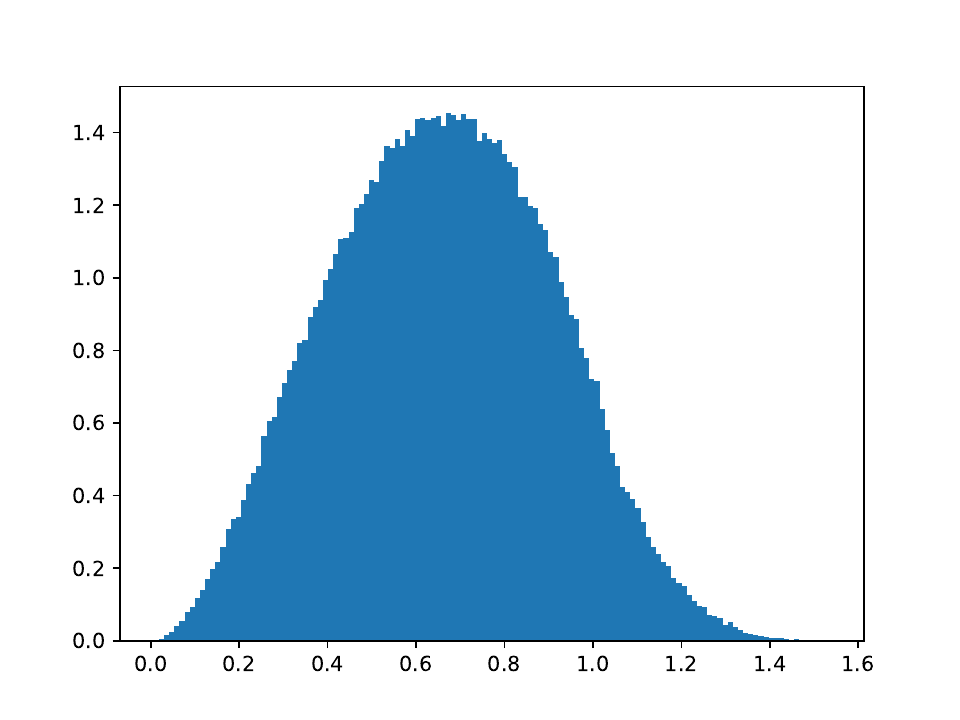}
         \caption{$n=50$}
         \label{fig:hist-EDB-50-6}
     \end{subfigure}
     \hfill
     \begin{subfigure}[b]{0.325\textwidth}
         \centering
         \includegraphics[width=\textwidth]{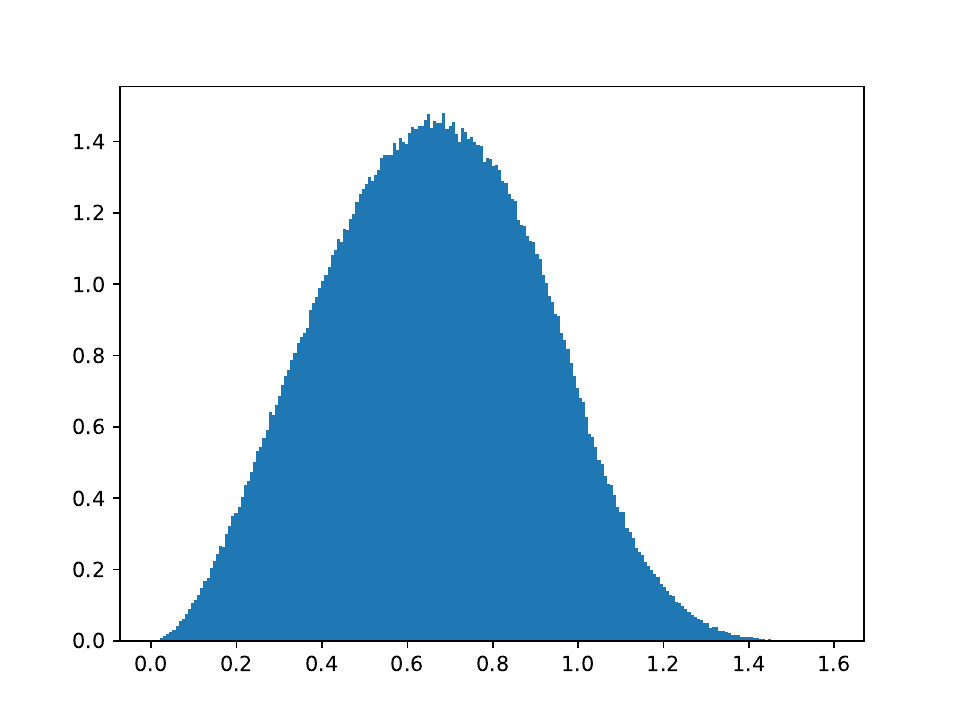}
         \caption{$n=100$}
         \label{fig:hist-EDB-100-6}
     \end{subfigure}
     \hfill
     \begin{subfigure}[b]{0.325\textwidth}
         \centering
         \includegraphics[width=\textwidth]{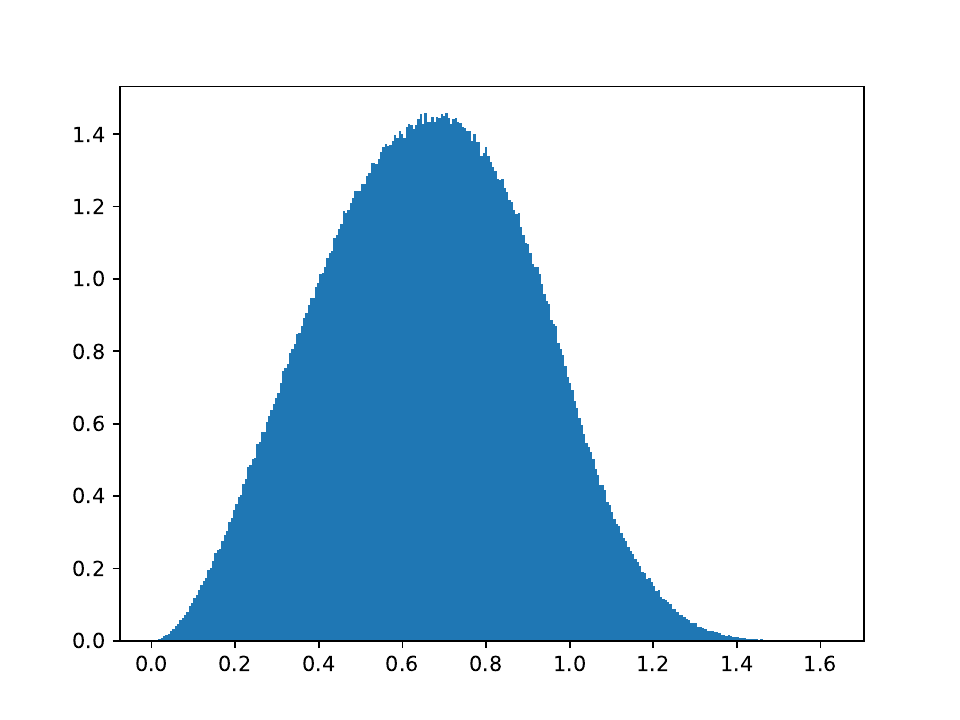}
         \caption{$n=150$}
         \label{fig:hist-EDB-150-6}
     \end{subfigure}
     \hfill
     \begin{subfigure}[b]{0.325\textwidth}
         \centering
         \includegraphics[width=\textwidth]{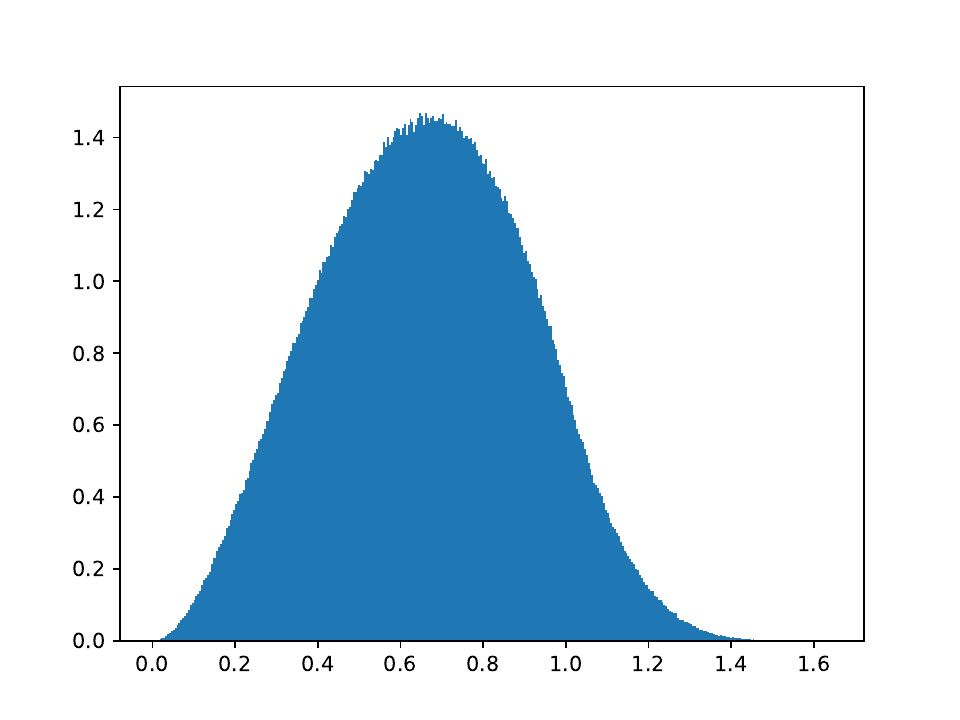}
         \caption{$n=200$}
         \label{fig:hist-EDB-200-6}
     \end{subfigure}
     \hfill
     \begin{subfigure}[b]{0.325\textwidth}
         \centering
         \includegraphics[width=\textwidth]{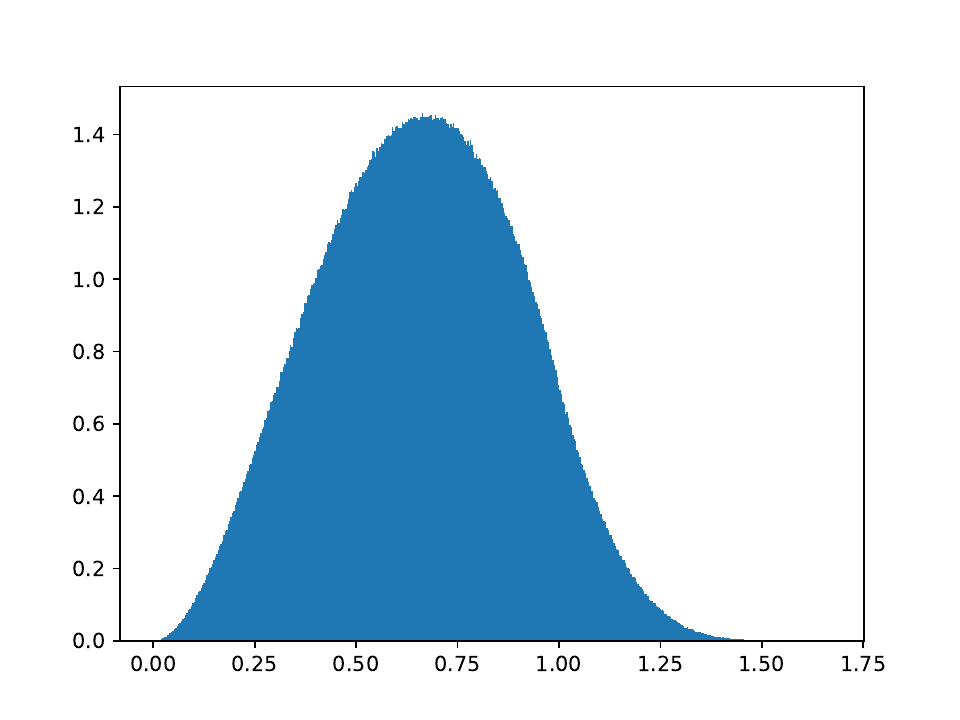}
         \caption{$n=300$}
         \label{fig:hist-EDB-300-6}
     \end{subfigure}
     \hfill
     \begin{subfigure}[b]{0.325\textwidth}
         \centering
         \includegraphics[width=\textwidth]{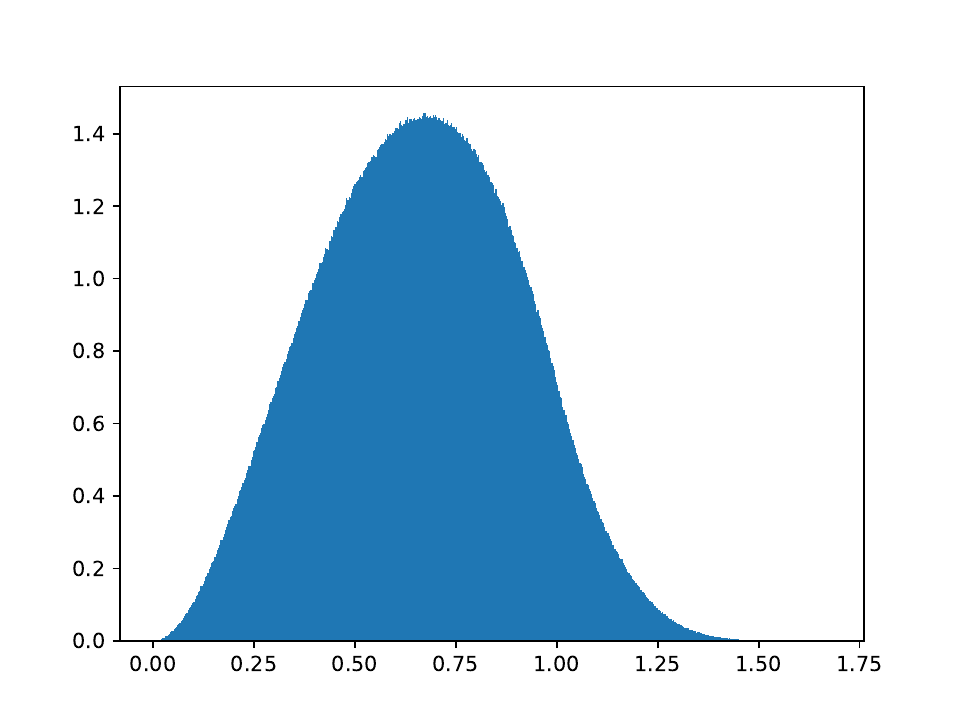}
         \caption{$n=400$}
         \label{fig:hist-EDB-400-6}
     \end{subfigure}
     \hfill
          \begin{subfigure}[b]{0.325\textwidth}
         \centering
         \includegraphics[width=\textwidth]{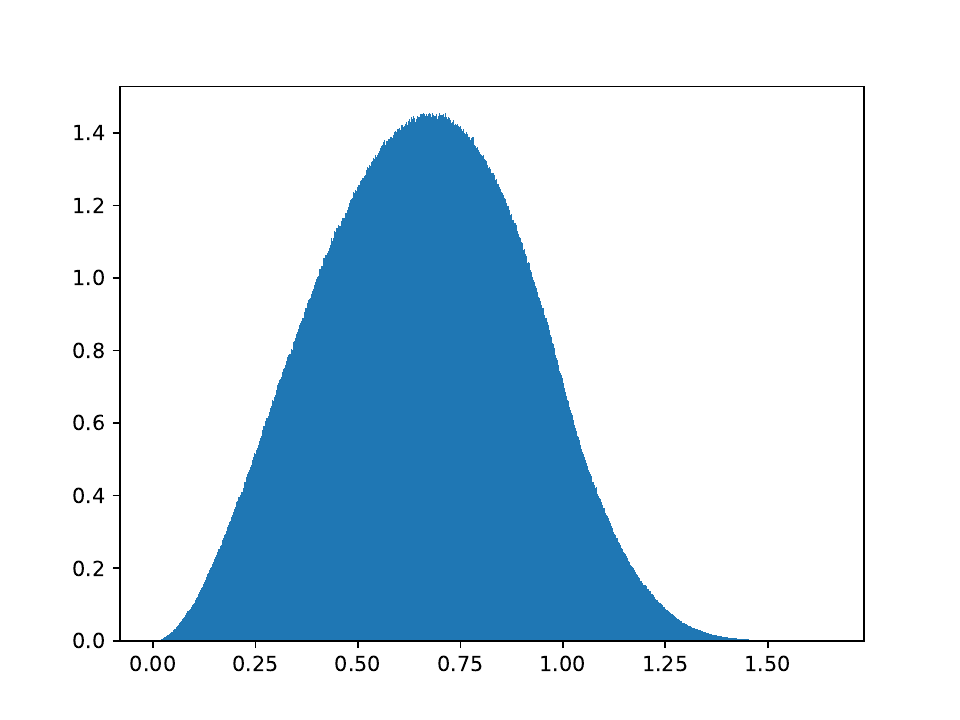}
         \caption{$n=500$}
         \label{fig:hist-EDB-500-6}
     \end{subfigure}
     \hfill
          \begin{subfigure}[b]{0.325\textwidth}
         \centering
         \includegraphics[width=\textwidth]{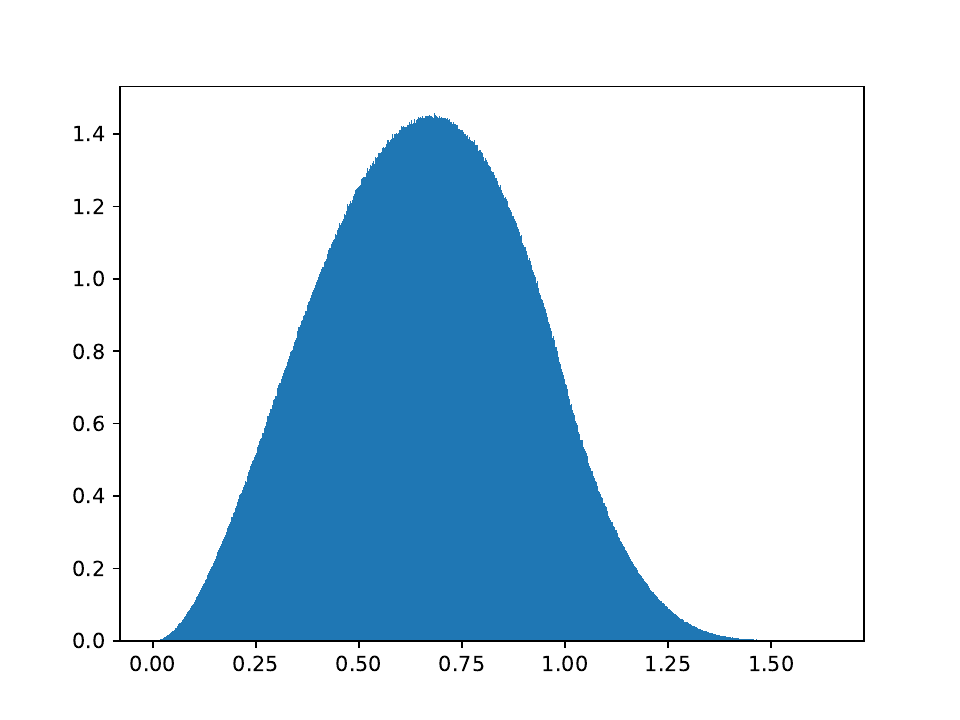}
         \caption{$n=600$}
         \label{fig:hist-EDB-600-6}
     \end{subfigure}
     \hfill
          \begin{subfigure}[b]{0.325\textwidth}
         \centering
         \includegraphics[width=\textwidth]{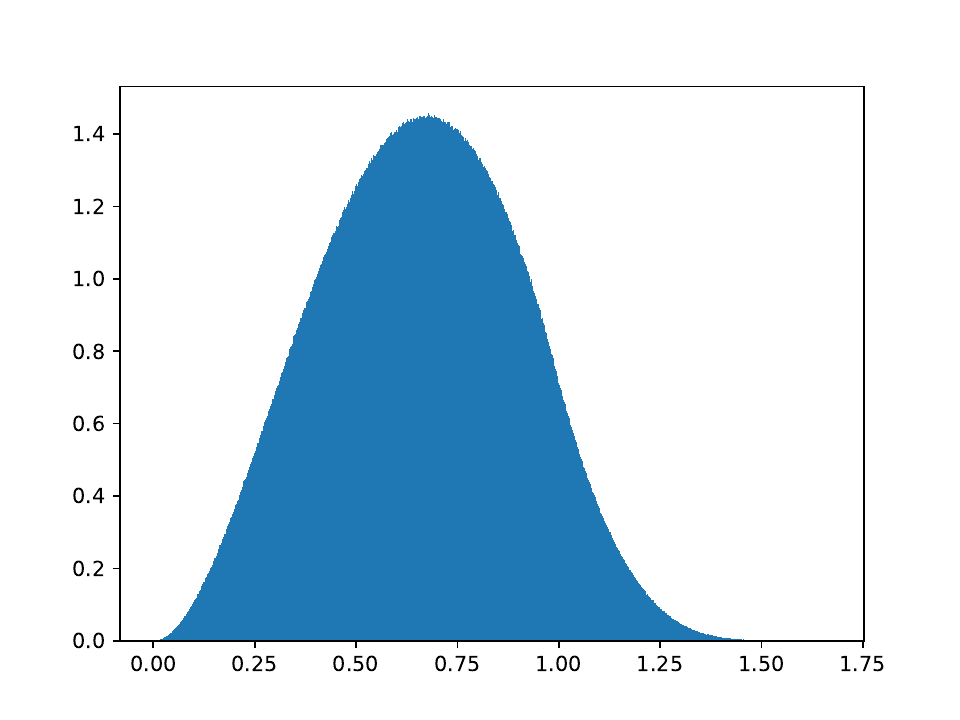}
         \caption{$n=700$}
         \label{fig:hist-EDB-700-6}
     \end{subfigure}
     \hfill
          \begin{subfigure}[b]{0.325\textwidth}
         \centering
         \includegraphics[width=\textwidth]{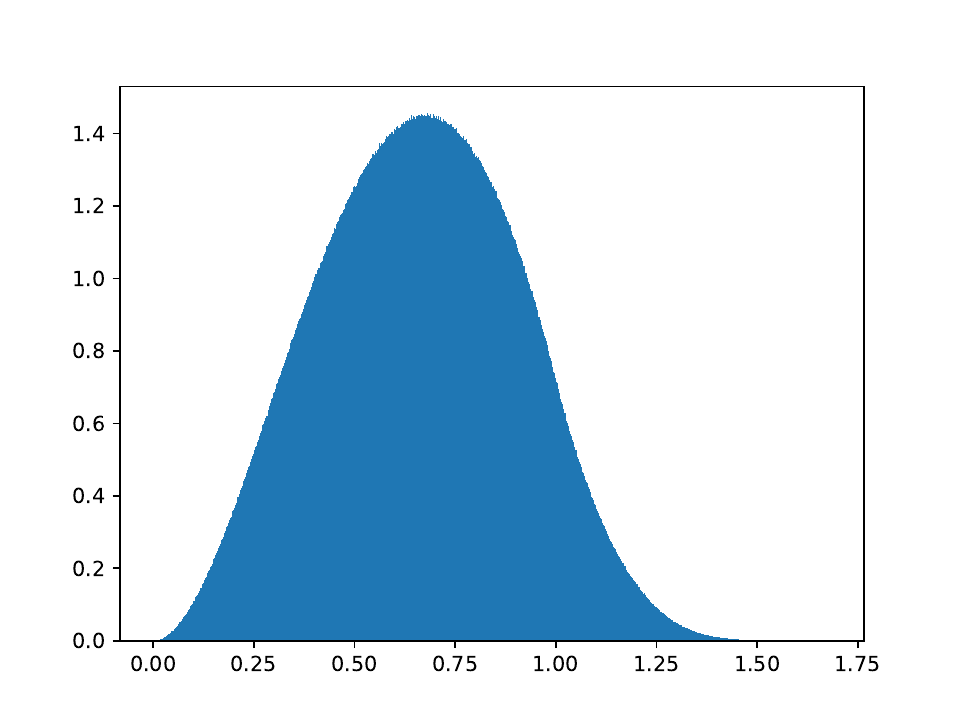}
         \caption{$n=800$}
         \label{fig:hist-EDB-800-6}
     \end{subfigure}
     \hfill
          \begin{subfigure}[b]{0.325\textwidth}
         \centering
         \includegraphics[width=\textwidth]{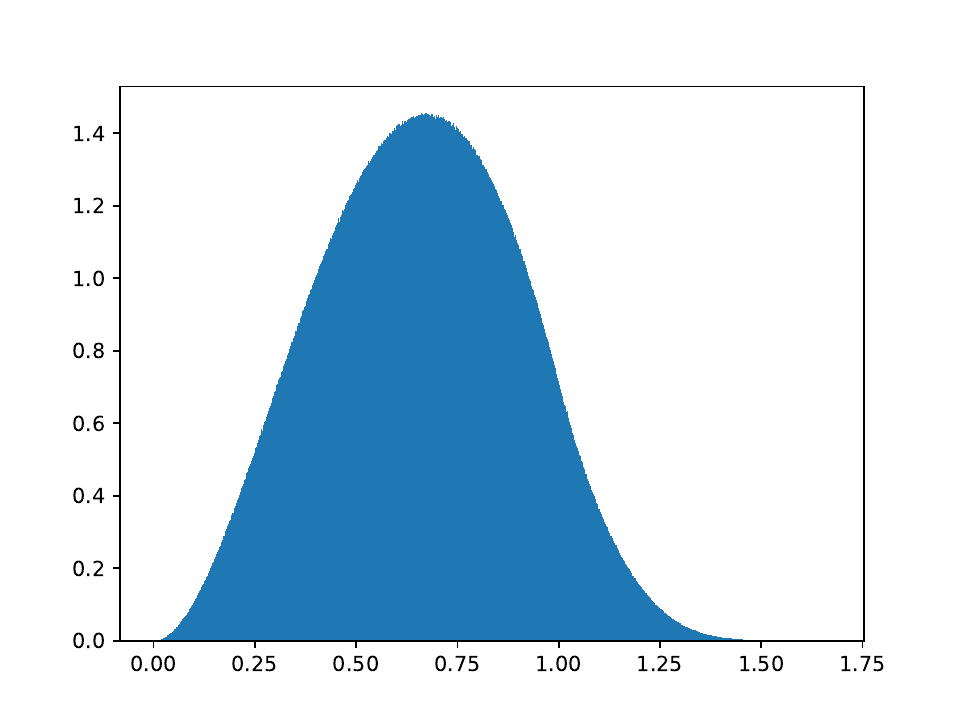}
         \caption{$n=900$}
         \label{fig:hist-EDB-900-6}
     \end{subfigure}
     \hfill
          \begin{subfigure}[b]{0.325\textwidth}
         \centering
         \includegraphics[width=\textwidth]{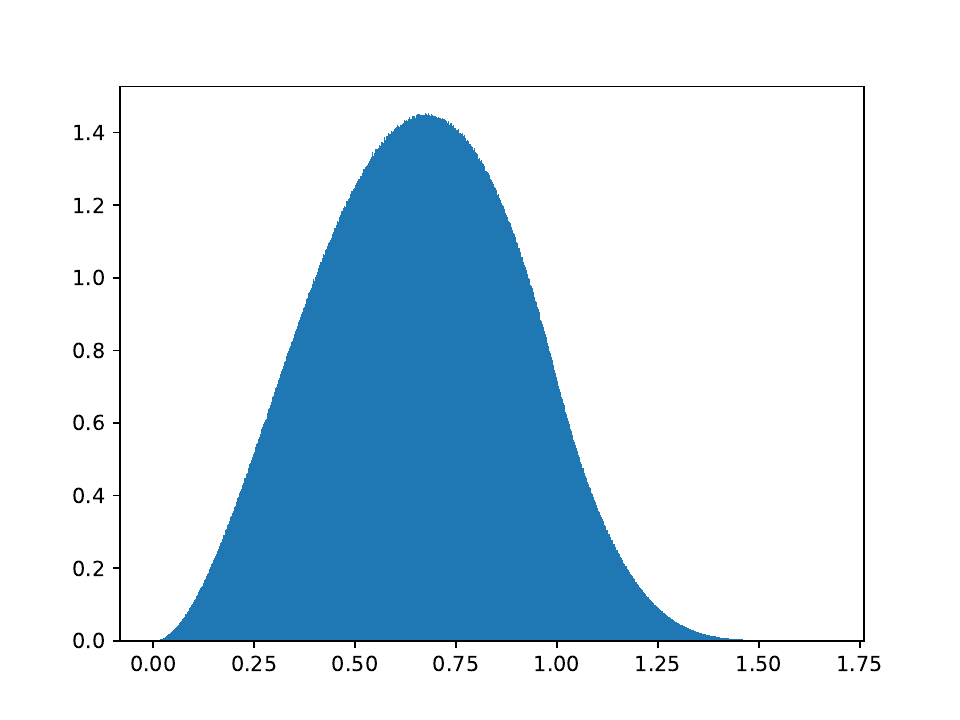}
         \caption{$n=1000$}
         \label{fig:hist-EDB-1000-6}
     \end{subfigure}
         \caption{Histograms of the empirical densities of baseline lengths for various $n$ and $k=6$.}
    \label{fig:ALL-EDBhist-6}
\end{figure}

In Figures~\ref{fig:ALL-EDBhist-4}, \ref{fig:ALL-EDBhist-5}, \ref{fig:ALL-EDBhist-6} we show histograms of empirical distributions of baseline lengths omitted from Section II.B.

\section{Simulation study of cross-correlation coverage}
\subsection{Introduction and setting}
This section contains a more detailed discussion of the simulations presented in Section III.C of the main paper. 
For the reader's convenience, we first recall the setting of our simulation experiment. We are interested in the behaviour of an idealized model of the communication network within a swarm of satellites acting as a~space-based interferometer. This network is generated based on a $k$-NN graph between the individual satellites. More concretely, the network is created as follows:

\begin{enumerate}
    \item Initially, all the $n$ satellites are assigned an energy budget $E_{\rm max} = q_{GG}(p)$, where $q_{GG}(p)$ is the quantile function associated with the distribution function identified in Section~II.C with finite-size correction described in Section~II.D and $p$ is a parameter of the simulation.
    \item We establish a \emph{pruned} $k$-NN communication network, which is created from $k$-NN graph between the satellites, with edges removed when their existence causes the satellite's power consumption to exceed the energy budget $E_{\rm max}$. We iteratively remove the most energetically expensive edge, until the edges no longer deplete the power budget.
    \item We set the energy per one cross-correlation computation as in Eq.~1 and assign computation jobs as described in Section~III.A.
\end{enumerate}

For a more detailed discussion of the setting and motivation of our study, we refer the reader to the main paper.

The choice of $E_{\rm max}$ and $\beta$ strongly influences the outcome of the simulation. Since we aim to explore disconnected networks, the quantile function $ q_{GG}(p)$ allows us to pick $E_{\rm max}$ in a way that will very probably induce network pruning and the intensity of this effect can be easily controlled by the parameter $p$ (note that pruning is independent of $\beta$). The parameter $\beta$ influences the intensity of \emph{competition} for energy between transmission and computational functions of satellites.

In our simulations, we focus on three  quantitative indicators:
\begin{enumerate}
    \item LSCC reduction factor $\rho_L = \frac{|\text{LSCC post-pruning}|}{|\text{LSCC pre-pruning}|}$,
    \item LSCC coverage factor $\alpha_L = \frac{|\text{jobs assigned within LSCC}|}{|\text{jobs available within LSCC}|}$,
    \item Coverage factor $\alpha = \frac{|\text{jobs assigned within LSCC}|}{|\text{jobs available within swarm}|}$.
\end{enumerate}

In this document, we present a more detailed account of our results and discuss the influence of various parameters on the aforementioned quantities.

\subsection{LSCC reduction factor}

Firstly, note that the LSCC reduction factor $\rho_L$ is \emph{independent} of the parameter~$\beta$, since the energy costs of initialisation and maintenance of the communication network are, in our model, given preference to those related to computation. In other words, the primary aim of every satellite is to establish its $k$-NN communication network, not to conserve power for possible computations.

In Sections II.C and II.D, we identified a closed-form expression for the probability distribution that is a good approximation of communication costs in the $k$-NN network. We use the quantile function $q_{\rm GG}(p) : (0,1) \rightarrow \mathbf{R}^+$ associated to this distribution (that is, the inverse of the cumulative distribution function) to set the parameter $E_{\rm max}$ that corresponds to the power budget of a~single satellite. As one would expect, the pruning has a more pronounced effect when $p$ is small. In this setting, the results also show significant random fluctuations. As the parameter $p$ increases, the LSCC reduction factor $\rho_L$ is being pushed towards $1$ (which would correspond to no reduction in the size of the LSCC) and results start to concentrate in the neighbourhood of~$1$.

We would also like to point out the effect of parameters $n$ and $k$. As we have seen throughout the main paper, the randomness in the network is more significant in the regime of small $n$. Similarly, networks with higher values of the parameter $k$ seem more robust, due to the presence of additional edges.

\begin{figure}[p]
     \centering
     \begin{subfigure}[b]{0.325\textwidth}
         \centering
         \includegraphics[width=\textwidth]{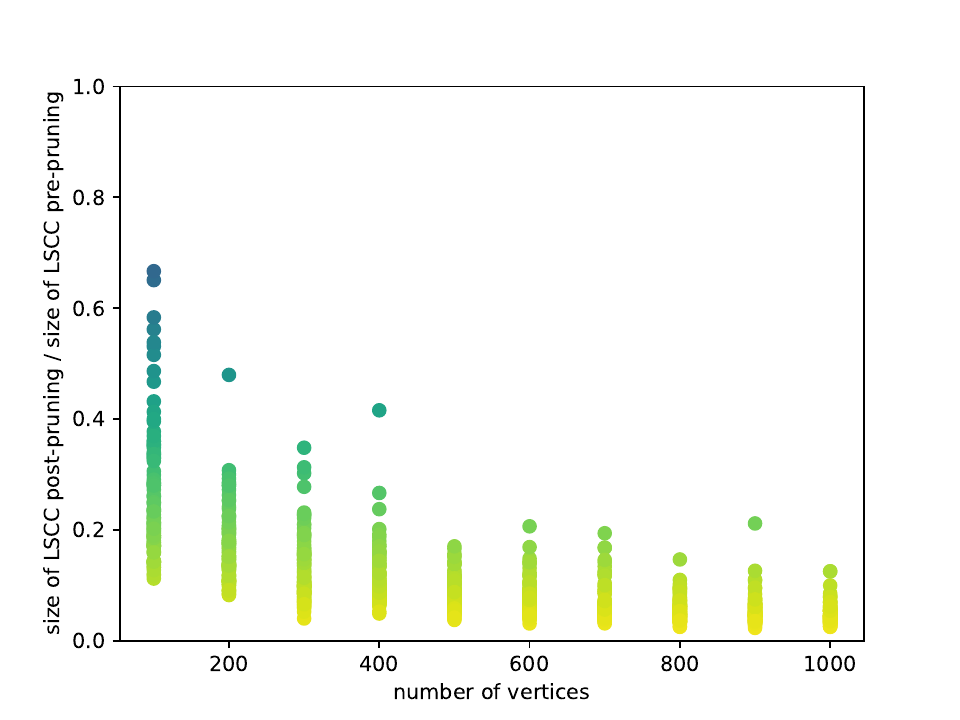}
         \caption{$k=4$, $p=0.1$}
         \label{fig:LSCCred.1-4}
     \end{subfigure}
     \hfill
     \begin{subfigure}[b]{0.325\textwidth}
         \centering\includegraphics[width=\textwidth]{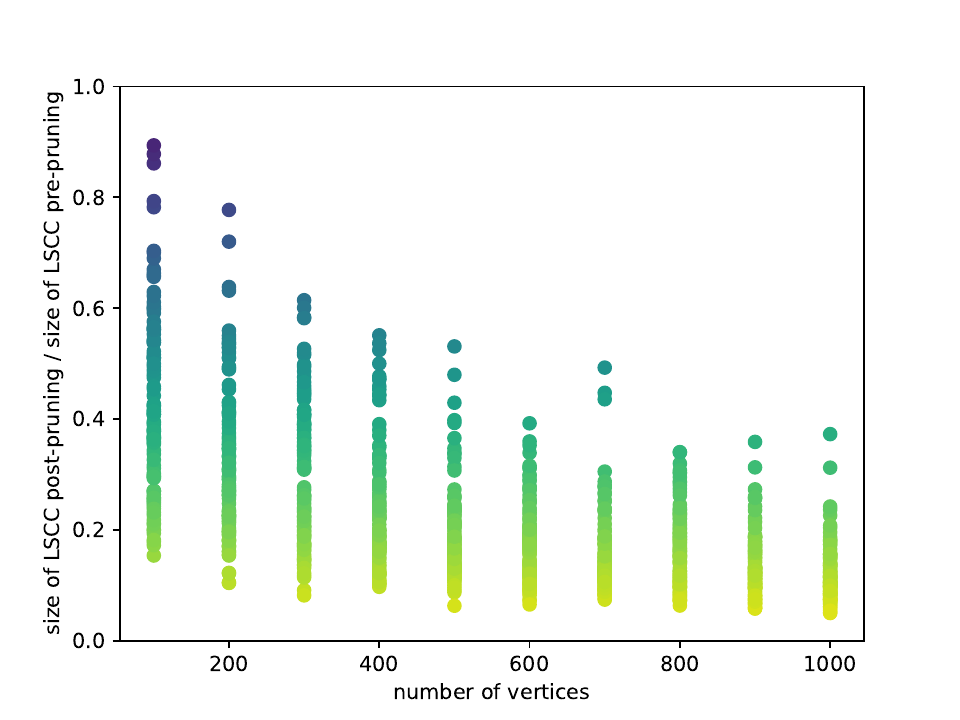}
         \caption{$k=4$, $p=0.2$}
         \label{fig:LSCCred.2-4}
     \end{subfigure}
     \hfill
     \begin{subfigure}[b]{0.325\textwidth}
         \centering\includegraphics[width=\textwidth]{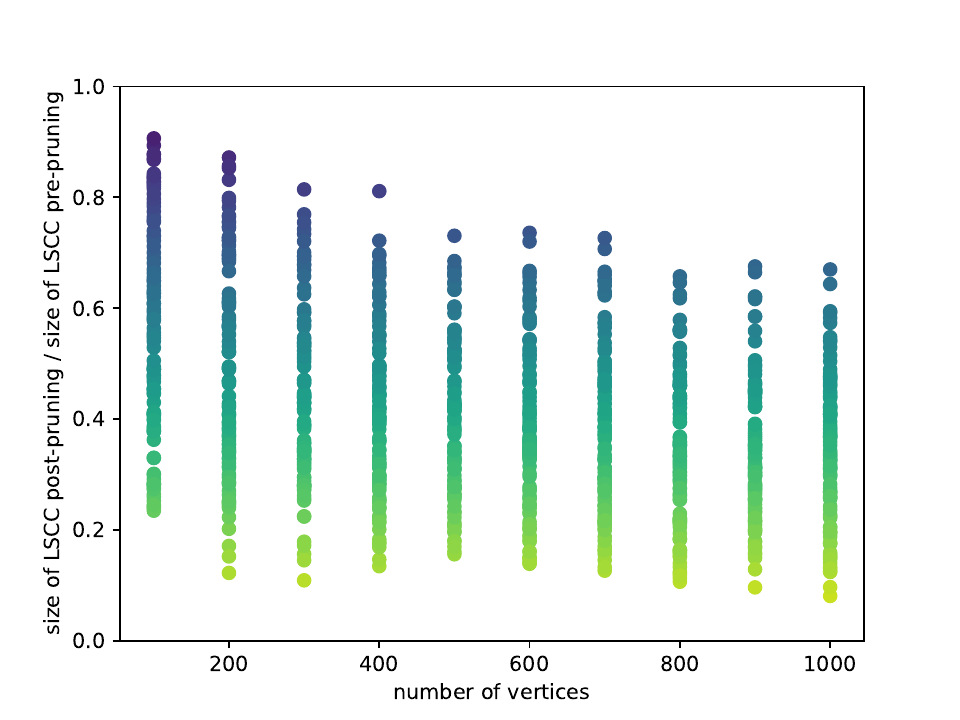}
         \caption{$k=4$, $p=0.3$}
         \label{fig:LSCCred.3-4}
     \end{subfigure}
     \hfill
    \begin{subfigure}[b]{0.325\textwidth}
         \centering\includegraphics[width=\textwidth]{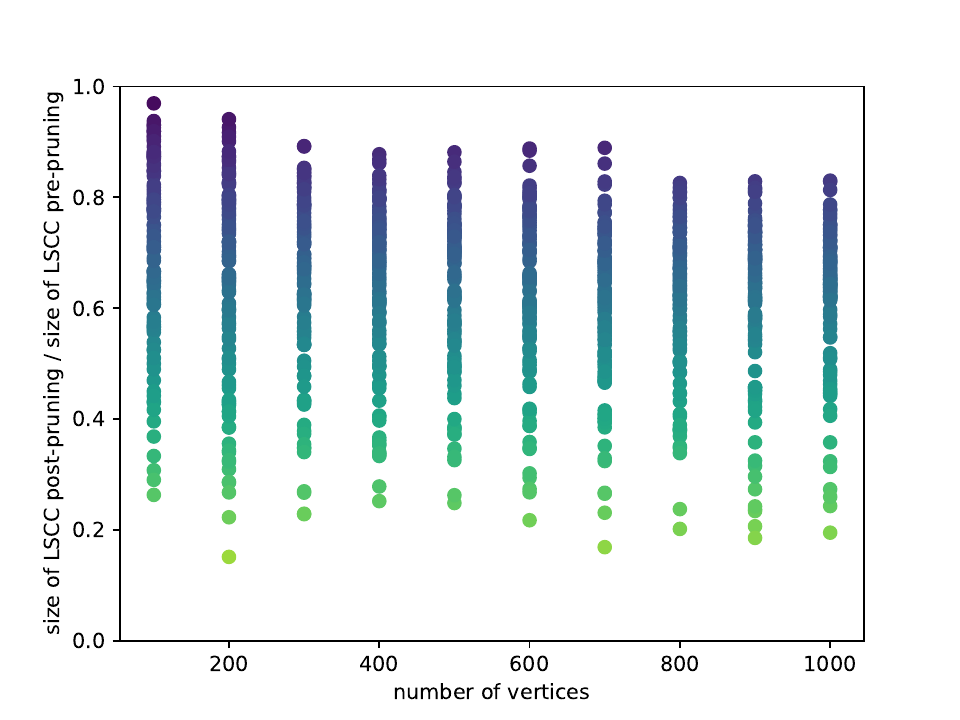}
         \caption{$k=4$, $p=0.4$}
         \label{fig:LSCCred.4-4}
     \end{subfigure}
     \hfill
     \begin{subfigure}[b]{0.325\textwidth}
         \centering\includegraphics[width=\textwidth]{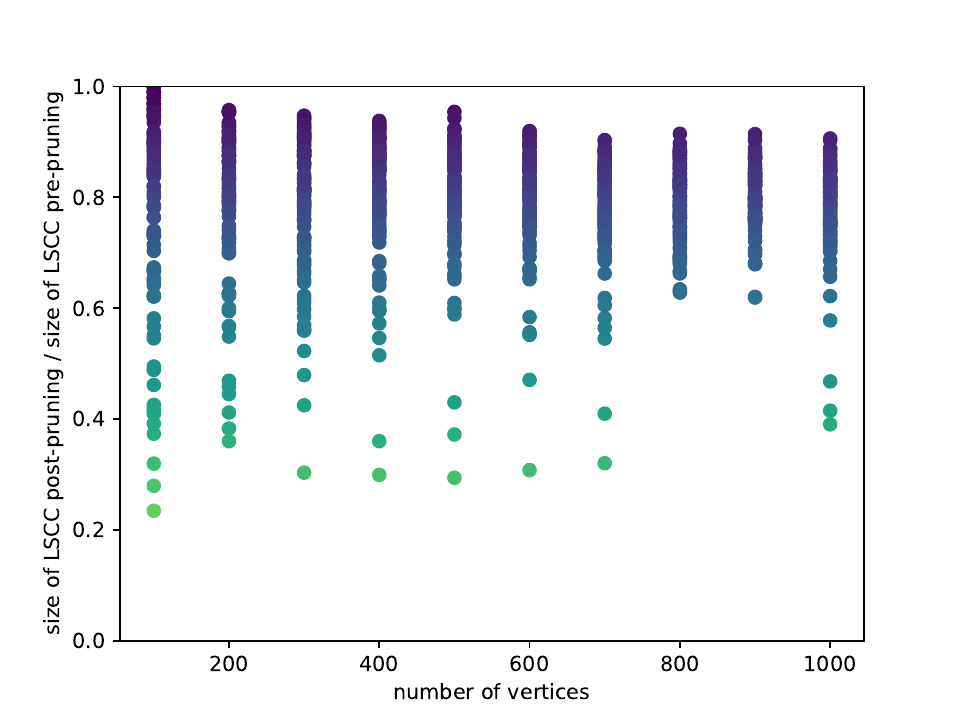}
         \caption{$k=4$, $p=0.5$}
         \label{fig:LSCCred.5-4}
     \end{subfigure}
     \hfill
     \begin{subfigure}[b]{0.325\textwidth}
         \centering\includegraphics[width=\textwidth]{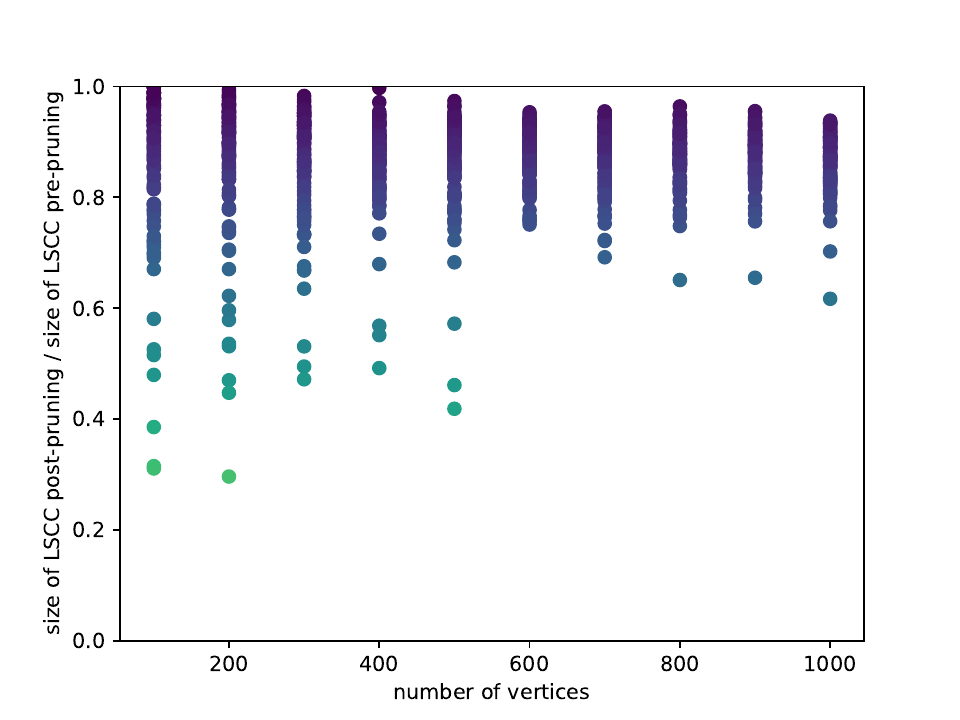}
         \caption{$k=4$, $p=0.6$}
         \label{fig:LSCCred.6-4}
     \end{subfigure}
     \hfill
     \begin{subfigure}[b]{0.325\textwidth}
         \centering\includegraphics[width=\textwidth]{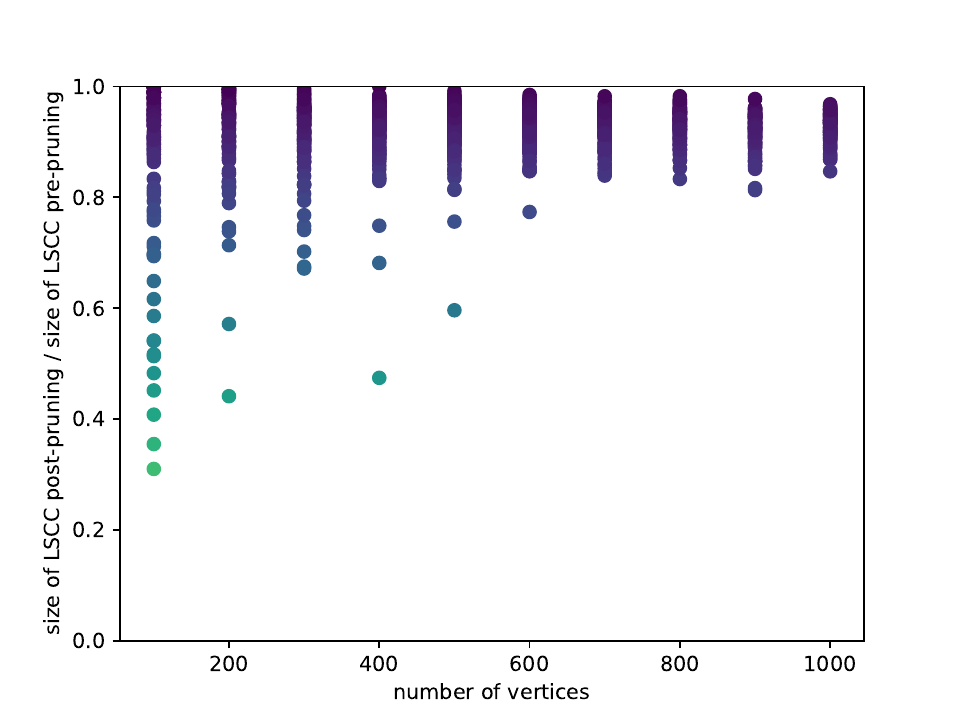}
         \caption{$k=4$, $p=0.7$}
         \label{fig:LSCCred.7-4}
     \end{subfigure}
     \hfill
     \begin{subfigure}[b]{0.325\textwidth}
         \centering\includegraphics[width=\textwidth]{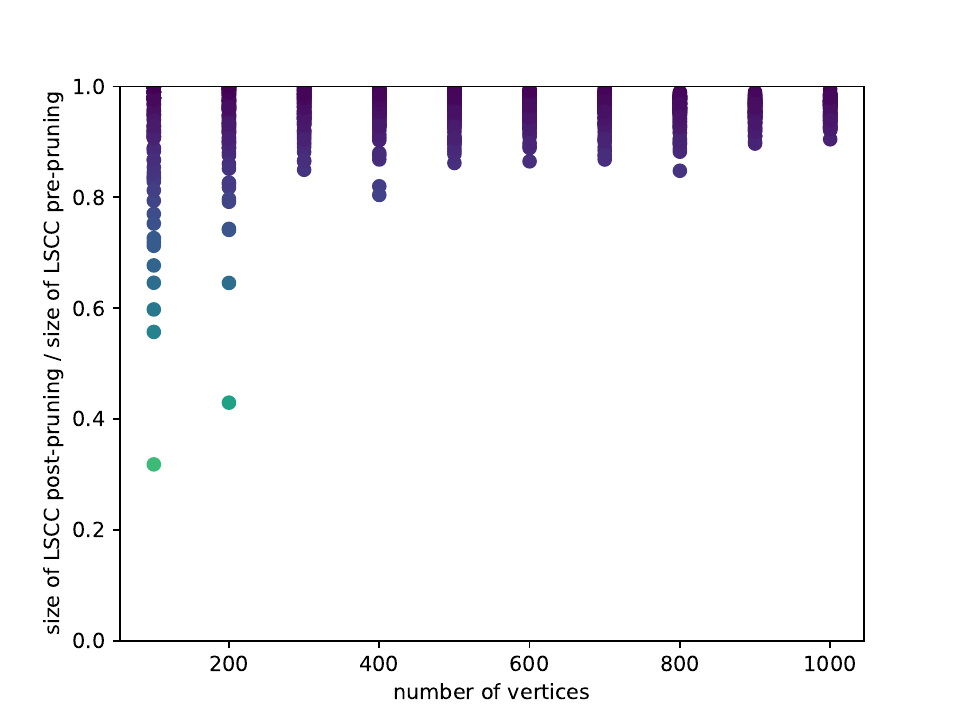}
         \caption{$k=4$, $p=0.8$}
         \label{fig:LSCCred.8-4}
     \end{subfigure}
     \hfill
     \begin{subfigure}[b]{0.325\textwidth}
         \centering\includegraphics[width=\textwidth]{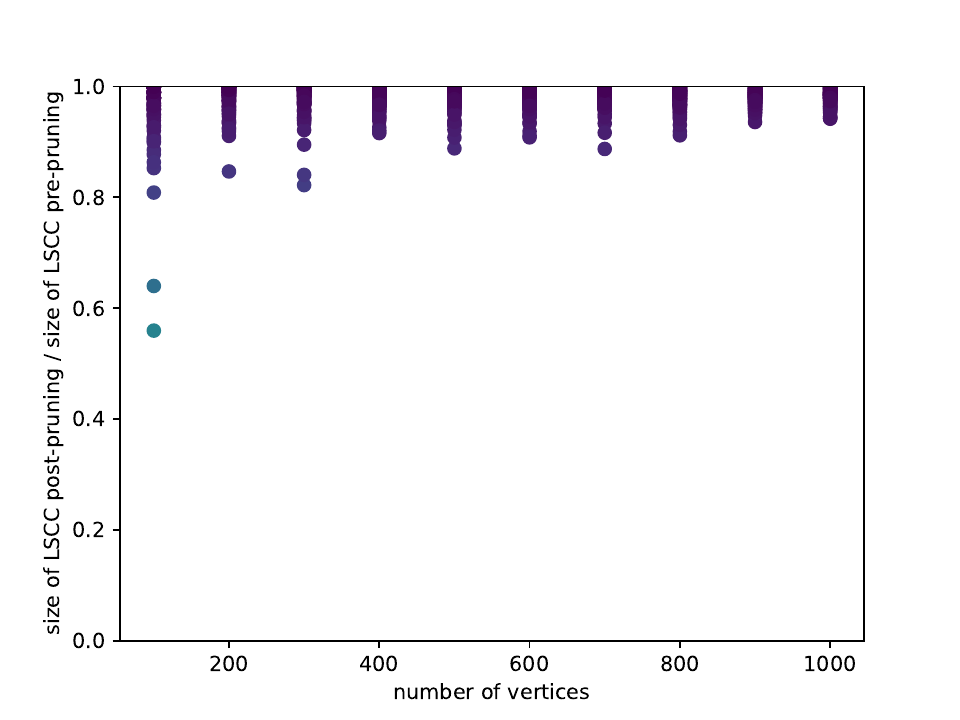}
         \caption{$k=4$, $p=0.9$}
         \label{fig:LSCCred.9-4}
     \end{subfigure}
     \hfill
     \begin{subfigure}[b]{0.325\textwidth}
         \centering\includegraphics[width=\textwidth]{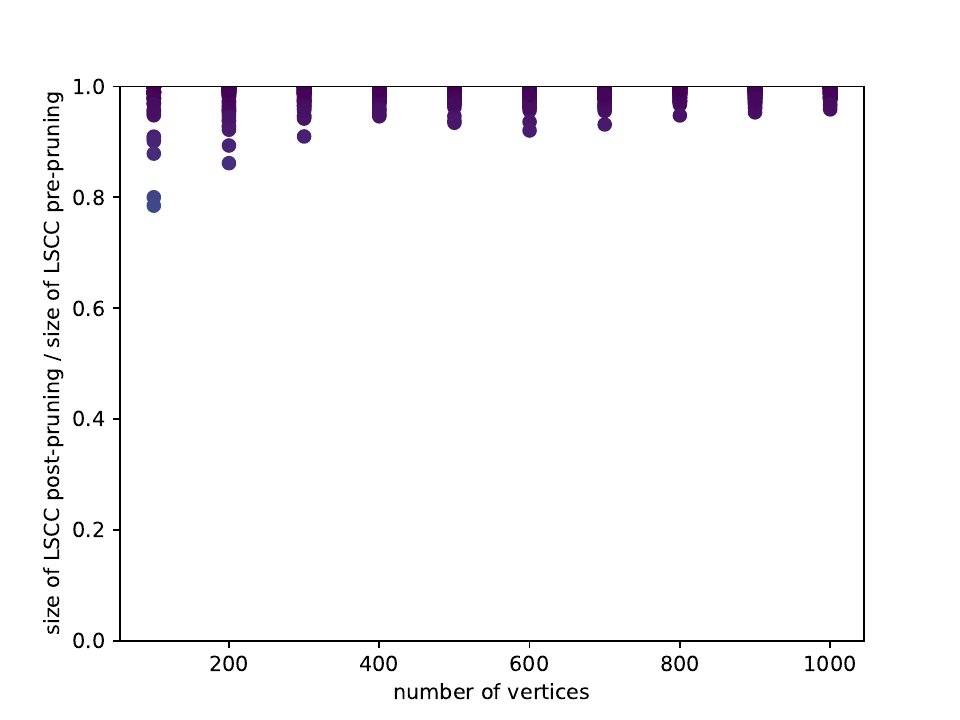}
         \caption{$k=4$, $p=0.95$}
         \label{fig:LSCCred.95-4}
     \end{subfigure}
     \hfill
     \begin{subfigure}[b]{0.325\textwidth}
         \centering\includegraphics[width=\textwidth]{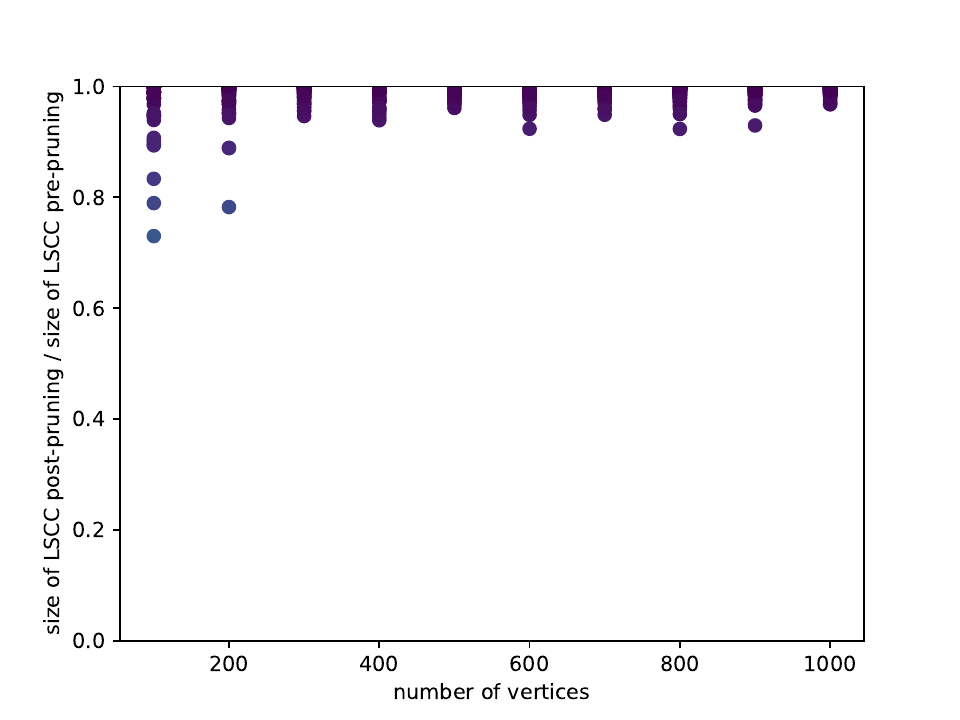}
         \caption{$k=4$, $p=0.97$}
         \label{fig:LSCCred.97-4}
     \end{subfigure}
     \hfill
     \begin{subfigure}[b]{0.325\textwidth}
         \centering\includegraphics[width=\textwidth]{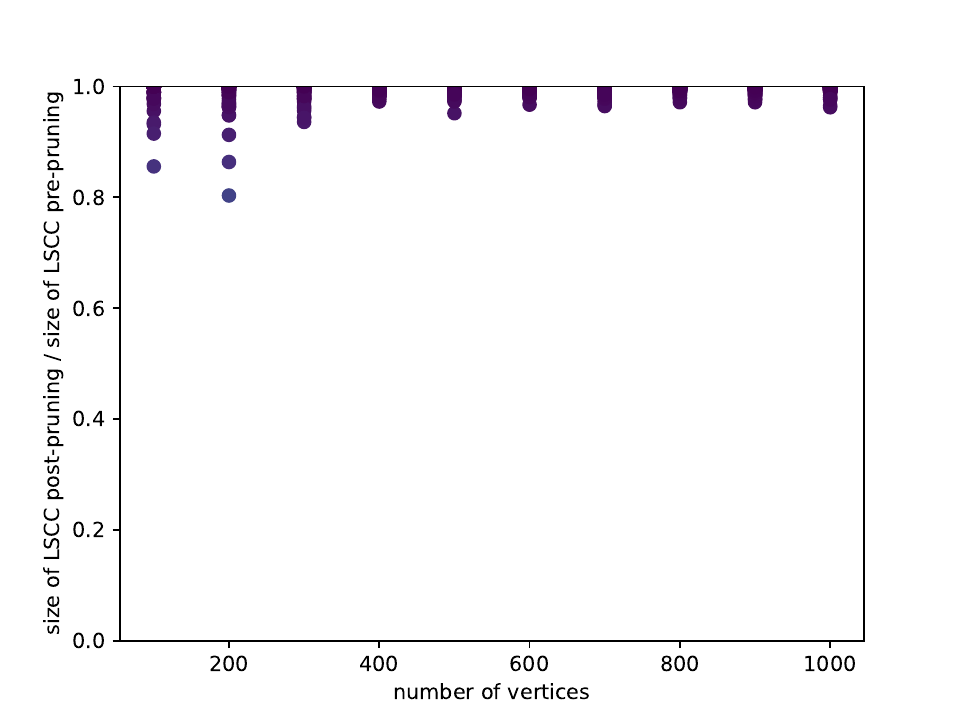}
         \caption{$k=4$, $p=0.99$}
         \label{fig:LSCCred.99-4}
     \end{subfigure}
     \hfill
     \caption[Scatter plots of LSSC reduction factor $\rho_L$ for $k=4$.]{Scatter plots of LSSC reduction factor $\rho_L$ obtained from simulations in the parameter regime $n\in(100, 1000)$ for $k=4$.}
    \label{fig:LSCCred-4-all}
\end{figure}

\begin{figure}[p]
     \centering
     \begin{subfigure}[b]{0.325\textwidth}
         \centering
         \includegraphics[width=\textwidth]{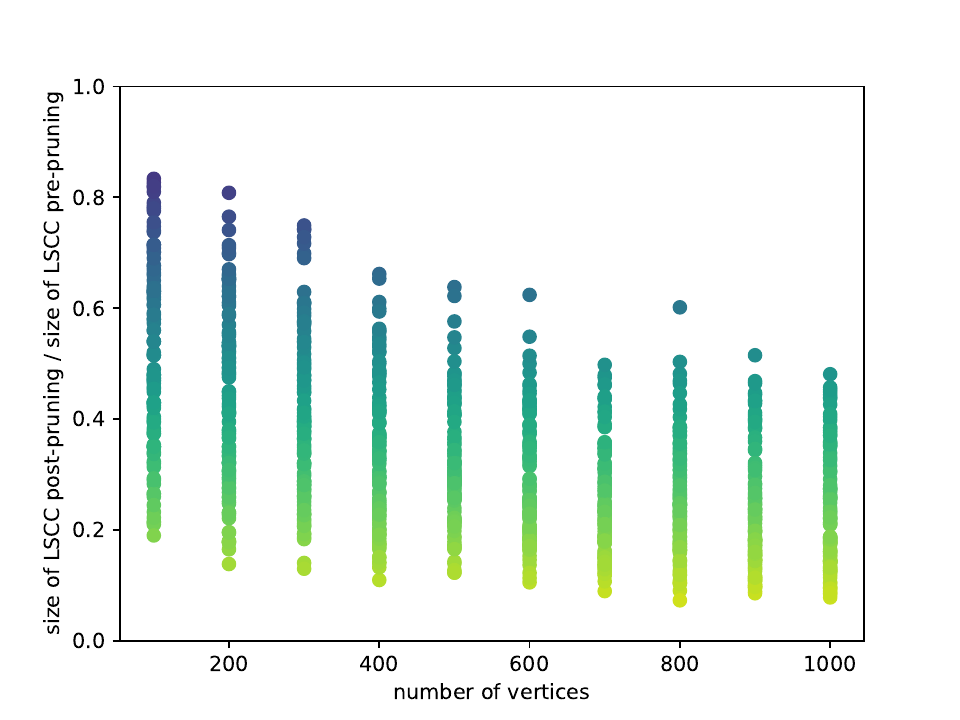}
         \caption{$k=5$, $p=0.1$}
         \label{fig:LSCCred.1-5}
     \end{subfigure}
     \hfill
     \begin{subfigure}[b]{0.325\textwidth}
         \centering\includegraphics[width=\textwidth]{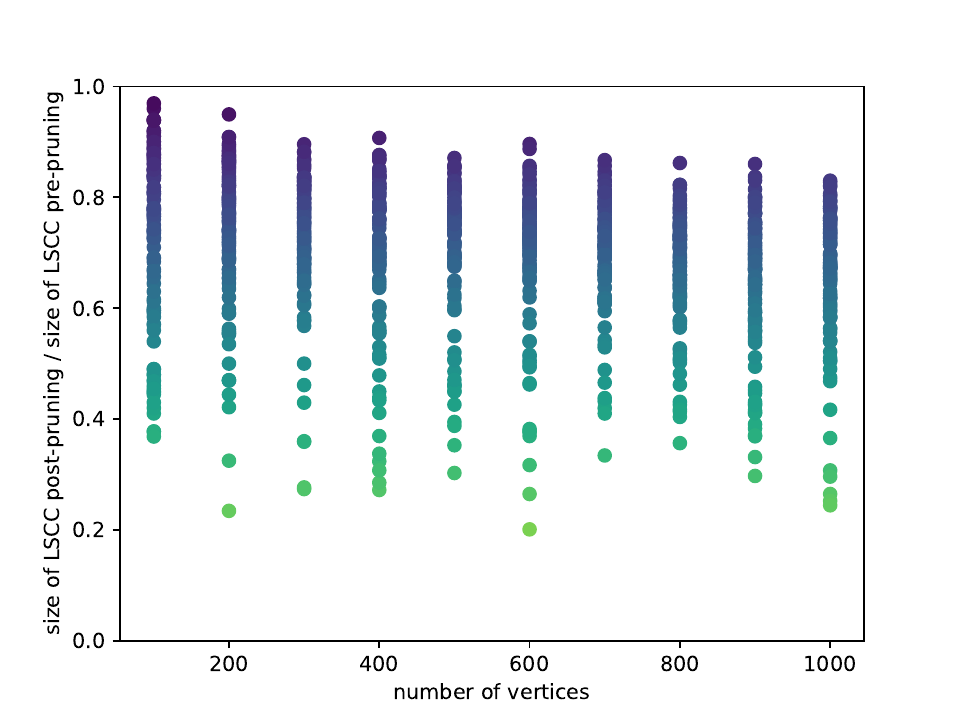}
         \caption{$k=5$, $p=0.2$}
         \label{fig:LSCCred.2-5}
     \end{subfigure}
     \hfill
     \begin{subfigure}[b]{0.325\textwidth}
         \centering\includegraphics[width=\textwidth]{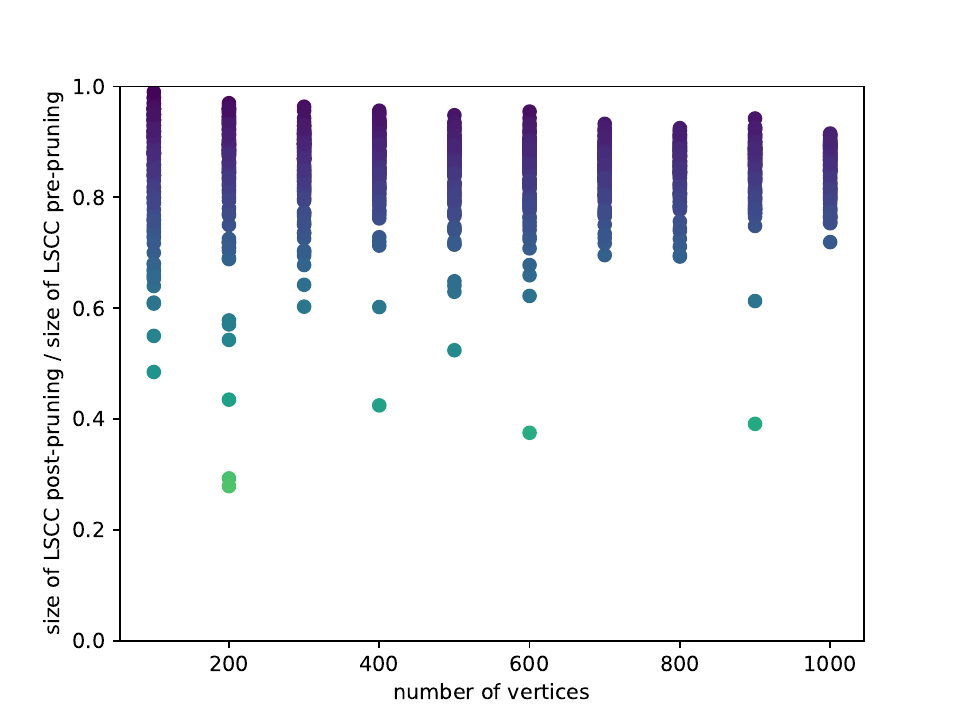}
         \caption{$k=5$, $p=0.3$}
         \label{fig:LSCCred.3-5}
     \end{subfigure}
     \hfill
    \begin{subfigure}[b]{0.325\textwidth}
         \centering\includegraphics[width=\textwidth]{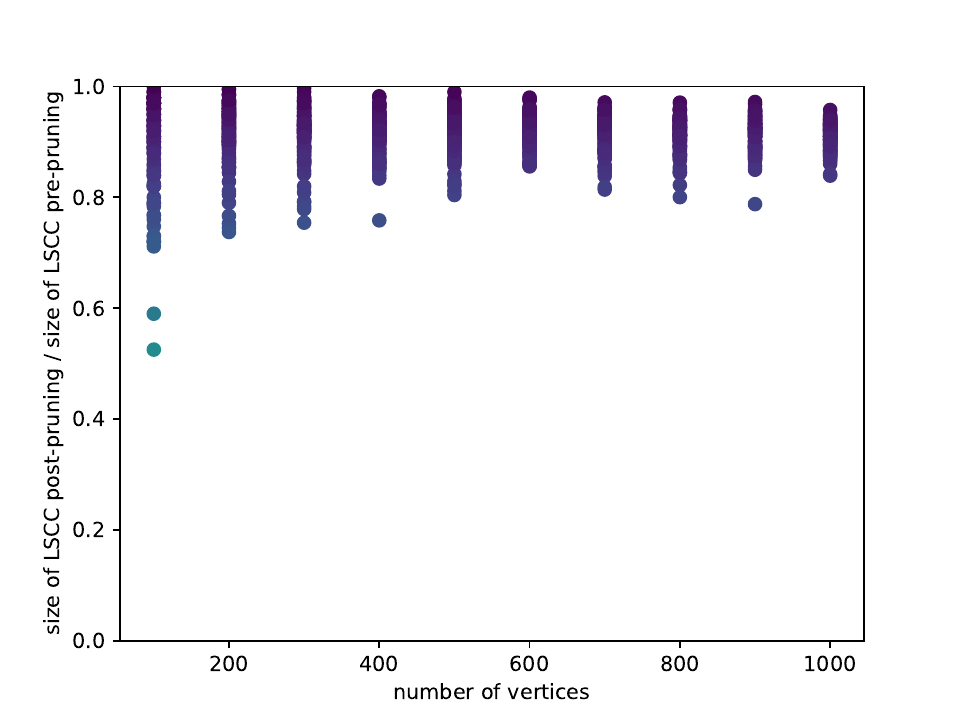}
         \caption{$k=5$, $p=0.4$}
         \label{fig:LSCCred.4-5}
     \end{subfigure}
     \hfill
     \begin{subfigure}[b]{0.325\textwidth}
         \centering\includegraphics[width=\textwidth]{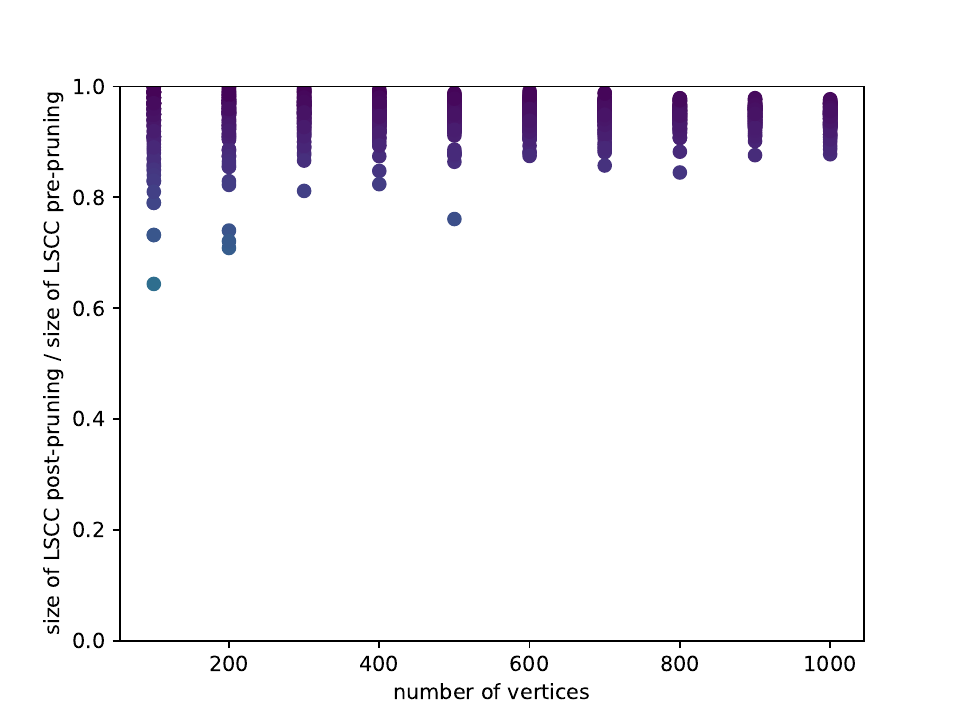}
         \caption{$k=5$, $p=0.5$}
         \label{fig:LSCCred.5-5}
     \end{subfigure}
     \hfill
     \begin{subfigure}[b]{0.325\textwidth}
         \centering\includegraphics[width=\textwidth]{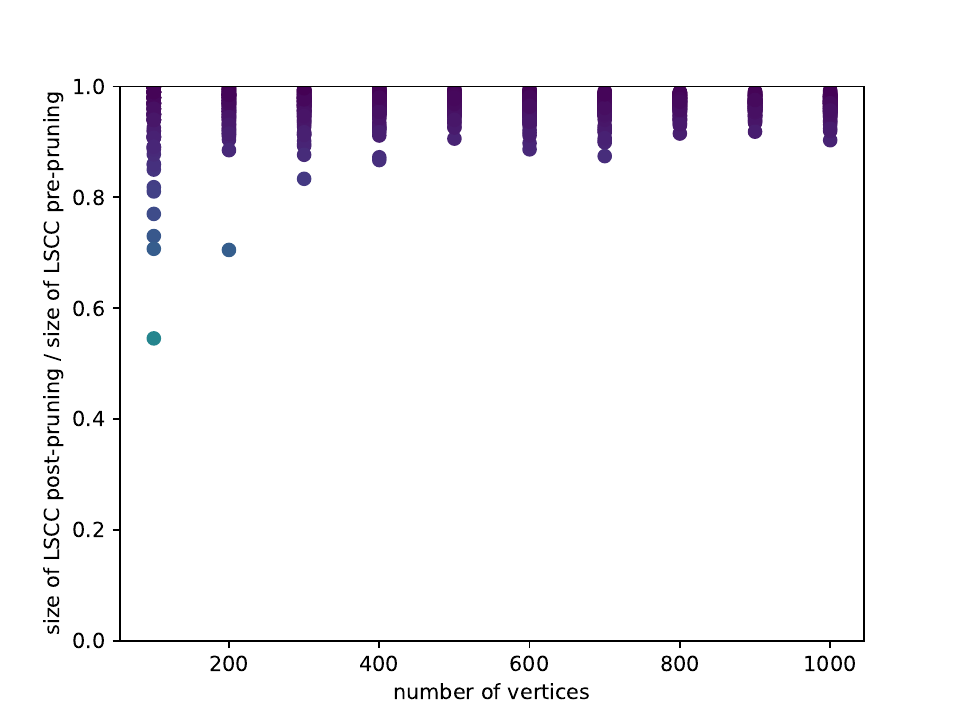}
         \caption{$k=5$, $p=0.6$}
         \label{fig:LSCCred.6-5}
     \end{subfigure}
     \hfill
     \begin{subfigure}[b]{0.325\textwidth}
         \centering\includegraphics[width=\textwidth]{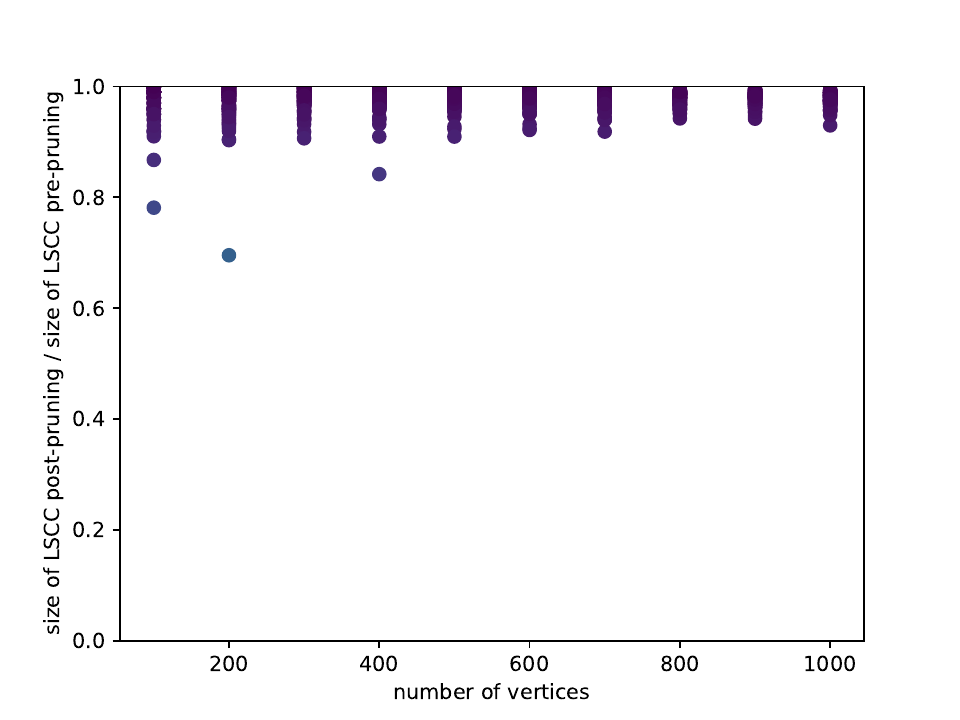}
         \caption{$k=5$, $p=0.7$}
         \label{fig:LSCCred.7-5}
     \end{subfigure}
     \hfill
     \begin{subfigure}[b]{0.325\textwidth}
         \centering\includegraphics[width=\textwidth]{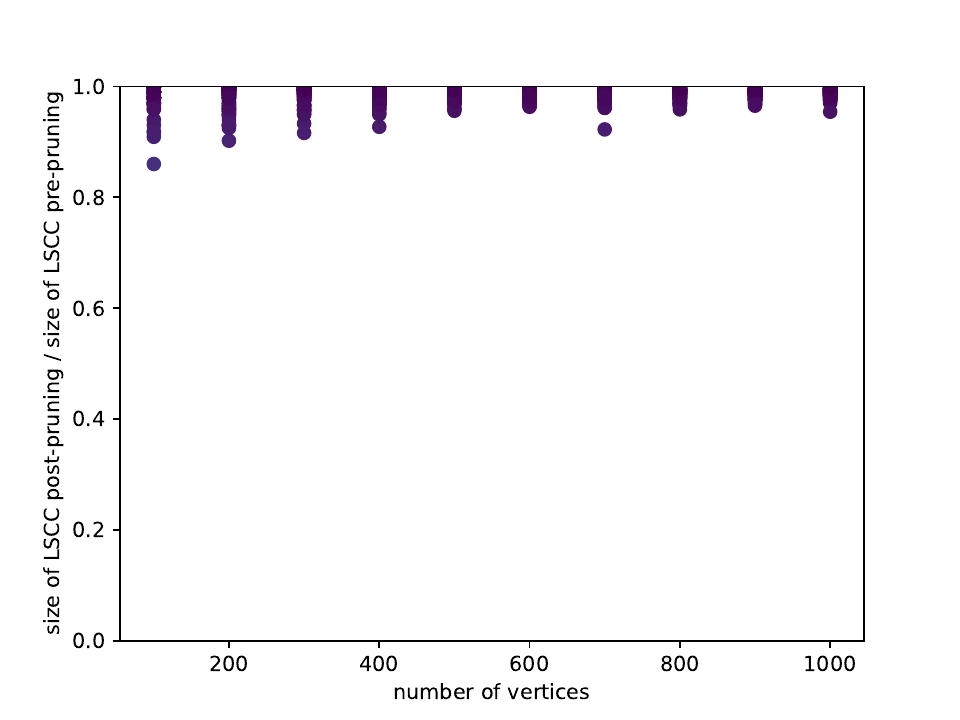}
         \caption{$k=5$, $p=0.8$}
         \label{fig:LSCCred.8-5}
     \end{subfigure}
     \hfill
     \begin{subfigure}[b]{0.325\textwidth}
         \centering\includegraphics[width=\textwidth]{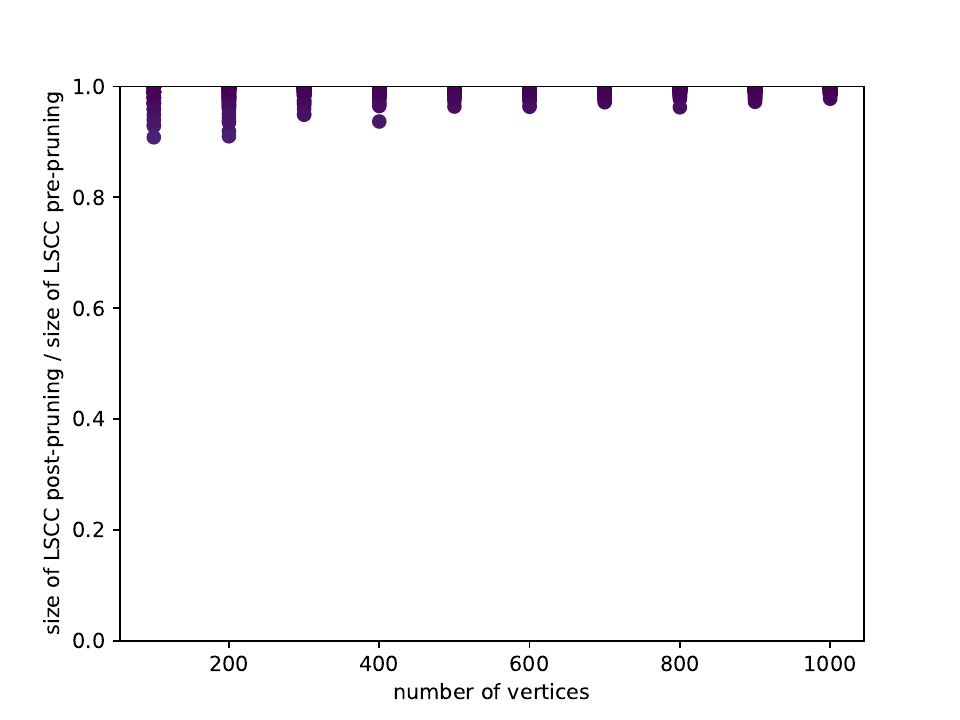}
         \caption{$k=5$, $p=0.9$}
         \label{fig:LSCCred.9-5}
     \end{subfigure}
     \hfill
     \begin{subfigure}[b]{0.325\textwidth}
         \centering\includegraphics[width=\textwidth]{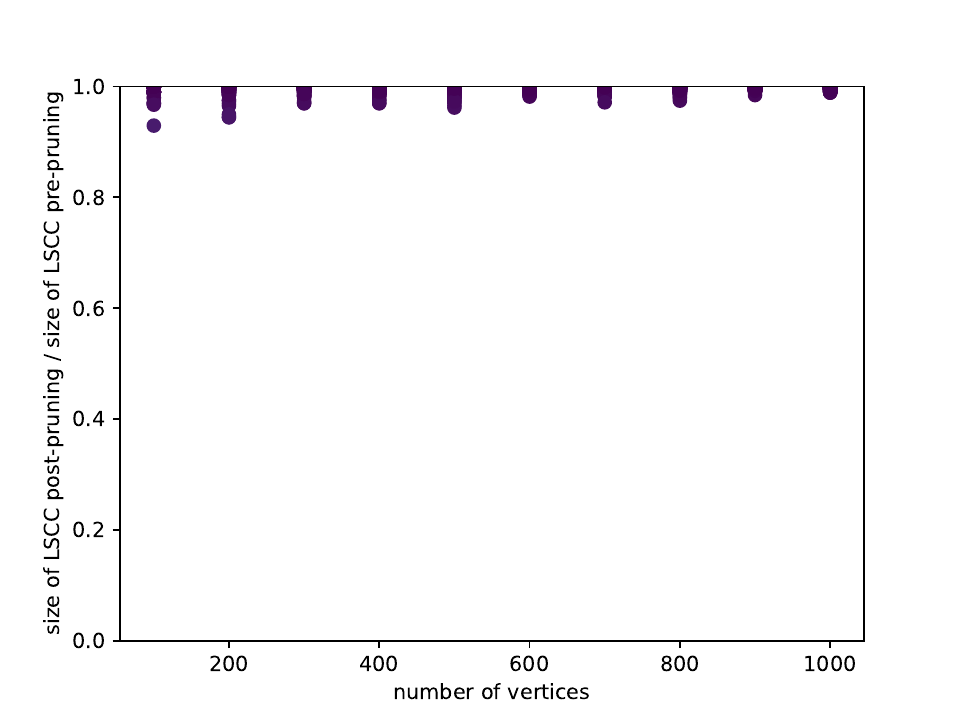}
         \caption{$k=5$, $p=0.95$}
         \label{fig:LSCCred.95}
     \end{subfigure}
     \hfill
     \begin{subfigure}[b]{0.325\textwidth}
         \centering\includegraphics[width=\textwidth]{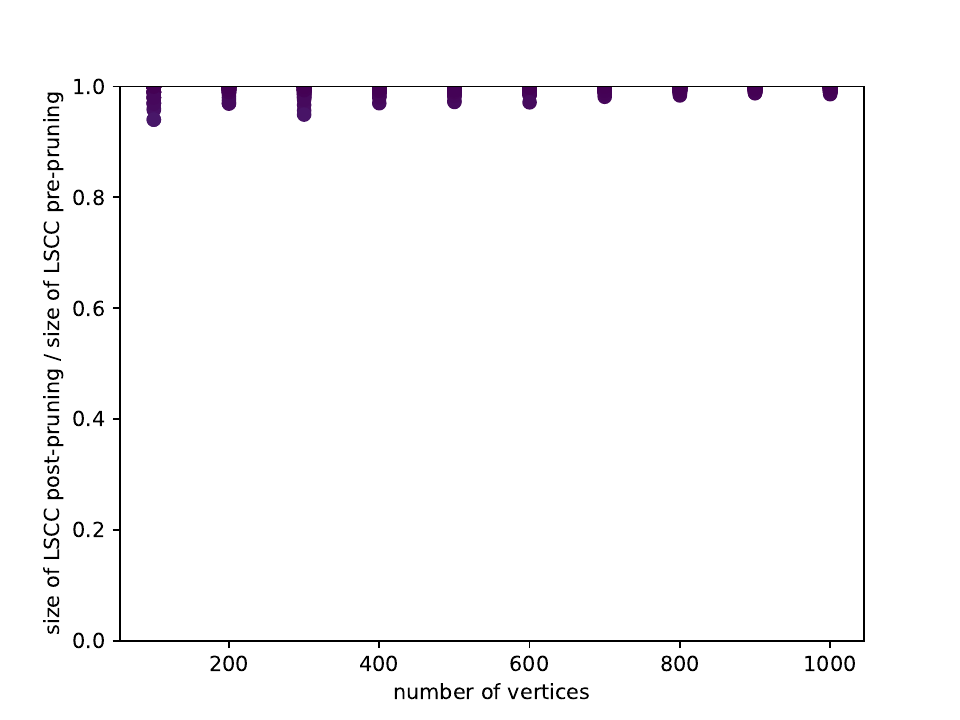}
         \caption{$k=5$, $p=0.97$}
         \label{fig:LSCCred.97-5}
     \end{subfigure}
     \hfill
     \begin{subfigure}[b]{0.325\textwidth}
         \centering\includegraphics[width=\textwidth]{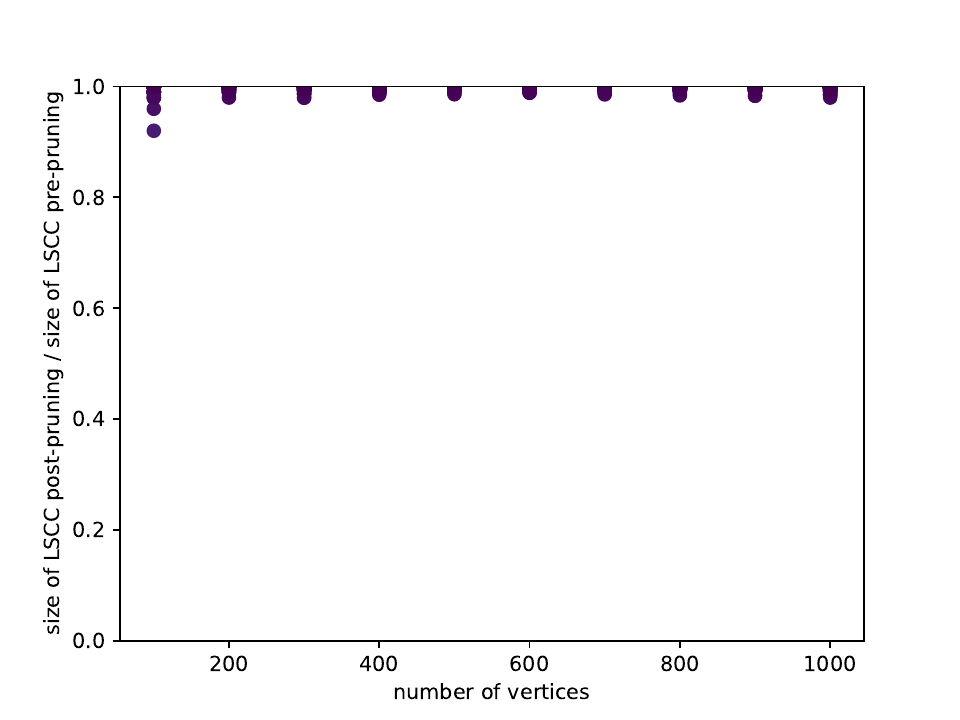}
         \caption{$k=5$, $p=0.99$}
         \label{fig:LSCCred.99-5}
     \end{subfigure}
     \hfill
     \caption[Scatter plots of LSSC reduction factor $\rho_L$ for $k=5$.]{Scatter plots of LSSC reduction factor $\rho_L$ obtained from simulations in the parameter regime $n\in(100, 1000)$ for $k=5$.}
    \label{fig:LSCCred-5-all}
\end{figure}

\begin{figure}[p]
     \centering
     \begin{subfigure}[b]{0.325\textwidth}
         \centering
         \includegraphics[width=\textwidth]{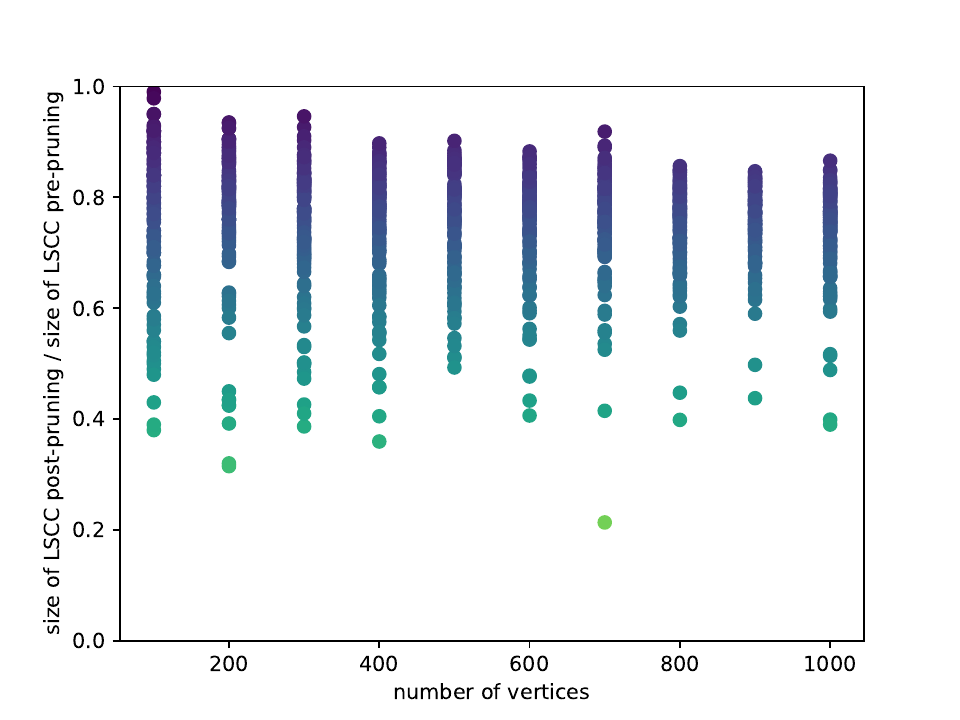}
         \caption{$k=6$, $p=0.1$}
         \label{fig:LSCCred.1-6}
     \end{subfigure}
     \hfill
     \begin{subfigure}[b]{0.325\textwidth}
         \centering\includegraphics[width=\textwidth]{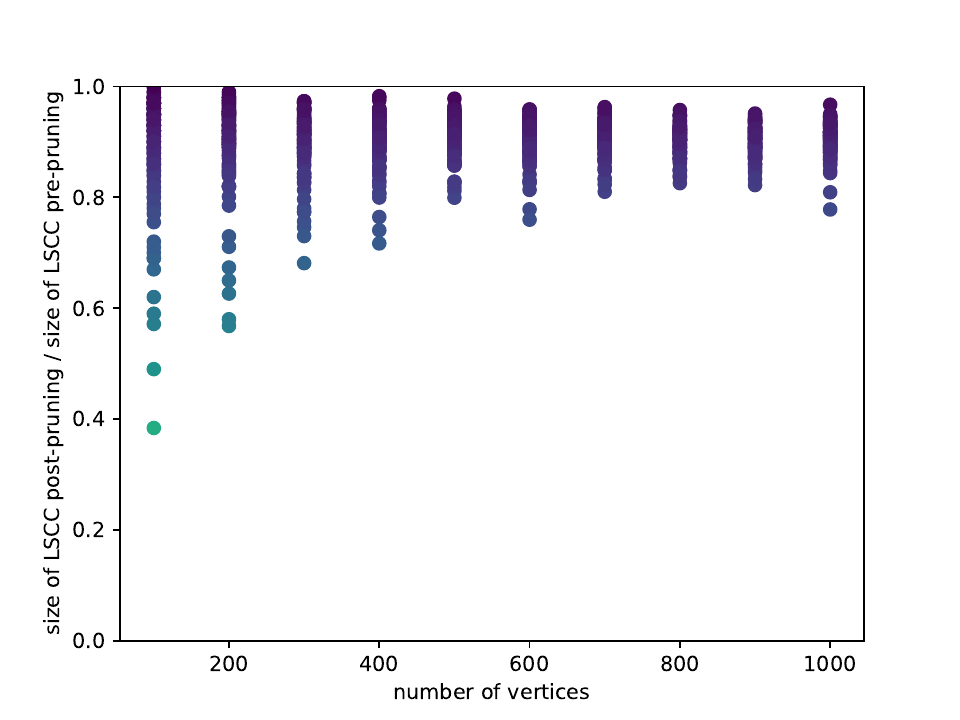}
         \caption{$k=6$, $p=0.2$}
         \label{fig:LSCCred.2-6}
     \end{subfigure}
     \hfill
     \begin{subfigure}[b]{0.325\textwidth}
         \centering\includegraphics[width=\textwidth]{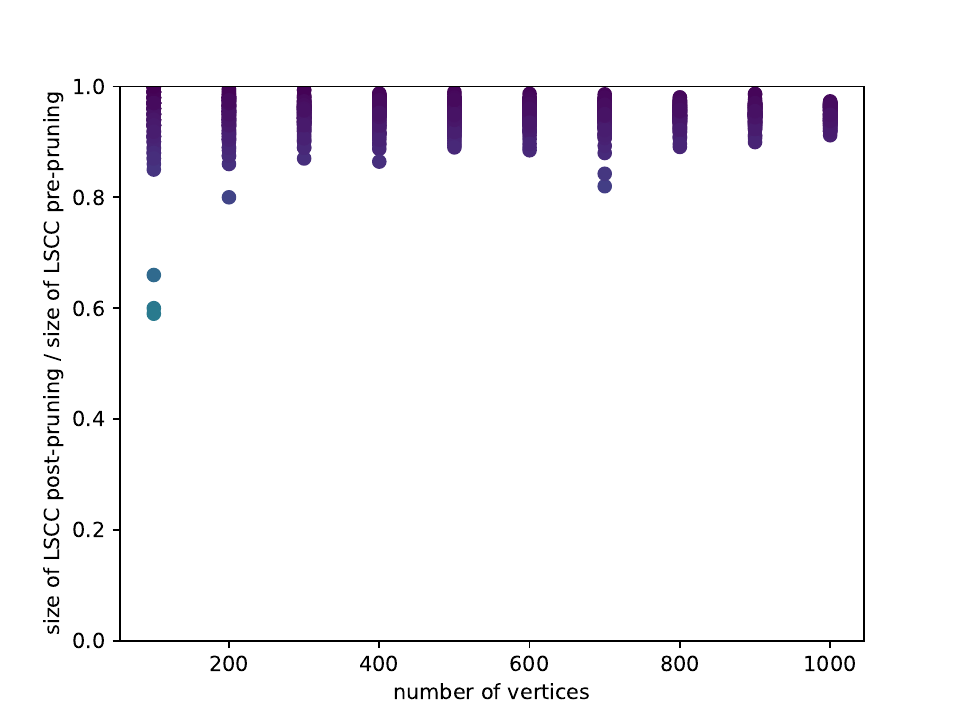}
         \caption{$k=6$, $p=0.3$}
         \label{fig:LSCCred.3-6}
     \end{subfigure}
     \hfill
    \begin{subfigure}[b]{0.325\textwidth}
         \centering\includegraphics[width=\textwidth]{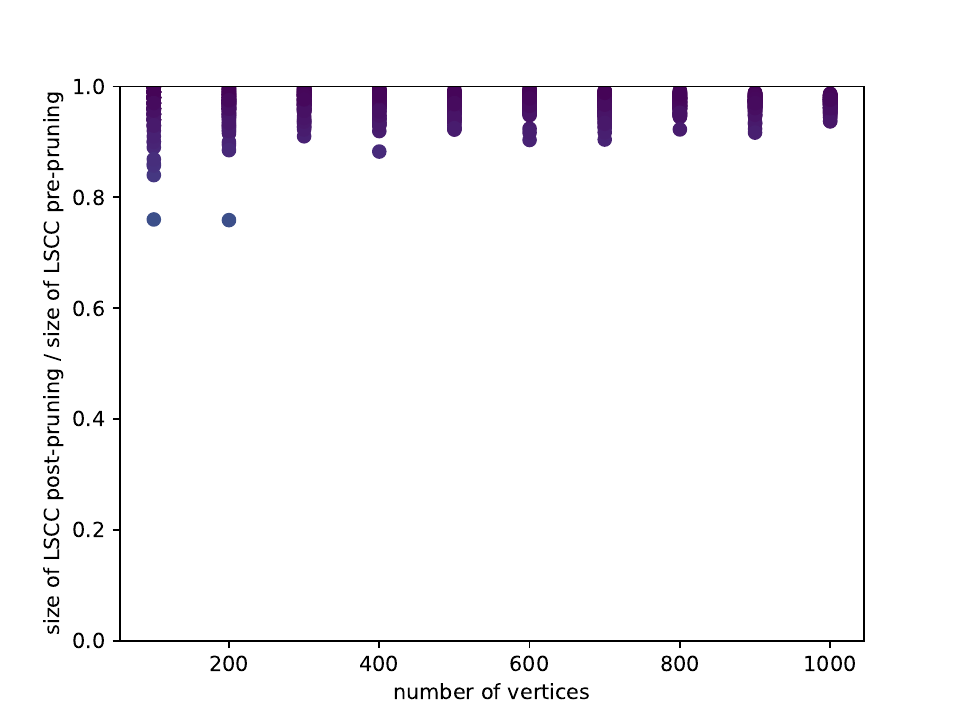}
         \caption{$k=6$, $p=0.4$}
         \label{fig:LSCCred.4-6}
     \end{subfigure}
     \hfill
     \begin{subfigure}[b]{0.325\textwidth}
         \centering\includegraphics[width=\textwidth]{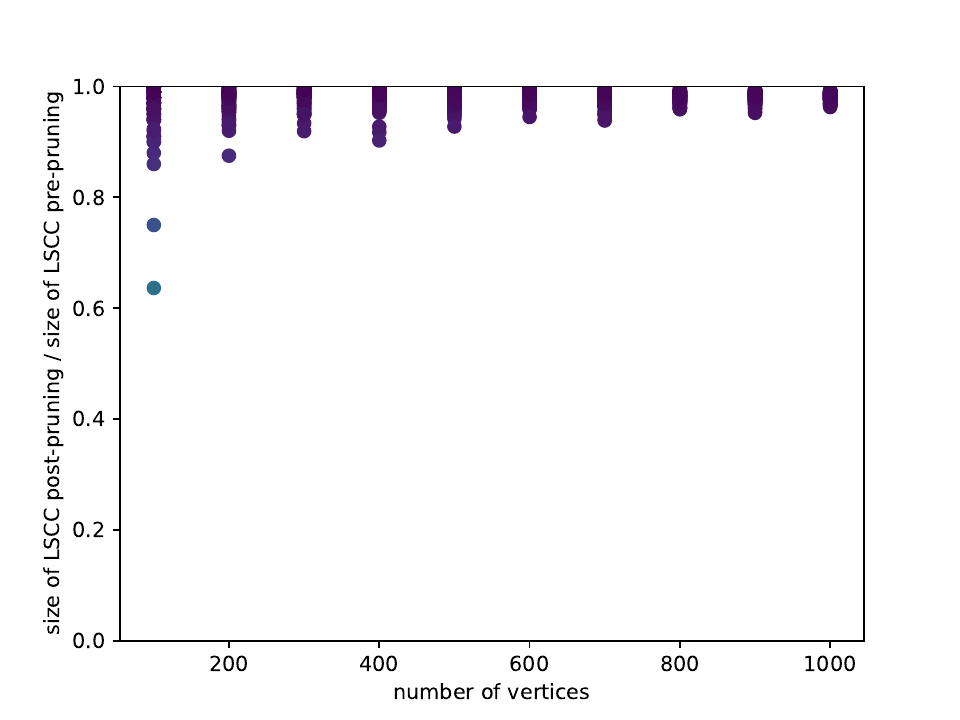}
         \caption{$k=6$, $p=0.5$}
         \label{fig:LSCCred.5-6}
     \end{subfigure}
     \hfill
     \begin{subfigure}[b]{0.325\textwidth}
         \centering\includegraphics[width=\textwidth]{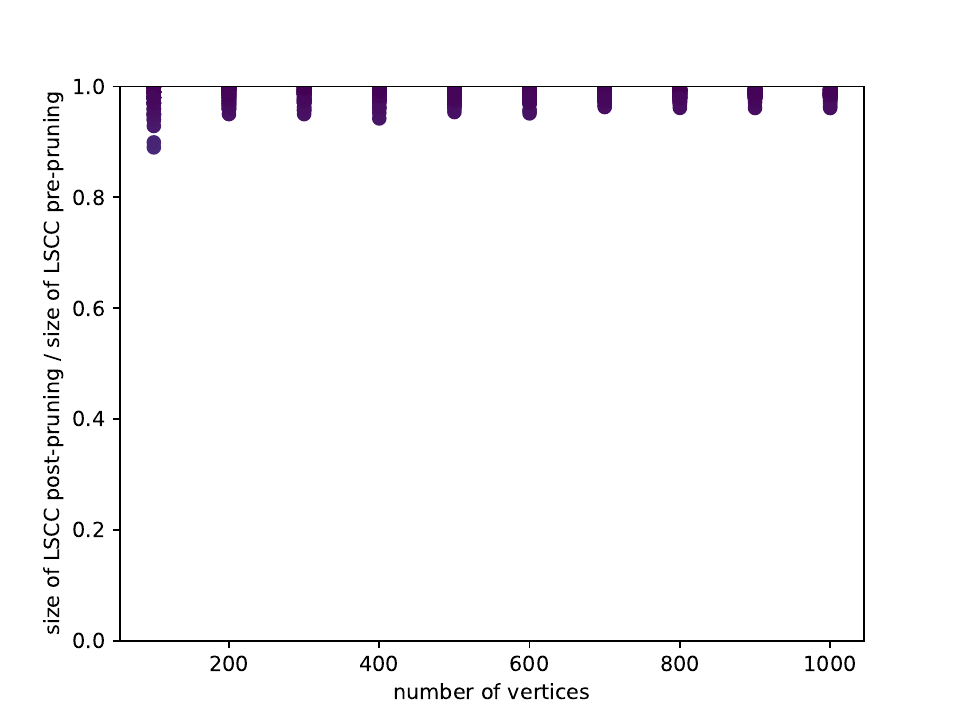}
         \caption{$k=6$, $p=0.6$}
         \label{fig:LSCCred.6-6}
     \end{subfigure}
     \hfill
     \begin{subfigure}[b]{0.325\textwidth}
         \centering\includegraphics[width=\textwidth]{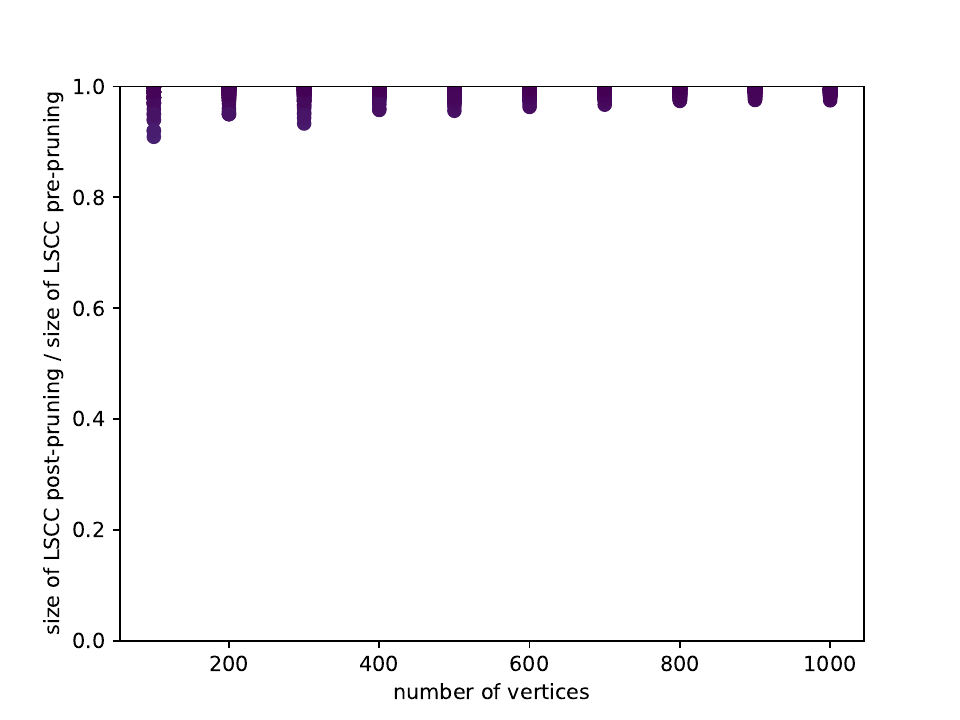}
         \caption{$k=6$, $p=0.7$}
         \label{fig:LSCCred.7-6}
     \end{subfigure}
     \hfill
     \begin{subfigure}[b]{0.325\textwidth}
         \centering\includegraphics[width=\textwidth]{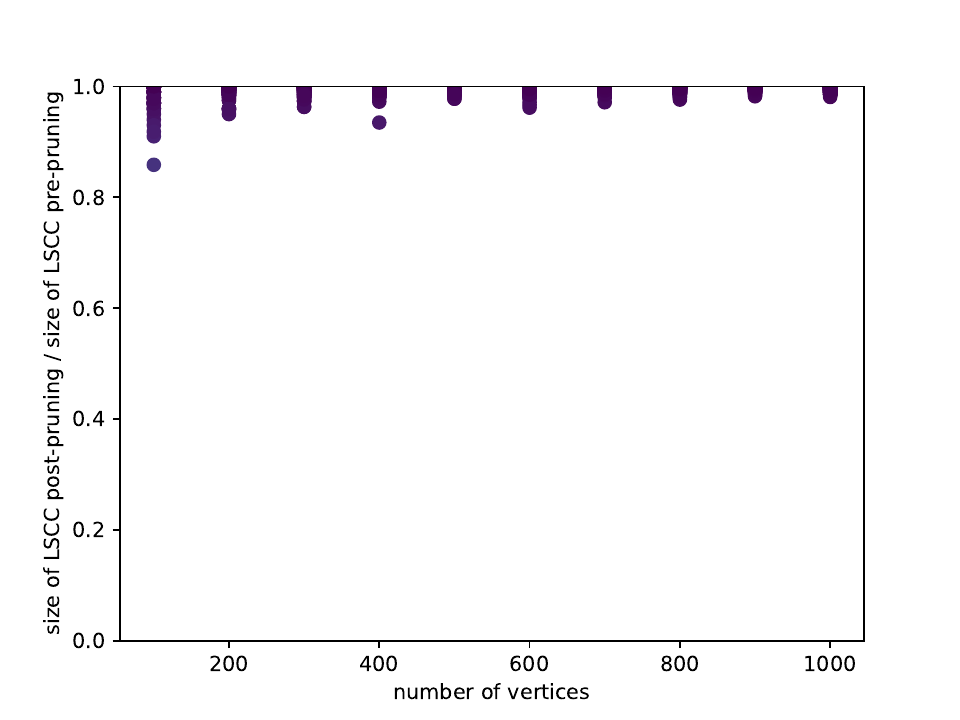}
         \caption{$k=6$, $p=0.8$}
         \label{fig:LSCCred.8-6}
     \end{subfigure}
     \hfill
     \begin{subfigure}[b]{0.325\textwidth}
         \centering\includegraphics[width=\textwidth]{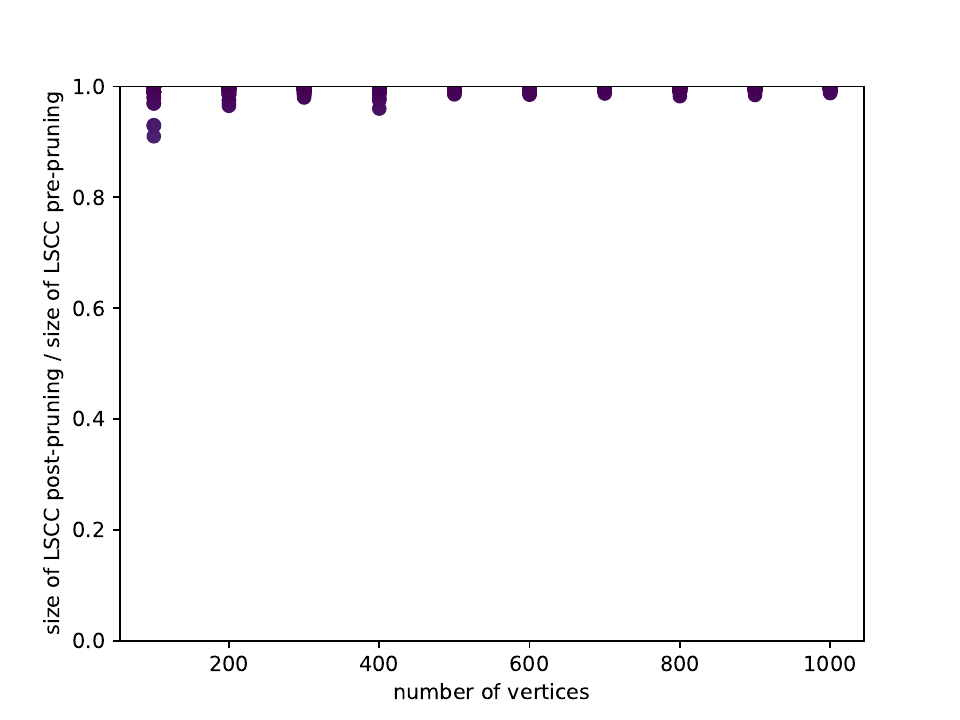}
         \caption{$k=6$, $p=0.9$}
         \label{fig:LSCCred.9-6}
     \end{subfigure}
     \hfill
     \begin{subfigure}[b]{0.325\textwidth}
         \centering\includegraphics[width=\textwidth]{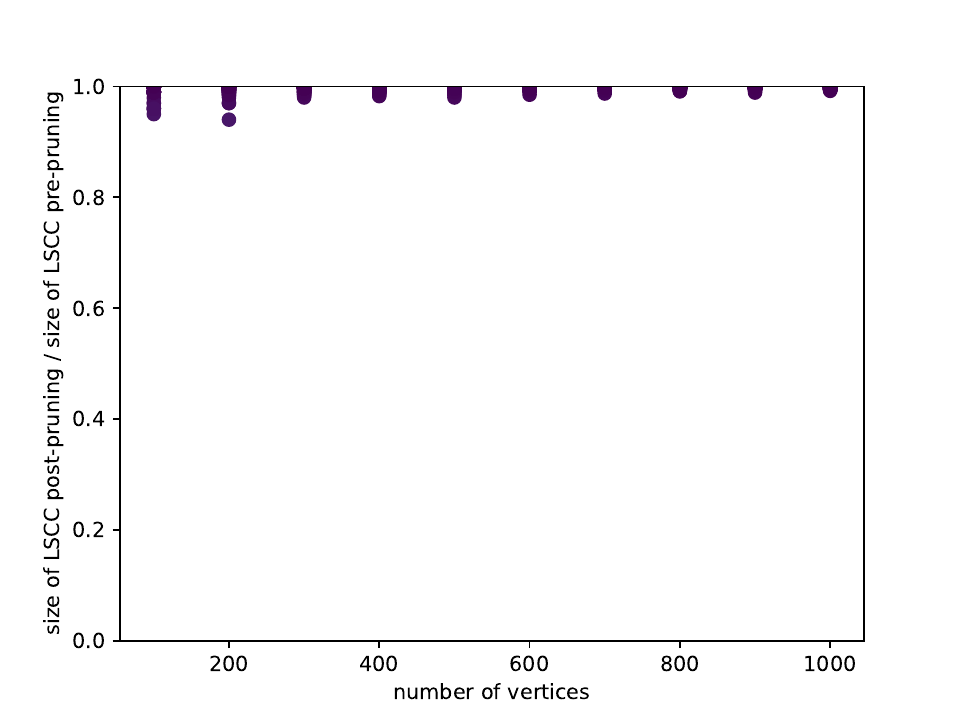}
         \caption{$k=6$, $p=0.95$}
         \label{fig:LSCCred.95-6}
     \end{subfigure}
     \hfill
     \begin{subfigure}[b]{0.325\textwidth}
         \centering\includegraphics[width=\textwidth]{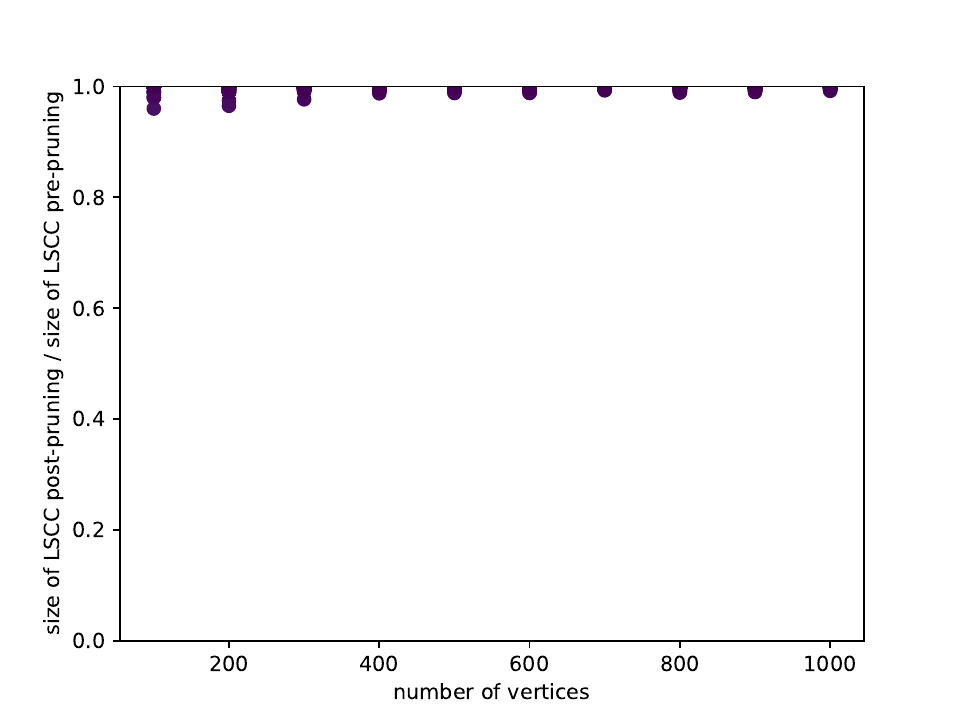}
         \caption{$k=6$, $p=0.97$}
         \label{fig:LSCCred.97-6}
     \end{subfigure}
     \hfill
     \begin{subfigure}[b]{0.325\textwidth}
         \centering\includegraphics[width=\textwidth]{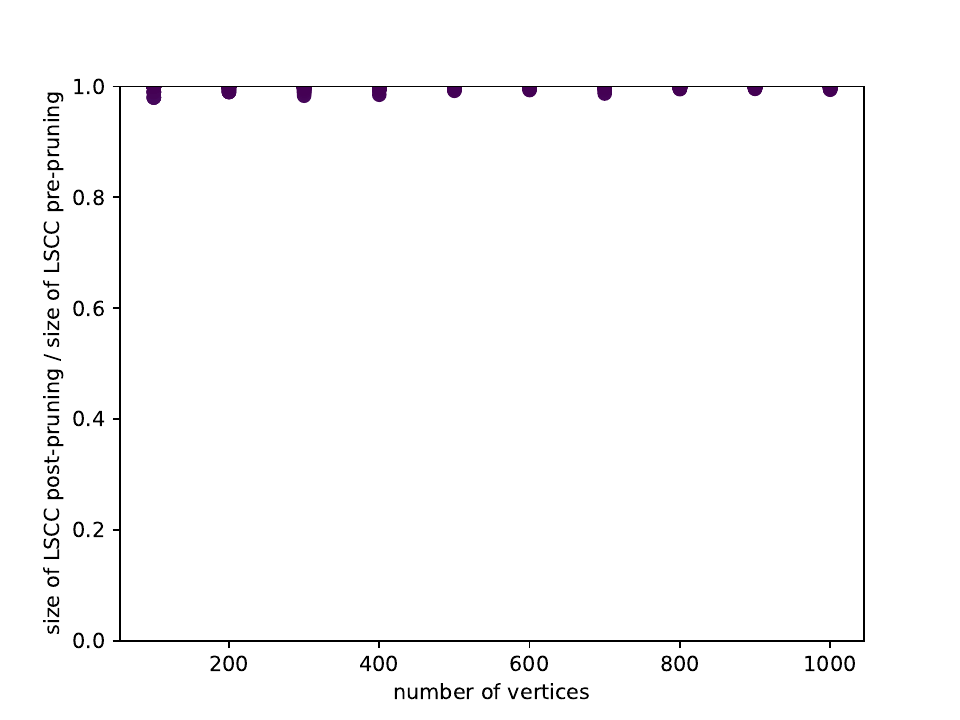}
         \caption{$k=6$, $p=0.99$}
         \label{fig:LSCCred.99-5-6}
     \end{subfigure}
     \hfill
     \caption[Scatter plots of LSSC reduction factor $\rho_L$ for $k=6$.]{Scatter plots of LSSC reduction factor $\rho_L$ obtained from simulations in the parameter regime $n\in(100, 1000)$ for $k=6$.}
     \label{fig:LSCCred-6-all}
\end{figure}

Some of our results are summarised in Figures~\ref{fig:LSCCred-4-all}, \ref{fig:LSCCred-5-all} and \ref{fig:LSCCred-6-all}.

\subsection{LSCC coverage factor}

The LSCC coverage factor $\alpha_L$, being a quantity related both to the size of the communication network and the allocation of computation jobs, depends on all the parameters $E_{\rm max}$ (or alternatively $p$), $\beta$, $n$ and $k$. 

\begin{figure}[p]
     \centering
     \begin{subfigure}[b]{0.325\textwidth}
         \centering\includegraphics[width=\textwidth]{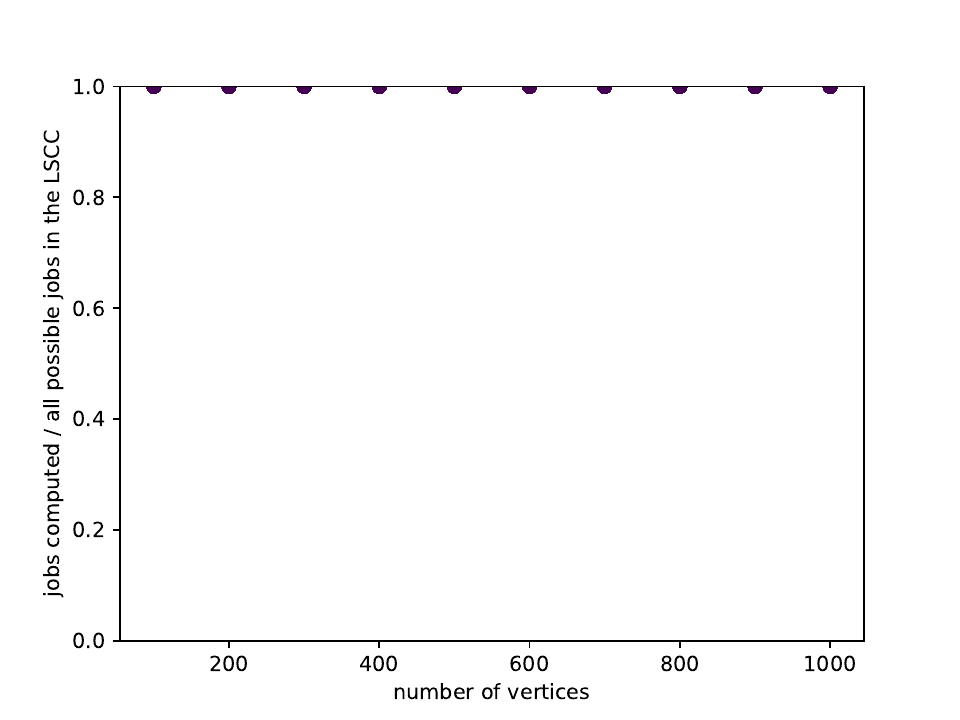}
         \caption{$k=4$, $p=0.1$, $\beta=0.1$}
         \label{fig:LSCCcov.1.1-4}
     \end{subfigure}
     \hfill
     \begin{subfigure}[b]{0.325\textwidth}
         \centering\includegraphics[width=\textwidth]{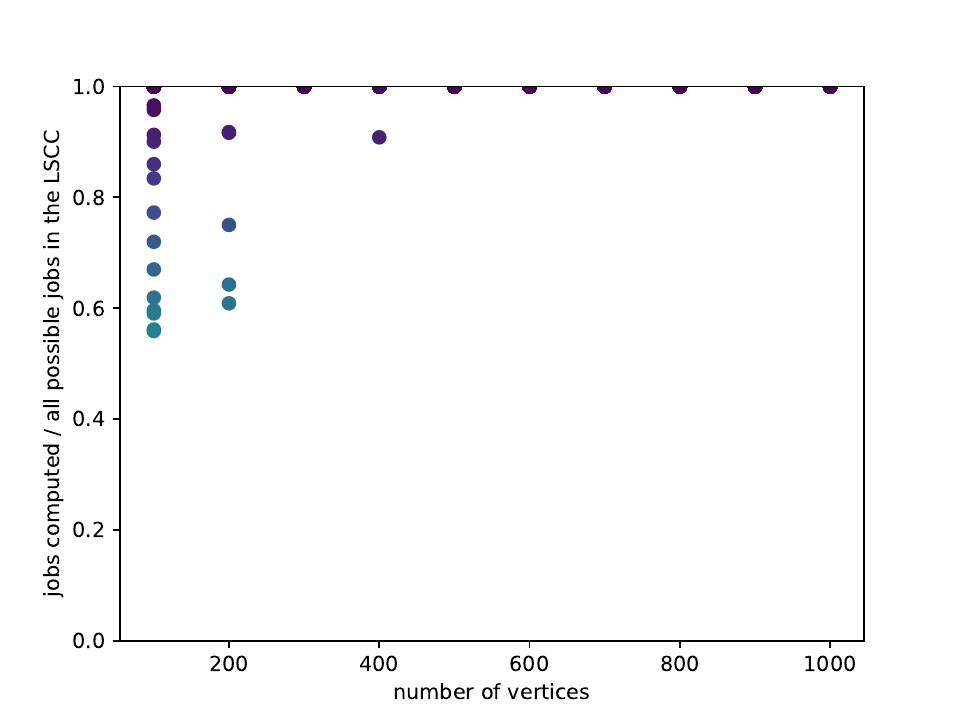}
         \caption{$k=4$, $p=0.1$, $\beta=0.7$}
         \label{fig:LSCCcov.1.7-4}
     \end{subfigure}
     \hfill
     \begin{subfigure}[b]{0.325\textwidth}
         \centering\includegraphics[width=\textwidth]{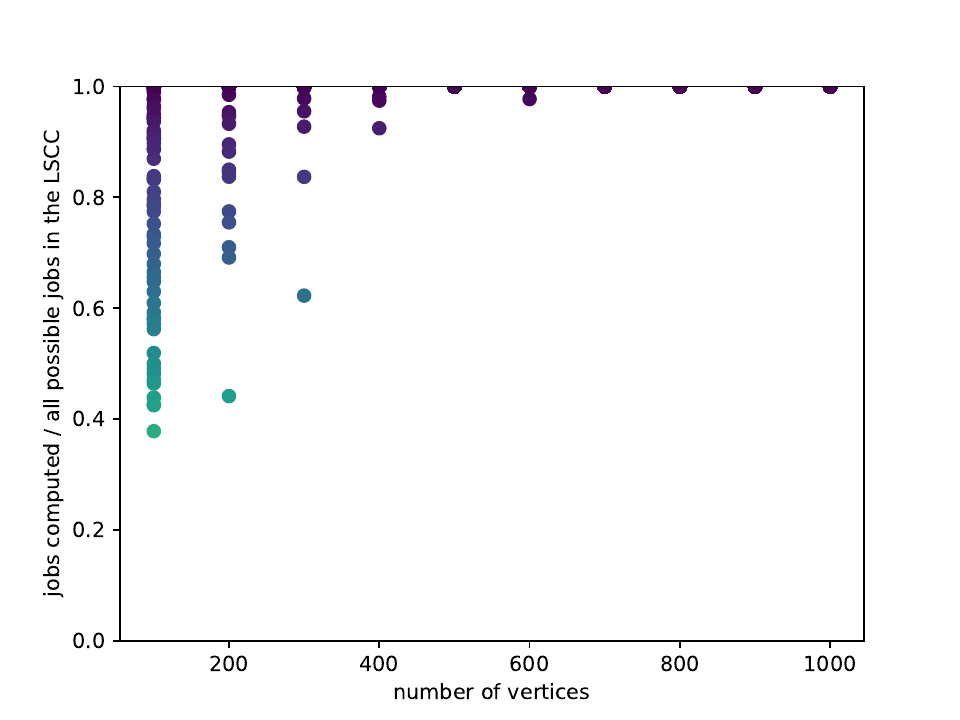}
         \caption{$k=4$, $p=0.1$, $\beta=1.0$}
         \label{fig:LSCCcov.1.0-4}
     \end{subfigure}
     \hfill
         \begin{subfigure}[b]{0.325\textwidth}
         \centering\includegraphics[width=\textwidth]{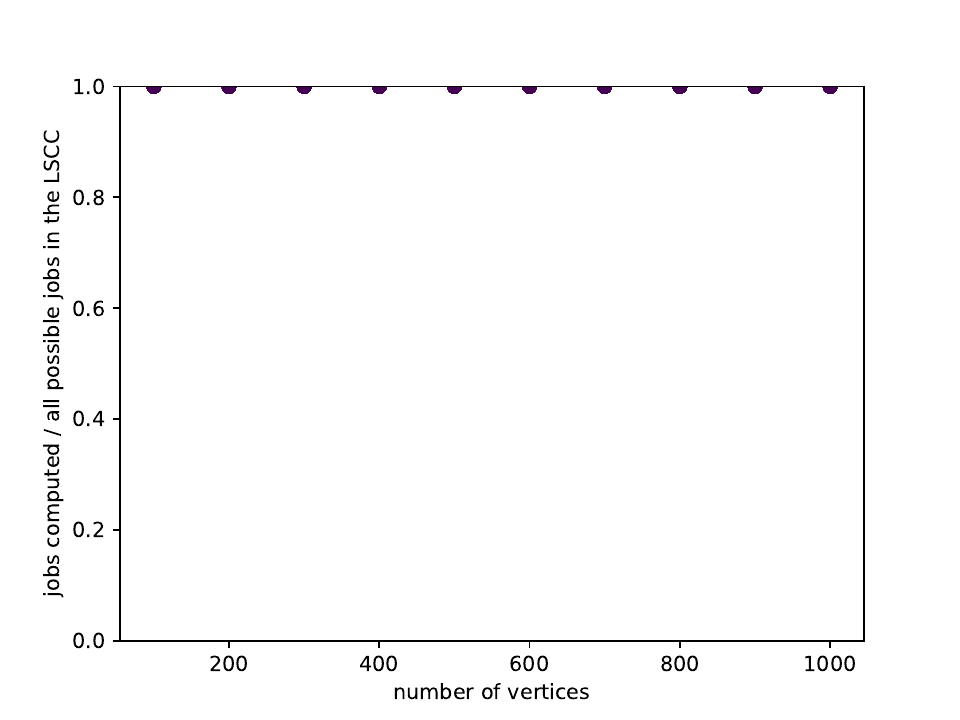}
         \caption{$k=4$, $p=0.2$, $\beta=0.1$}
         \label{fig:LSCCcov.2.1-4}
     \end{subfigure}
     \hfill
     \begin{subfigure}[b]{0.325\textwidth}
         \centering\includegraphics[width=\textwidth]{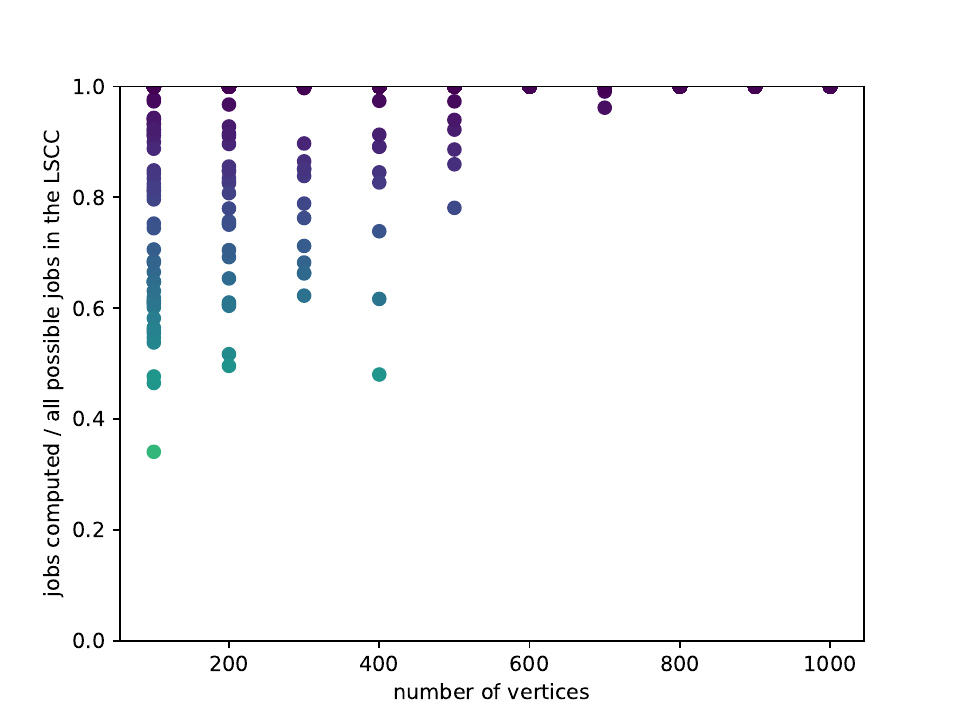}
         \caption{$k=4$, $p=0.2$, $\beta=0.6$}
         \label{fig:LSCCcov.2.6-4}
     \end{subfigure}
     \hfill
     \begin{subfigure}[b]{0.325\textwidth}
         \centering\includegraphics[width=\textwidth]{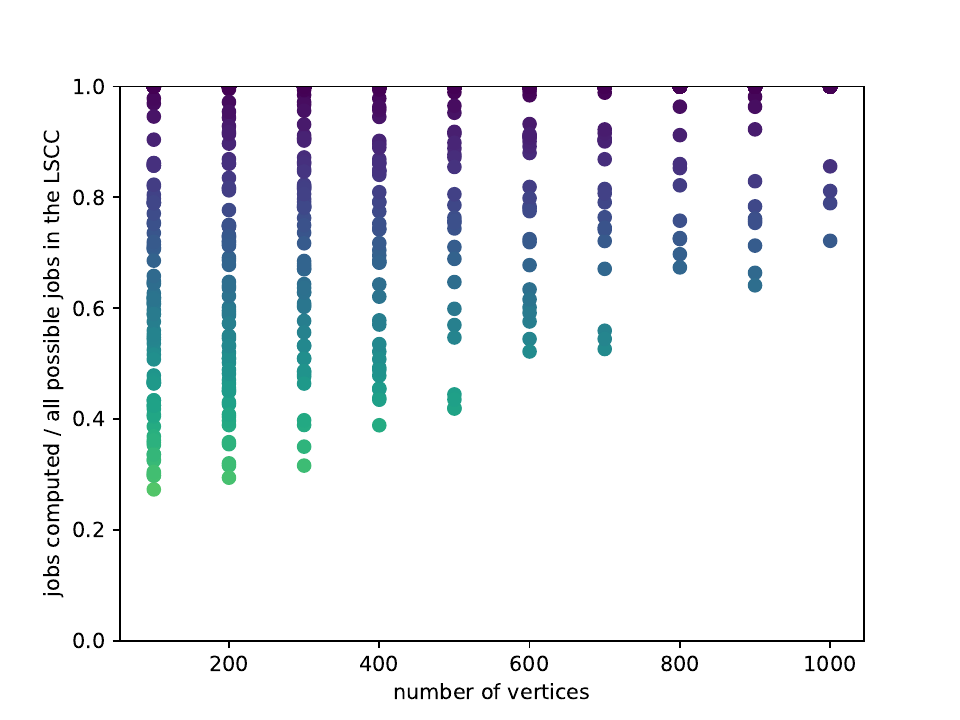}
         \caption{$k=4$, $p=0.2$, $\beta=1.0$}
         \label{fig:LSCCcov.2.0-4}
     \end{subfigure}
     \hfill
          \begin{subfigure}[b]{0.325\textwidth}
         \centering\includegraphics[width=\textwidth]{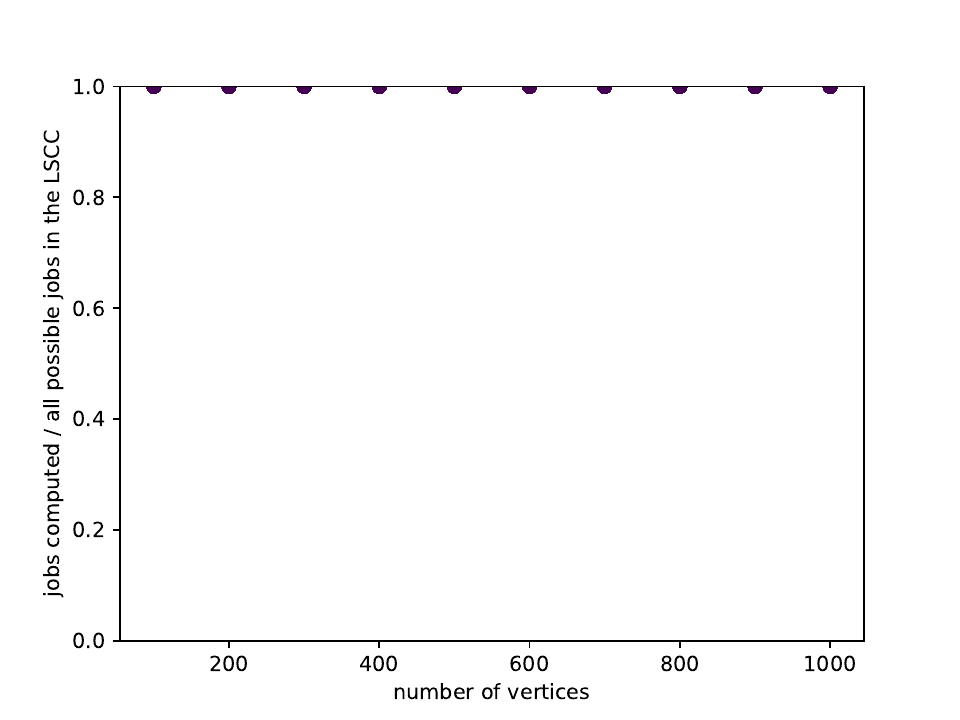}
         \caption{$k=4$, $p=0.7$, $\beta=0.1$}
         \label{fig:LSCCcov.7.1-4}
     \end{subfigure}
     \hfill
          \begin{subfigure}[b]{0.325\textwidth}
         \centering\includegraphics[width=\textwidth]{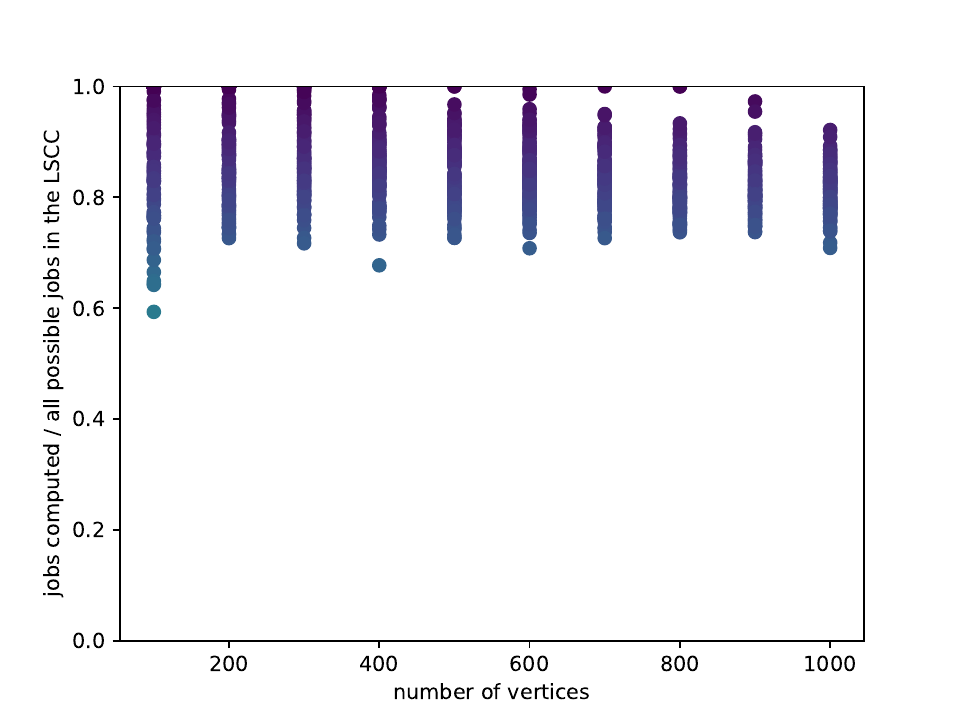}
         \caption{$k=4$, $p=0.7$, $\beta=0.4$}
         \label{fig:LSCCcov.7.4-4}
     \end{subfigure}
     \hfill
     \begin{subfigure}[b]{0.325\textwidth}
         \centering\includegraphics[width=\textwidth]{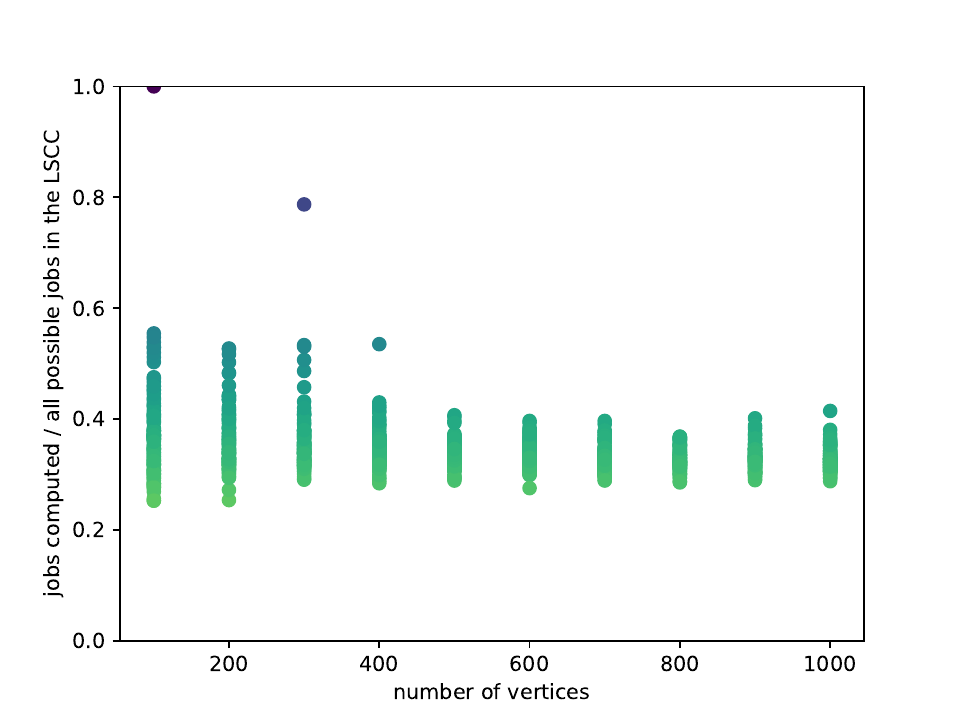}
         \caption{$k=4$, $p=0.7$, $\beta=1.0$}
         \label{fig:LSCCcov.7.0-4}
     \end{subfigure}
     \hfill
          \begin{subfigure}[b]{0.325\textwidth}
         \centering\includegraphics[width=\textwidth]{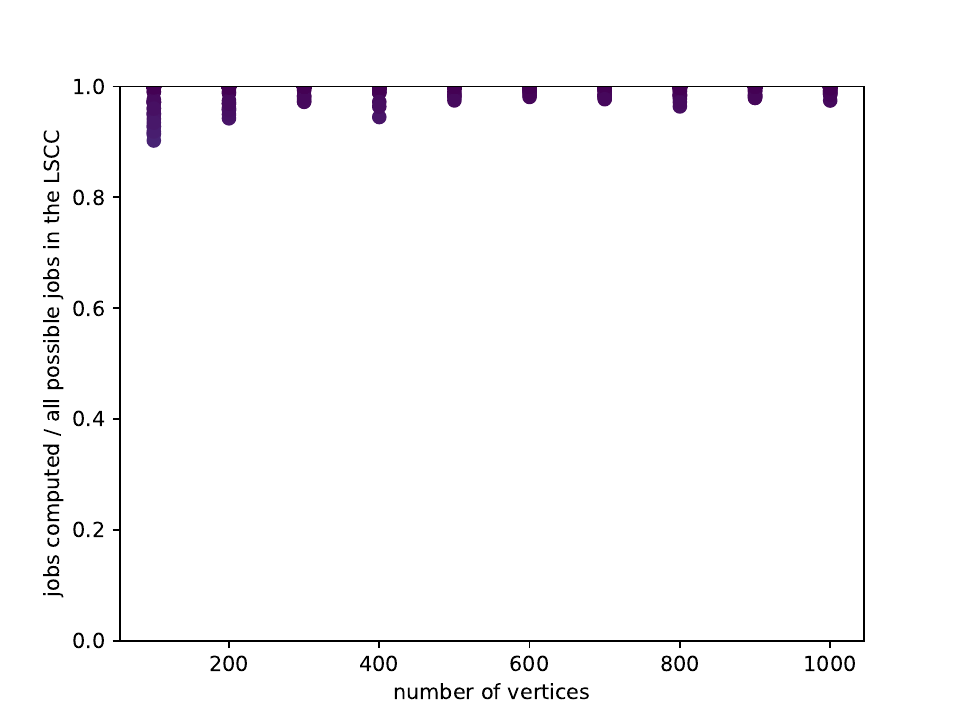}
         \caption{$k=4$, $p=0.99$, $\beta=0.5$}
         \label{fig:LSCCcov.99.5-4}
     \end{subfigure}
     \hfill
          \begin{subfigure}[b]{0.325\textwidth}
         \centering\includegraphics[width=\textwidth]{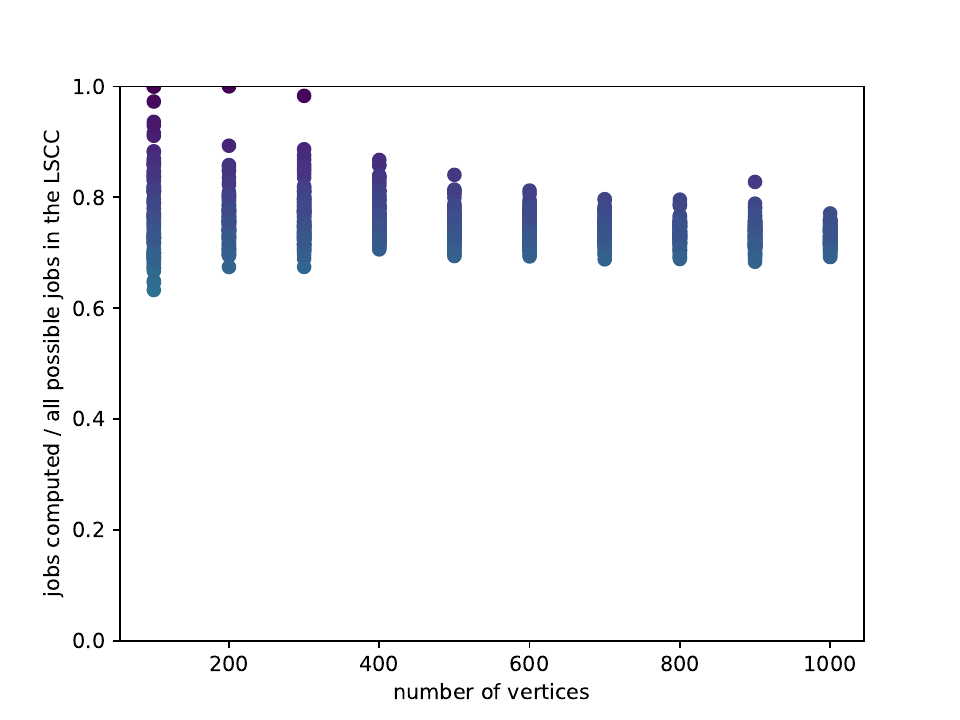}
         \caption{$k=4$, $p=0.99$, $\beta=0.7$}
         \label{fig:LSCCcov.99.7-4}
     \end{subfigure}
     \hfill
          \begin{subfigure}[b]{0.325\textwidth}
         \centering\includegraphics[width=\textwidth]{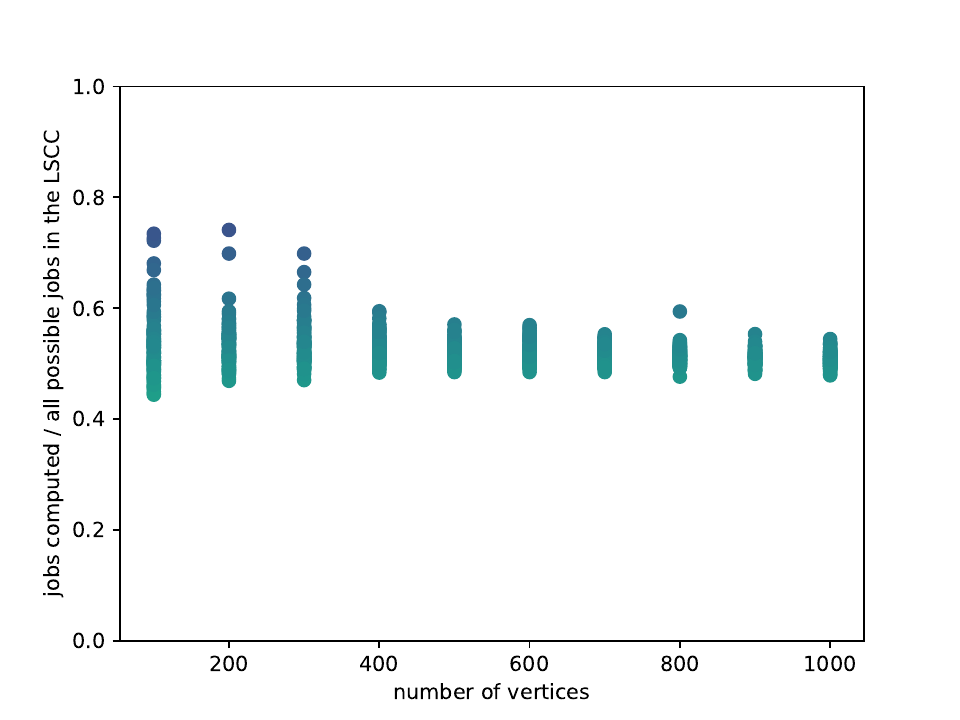}
         \caption{$k=4$, $p=0.99$, $\beta=1.0$}
         \label{fig:LSCCcov.99.0-4}
     \end{subfigure}
     \hfill
     \caption[Scatter plots of LSSC coverage factor $\alpha_L$ for $k=4$]{Scatter plots of LSSC coverage factor $\alpha_L$ obtained from simulations in the parameter regime $n\in(100, 1000)$ for $k=4$. Each row has $p$ fixed and $\beta$ is chosen to showcase different possible behaviour. }
     \label{fig:LSCCcov-4-all}
\end{figure}

The behaviour of $\alpha_L$ might be slightly misleading. For example, for $p=0.1$, $k=4$, and $\beta$ low enough (see Figure~\ref{fig:LSCCcov.1.1-4}), $\alpha_L$ is concentrated at the value 1. When interpreting these results, one must realise that this is because the LSCC itself is rather small, and therefore the assignment of all the jobs available within the LSCC, which are furthermore energetically cheap due to $\beta$ being small, is almost trivial. If $\beta$ is close to one, then even on a small LSCC there might not be enough energy to carry out all the computational jobs. As $p$ becomes slightly larger, the results are quite spread out and the influence of $\beta$ becomes significant. Finally, as $p$ becomes closer to 1, $\beta$ becomes the main controlling parameter.

\begin{figure}[p]
     \centering
     \begin{subfigure}[b]{0.325\textwidth}
         \centering\includegraphics[width=\textwidth]{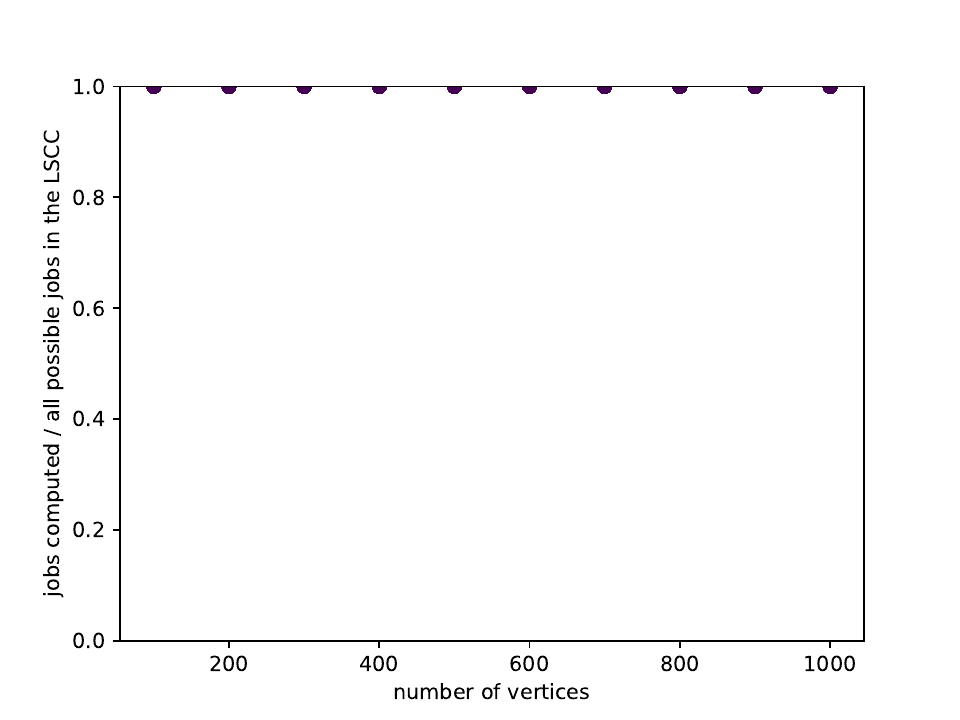}
         \caption{$k=5$, $p=0.1$, $\beta=0.1$}
         \label{fig:LSCCcov.1.1-5}
     \end{subfigure}
     \hfill
     \begin{subfigure}[b]{0.325\textwidth}
         \centering\includegraphics[width=\textwidth]{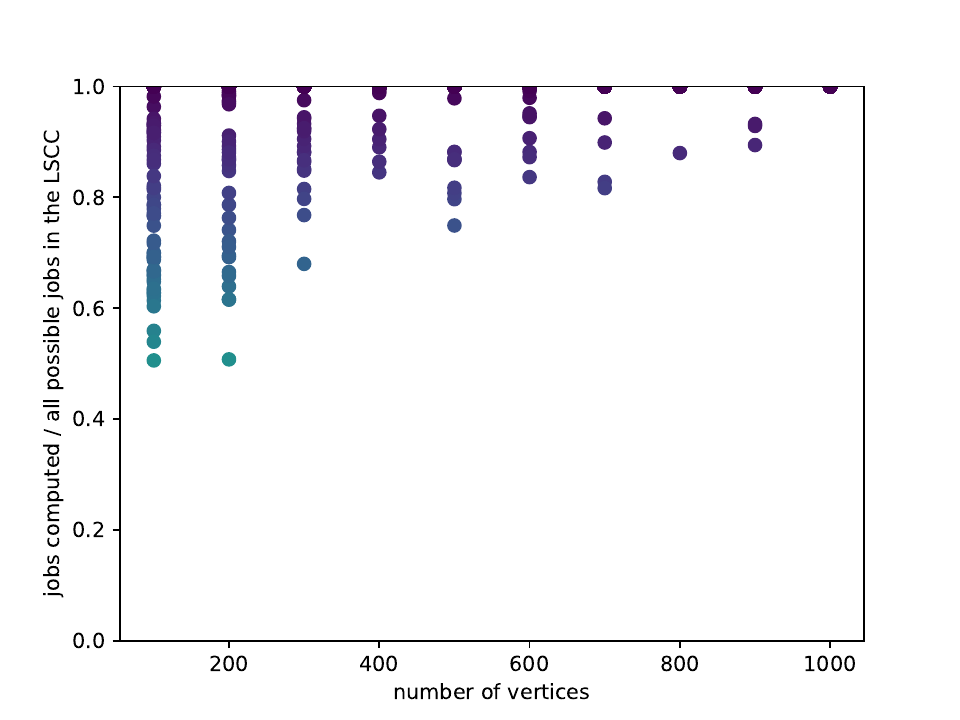}
         \caption{$k=5$, $p=0.1$, $\beta=0.4$}
         \label{fig:LSCCcov.1.4-5}
     \end{subfigure}
     \hfill
     \begin{subfigure}[b]{0.325\textwidth}
         \centering\includegraphics[width=\textwidth]{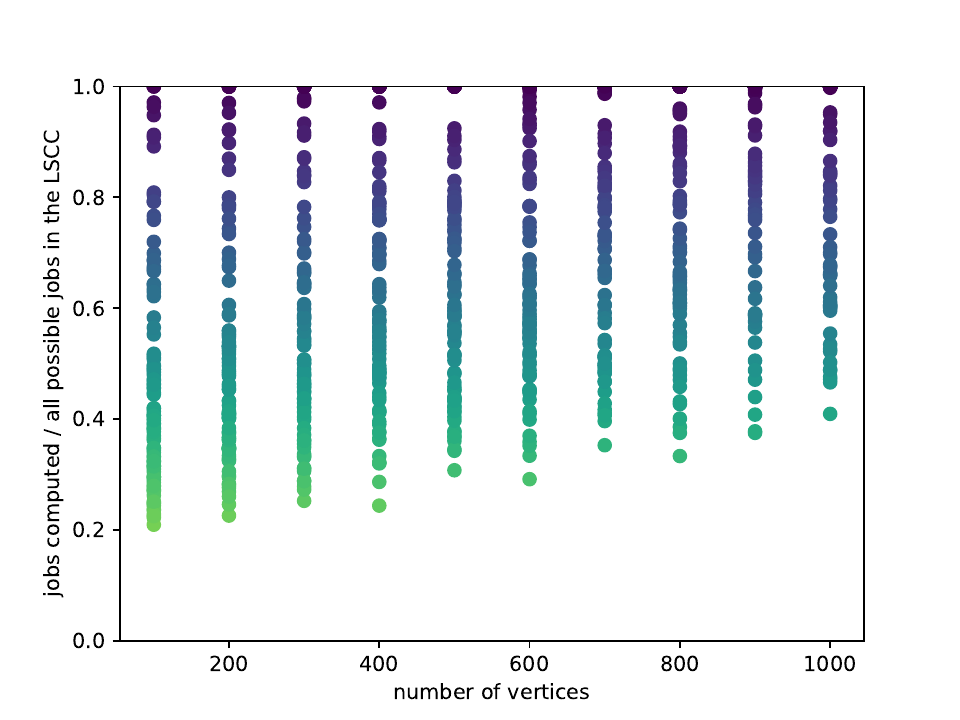}
         \caption{$k=5$, $p=0.1$, $\beta=1.0$}
         \label{fig:LSCCcov.1.0-5}
     \end{subfigure}
     \hfill
         \begin{subfigure}[b]{0.325\textwidth}
         \centering\includegraphics[width=\textwidth]{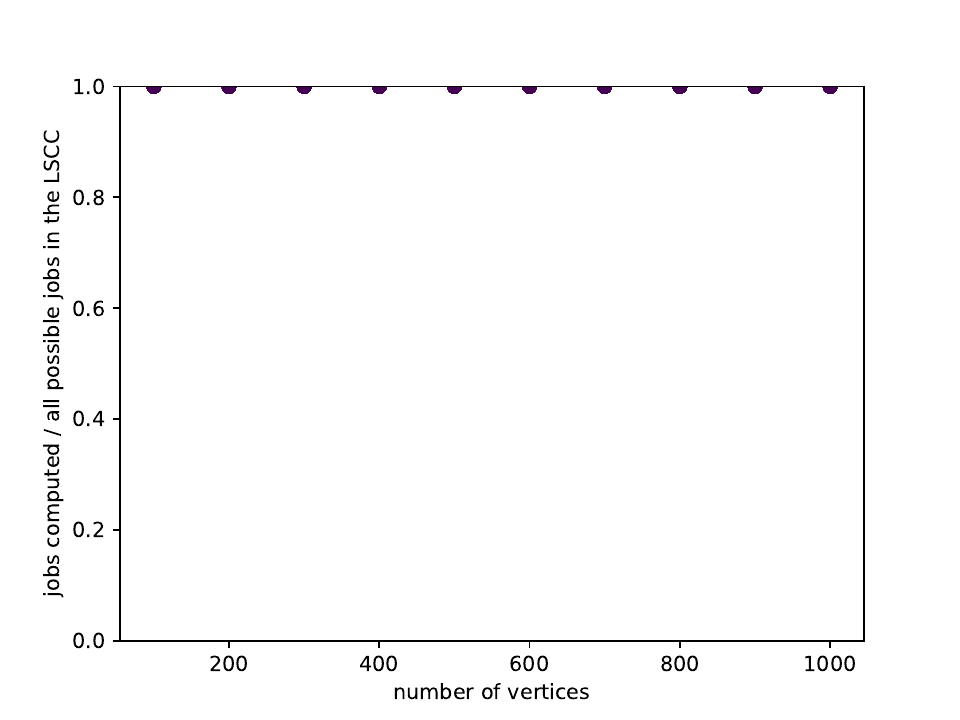}
         \caption{$k=5$, $p=0.3$, $\beta=0.1$}
         \label{fig:LSCCcov.3.1-5}
     \end{subfigure}
     \hfill
     \begin{subfigure}[b]{0.325\textwidth}
         \centering\includegraphics[width=\textwidth]{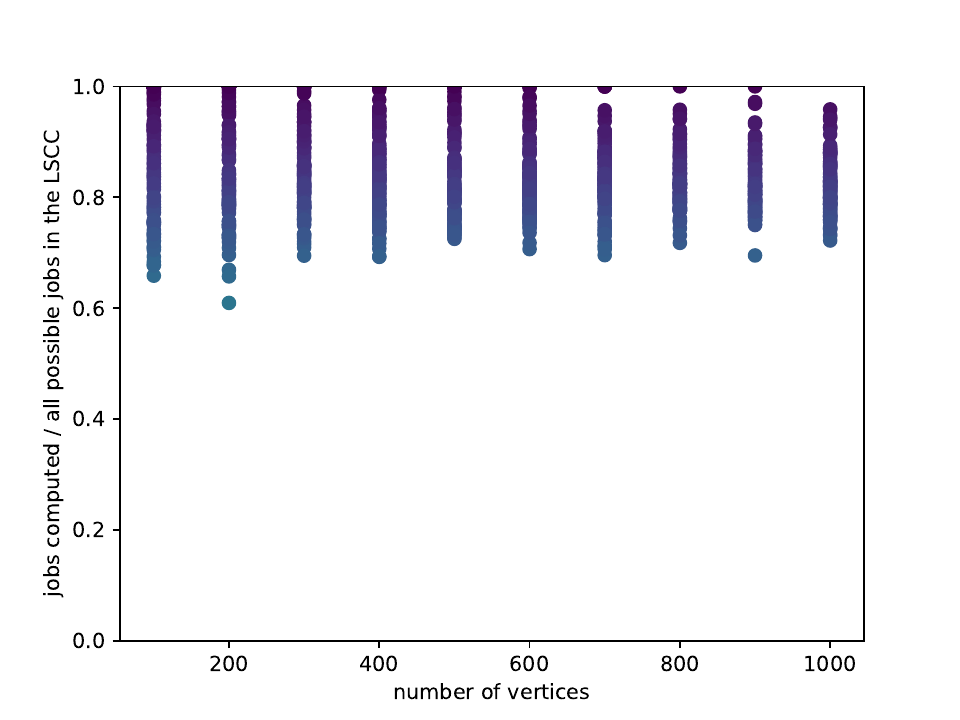}
         \caption{$k=5$, $p=0.3$, $\beta=0.3$}
         \label{fig:LSCCcov.3.3-5}
     \end{subfigure}
     \hfill
     \begin{subfigure}[b]{0.325\textwidth}
         \centering\includegraphics[width=\textwidth]{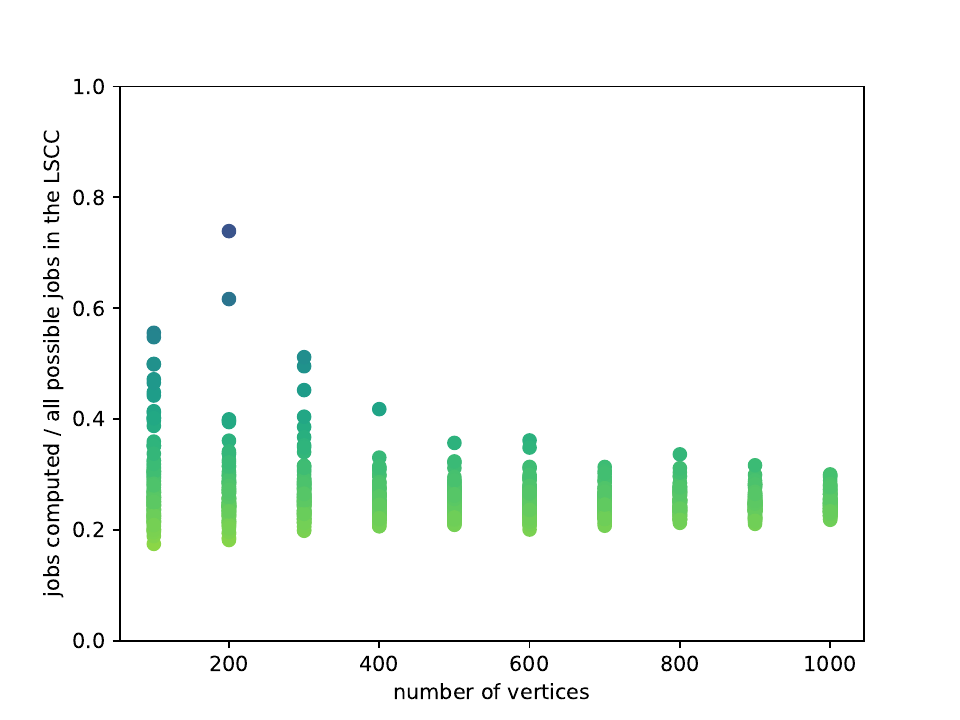}
         \caption{$k=5$, $p=0.3$, $\beta=1.0$}
         \label{fig:LSCCcov.3.0-5}
     \end{subfigure}
     \hfill
          \begin{subfigure}[b]{0.325\textwidth}
         \centering\includegraphics[width=\textwidth]{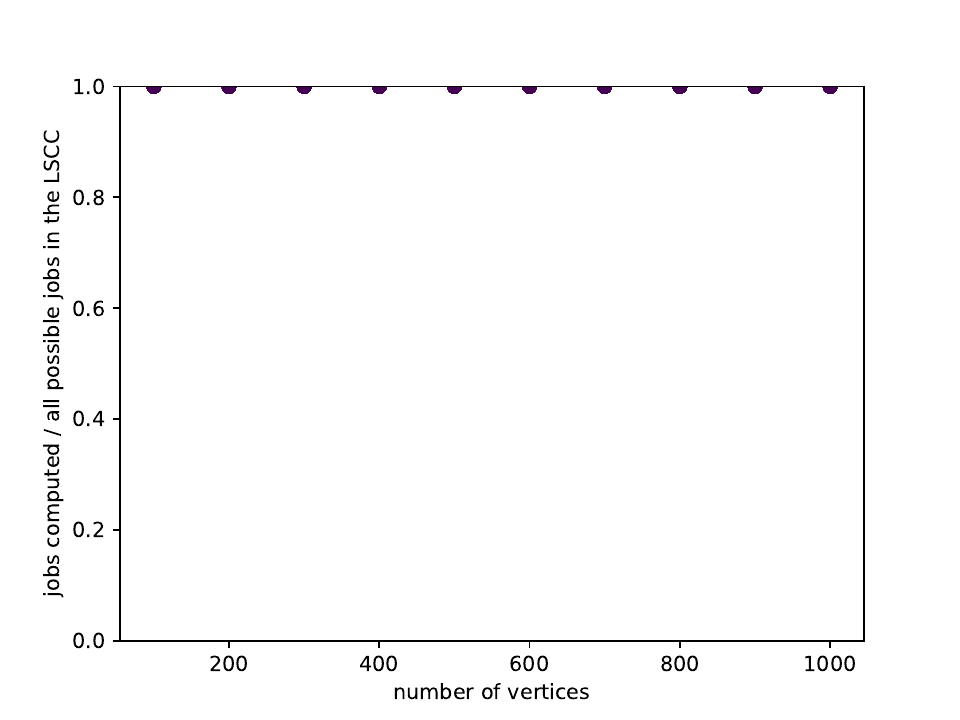}
         \caption{$k=5$, $p=0.7$, $\beta=0.1$}
         \label{fig:LSCCcov.7.1-5}
     \end{subfigure}
     \hfill
          \begin{subfigure}[b]{0.325\textwidth}
         \centering\includegraphics[width=\textwidth]{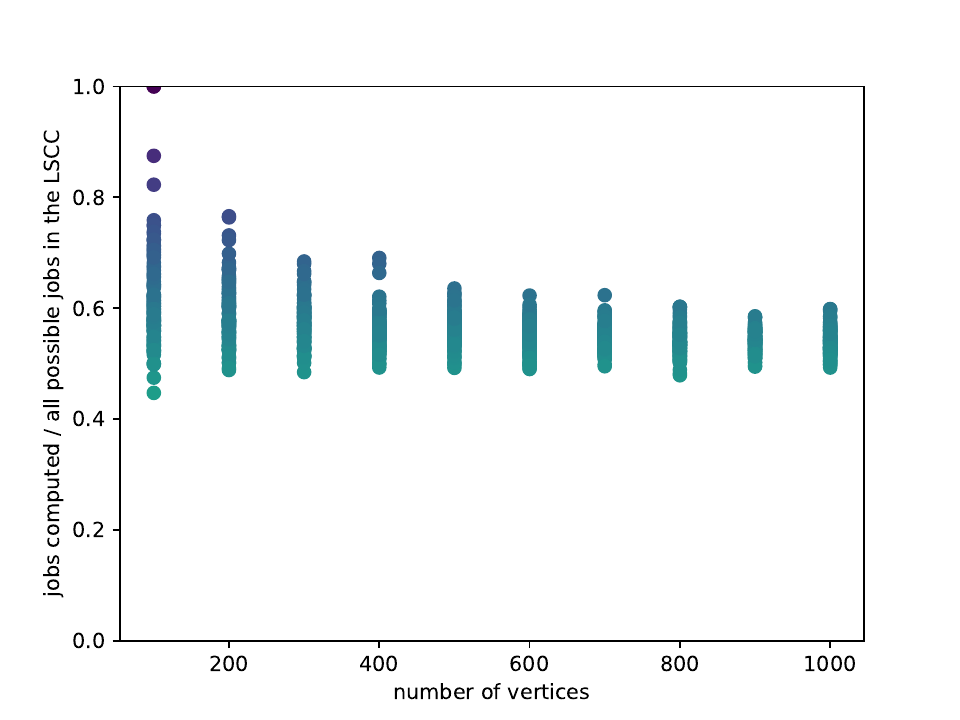}
         \caption{$k=5$, $p=0.7$, $\beta=0.5$}
         \label{fig:LSCCcov.7.5-5}
     \end{subfigure}
     \hfill
     \begin{subfigure}[b]{0.325\textwidth}
         \centering\includegraphics[width=\textwidth]{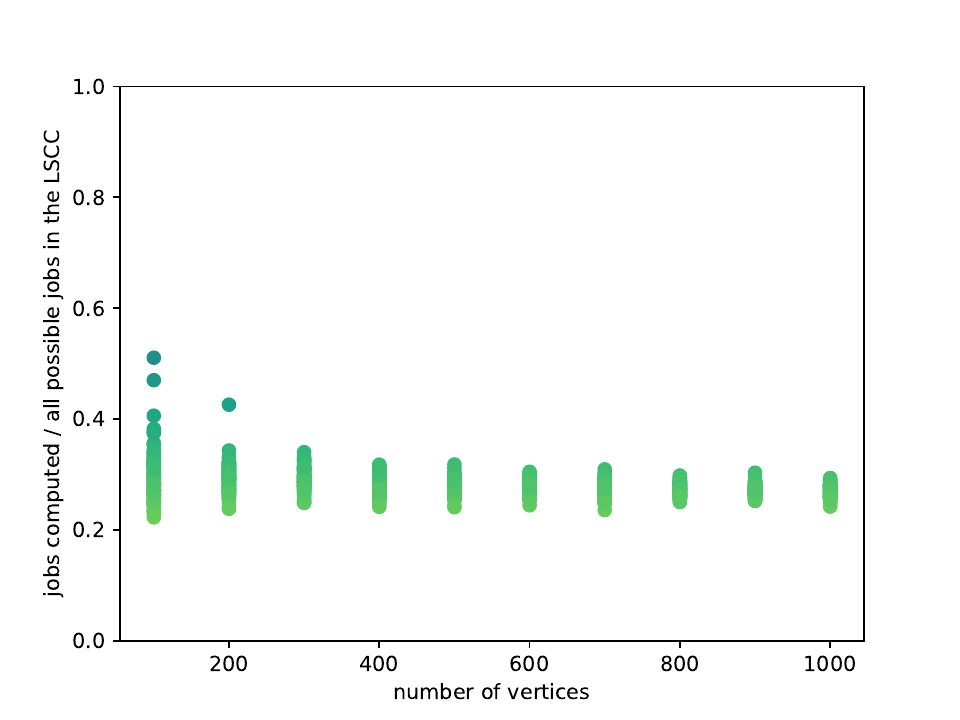}
         \caption{$k=5$, $p=0.7$, $\beta=1.0$}
         \label{fig:LSCCcov.7.0-5}
     \end{subfigure}
     \hfill
          \begin{subfigure}[b]{0.325\textwidth}
         \centering\includegraphics[width=\textwidth]{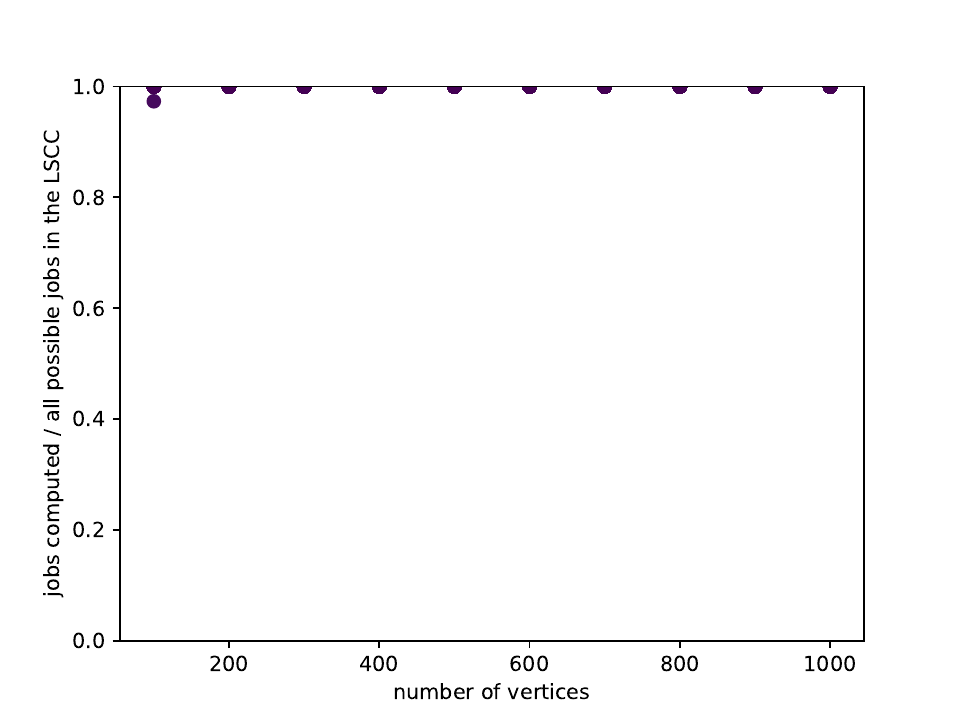}
         \caption{$k=5$, $p=0.99$, $\beta=0.4$}
         \label{fig:LSCCcov.99.4-5}
     \end{subfigure}
     \hfill
          \begin{subfigure}[b]{0.325\textwidth}
         \centering\includegraphics[width=\textwidth]{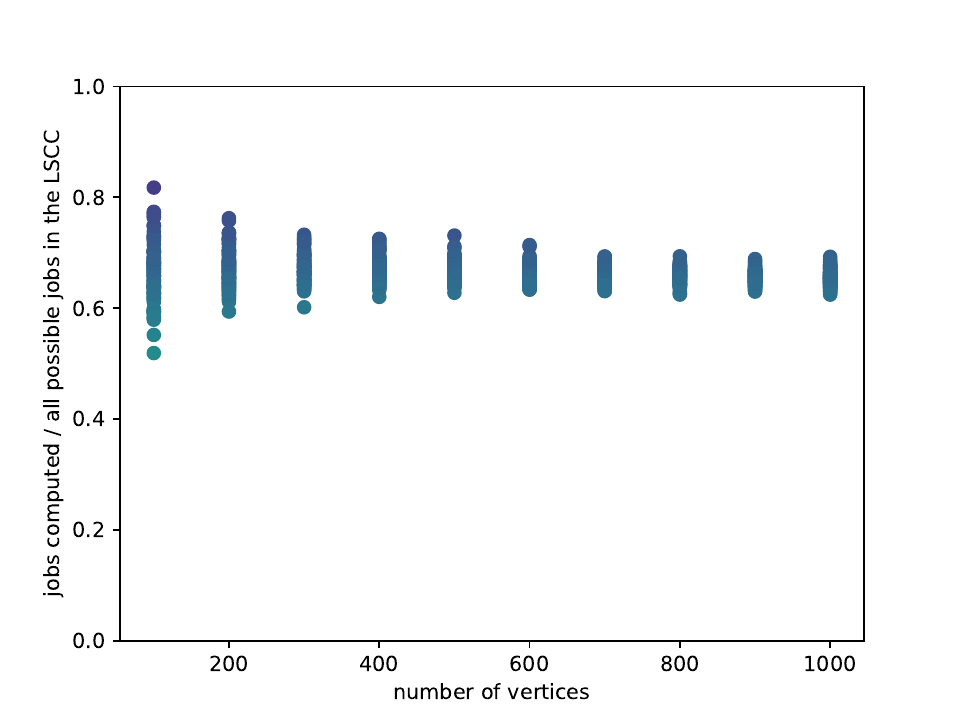}
         \caption{$k=5$, $p=0.99$, $\beta=0.7$}
         \label{fig:LSCCcov.99.7-5}
     \end{subfigure}
     \hfill
          \begin{subfigure}[b]{0.325\textwidth}
         \centering\includegraphics[width=\textwidth]{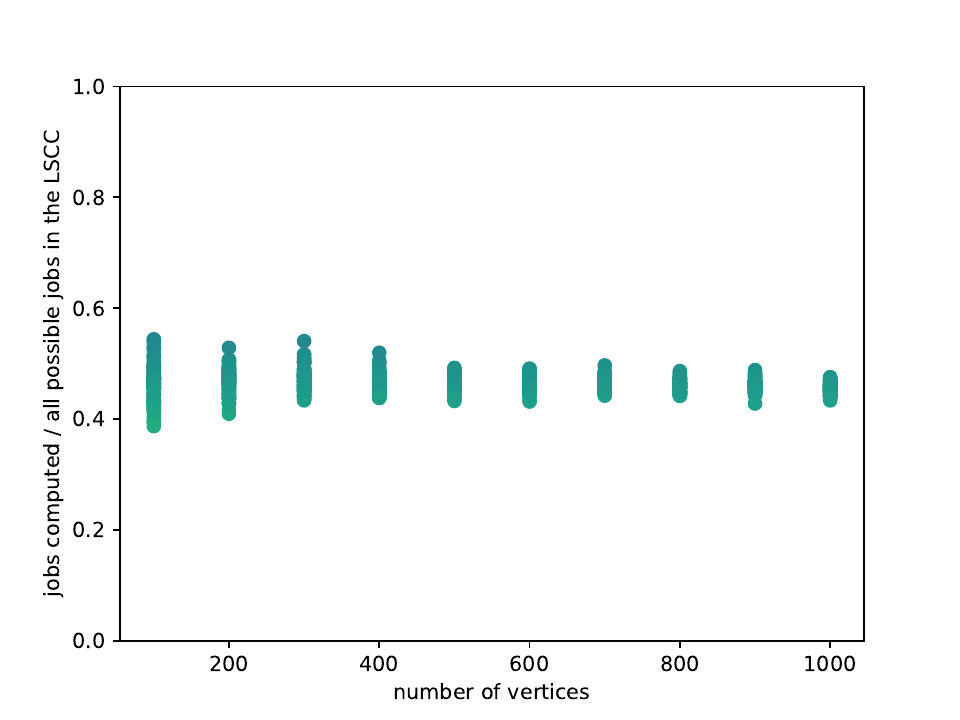}
         \caption{$k=5$, $p=0.99$, $\beta=1.0$}
         \label{fig:LSCCcov.99.0-5}
     \end{subfigure}
     \hfill
     \caption[Scatter plots of LSSC coverage factor $\alpha_L$ for $k=5$]{Scatter plots of LSSC coverage factor $\alpha_L$ obtained from simulations in the parameter regime $n\in(100, 1000)$ for $k=5$. Each row has $p$ fixed and $\beta$ is chosen to showcase different possible behaviour. }
     \label{fig:LSCCcov-5-all}
\end{figure}

\begin{figure}[p]
     \centering
     \begin{subfigure}[b]{0.325\textwidth}
         \centering\includegraphics[width=\textwidth]{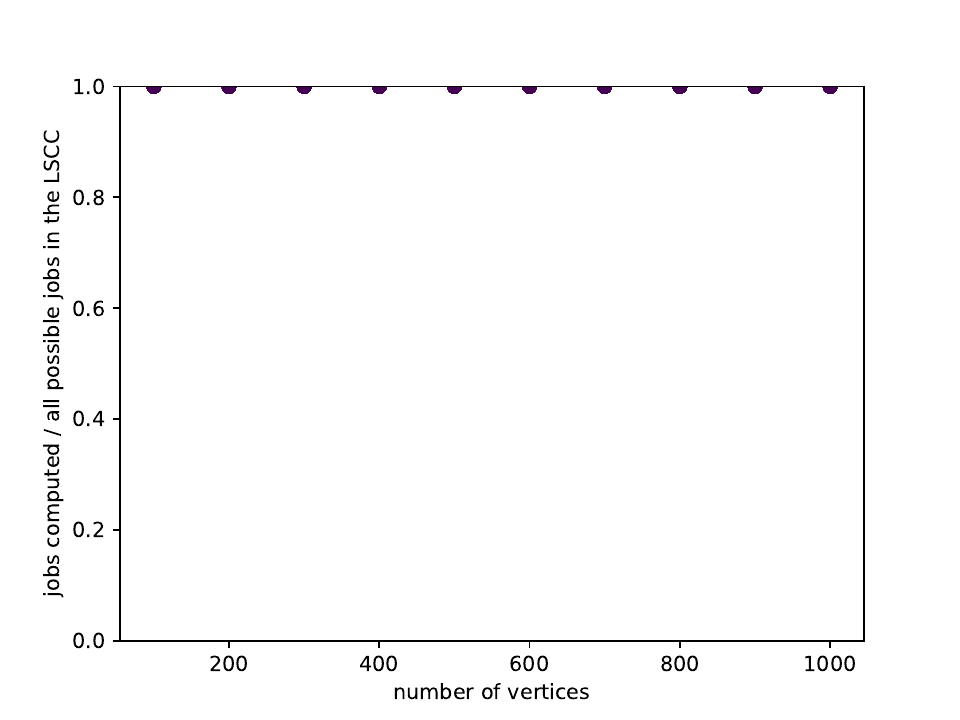}
         \caption{$k=6$, $p=0.1$, $\beta=0.1$}
         \label{fig:LSCCcov.1.1-6}
     \end{subfigure}
     \hfill
     \begin{subfigure}[b]{0.325\textwidth}
         \centering\includegraphics[width=\textwidth]{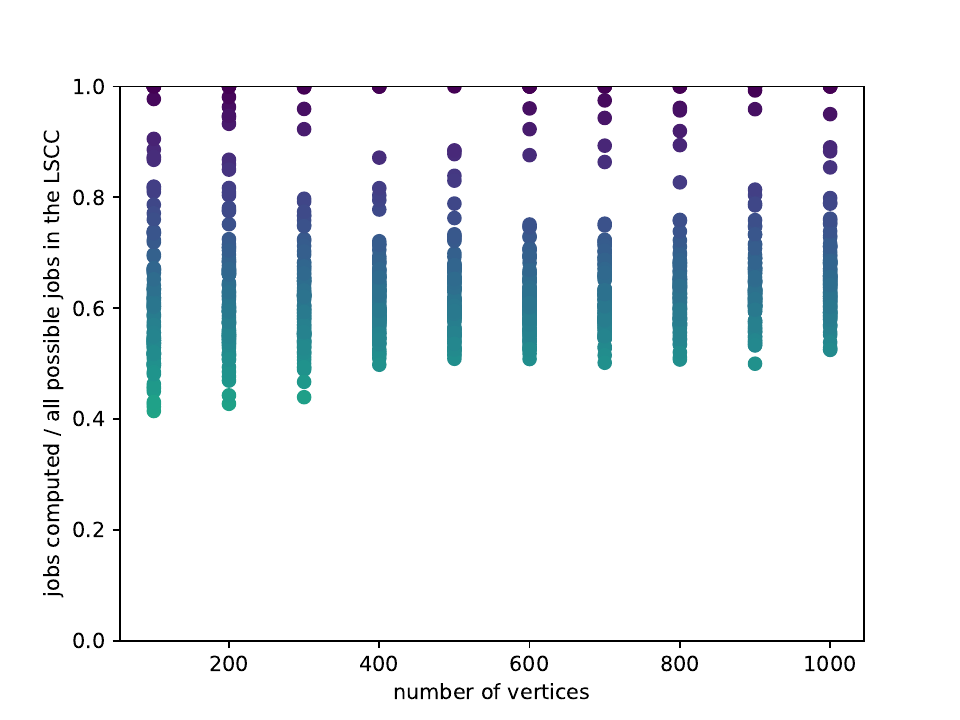}
         \caption{$k=6$, $p=0.1$, $\beta=0.4$}
         \label{fig:LSCCcov.1.4-6}
     \end{subfigure}
     \hfill
     \begin{subfigure}[b]{0.325\textwidth}
         \centering\includegraphics[width=\textwidth]{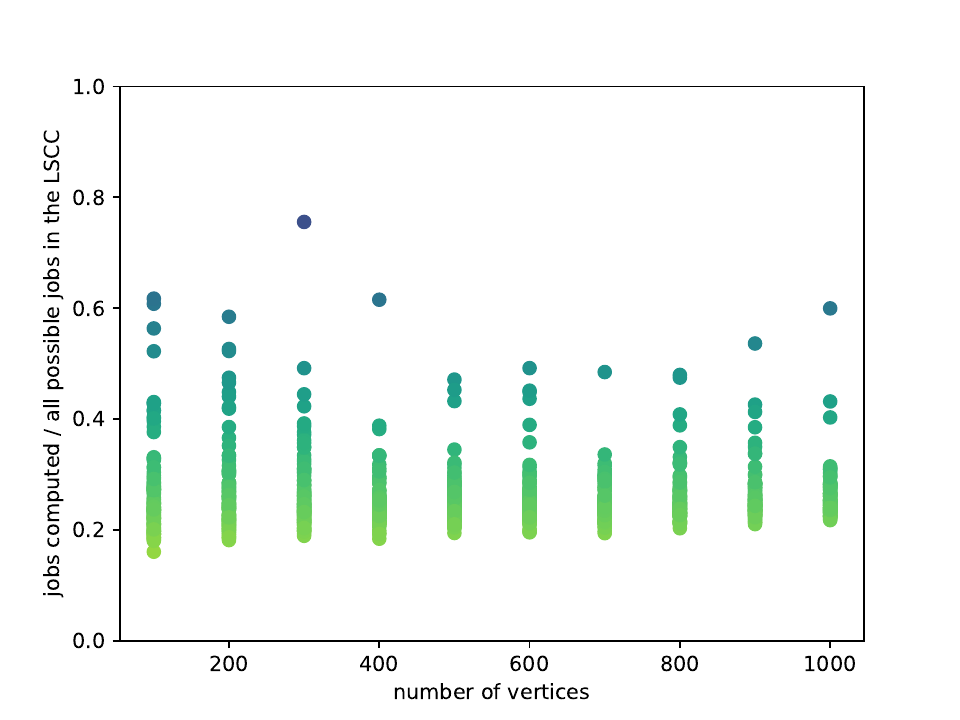}
         \caption{$k=6$, $p=0.1$, $\beta=1.0$}
         \label{fig:LSCCcov.1.0-6}
     \end{subfigure}
     \hfill
         \begin{subfigure}[b]{0.325\textwidth}
         \centering\includegraphics[width=\textwidth]{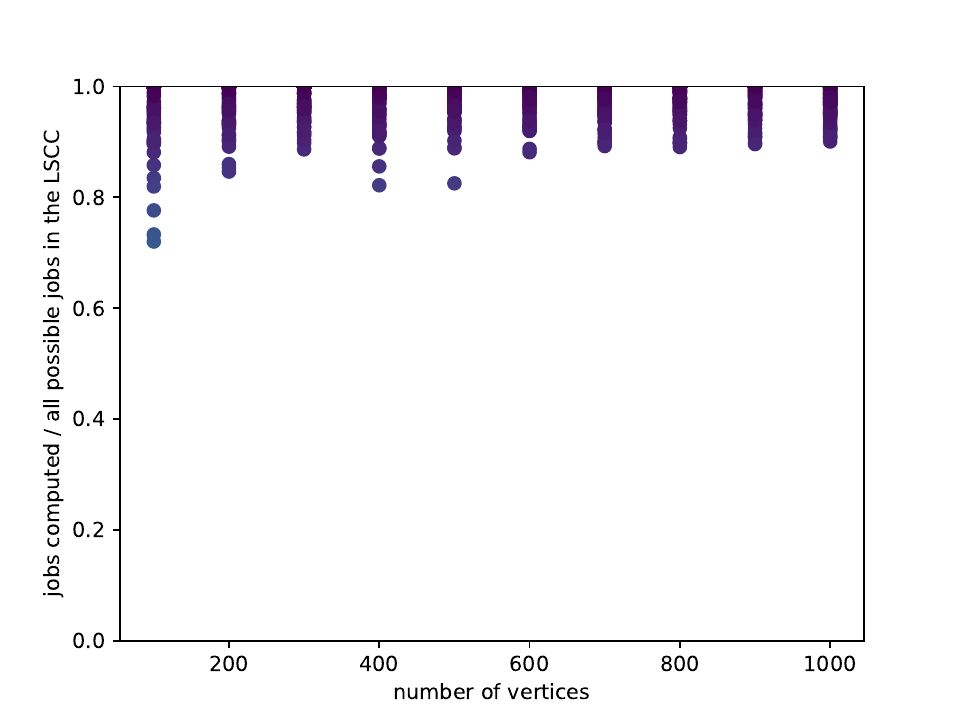}
         \caption{$k=6$, $p=0.3$, $\beta=0.2$}
         \label{fig:LSCCcov.3.2-6}
     \end{subfigure}
     \hfill
     \begin{subfigure}[b]{0.325\textwidth}
         \centering\includegraphics[width=\textwidth]{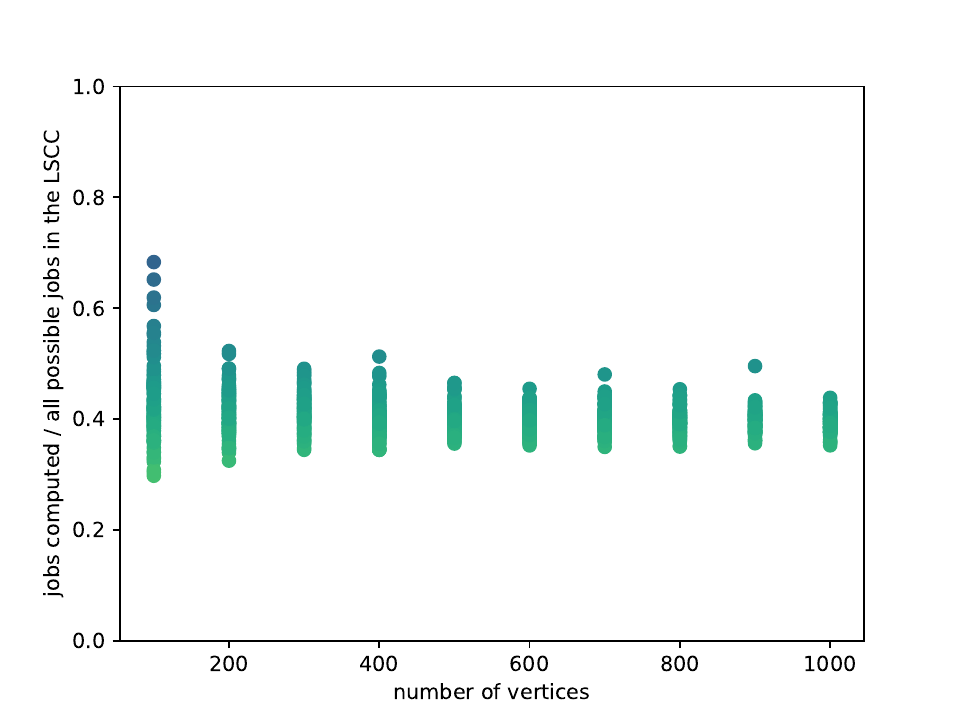}
         \caption{$k=6$, $p=0.3$, $\beta=0.5$}
         \label{fig:LSCCcov.3.5-6}
     \end{subfigure}
     \hfill
     \begin{subfigure}[b]{0.325\textwidth}
         \centering\includegraphics[width=\textwidth]{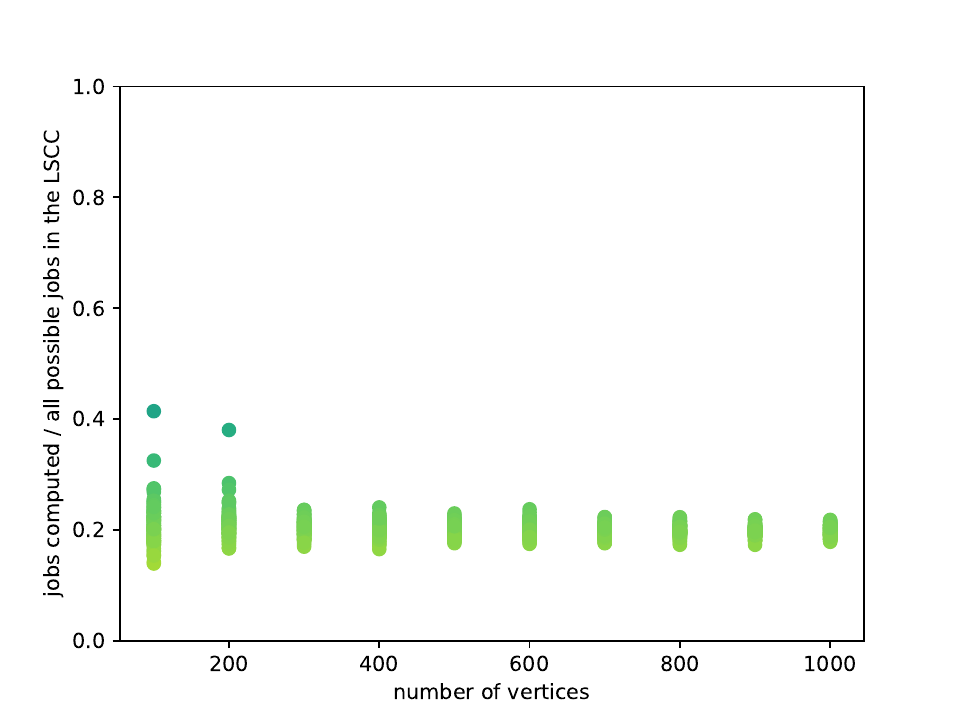}
         \caption{$k=6$, $p=0.3$, $\beta=1.0$}
         \label{fig:LSCCcov.3.0-6}
     \end{subfigure}
     \hfill
          \begin{subfigure}[b]{0.325\textwidth}
         \centering\includegraphics[width=\textwidth]{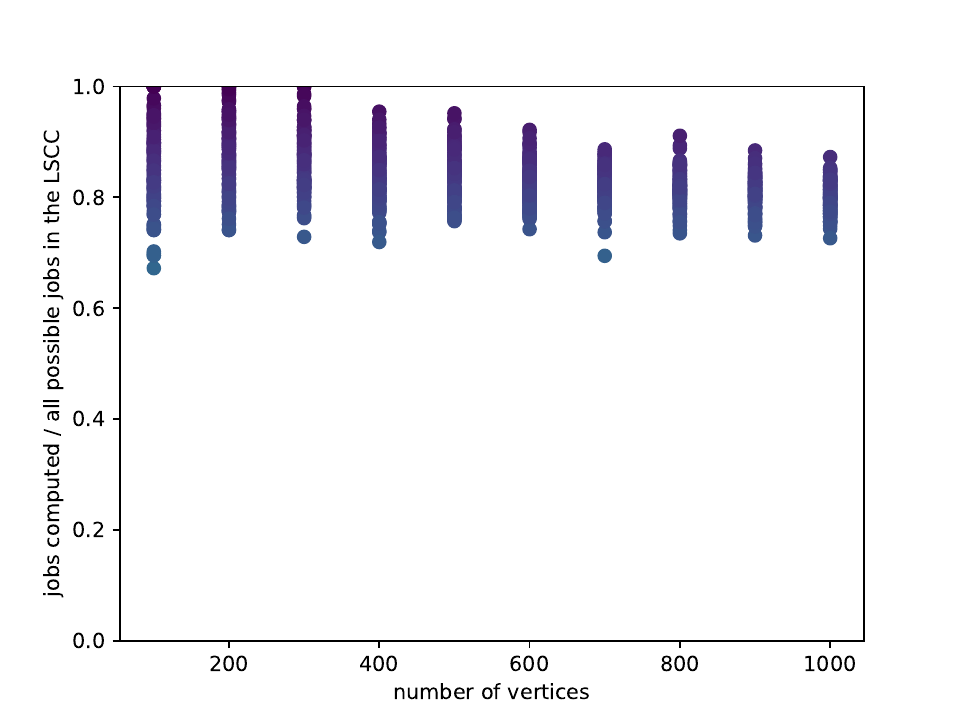}
         \caption{$k=6$, $p=0.7$, $\beta=0.3$}
         \label{fig:LSCCcov.7.3-6}
     \end{subfigure}
     \hfill
          \begin{subfigure}[b]{0.325\textwidth}
         \centering\includegraphics[width=\textwidth]{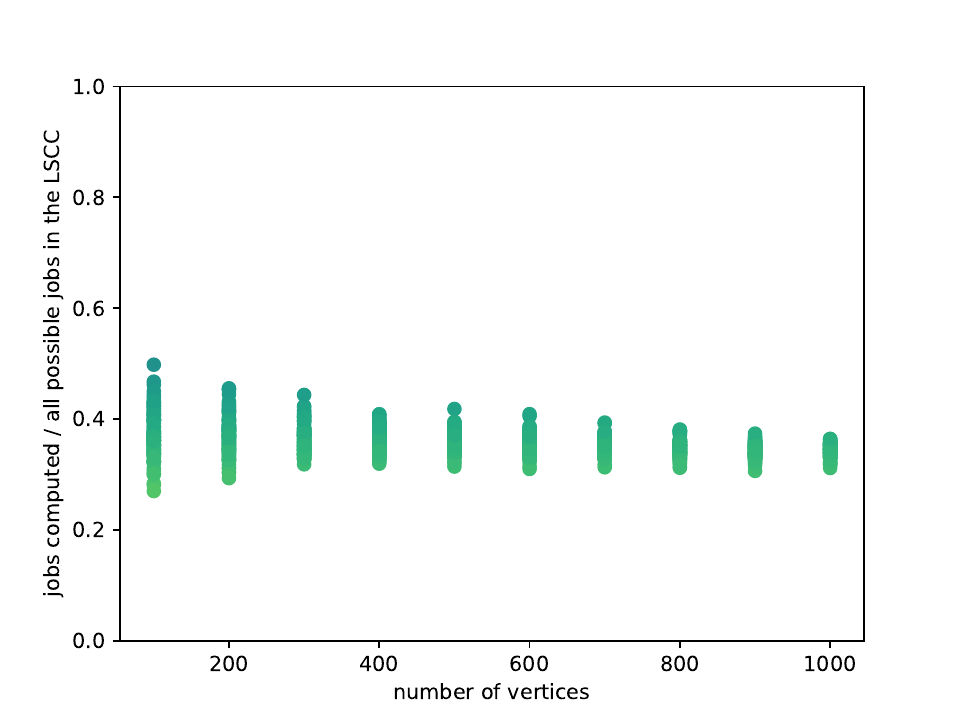}
         \caption{$k=6$, $p=0.7$, $\beta=0.7$}
         \label{fig:LSCCcov.7.7-6}
     \end{subfigure}
     \hfill
     \begin{subfigure}[b]{0.325\textwidth}
         \centering\includegraphics[width=\textwidth]{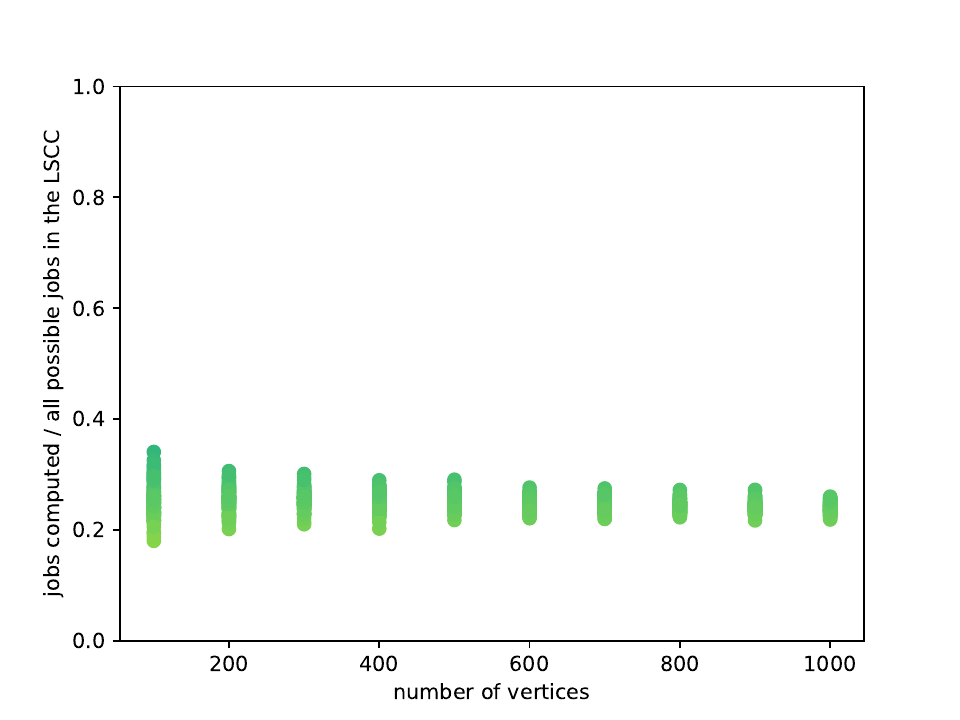}
         \caption{$k=6$, $p=0.7$, $\beta=1.0$}
         \label{fig:LSCCcov.7.0-6}
     \end{subfigure}
     \hfill
          \begin{subfigure}[b]{0.325\textwidth}
         \centering\includegraphics[width=\textwidth]{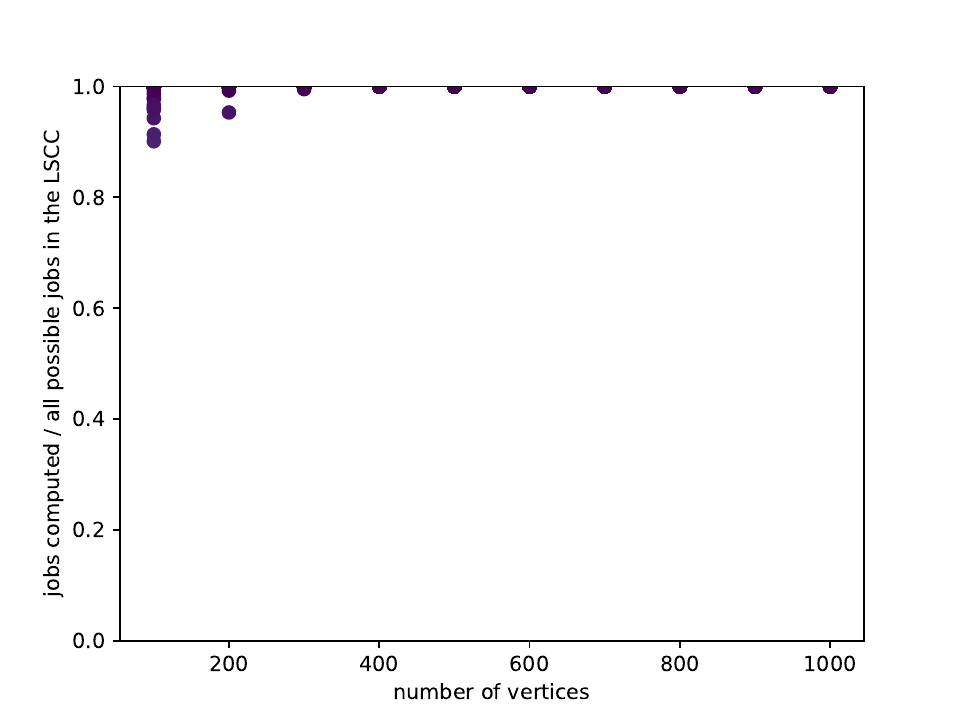}
         \caption{$k=6$, $p=0.99$, $\beta=0.4$}
         \label{fig:LSCCcov.99.4-6}
     \end{subfigure}
     \hfill
          \begin{subfigure}[b]{0.325\textwidth}
         \centering\includegraphics[width=\textwidth]{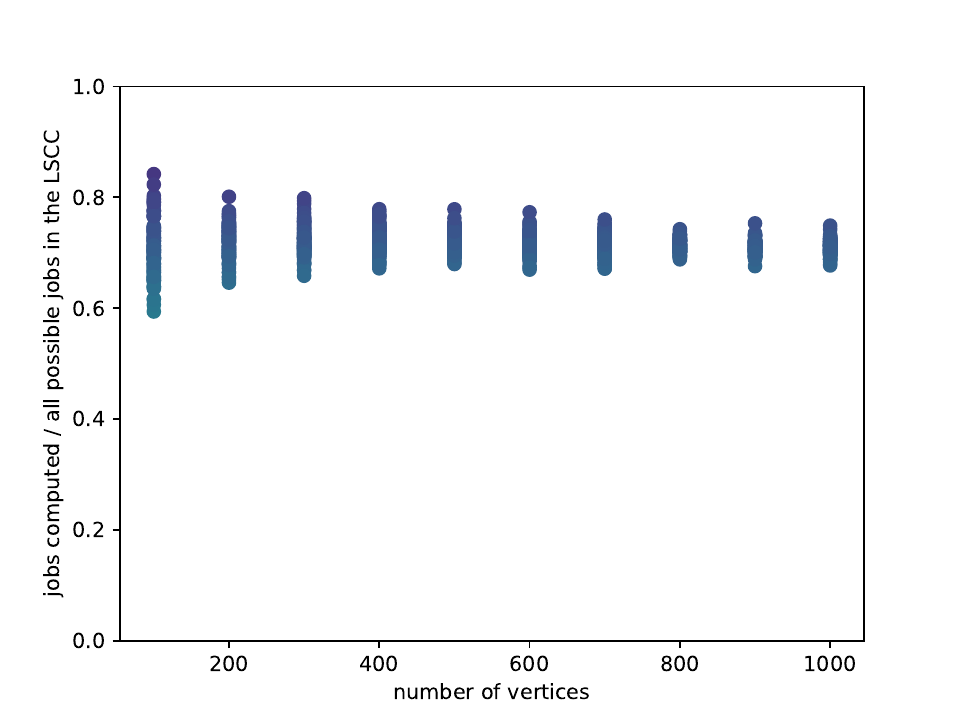}
         \caption{$k=6$, $p=0.99$, $\beta=0.6$}
         \label{fig:LSCCcov.99.6-6}
     \end{subfigure}
     \hfill
          \begin{subfigure}[b]{0.325\textwidth}
         \centering\includegraphics[width=\textwidth]{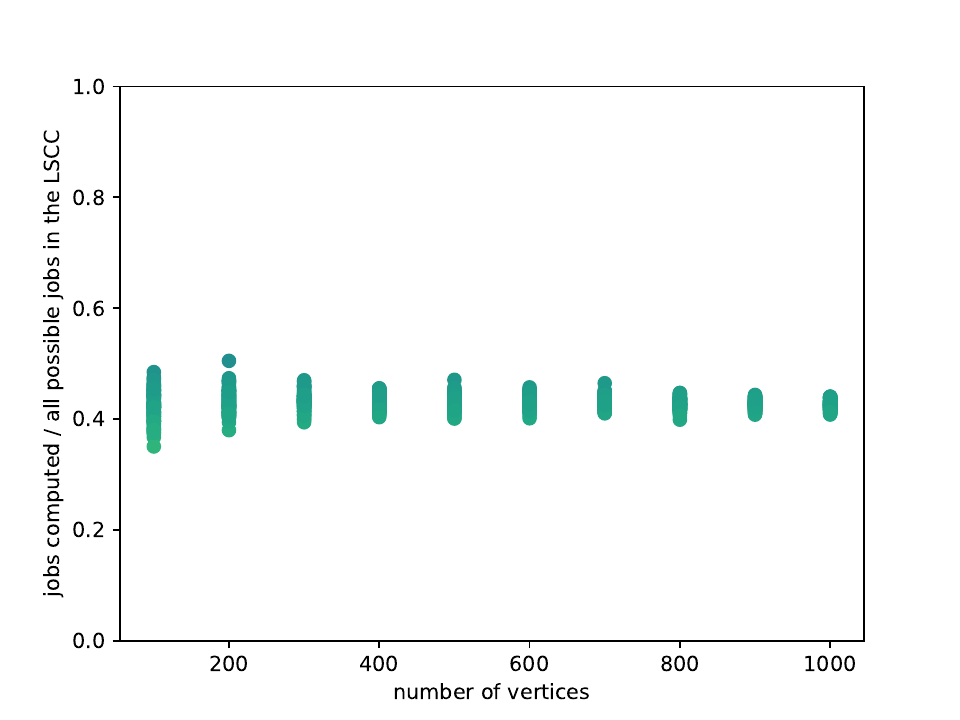}
         \caption{$k=6$, $p=0.99$, $\beta=1.0$}
         \label{fig:LSCCcov.99.0-6}
     \end{subfigure}
     \hfill
     \caption[Scatter plots of LSSC coverage factor $\alpha_L$ for $k=6$]{Scatter plots of LSSC coverage factor $\alpha_L$ obtained from simulations in the parameter regime $n\in(100, 1000)$ for $k=6$. Each row has $p$ fixed and $\beta$ is chosen to showcase different possible behaviour. }
     \label{fig:LSCCcov-6-all}
\end{figure}

Some of our results are plotted in Figures~\ref{fig:LSCCcov-4-all}, \ref{fig:LSCCcov-5-all} and \ref{fig:LSCCcov-6-all}.

\subsection{Coverage factor}

Unlike the LSCC coverage factor $\alpha_L$, the (global) coverage factor $\alpha$ is easy to interpret. The coverage $\alpha$ increases with $p$ and decreases with $\beta$, in line with intuitive expectations. Regimes of intermediate $\beta$ and $p$ can roughly be divided into 2 classes: regimes where the coverage $\alpha$ concentrates around some intermediate value and regimes dominated by inherent randomness, with values of $\alpha$ significantly spread out.

A somewhat counterintuitive phenomenon in the data is that for small values of $p$ and $n$, the coverage is higher than naïvely expected. This is mostly due to the network being small, and corresponds to the similarly counterintuitive behaviour of $\alpha_L$ in this parameter regime explained in the previous section. 

\begin{figure}[p]
     \centering
     \begin{subfigure}[b]{0.325\textwidth}
         \centering\includegraphics[width=\textwidth]{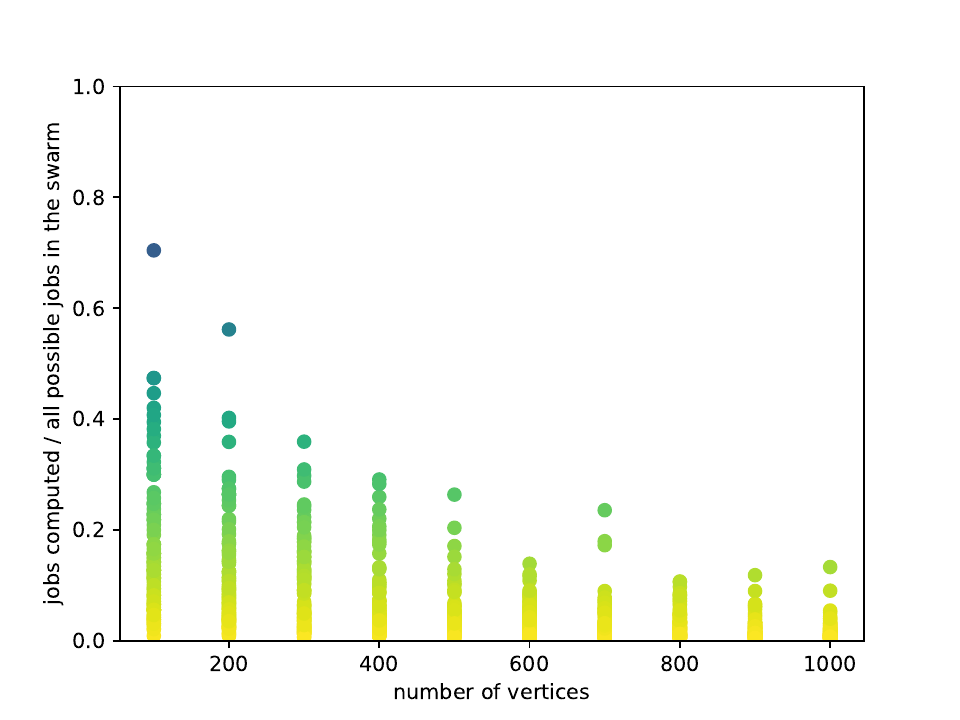}
         \caption{$k=4$, $p=0.2$, $\beta=0.1$}
         \label{fig:ALLcov.2.1-4}
     \end{subfigure}
     \hfill
     \begin{subfigure}[b]{0.325\textwidth}
         \centering\includegraphics[width=\textwidth]{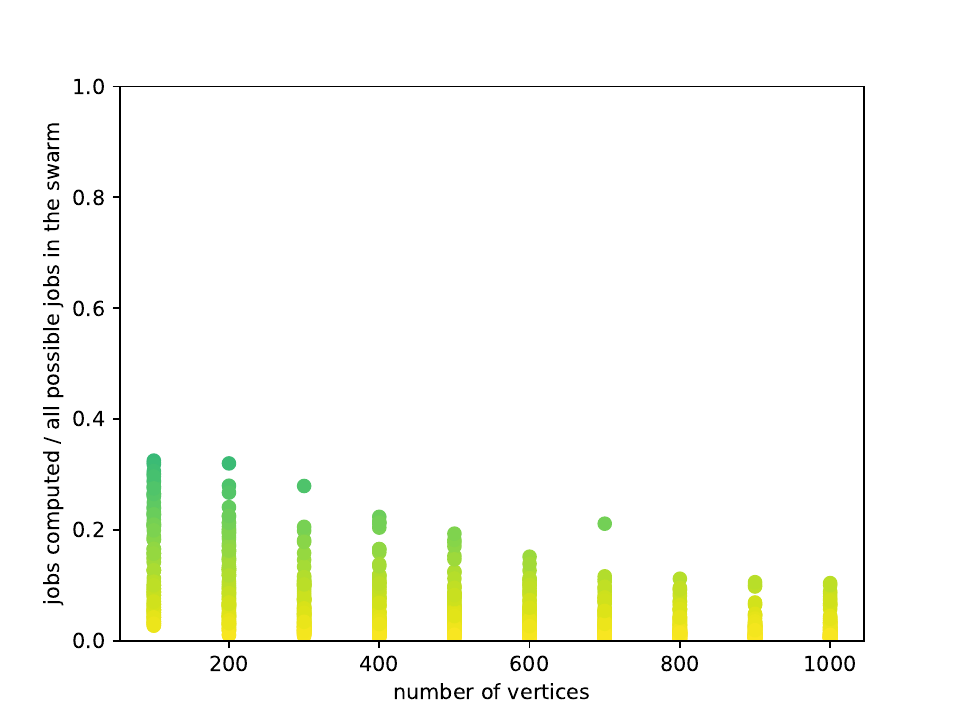}
         \caption{$k=4$, $p=0.2$, $\beta=0.5$}
         \label{fig:ALLcov.2.5-4}
     \end{subfigure}
     \hfill
     \begin{subfigure}[b]{0.325\textwidth}
         \centering\includegraphics[width=\textwidth]{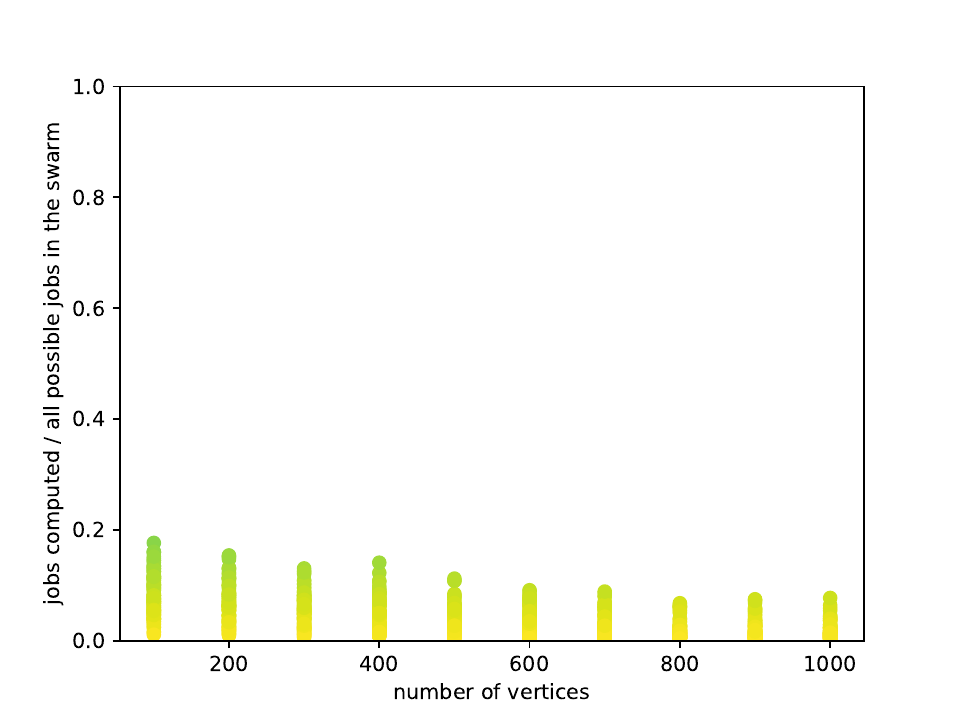}
         \caption{$k=4$, $p=0.2$, $\beta=1.0$}
         \label{fig:ALLcov.2.0-4}
     \end{subfigure}
     \hfill
         \begin{subfigure}[b]{0.325\textwidth}
         \centering\includegraphics[width=\textwidth]{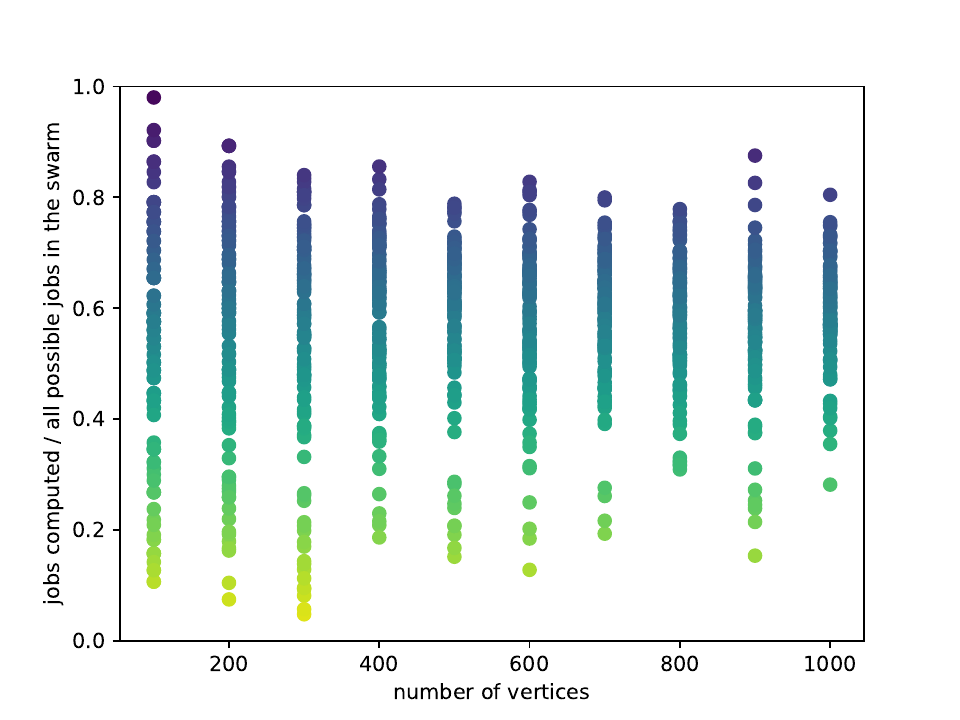}
         \caption{$k=4$, $p=0.5$, $\beta=0.1$}
         \label{fig:ALLcov.5.1-4}
     \end{subfigure}
     \hfill
     \begin{subfigure}[b]{0.325\textwidth}
         \centering\includegraphics[width=\textwidth]{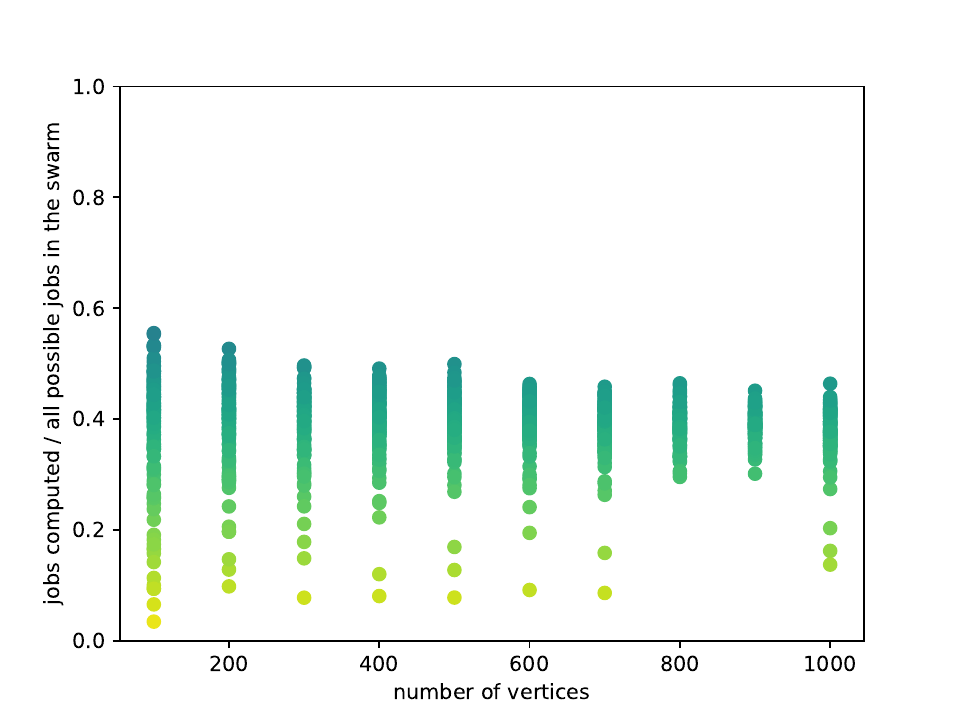}
         \caption{$k=4$, $p=0.5$, $\beta=0.5$}
         \label{fig:ALLcov.5.5-4}
     \end{subfigure}
     \hfill
     \begin{subfigure}[b]{0.325\textwidth}
         \centering\includegraphics[width=\textwidth]{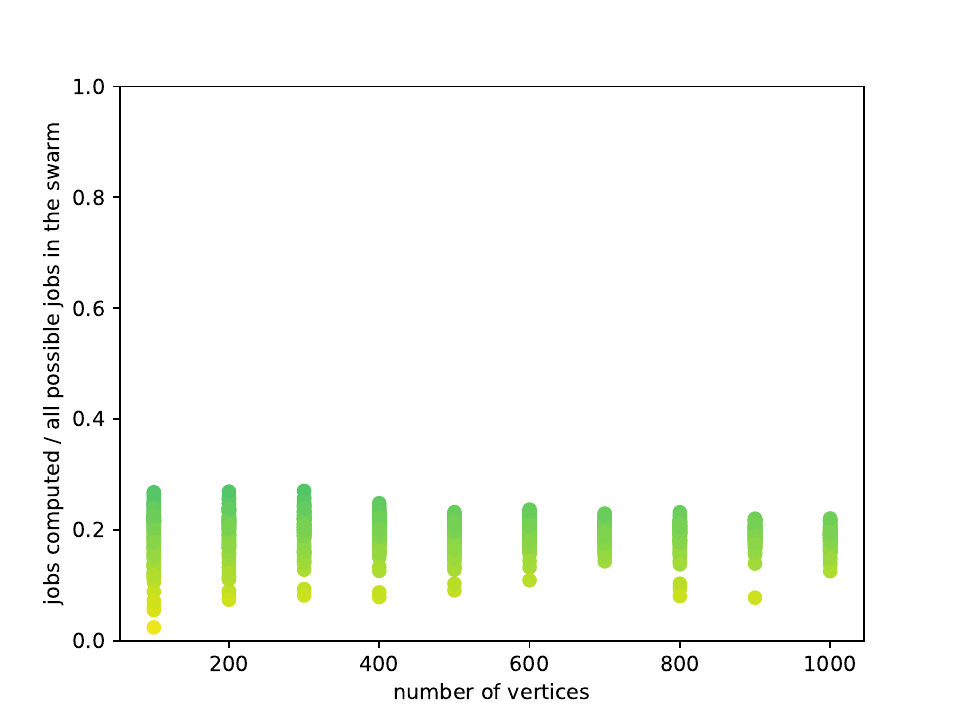}
         \caption{$k=4$, $p=0.5$, $\beta=1.0$}
         \label{fig:ALLcov.5.0-4}
     \end{subfigure}
     \hfill
          \begin{subfigure}[b]{0.325\textwidth}
         \centering\includegraphics[width=\textwidth]{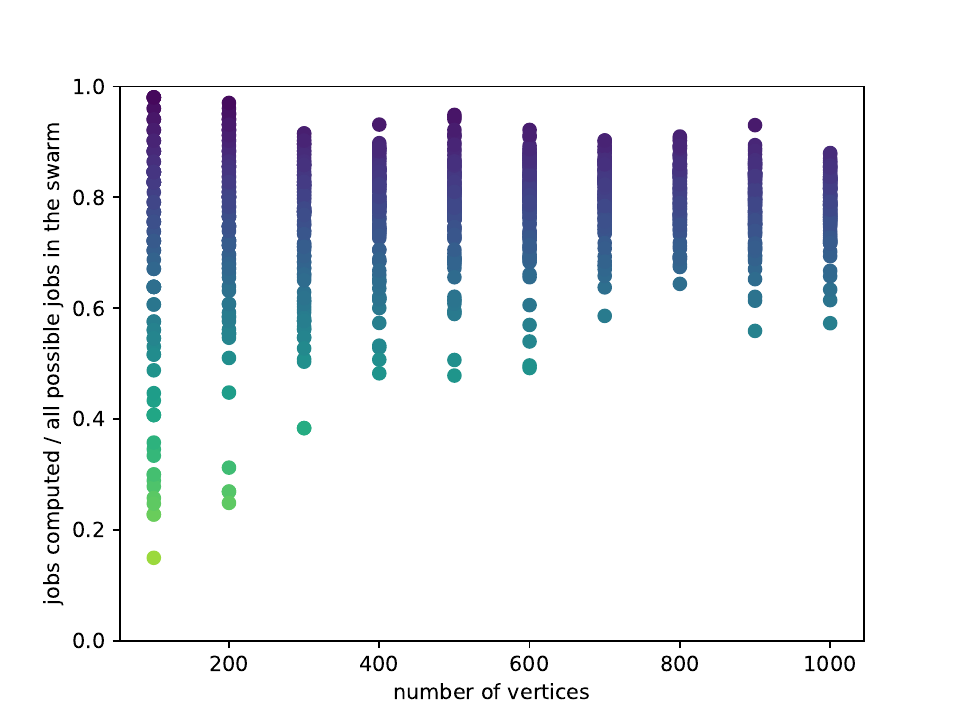}
         \caption{$k=4$, $p=0.7$, $\beta=0.1$}
         \label{fig:ALLcov.7.1-4}
     \end{subfigure}
     \hfill
          \begin{subfigure}[b]{0.325\textwidth}
         \centering\includegraphics[width=\textwidth]{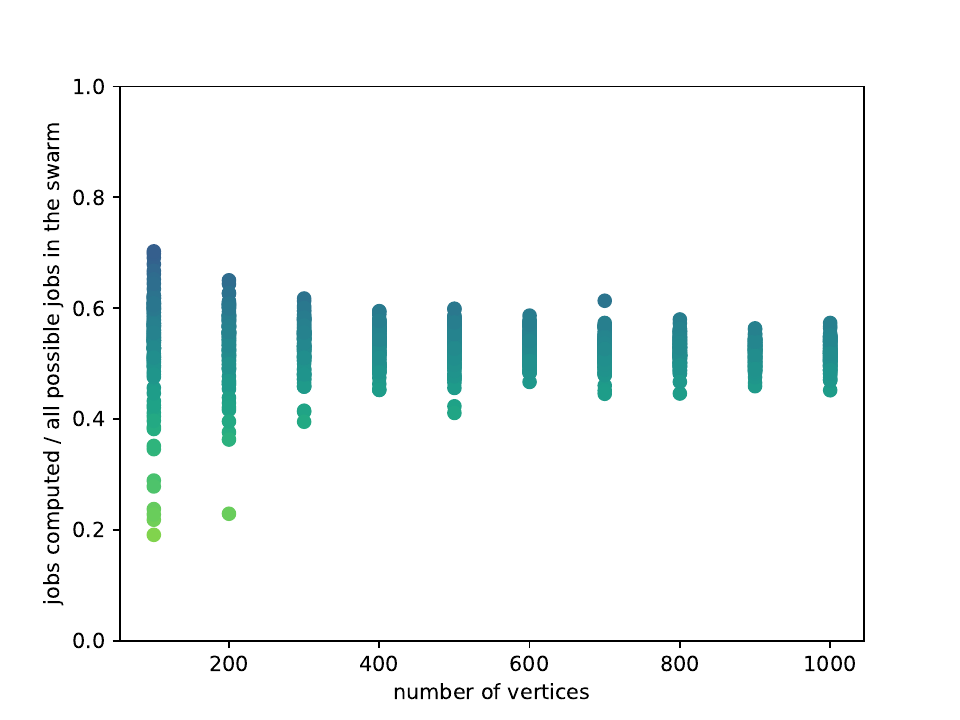}
         \caption{$k=4$, $p=0.7$, $\beta=0.5$}
         \label{fig:ALLcov.7.5-4}
     \end{subfigure}
     \hfill
     \begin{subfigure}[b]{0.325\textwidth}
         \centering\includegraphics[width=\textwidth]{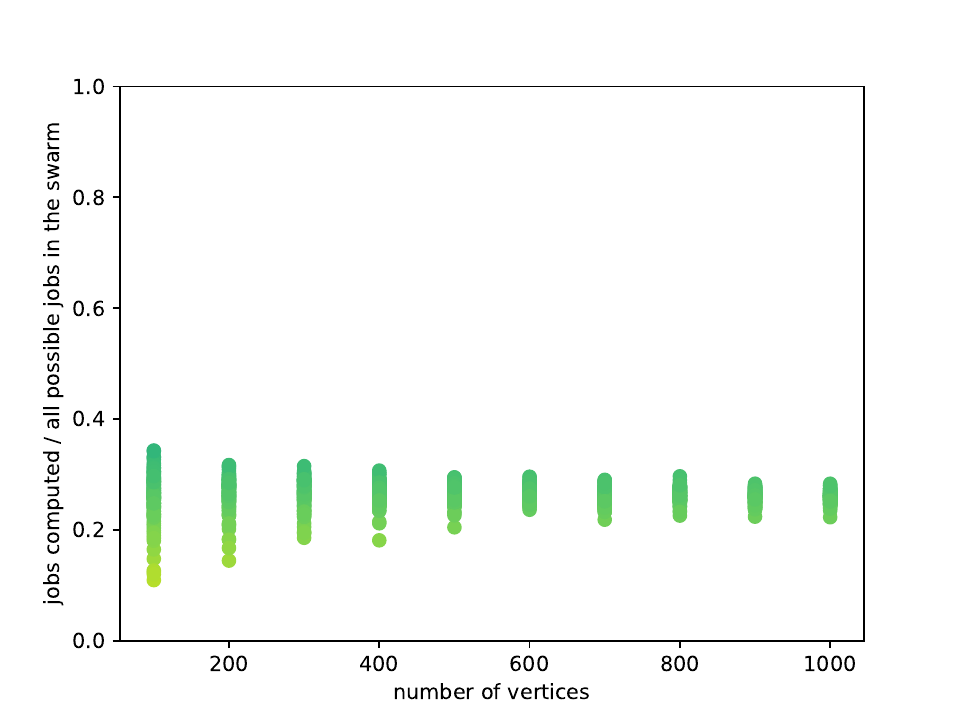}
         \caption{$k=4$, $p=0.7$, $\beta=1.0$}
         \label{fig:ALLcov.7.0-4}
     \end{subfigure}
     \hfill
          \begin{subfigure}[b]{0.325\textwidth}
         \centering\includegraphics[width=\textwidth]{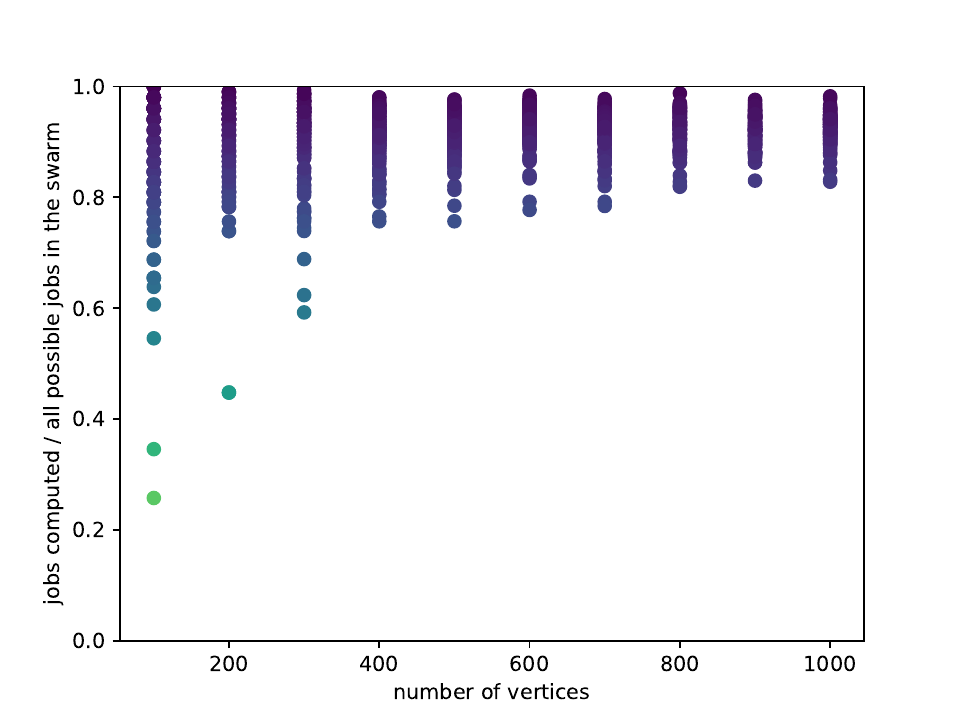}
         \caption{$k=4$, $p=0.99$, $\beta=0.1$}
         \label{fig:ALLcov.9.1-4}
     \end{subfigure}
     \hfill
          \begin{subfigure}[b]{0.325\textwidth}
         \centering\includegraphics[width=\textwidth]{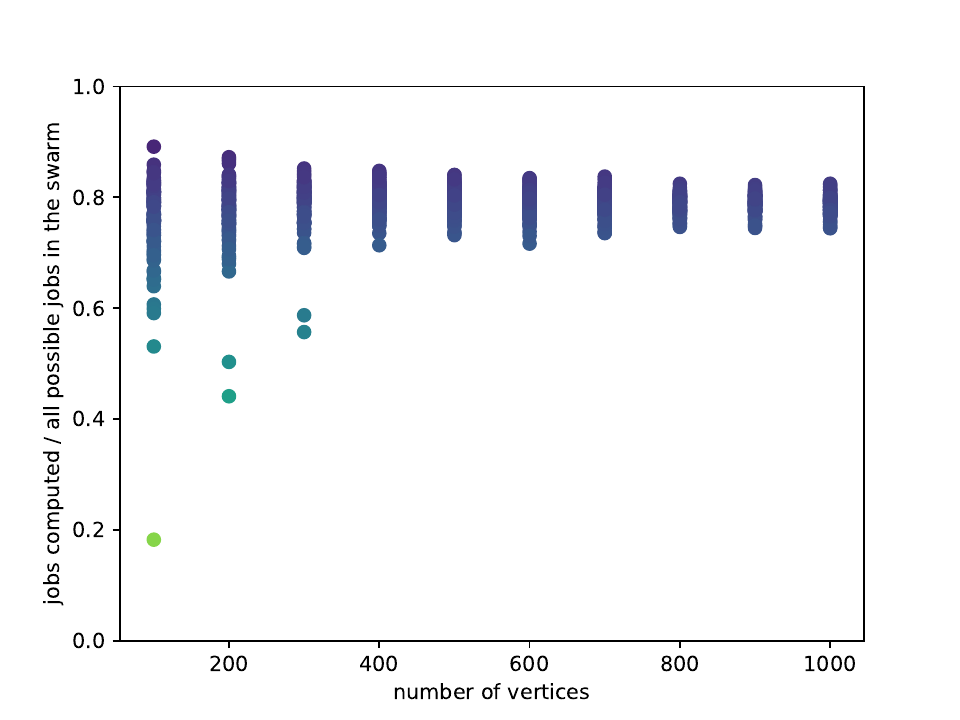}
         \caption{$k=4$, $p=0.99$, $\beta=0.1$}
         \label{fig:ALLcov.9.6-4}
     \end{subfigure}
     \hfill
          \begin{subfigure}[b]{0.325\textwidth}
         \centering\includegraphics[width=\textwidth]{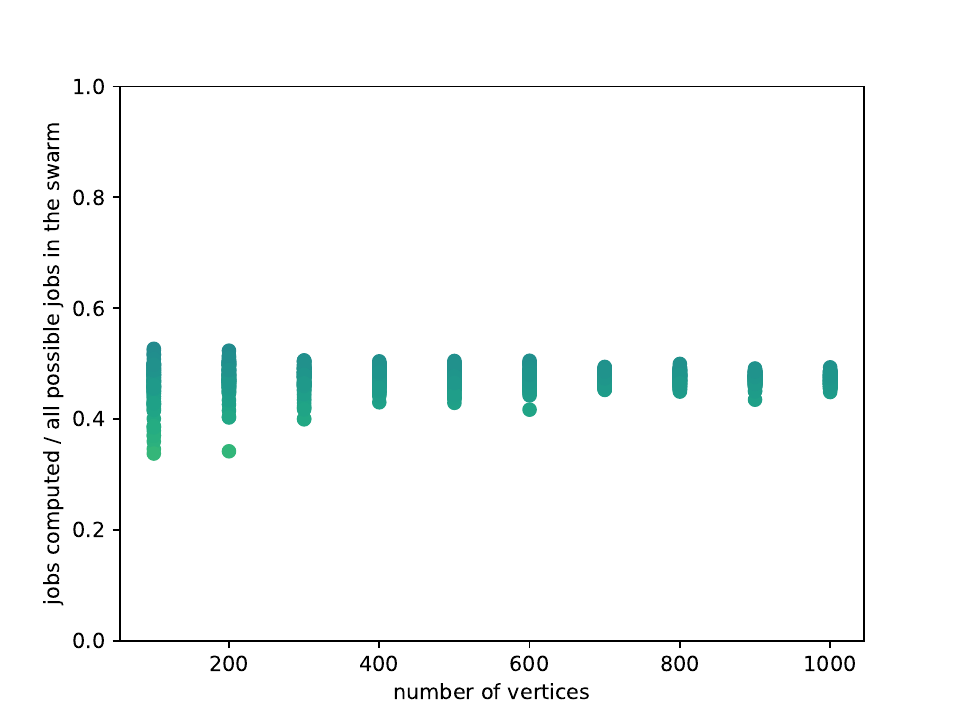}
         \caption{$k=4$, $p=0.99$, $\beta=1.0$}
         \label{fig:ALLcov.9.0-4}
     \end{subfigure}
     \hfill
     \caption[Scatter plots of coverage factor $\alpha$ for $k=4$]{Scatter plots of coverage factor $\alpha$ obtained from simulations in the parameter regime $n\in(100, 1000)$ for $k=4$. Each row has $p$ fixed and $\beta$ is chosen to showcase different possible behaviour. }
     \label{fig:ALLcov-4-all}
\end{figure}

\begin{figure}[p]
     \centering
     \begin{subfigure}[b]{0.325\textwidth}
         \centering\includegraphics[width=\textwidth]{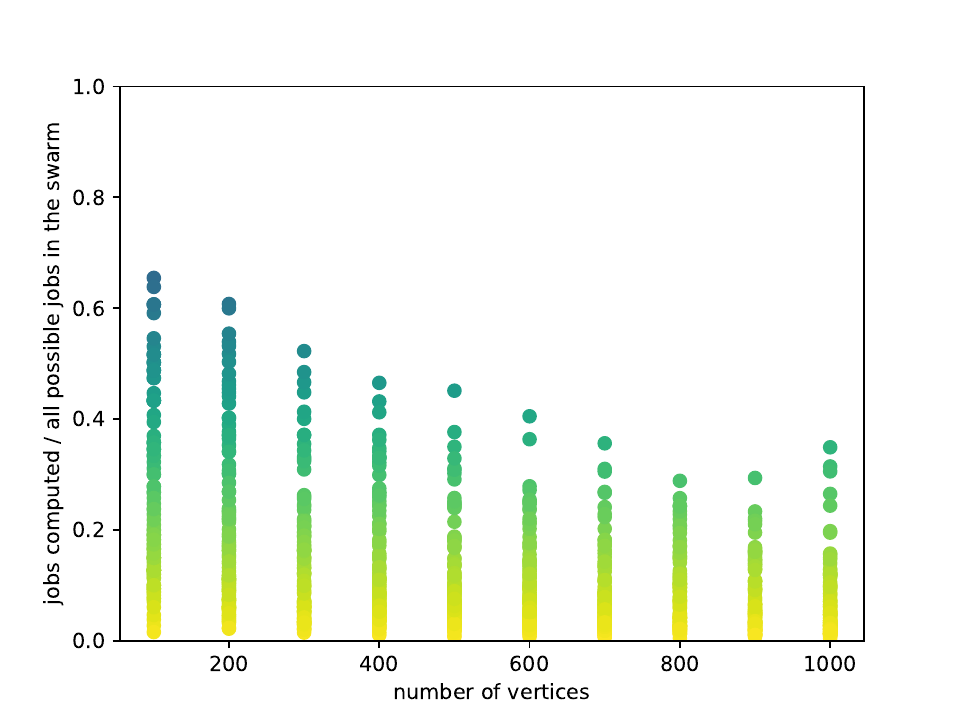}
         \caption{$k=5$, $p=0.1$, $\beta=0.1$}
         \label{fig:ALLcov.1.1-5}
     \end{subfigure}
     \hfill
     \begin{subfigure}[b]{0.325\textwidth}
         \centering\includegraphics[width=\textwidth]{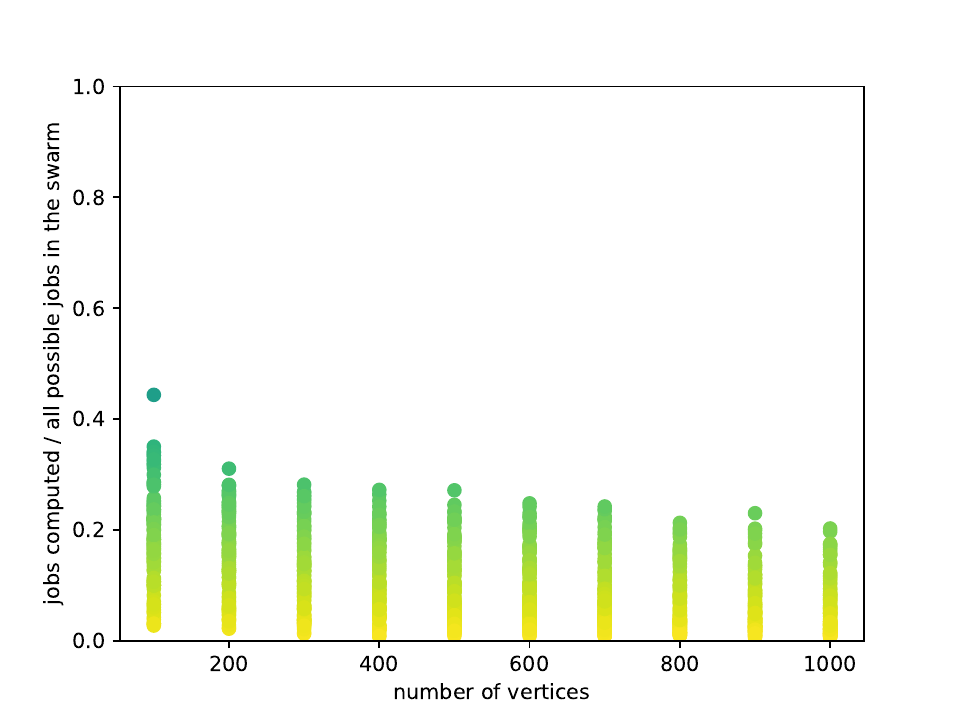}
         \caption{$k=5$, $p=0.1$, $\beta=0.5$}
         \label{fig:ALLcov.1.5-5}
     \end{subfigure}
     \hfill
     \begin{subfigure}[b]{0.325\textwidth}
         \centering\includegraphics[width=\textwidth]{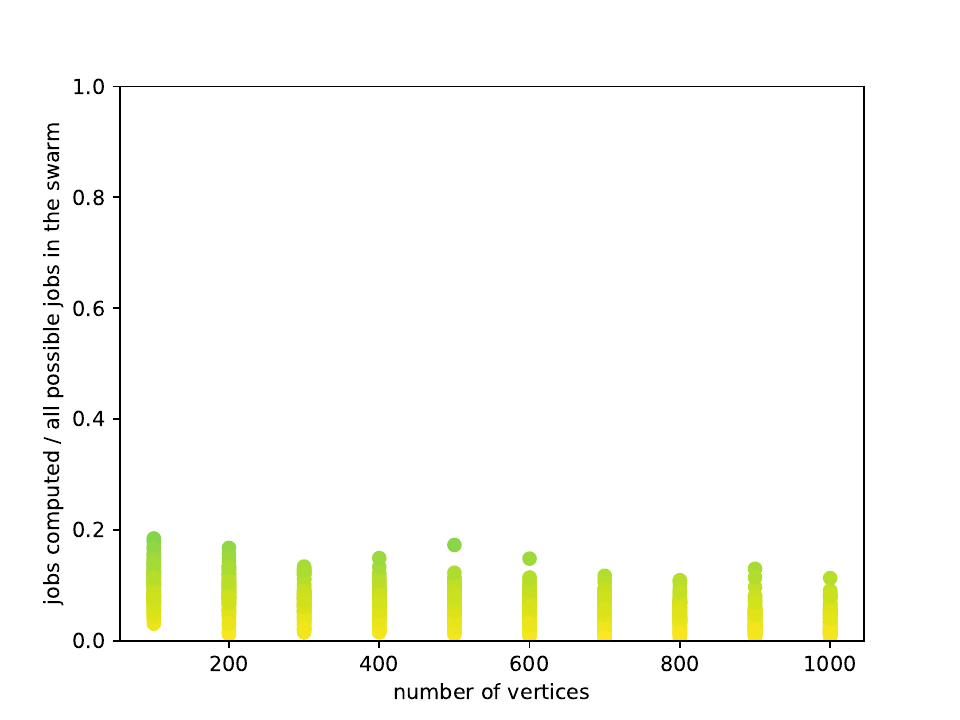}
         \caption{$k=5$, $p=0.1$, $\beta=1.0$}
         \label{fig:ALLcov.1.0-5}
     \end{subfigure}
     \hfill
         \begin{subfigure}[b]{0.325\textwidth}
         \centering\includegraphics[width=\textwidth]{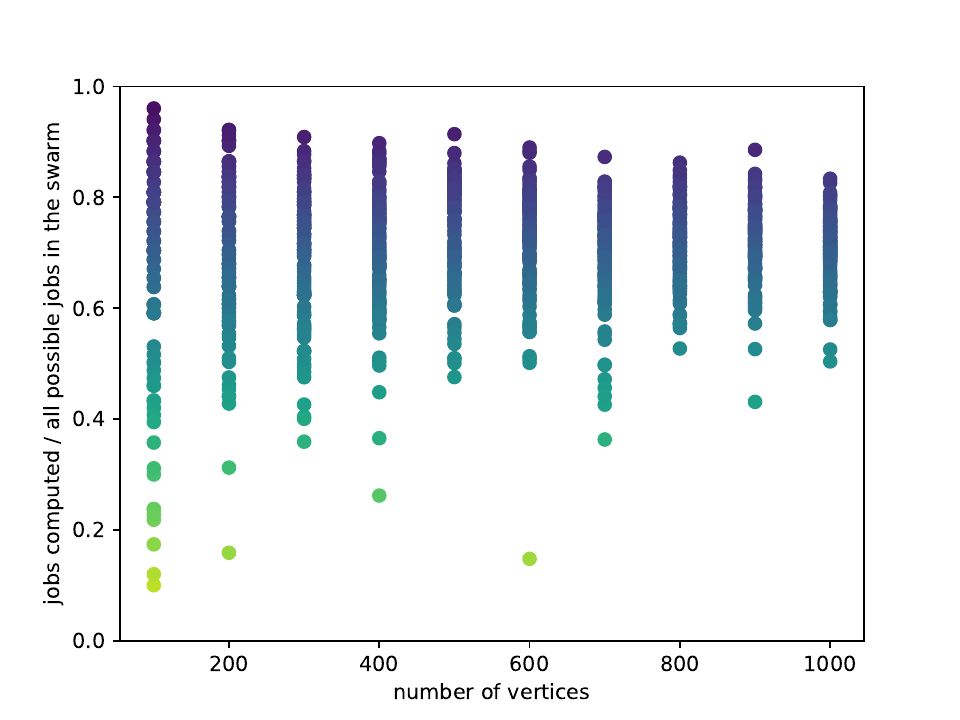}
         \caption{$k=5$, $p=0.3$, $\beta=0.1$}
         \label{fig:ALLcov.3.1-5}
     \end{subfigure}
     \hfill
     \begin{subfigure}[b]{0.325\textwidth}
         \centering\includegraphics[width=\textwidth]{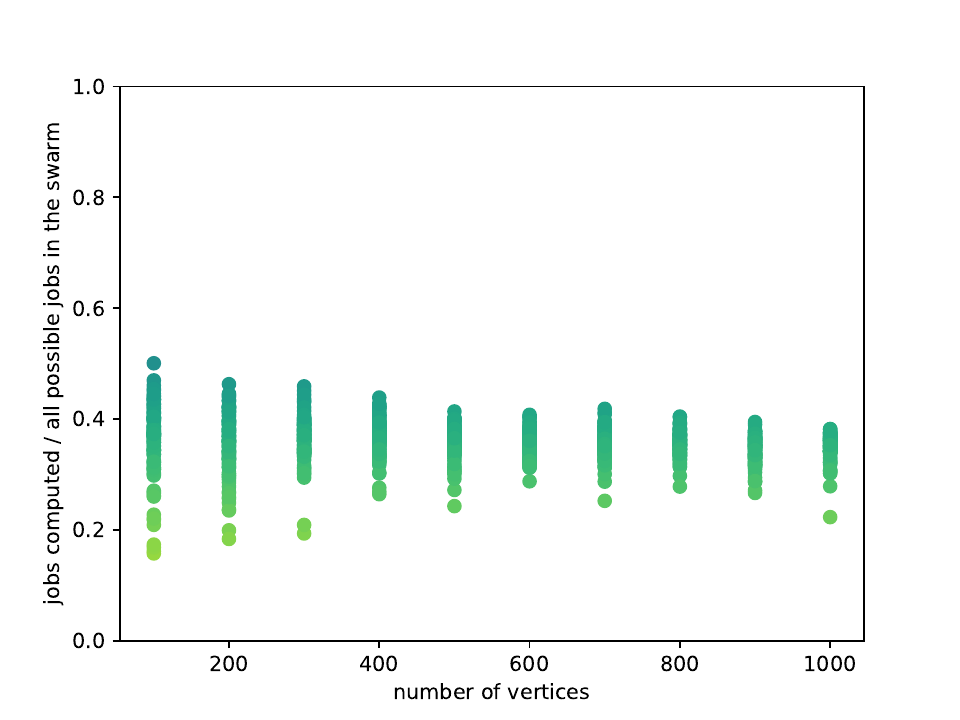}
         \caption{$k=5$, $p=0.3$, $\beta=0.5$}
         \label{fig:ALLcov.3.5-5}
     \end{subfigure}
     \hfill
     \begin{subfigure}[b]{0.325\textwidth}
         \centering\includegraphics[width=\textwidth]{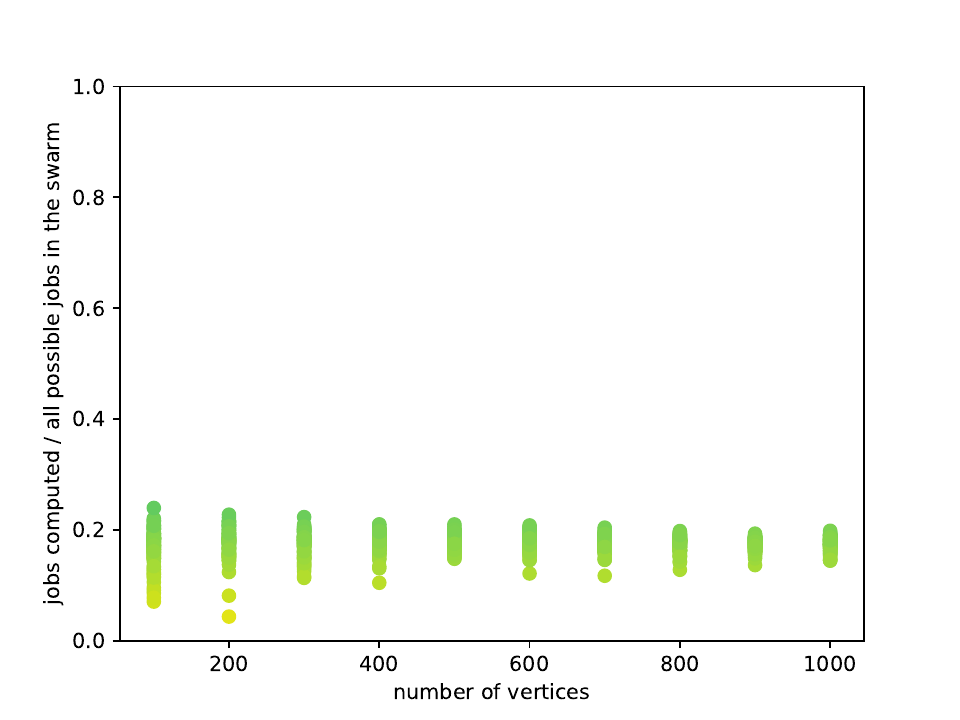}
         \caption{$k=5$, $p=0.3$, $\beta=1.0$}
         \label{fig:ALLcov.3.0-5}
     \end{subfigure}
     \hfill
          \begin{subfigure}[b]{0.325\textwidth}
         \centering\includegraphics[width=\textwidth]{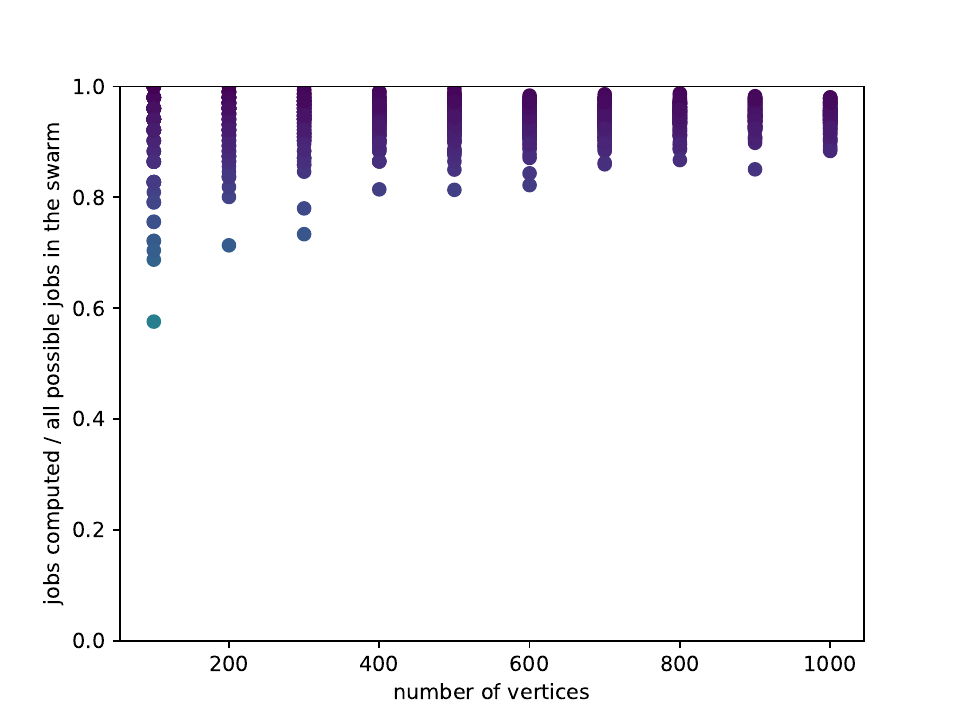}
         \caption{$k=5$, $p=0.7$, $\beta=0.1$}
         \label{fig:ALLcov.7.1-5}
     \end{subfigure}
     \hfill
          \begin{subfigure}[b]{0.325\textwidth}
         \centering\includegraphics[width=\textwidth]{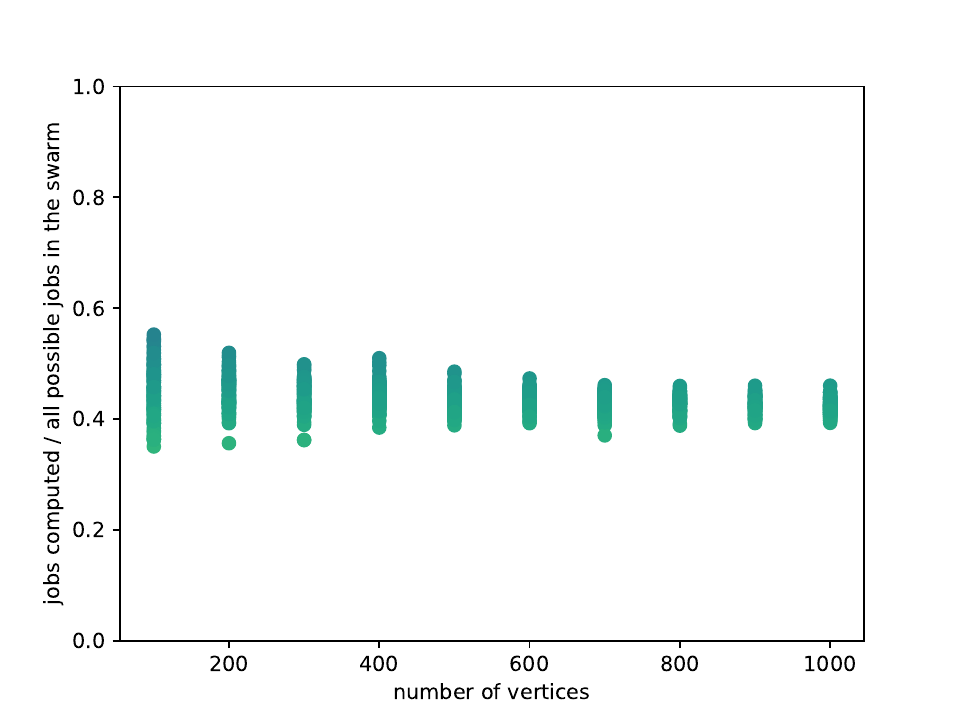}
         \caption{$k=5$, $p=0.7$, $\beta=0.6$}
         \label{fig:ALLcov.7.6-5}
     \end{subfigure}
     \hfill
     \begin{subfigure}[b]{0.325\textwidth}
         \centering\includegraphics[width=\textwidth]{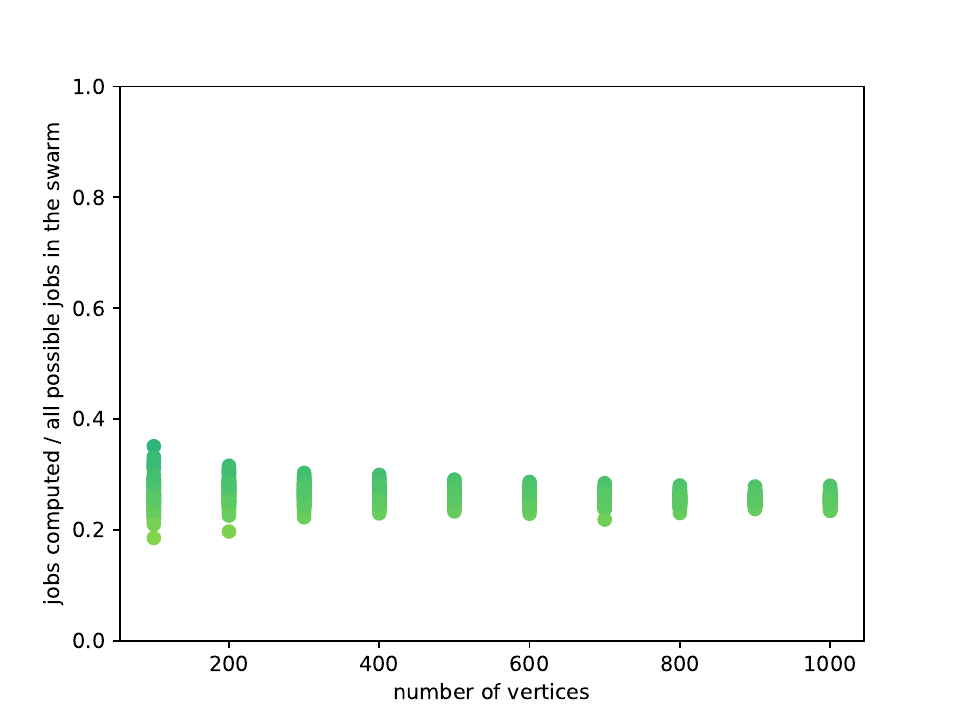}
         \caption{$k=5$, $p=0.7$, $\beta=1.0$}
         \label{fig:ALLcov.7.0-5}
     \end{subfigure}
     \hfill
          \begin{subfigure}[b]{0.325\textwidth}
         \centering\includegraphics[width=\textwidth]{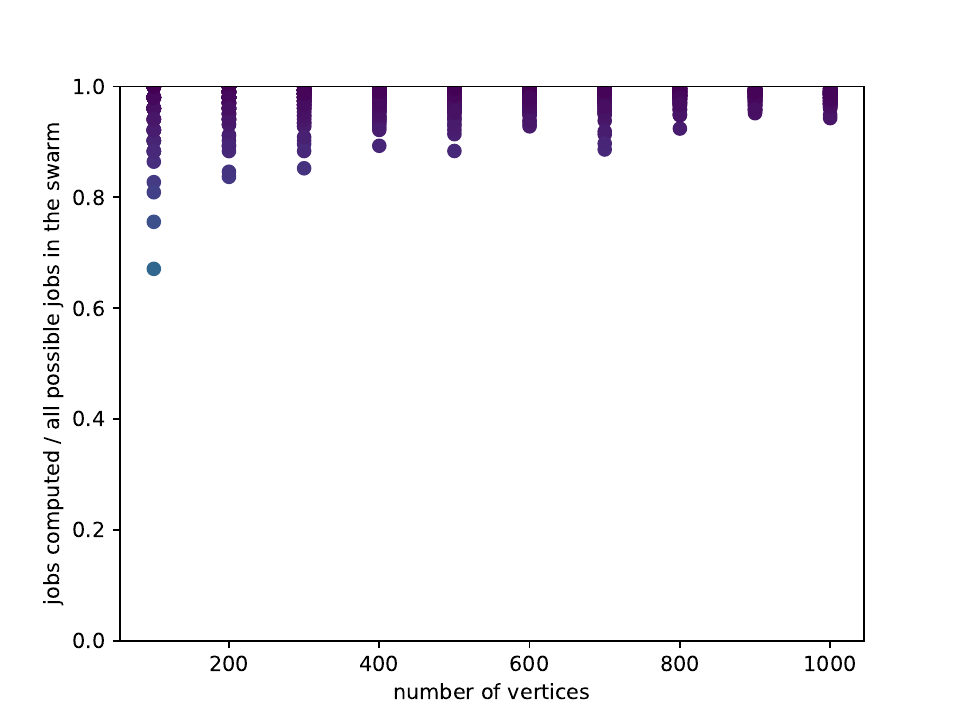}
         \caption{$k=5$, $p=0.99$, $\beta=0.1$}
         \label{fig:ALLcov.99.1-5}
     \end{subfigure}
     \hfill
          \begin{subfigure}[b]{0.325\textwidth}
         \centering\includegraphics[width=\textwidth]{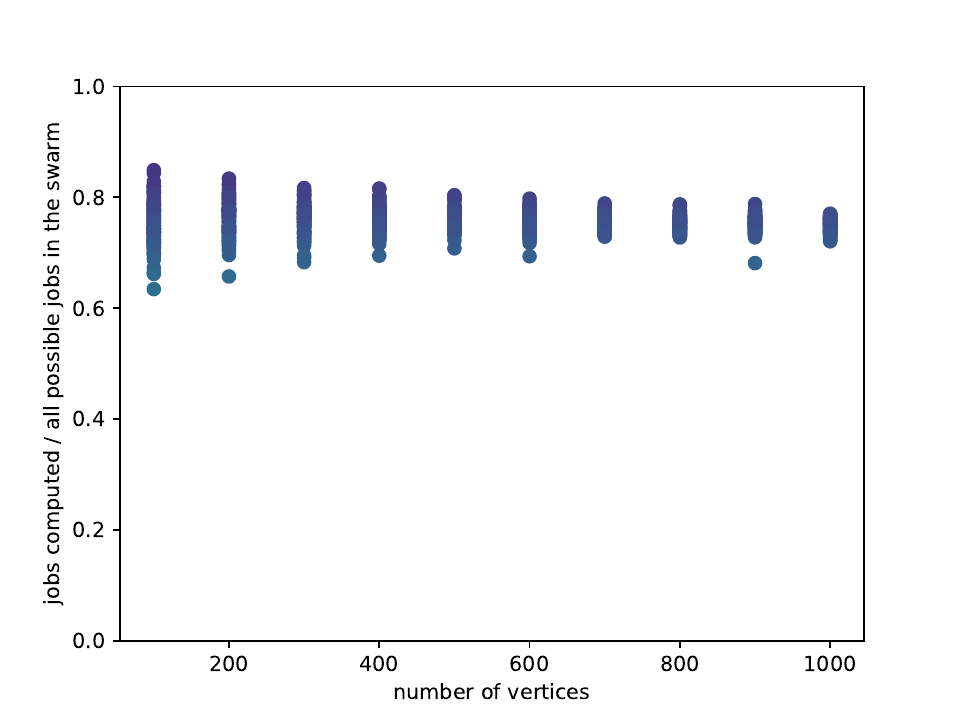}
         \caption{$k=5$, $p=0.99$, $\beta=0.6$}
         \label{fig:ALLcov.99.6-5}
     \end{subfigure}
     \hfill
          \begin{subfigure}[b]{0.325\textwidth}
         \centering\includegraphics[width=\textwidth]{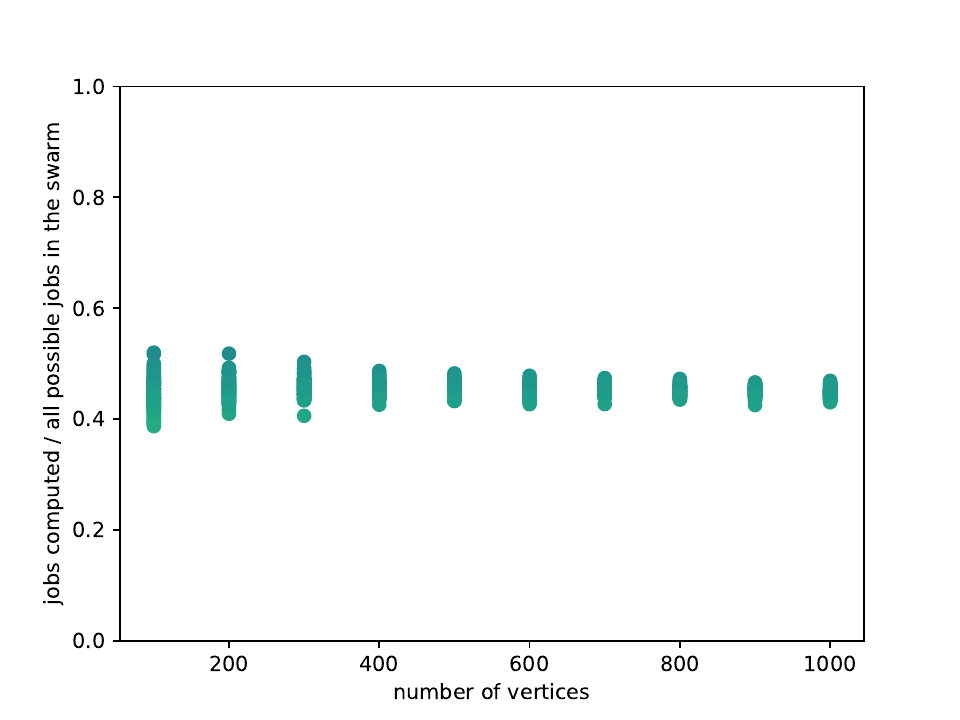}
         \caption{$k=5$, $p=0.99$, $\beta=1.0$}
         \label{fig:ALLcov.99.0-5}
     \end{subfigure}
     \hfill
     \caption[Scatter plots of coverage factor $\alpha$ for $k=5$]{Scatter plots of coverage factor $\alpha$ obtained from simulations in the parameter regime $n\in(100, 1000)$ for $k=5$. Each row has $p$ fixed and $\beta$ is chosen to showcase different possible behaviour. }     \label{fig:ALLcov-5-all}
\end{figure}

\begin{figure}[p]
     \centering
     \begin{subfigure}[b]{0.325\textwidth}
         \centering\includegraphics[width=\textwidth]{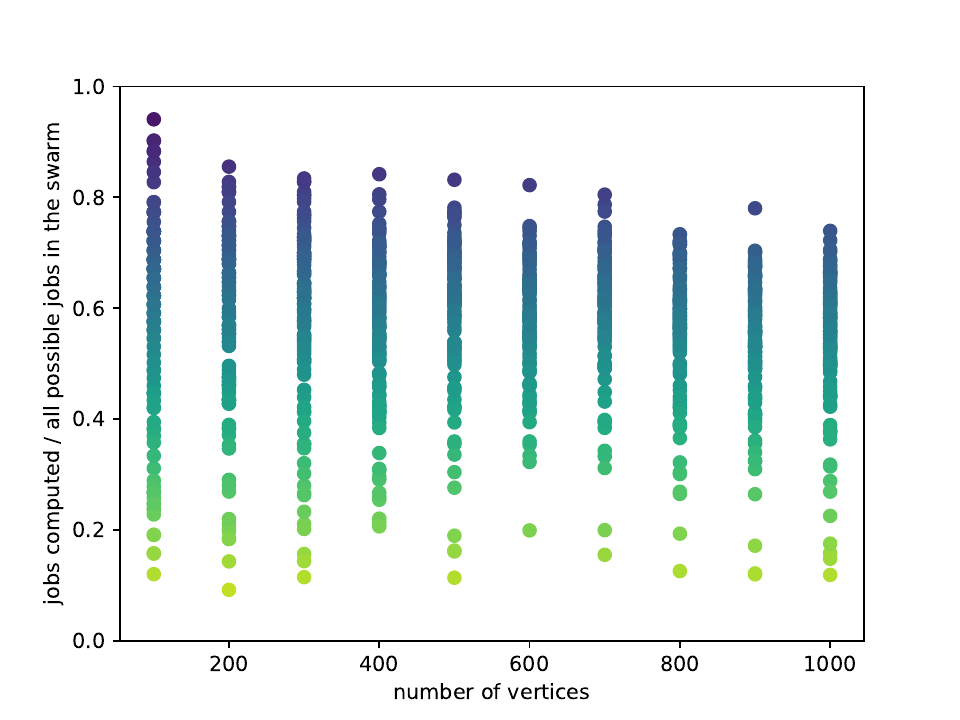}
         \caption{$k=6$, $p=0.1$, $\beta=0.1$}
         \label{fig:ALLcov.1.1-6}
     \end{subfigure}
     \hfill
     \begin{subfigure}[b]{0.325\textwidth}
         \centering\includegraphics[width=\textwidth]{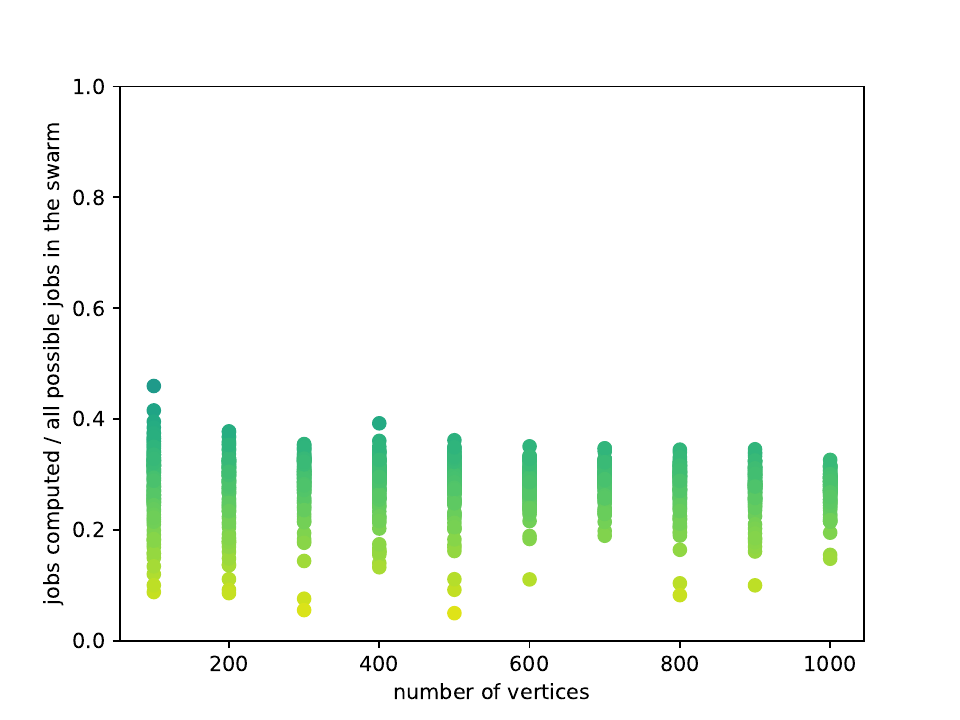}
         \caption{$k=6$, $p=0.1$, $\beta=0.5$}
         \label{fig:ALLcov.1.5-6}
     \end{subfigure}
     \hfill
     \begin{subfigure}[b]{0.325\textwidth}
         \centering\includegraphics[width=\textwidth]{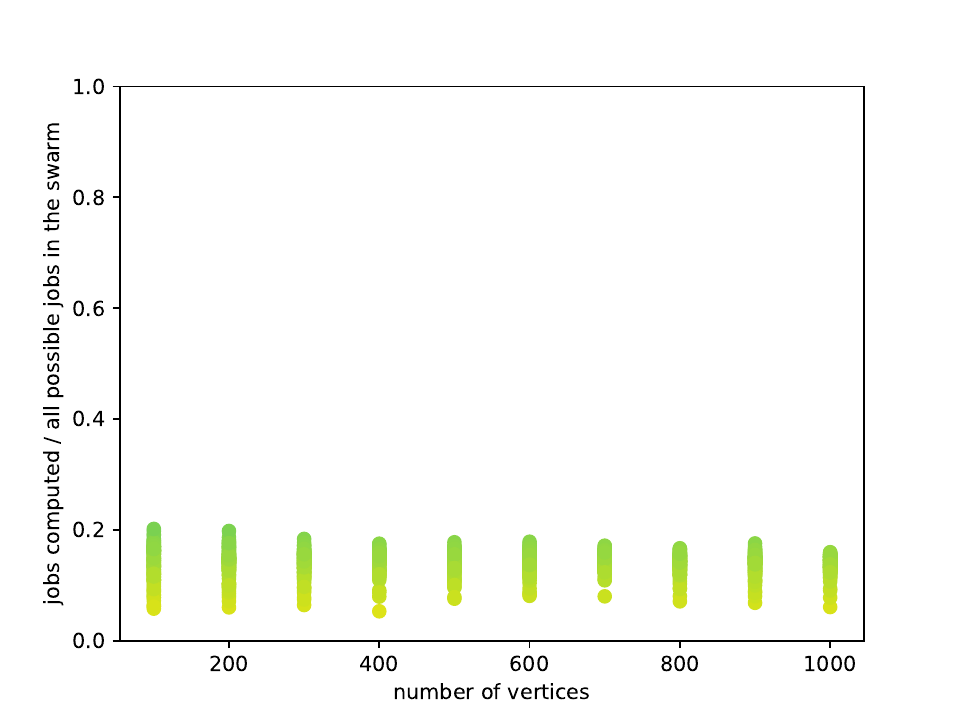}
         \caption{$k=6$, $p=0.1$, $\beta=0.1$}
         \label{fig:ALLcov.1.0-6}
     \end{subfigure}
     \hfill
         \begin{subfigure}[b]{0.325\textwidth}
         \centering\includegraphics[width=\textwidth]{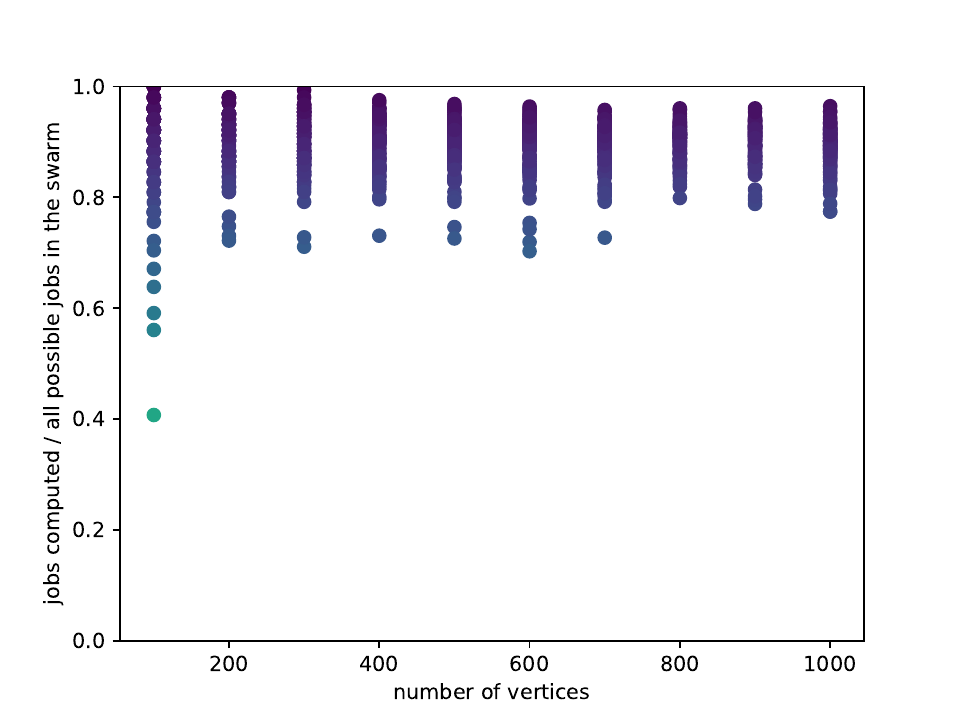}
         \caption{$k=6$, $p=0.3$, $\beta=0.1$}
         \label{fig:ALLcov.3.1-6}
     \end{subfigure}
     \hfill
     \begin{subfigure}[b]{0.325\textwidth}
         \centering\includegraphics[width=\textwidth]{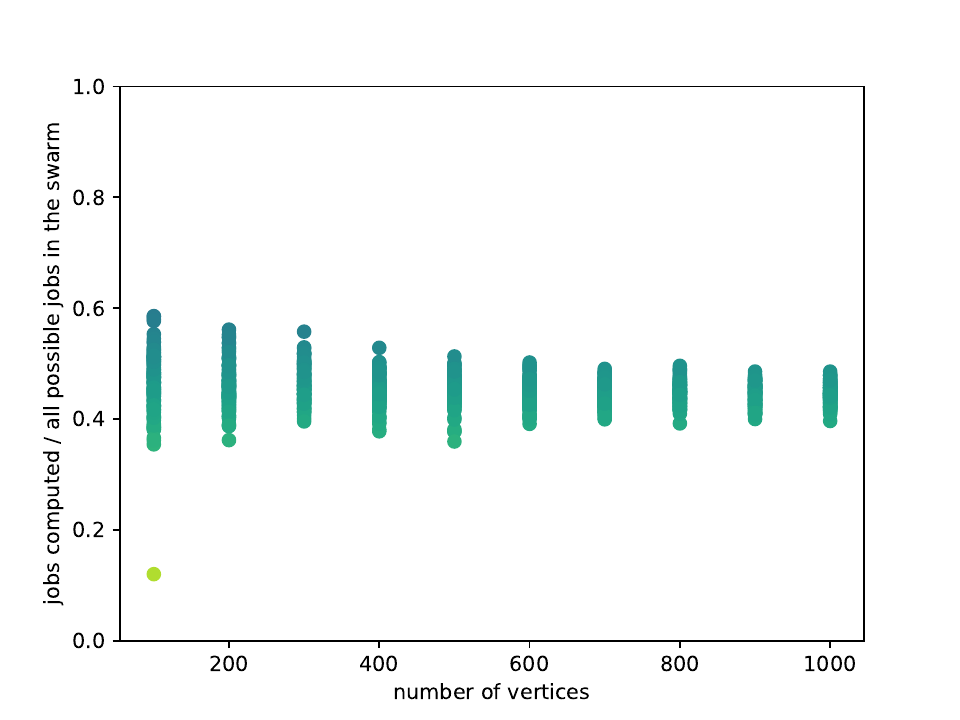}
         \caption{$k=6$, $p=0.3$, $\beta=0.4$}
         \label{fig:ALLcov.3.4-6}
     \end{subfigure}
     \hfill
     \begin{subfigure}[b]{0.325\textwidth}
         \centering\includegraphics[width=\textwidth]{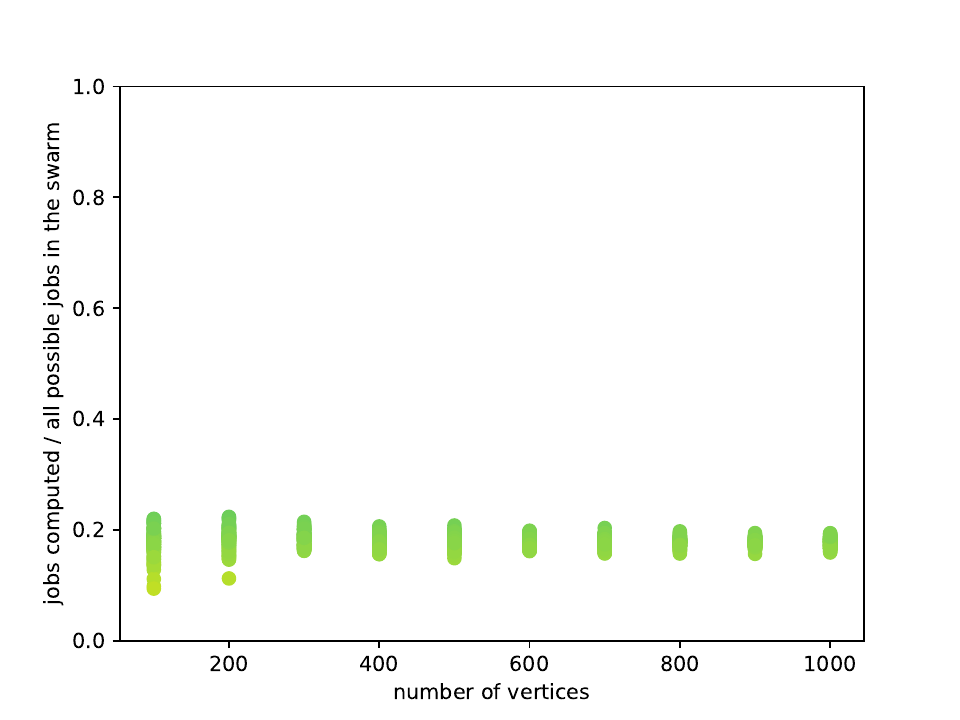}
         \caption{$k=6$, $p=0.3$, $\beta=1.0$}
         \label{fig:ALLcov.3.0-6}
     \end{subfigure}
     \hfill
          \begin{subfigure}[b]{0.325\textwidth}
         \centering\includegraphics[width=\textwidth]{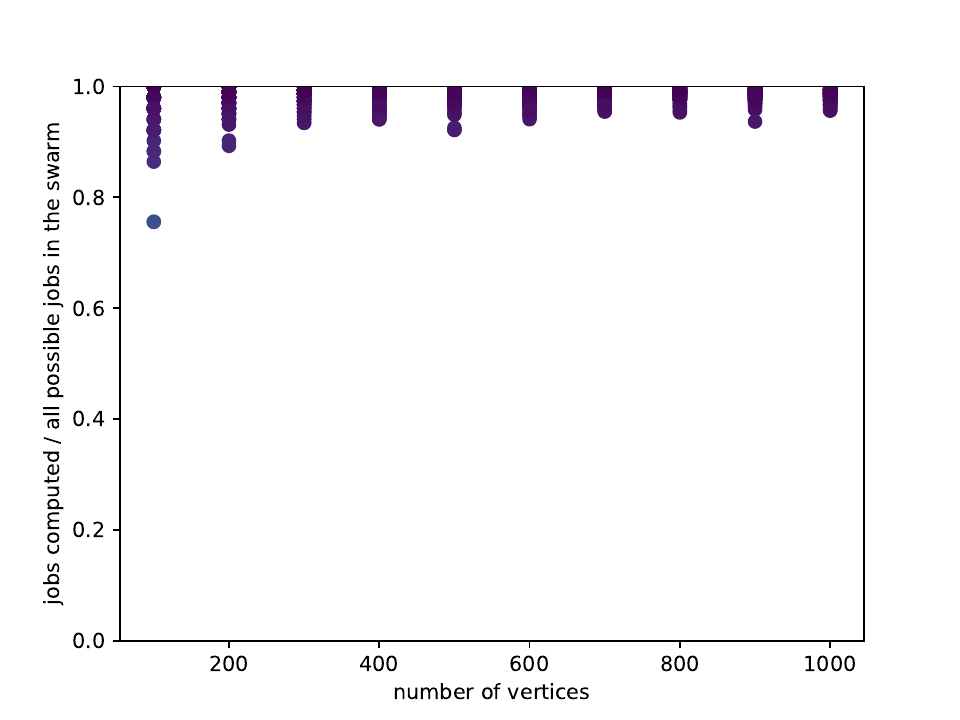}
         \caption{$k=6$, $p=0.8$, $\beta=0.1$}
         \label{fig:ALLcov.8.1-6}
     \end{subfigure}
     \hfill
          \begin{subfigure}[b]{0.325\textwidth}
         \centering\includegraphics[width=\textwidth]{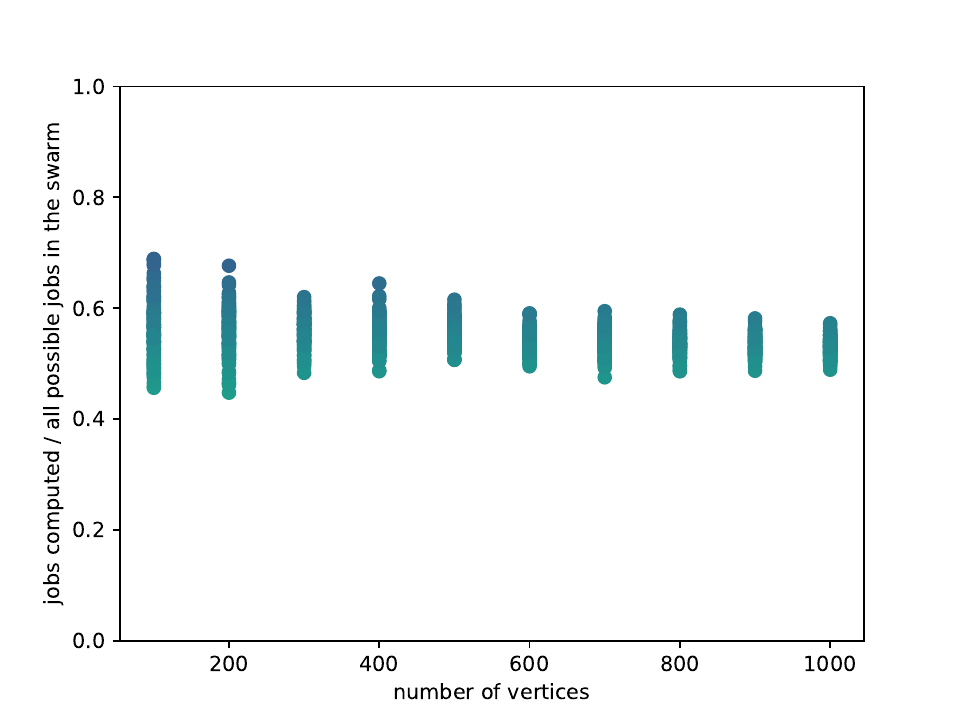}
         \caption{$k=6$, $p=0.8$, $\beta=0.5$}
         \label{fig:ALLcov.8.5-6}
     \end{subfigure}
     \hfill
     \begin{subfigure}[b]{0.325\textwidth}
         \centering\includegraphics[width=\textwidth]{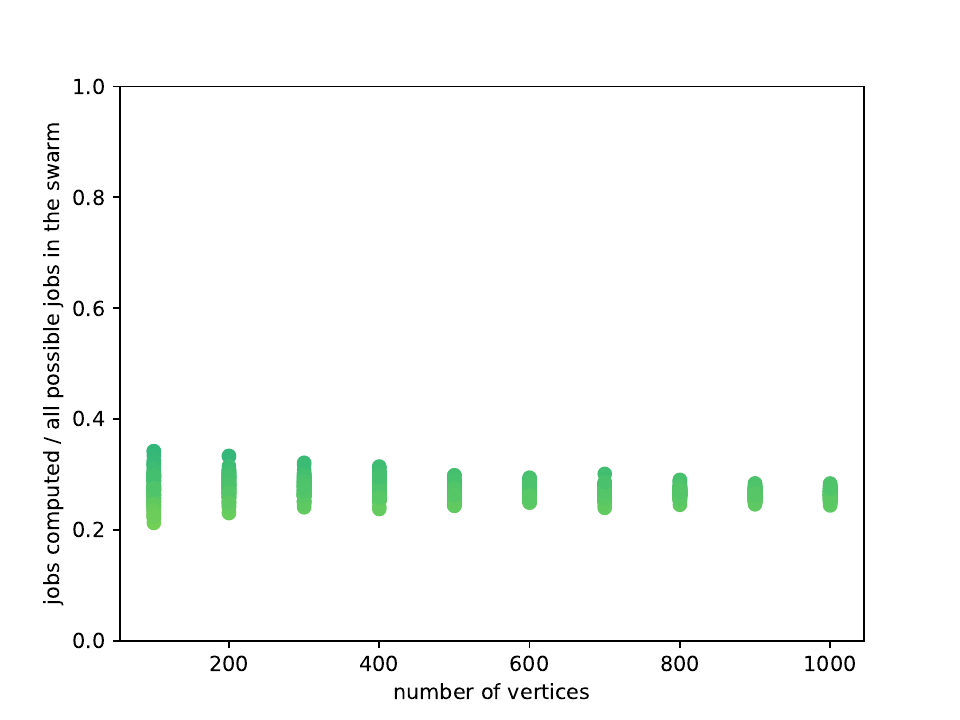}
         \caption{$k=6$, $p=0.8$, $\beta=1.0$}
         \label{fig:ALLcov.8.0-6}
     \end{subfigure}
     \hfill
          \begin{subfigure}[b]{0.325\textwidth}
         \centering\includegraphics[width=\textwidth]{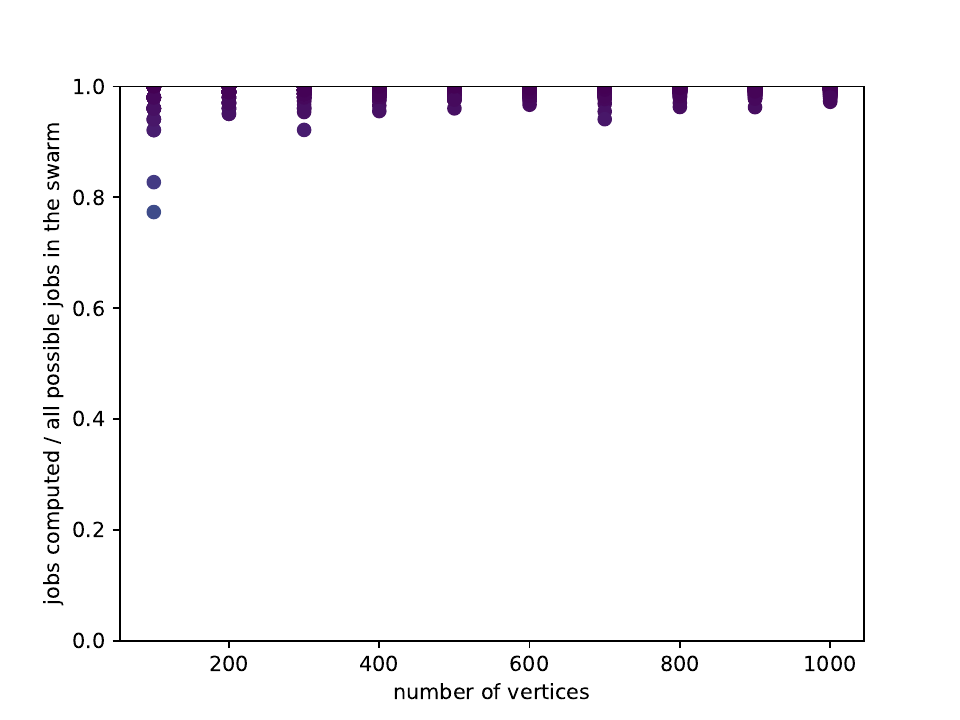}
         \caption{$k=6$, $p=0.99$, $\beta=0.1$}
         \label{fig:ALLcov.99.1-6}
     \end{subfigure}
     \hfill
          \begin{subfigure}[b]{0.325\textwidth}
         \centering\includegraphics[width=\textwidth]{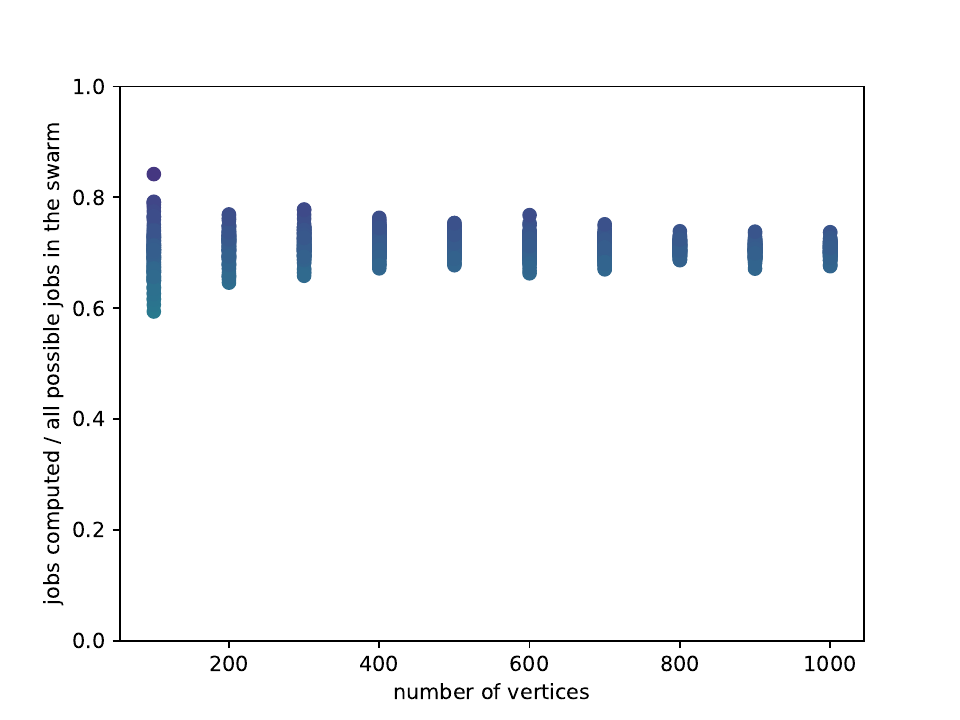}
         \caption{$k=6$, $p=0.99$, $\beta=0.6$}
         \label{fig:ALLcov.99.6-6}
     \end{subfigure}
     \hfill
          \begin{subfigure}[b]{0.325\textwidth}
         \centering\includegraphics[width=\textwidth]{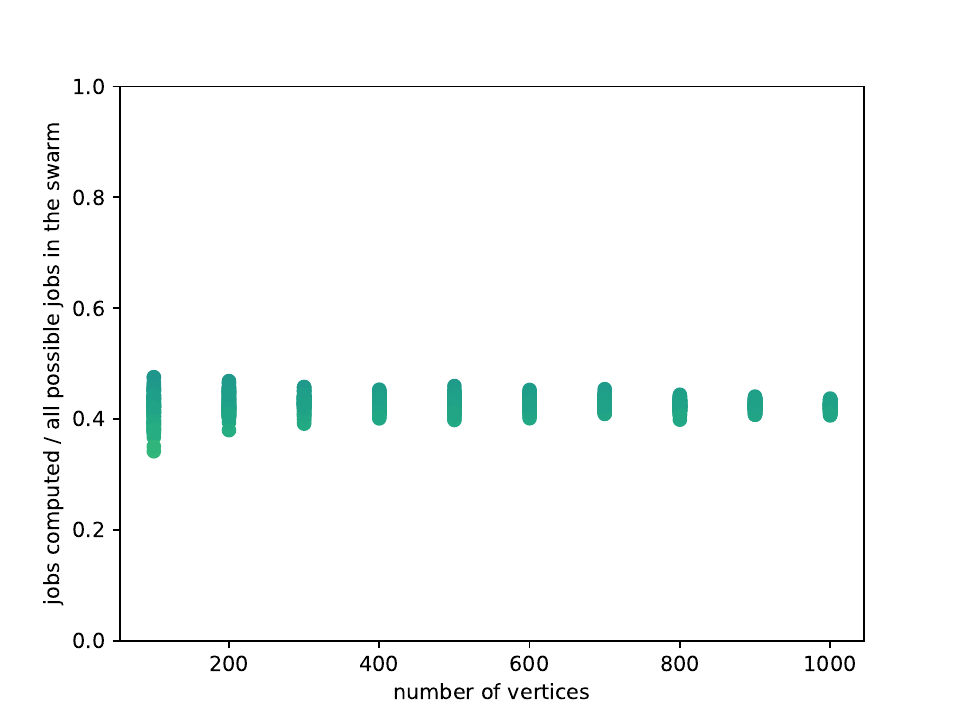}
         \caption{$k=6$, $p=0.99$, $\beta=1.0$}
         \label{fig:ALLcov.99.0-6}
     \end{subfigure}
     \hfill
     \caption[Scatter plots of coverage factor $\alpha$ for $k=6$]{Scatter plots of coverage factor $\alpha$ obtained from simulations in the parameter regime $n\in(100, 1000)$ for $k=6$. Each row has $p$ fixed and $\beta$ is chosen to showcase different possible behaviour. }
     \label{fig:ALLcov-6-all}
\end{figure}

These results are illustrated in Figures~\ref{fig:ALLcov-4-all}, \ref{fig:ALLcov-5-all} and \ref{fig:ALLcov-6-all}.

\subsection{Additional plots}

Additional plots are available via \href{https://doi.org/10.5281/zenodo.8433181}{DOI:10.5281/zenodo.8433181}.

The naming scheme of the files is as follows:
\begin{center}
\begin{verbatim}
    fig_{descriptor}_{k}_{p}_{beta}.pdf
\end{verbatim}
\end{center}
where \verb+{descriptor}+ can be \verb+fracLSCC+ for plots related to LSCC reduction parameter $\rho_L$, \verb+coverLSCC+ for LSSC coverage $\alpha_L$ and \verb+coverALL+ for coverage parameter $\alpha$. The placeholder \verb+{k}+ corresponds to the value of $k$ chosen in the simulation, \verb+p+ denotes the chosen value of $p$ and \verb+{beta}+ denotes the chosen value of $\beta$. For $\rho_L$, which is independent of $\beta$, the naming convention uses $\beta = 1$.

For example, \verb+fig_coverALL_5-0.980-0.700.pdf+ corresponds to the plot of the coverage parameter $\alpha$ for networks with $k=5$, $p=0.98$ and $\beta = 0.7$.

\bibliographystyle{IEEEtaes.bst}





